\documentclass[10pt,one column]{book}
\usepackage{amsmath}
\usepackage{amssymb}
\usepackage{graphicx}
\usepackage{float}
\usepackage{array}
\newcolumntype{P}[1]{>{\centering\arraybackslash}p{#1}}
\newcolumntype{M}[1]{>{\centering\arraybackslash}m{#1}}
\usepackage{color}
\usepackage{multirow}
\usepackage{graphicx}
\usepackage[utf8]{inputenc}

\newcommand{\bi}{\begin{itemize}}

\newcommand{\be}{\begin{enumerate}}
\newcommand{\ee}{\end{enumerate}}

\newcommand{\ed}{\end{description}}
\newcommand{\bc}{\begin{center}}
\newcommand{\ec}{\end{center}}
\newcommand{\bt}{\begin{tabbing}}
\newcommand{\et}{\end{tabbing}}
\newcommand{\bfig}{\begin{figure}}
\newcommand{\efig}{\end{figure}}
\newcommand{\beq}{\begin{equation}}
\newcommand{\beqarr}{\begin{eqnarray}}
\newcommand{\beqarrn}{\begin{eqnarray*}}
\newcommand{\eeq}{\end{equation}}
\newcommand{\eeqarr}{\end{eqnarray}}
\newcommand{\eeqarrn}{\end{eqnarray*}}
\newcommand{\bflr}{\begin{flushright}\vspace{-0.2in}}
\newcommand{\eflr}{\end{flushright}}
\newcommand{\bsub}{\begin{subequations}}
\newcommand{\esub}{\end{subequations}}
\newcommand{\barr}{\begin{array}}
\newcommand{\earr}{\end{array}}

\def\x{{\mathbf x}}
\def\L{{\cal L}}
\begin{document}

\title{Stochastic processes in classical and quantum physics and engineering\author{Harish Parthasarathy\\ECE division, NSUT}}\maketitle

{\bf Preface}

This book covers a wide range of problems involving the applications of stochastic processes, stochastic calculus, large deviation theory, group representation theory and quantum statistics to diverse fields in dynamical systems, electromagnetics, statistical signal processing, quantum information theory, quantum neural network theory, quantum filtering theory, quantum electrodynamics, quantum general relativity, string theory, problems in biology and classical and quantum fluid dynamics. The selection of the problems has been based on courses taught by the author to undergraduates and postgraduates in electronics and communication engineering. The theory of stochastic processes has been applied to problems of system identification based on time series models of the process concerned to derive time and order recursive parameter estimation followed by statistical performance analysis. The theory of matrices has been applied to problems such as quantum hypothesis testing between two states including a proof of the quantum Stein Lemma involving asymptotic formulas for the error probabilities involved in discriminating between a sequence of ordered pairs of tensor product states. This problem is the quantum generalization of the large deviation result regarding discriminating between to probability distributions from iid measurements. The error rates are obtained in terms of the quantum relative entropy between the two states. Classical neural networks are used to match an input-output signal pair. This means that the weights of the network are trained so that it takes as input the true system input and reproduces a good approximaition of the system output. But if the input and output signals are stochastic processes characterized by probability densities and not by signal values, then the network must take as input the input probability distribution and reproduce a good approximation of the output probability distribution. This is a very hard task since a probability distribution is specified by an infinite number of values. A quantum neural network solves this problem easily since Schrodinger's equation evolves a wave function whose magnitude square is always a probability density and hence by controlling the potential in this equation using the input pdf (probability distribution function) and the error between the desired output pdf and that generated by Schrodinger's equation, we are able to track a time varying output pdf successfully during the training process. This has applications to the problem of transforming speech data to EEG data. Speech is characterized by a pdf and so is EEG. Further the speech pdf is correlated to the EEG pdf because if the brain acquires a disease, the EEG signal will get distorted and correspondingly the speech signal will also get slurred. Thus, the quantum neural network based on Schrodinger's equation can exploit this correlation to get a fairly good approximation of the EEG pdf given the speech pdf. These issues have been discussed in this book. Other problems include a proof of the celebrated Lieb's inequality which is used in proving convexity and concavity properties of various kinds of quantum entropies and quantum relative entropies that are used in establishing coding theorems in quantum information theory like what is the maximum rate at which classical information can be transmitted over a Cq channel so that a decoder with error probability converging to zero can be constructed. Some of the inequalities discussed in this book find applications to more advanced problems in quantum information theory that have not been discussed here like how to generate a sequence of maximally entangled states having maximal dimension asymptotically from a sequence of tensor product states using local quantum operations with one way or two way classical communication using a sequence of TPCP maps or conversely, how to generate a sequence tensor product states optimally from a sequence of maximally entangled states by a sequence of TPCP maps with the maximally entangled states having minimal dimension. Or how to generate a sequence of maximal shared randomness between two users a sequence of quantum operations on a sequence of maximally entangled states. The performance of such algorithms are characterized by the limit of the logarithm of the length divided by the number of times that the tensor product is taken. These quantities are related to the notion of information rate. Or, the maximal rate at which information can be transmitted from one user to another when they share a partially entangled state. This book also covers problems involving quantization of the action functionals in general relativity and expressing the resulting approximate Hamiltonian as perturbations to harmonic oscillator Hamiltonians. It also covers topics such as writing down the Schrodinger equation for a mixed state of a particle moving in a potential coupled to a noisy bath in terms of the Wigner distribution function for position and momentum. The utility of this lies in the fact that the position and momentum observables in quantum mechanics do not commute and hence by the Heisenberg uncertainty principle, they do not have a joint probability density unlike the classical case of a particle moving in a potential with fluctuation and dissipation forces present where we have a Fokker-Planck equation for the joint pdf of the position and momentum. The Wigner distribution being complex, is not a joint probability distribution function but its marginals are and it is the closest that one can come in quantum mechanics for making an analogy with classical joint probabilities and we show in this book that its dynamics is the same a that of a classical Fokker-Planck equation but with small quantum correction terms that are represented by a power series in Planck's constant. We then also apply this formalism to Belavkin's quantum filtering theory based on the Hudson-Parthasarathy quantum stochastic calculus by showing that when this equation for a particle moving in a potential with damping and noise is described in terms of the Wigner distribution function, then it just the same as the Kushner-Kallianpur stochastic filter but with quantum correction terms expressed as a power series in Planck's constant. This book also covers some aspects of classical large deviation theory especially in the derivation of the rate function of the empirical distribution of a Markov process and applied to the Brownian motion and the Poisson process. This problem has applications to stochastic control where we desire to control the probability of deviation of the empirical distribution of the output process of a stochastic system from a desired probability distribution by incorporating control terms in the stochastic differential equation.
This book also discusses some aspects of string theory especially the problem of how field equations of physics get modified when string theory is taken into account and conformal invariance is imposed. How supersymmetry breaking stochastic terms in the Lagrangian can be reduced by feedback stochastic control etc. Some problem deal with modeling blood flow, breathing mechanisms etc using stochastic ordinary and partial differential equations and how to control them by applying large deviation theory. One of  the problems deals with analysis of the motion of a magnet through a tube with a coil wound around it in a graviational field taking into account electromagnetic induction and back reaction. This problem is result of discussions with one of my colleagues. The book discusses quantization of this motion when the electron and electromagnetic fields are also quantized. In short, this book will be useful to research workers in applied mathematics, physics and quantum information theory who wish to explore the range of applications of classical and quantum stochastics to problems of physics and engineering.
\bigskip

{\bf Table of Contents}

{\bf chapter 1}

Stability theory of nonlinear systems based on Taylor approximations of the nonlinearity

1.1 Stochastic perturbation of nonlinear dynamical systems:Linearized analysis

1.2 Large deviation analysis of the stability problem

1.3 LDP for Yang-Mills gauge fields and LDP for the quantized metric tensor of space-time

1.4 Strings in background fields
\bigskip

{\bf chapter 2}

2.1 String theory and large deviations

2.2 Evaluating quantum string amplitudes using functional integration

2.3 LDP for Poisson fields

2.4 LDP for strings perturbed by mixture of Gaussian and Poisson noise
\bigskip

{\bf chapter 3}

ordinary, partial and stochastic differential equation models for biological systems,quantum gravity, Hawking radiation, quantum fluids,statistical signal processing, group representations, large deviations in statistical physics

3.1 A lecture on biological aspects of virus attacking the immune system

3.2 More on quantum gravity

3.3 Large deviation properties of virus combat with antibodies

3.4 Large deviations problems in quantum fluid dynamics

3.5 Lecture schedule for statistical signal processing

3.6 Basic concepts in group theory and group representation theory

3.7 Some remarks on quantum gravity using the ADM action

3.8 Diffusion exit problem

3.9 Large deviations for the Boltzmann kinetic transport equation

3.10 Orbital integrals for $SL(2,\Bbb R)$ with applications to statistical image processing.

3.11 Loop Quantum Gravity (LQG)

3.12 Group representations in relativistic image processing

3.13 Approximate solution to the wave equation in the vicinity of the event horizon of a Schwarchild black-hole

3.14 More on group theoretic image processing

3.15 A problem in quantum information theory.

3.16 Hawking radiation

{\bf chapter 4}

Research Topics on representations with image processing applications,breathing dynamics

4.1 Remarks on induced representations and Haar integrals.

4.2 On the dynamics of breathing in the presence of a mask

4.3 Plancherel formula for $SL(2,\Bbb R)$

4.4 Some applications of the Plancherel formula for a Lie group to image processing

{\bf chapter 5}

Some problems in quantum information theory

5.1 Quantum information theory and quantum blackhole physics

5.2 Harmonic analysis on the Schwarz space of $G=SL(2,\Bbb R)$

5.3 Monotonicity of quantum entropy under pinching

{\bf chapter 6}

6.1 Lectures on Statistical Signal Processing

6.2 Large deviation analysis of the RLS algorithm

6.3 Problem suggested by my colleague Prof. Dhananjay

6.4 Fidelity between two quantum states

6.5 Stinespring's representation of a TPCP map, ie, of a quantum operation

6.6 Approximate computation of the matrix elements of the principal and discrete series for $G=SL(2,\Bbb R)$

6.7 Proof of the converse part of the Cq coding theorem

6.8 Proof of the direct part of the Cq coding theorem

6.9 Induced representations

6.10 Estimating non-uniform transmission line parameters and line voltage and current using the extended Kalman filter

6.11 Multipole expansion of the magnetic field produced by a time varying current distribution

{\bf chapter 7}

Some aspects of stochastic processes

7.1 Problem suggested by Prof.Dhananjay

7.2 Converse of the Shannon coding theorem

{\bf chapter 8}

Problems in electromagnetics and gravitation

{\bf chapter 9}

Some more aspects of quantum information theory

9.1 An expression for state detection error.

9.2 Operator convex functions and operator monotone functions

9.3 A selection of matrix inequalities

9.4 Matrix Holder Inequality

{\bf chapter 10}

10.1 Lecture Plan for Statistical Signal Processing

10.2 End-Semester Exam, duration:four hours, Pattern Recognition

10.3 Quantum Stein's theorem

10.4 Question paper, short test, Statistical signal processing, M.Tech

10.5 Question paper on statistical signal processing, long test

10.5.1 Atom excited by a quantum field

10.5.2 Estimating the power spectral density of a stationary Gaussian process using an AR model:Performance analysis of the algorithm

10.5.3 Statistical performance analysis of the MUSIC and ESPRIT algorithms

10.6 Quantum neural networks for estimating EEG pdf from speech pdf based on Schrodinger's equation

{\bf Reference}:Paper by Vijay Upreti, Harish Parthasarathy, Vijayant Agarwal and Guguloth Sagar.

section{Statistical Signal Processing, M.Tech, Long Test\author{Max marks:50}}

10.7 Monotonocity of quantum relative Renyi entropy

{\bf chapter 11}

My commentary on the Ashtavakra Gita

{\bf chapter 12}

LDP-UKF based controller

{\bf chapter 13}

Abstracts for a book on quantum signal processing using quantum field theory

{\bf chapter 14}

Quantum neural networks and learning using mixed state dynamics for open quantum systems

{\bf chapter 15}

Quantum gate design, magnet motion, quantum filtering, Wigner distributions

15.1 Designing a quantum gate using QED with a control c-number four potential

15.2 On the motion of a magnet through a pipe with a coil under gravity and the reaction force exerted by the induced current through the coil on the magnet

15.3 Quantum filtering theory applied to quantum field theory for tracking a given probability density function

15.4 Lecture on the role of statistical signal processing in general relativity and quantum mechanics

15.5 Problems in classical and quantum stochastic calculus

15.6 Some remarks on the equilibrium Fokker-Planck equation for the Gibbs distribution

15.7 Wigner-distributions in quantum mechanics

{\bf Construction of the quantum Fokker-Planck equation for a specific choice of the Lindblad operator}

{\bf Problems in quantum corrections to classical theories in probability theory and in mechanics}

15.8 Motion of an element having both an electric and a magnetic dipole moment in an external electromagnetic field plus a coil

{\bf chapter 16}

Large deviations for Markov chains, quantum master equation, Wigner distribution for the Belavkin filter

16.1 Large deviation rate function for the empirical distribution of a Markov chain

16.2 Quantum Master Equation in general relativity

16.3 Belavkin filter for the Wigner distirbution function

{\bf chapter 17}

String and field theory with large deviations stochastic algorithm convergence

17.1 String theoretic corrections to field theories

17.2 Large deviations problems in string theory

17.3 Convergence analysis of the LMS algorithm

17.4 String interacting with scalar field, gauge field and a gravitational field
\newpage

\chapter{Stability theory of nonlinear systems based on Taylor approximations of the nonlinearity}

\section{Stochastic perturbation of nonlinear dynamical systems:Linearized analysis}

Consider the nonlinear system defined by the differential equation
$$
dx(t)/dt=f(x(t)), t\geq 0
$$
where $f:\Bbb R^n\rightarrow\Bbb R^n$ is a continuously differentiable mapping. Let $x_0$ be a fixed point of this system, ie
$$
f(x_0)=0
$$
Denoting $\delta x(t)$ to be a small fluctuation of the state $x(t)$ around the stability point $x_0$, we get on first order Taylor expansion, ie, linearization,
$$
d\delta x(t)/dt=f'(x_0)\delta x(t)+O(|\delta x(t)|^2)
$$
If we neglect $O(|\delta x|^2)$ terms, then we obtain the approximate solution to this differential equation as
$$
\delta x(t)=exp(tA)\delta x(0), t\geq 0
$$
where 
$$
A=f'(x_0)
$$
is the Jacobian matrix of $f$ at the equilibrium point $x_0$ and hence it follows that this linearized system is asymptotically stable under small perturbations in all directions iff all the eigenvalues of $A$ have negative real parts. By asymptotically stable, we meant that $lim_{t\rightarrow\infty}\delta x(t)=0$. The linearized system is critically stable at $x_0$ if all the eigenvalues of $A$ have non-positive real part, ie, an eigenvalue being purely imaginary is also allowed. By critically stable, we mean that
$\delta x(t), t\geq 0$ remains bounded in time.
\bigskip

Stochastic perturbations: Consider now the stochastic dynamical system obtained by adding a source term on the rhs of the above dynamical system
$$
dx(t)/dt=f(x(t))+w(t)
$$
where $w(t)$ is stationary Gaussian noise with mean zero and autcorrelation
$$
R_w(\tau)=\Bbb E(w(t+\tau)w(t))
$$
Now it is meaningless to speak of a fixed point, so we choose a trajectory $x_0(t)$ that solves the noiseless problem
$$
dx_0(t)/dt=f(x_0(t))
$$
Let $\delta x(t)$ be a small perturbation around $x_0(t)$ so that $x(t)=x_0(t)+\delta x(t) $ solves the noisy problem. Then, assuming $w(t)$ to be small and $\delta x(t)$ to be of the "same order or magnitude" as $w(t)$, we get approximately upto linear orders in $w(t)$, the linearized equation
$$
d\delta x(t)/dt=A\delta x(t)+w(t), A=f'(x_0)
$$
which has solution
$$
\delta x(t)=\Phi(t)\delta x(0)+\int_0^t\Phi(t-\tau)w(\tau)d\tau, t\geq 0,
$$
where
$$
\Phi(t)=exp(tA)
$$
Assuming $\delta x(0)$ to be a nonrandom initial perturbation, we get
$$
\Bbb E(\delta x(t))=\Phi(t)\delta x(0)
$$
which converges to zero as $t\rightarrow\infty$ provided that all the eigenvalues of $A$ have negative real part. Moreover, in this case,
$$
\Bbb E(|\delta x(t)|^2)=Tr\int_0^t\int_0^t\Phi(t-t_1)R_w(t_1-t_2)\Phi(t-t_2)^Tdt_1dt_2
$$
$$
=Tr\int_0^t\int_0^t\Phi(t_1)R_w(t_1-t_2)\Phi(t_2)^Tdt_1dt_2
$$
which converges as $t\rightarrow\infty$ to
$$
P_x=Tr\int_0^{\infty}\int_0^{\infty}\Phi(t_1)R_w(t_1-t_2)\Phi(t_2)^Tdt_1dt_2
$$
Suppose now that $A$ is diagonalizable with spectral representation
$$
A=-\sum_{k=1}^r\lambda(k)E_k
$$
Then
$$
\Phi(t)=\sum_{k=1}^rexp(-t\lambda(k))E_k
$$
and we get
$$,
P_x=\sum_{k,m=1}^r\int_0^{\infty}\int_0^{\infty}Tr(E_k.R_w(t_1-t_2).E_m)exp(-\lambda(k)t_1-\lambda(m)t_2)dt_1dt_2
$$
A sufficient condition for this to be finite is that $|R_w(t)|$ remain bounded in which case we get when
$$
sup_t|R_w(t)|\leq B<\infty\forall t
$$
that
$$
P_x\leq B\sum_{k,m=1}^r(Re(\lambda(k)+\lambda(m)))^{-1}\leq 2Br^2/\lambda_R(min)
$$
where 
$$
\lambda_R(min)=min_kRe(\lambda(k))
$$
\bigskip

\section{Large deviation analysis of the stability problem}

Let $x(t)$ be a diffusion process
starting at some point $x_0$ within an open set $V$ with boundary $S=\partial V$. If the noise level is low say parametrized by the parameter $\epsilon$, then the diffusion equation can be expressed as
$$
dx(t)=f(x(t))dt+\sqrt{\epsilon}g(x(t))dB(t)
$$
Let $P_{\epsilon}(x)$ denote the probability of a sample path of the process that starts at $x_0$ end ends at the stop time $\tau_{\epsilon}=t$ at which the process first hits the boundary $S$. Then, from basic LDP, we know that the approximate probability of such a path is given by 
$$
exp(-\epsilon^{-1}inf_{y\in S}V(x_0,y,t))
$$
where 
$$
V(x,t)=inf_{u:dx(s)/ds=f(x(s))+g(x(s))u(s), 0\leq s\leq t, x(0)=x_0}\int_0^tu(s)^2ds/2
$$
and
$$
V(x_0,y,t)=inf_{x:x(0)=x_0, x(t)=y}V(x_0,x,t)
$$
The greater the probability of the process hitting $S$, the lesser the mean value of the hitting time. This is valid with an inverse proportion. Thus, we can say that
$$
\tau_{\epsilon}\approx exp(V/\epsilon), \epsilon\rightarrow 0
$$
where
$$
V=inf_{y\in S,t\geq 0}V(x_0,y,t)
$$
More precisely, we have the LDP
$$
lim_{\epsilon\rightarrow 0}P(exp((V-\delta)/\epsilon)<\tau_{\epsilon}<exp((V+\delta)/\epsilon))=1
$$
Note that $\tau_{\epsilon}=exp(V/\epsilon)$ for a given path $x$ implies that the probability of $x$ $P_{\epsilon}(x)=exp(-V/\epsilon)$. This is because of the relative frequency interpretation of probability given by Von Mises or equivalently because of ergodic theory: If the path hits the boundary $N(\epsilon)$ times in a total of $N_0$
 trials, with the time taken to hit being $\tau_{\epsilon}$ for each path, then the probability of hitting the boundary is $N(\epsilon)/N$. However, since it takes a time
$\tau_{\epsilon}$ for each trial hit, it follows that the number of hits in time $T$
is $T/\tau_{\epsilon}$. Thus if each trial run in the set of $N$ trials is of duration $T$, then the probability of hit is
$$
P_{\epsilon}=N(\epsilon)/N=T/(N.\tau_{\epsilon})
$$
and if we assume that that $T,N$ are fixed, or equivalently that $T,N\rightarrow\infty$ with $N/T\rightarrow$ to a finite nonzero constant, then we deduce from the above,
$$
\tau_{\epsilon}\approx exp(V/\epsilon)
$$
in the sense that $\epsilon.log(\tau_{\epsilon})$ converges to $V$ as $\epsilon\rightarrow 0$. 
\bigskip

\section{LDP for Yang-Mills gauge fields and LDP for the quantized metric tensor of space-time}

Reference: Thomas Thiemann, "Modern canonical quantum general relativity", Cambridge university press.

Let $L(q_{ab},q_{ab}',N,N^a)$ denote the ADM Lagrangian of space-time. Here, $N^a$ defines the diffeomorphism constraint, $N$ the Hamiltonian constraint. The canonical momenta are
$$
P^{ab}=\partial L/\partial q_{ab}'
$$
The Hamiltonian is then
$$
H=P^{ab}q_{ab}'-L
$$
Note that the time derivatives $N', N^{a'}$ of $N$ and $N^a$ do not appear in $L$. 
QGTR is therefore described by a constrained Hamiltonian with the constraints being
$$
P^N=\partial L/\partial N'=0,
$$
leading to the equation of motion
$$
H=\partial L/\partial N=0
$$
which is the Wheeler-De-Witt equation, the analog of Schrodinger's equation in particle non-relativistic mechanics. It should be interpreted as
$$
H(N)\psi=0\forall N
$$
where
$$
H(N)=\int N.Hd^3x
$$

The Large deviation problem: If there is a small random perturbation in the initial conditions of the Einstein field equations, then what will be the statistics of the perturbed spatial metric $q_{ab}$ after time $t$ ? This is the classical LDP. The quantum LDP is that if there is a small random perturbation in the initial wave function of a system and also small random Hamiltonian perturbations as well as small random time varying perturbations in the potentials appearing in the Hamiltonian, then what will be the statistics of quantum averaged observables ?
\bigskip

The Yang-Mills Lagrangian in curved space-time: Let $A_{\mu}=A_{\mu}^a\tau_a$ denote the 
Lie algebra valued Yang-Mills connection potential field. Let 
$$
D_{\mu}=\partial_{\mu}+\Gamma_{\mu}
$$
denote the spinor gravitational covariant derivative. Then the Yang-Mills field tensor is given by
$$
F_{\mu\nu}=D_{\mu}A_{\nu}-D_{\nu}A_{\mu}+[A_{\mu},A_{\nu}]
$$
The action functional of the Yang-Mills field in the background gravitational field is given by
$$
S_{YM}(A)=\int g^{\mu\alpha}g^{\nu\beta}Tr(F_{\mu\nu}F_{\alpha\beta})\sqrt{-g}d^4x
$$

Problem: Derive the canonical momentum fields for this curved space-time Yang-Mills Lagrangian and hence obtain a formula for the canonical Hamiltonian density. 
\bigskip

\section{String in background fields}

The Lagrangian is
$$
L(X,X_{,\alpha},\alpha=1,2)=(1/2)g_{\mu\nu}(X)h^{\alpha\beta}X^{\mu}_{,\alpha}X^{\nu}_{,\beta}
$$
$$
+\Phi(X)+H_{\mu\nu}(X)\epsilon^{\alpha\beta}X^{\mu}_{,\alpha}X^{\nu}_{,\beta}
$$
where a conformal transformation has been applied to bring the string world sheet metric
$h_{\alpha\beta}$ to the canonical form $diag[1,-1]$. Here, $\Phi(X)$ is a scalar field.
The string action functional is
$$
S[X]=\int Ld\tau d\sigma
$$
The string field equations
$$
\partial_{\alpha}\partial L/\partial X^{\mu}_{,\alpha}-\partial L/\partial X^{\mu}=0
$$
are
$$
\partial_{\alpha}(g_{\mu\nu}(X)h^{\alpha\beta}X^{\nu}_{,\beta})
+\epsilon^{\alpha\beta}(H_{\mu\nu}(X)X^{\nu}_{,\beta})_{,\alpha}=
(1/2)g_{\rho\sigma,\mu}(X)h^{ab}X^{\rho}_{,a}X^{\sigma}_{,b}+\Phi_{,\mu}(X)
+H_{\rho\sigma,\mu}(X)\epsilon^{ab}X^{\rho}_{,a}X^{\sigma}_{,b}
$$
or equivalently,
$$
g_{\mu\nu,\rho}(X)h^{ab}X^{\rho}_{,a}X^{\nu}_{,b}+g_{\mu\nu}(X)h^{ab}X^{\nu}_{,ab}
+\epsilon^{ab}H_{\mu\nu,\rho}(X)X^{\rho}_{,a}X^{\nu}_{,b}
+\epsilon^{ab}H_{\mu\nu}(X)X^{\nu}_{,ab}
$$
$$
=(1/2)g_{\rho\sigma,\mu}(X)h^{ab}X^{\rho}_{,a}X^{\sigma}_{,b}+\Phi_{,\mu}(X)
+H_{\rho\sigma,\mu}(X)\epsilon^{ab}X^{\rho}_{,a}X^{\sigma}_{,b}
$$
These differential equations can be put in the form
$$
\eta^{ab}X^{\mu}_{,ab}=\delta.F^{\mu}(X^{\rho},X^{\rho}_{,a},X^{\rho}_{,ab})
$$
where $\delta$ is a small perturbation parameter and $((\eta^{ab}))=diag[1,-1]$
The problem is to derive the correction to string propagator caused by the nonlinear perturbation term. The propagator is
$$
D^{\mu\nu}(\tau,\sigma|\tau',\sigma')=<0|T(X^{\mu}(\tau,\sigma).X^{\nu}(\tau',\sigma'))|0>
$$
and we see that it satisfies the following fundamental propagator equations: 
let
$$
x=(\tau,\sigma),x'=(\tau',\sigma'), \partial_0=\partial_{\tau},\partial_1=\partial_{\sigma}
$$
Then,
$$
\partial_0D^{\mu\nu}(x|x')=\delta(x^0-x^{'0})<[X^{\mu}(x),X^{\nu}(x')]>
$$
$$
+<T(X^{\mu}_{,0}(x).X^{\nu}(x'))>
$$
$$
=<T(X^{\mu}_{,0}(x).X^{\nu}(x')>
$$
and likewise
$$
\partial_0^2D^{\mu\nu}(x|x')=\delta(x^0-x^{'0})<[X^{\mu}_{,0}(x),X^{\nu}(x')]>
$$
$$
+<T(X^{\mu}_{,00}(x).X^{\nu}(x'))>
$$
$$
=\delta(x^0-x^{'0})<[X^{\mu}_{,0}(x),X^{\nu}(x')]>
$$
$$
+\partial_1^2D^{\mu\nu}(x|x')+\delta.<T(F^{\mu}(X,X_{,a}X_{,b})(x).X^{\nu}(x'))>
$$
or equivalently
$$
\eta^{ab}\partial_a\partial_bD(x|x')=
$$
$$
\delta.<T(F^{\mu}(X,X_{,a}X_{,b})(x).X^{\nu}(x'))>
$$
$$
+\delta(x^0-x^{'0})<[X^{\mu}_{,0}(x),X^{\nu}(x')]>---(1)
$$
Now consider the canonical momenta
$$
P_{\mu}(x)=\partial L/\partial X^{\mu}_{,0}=g_{\mu\nu}(X(x))X^{\nu}_{,0}(x)
$$
$$
-2H_{\mu\nu}(X(x))X^{\nu}_{,1}(x)
$$
Using the canonical equal time commutation relations
$$
[X^{\mu}(\tau,\sigma),X^{\nu}(\tau,\sigma')]=0
$$
so that
$$
[X^{\mu}(\tau,\sigma),X^{\nu}_{,1}(\tau,\sigma')]=0
$$
and
$$
[X^{\mu}(\tau,\sigma),P_{\nu}(\tau,\sigma')]=i\delta^{\nu}_{\mu}\delta(\sigma-\sigma')
$$
we therefore deduce that
$$
[X^{\mu}(\tau,\sigma),g_{\nu\rho}(X(\tau,\sigma'))X^{\rho}_{,0}(\tau,\sigma')]=i\delta^{\nu}_{\mu}\delta(\sigma-\sigma')
$$
or equivalently,
$$
[X^{\mu}(\tau,\sigma),X^{\nu}_{,0}(\tau,\sigma')]=ig^{\mu\nu}(X(\tau,\sigma))\delta(\sigma-\sigma')
$$
and therefore (1) can be expressed as
$$
\square D^{\mu\nu}(x|x')=\eta^{ab}\partial_a\partial_bD^{\mu\nu}(x|x')=
$$
$$
\delta.<T(F^{\mu}(X,X_{,a}X_{,b})(x).X^{\nu}(x'))>
$$
$$
-i<g^{\mu\nu}(X(x))>\delta^2(x-x')---(2)
$$
Now suppose we interpret the quantum string field $X^{\mu}$ as being the sum of a large classical component $X^{\mu}_0(\tau,\sigma)$ and a small quantum fluctuating component
$\delta X^{\mu}(\tau,\sigma)$, then we can express the above exact relation as
$$
D^{\mu\nu}(x|x')=X_0^{\mu}(x)X_0^{\nu}(x')+<T(\delta X^{\mu}(x).\delta X^{\nu}(x'))>
$$
and further we have approximately upto first order of smallness
$$
F^{\mu}(X,X',X'')=F^{\mu}(X_0,X_{0,a},X_{0,b})+F^{\mu}_{,1\rho}(X_0,X_0',X_0'')\delta X^{\rho}+F^{\mu}_{,2\rho a}(X,X',X'')\delta X^{\rho}_{,a}+F^{\mu}_{,3\rho ab}(X,X',X'')\delta X^{\rho}_{,ab}
$$
and hence upto second orders of smallness, (2) becomes
$$
\square(X_0^{\mu}(x)X_0^{\nu}(x'))+\square<T(\delta X^{\mu}(x).\delta X^{\nu}(x'))>
$$
$$
=\delta F^{\mu}(X_0,X_{0,a},X_{0,b})(x)X_0^{\nu}(x')
$$
$$
+\delta.F^{\mu}_{,1\rho}(X_0,X_0',X_0'')(x)<T(\delta X^{\rho}(x)\delta X^{\nu}(x'))>
$$
$$
+\delta.F^{\mu}_{,2\rho a}(X,X',X'')(x)<T(\delta X^{\rho}_{,a}(x).\delta X^{\nu}(x'))>
$$
$$
+\delta.F^{\mu}_{,3\rho ab}(X,X',X'')(x)<T(\delta X^{\rho}_{,ab}(x).\delta X^{\nu}(x'))>
$$
$$
-i<g^{\mu\nu}(X(x))>\delta^2(x-x')
$$
To proceed further, we require the dynamics of the quantum fluctuation part $\delta X^{\mu}$. To get this, we apply first order perturbation theory to the system
$$
\eta^{ab}X^{\mu}_{,ab}=\delta.F^{\mu}(X^{\rho},X^{\rho}_{,a},X^{\rho}_{,ab})
$$
to get
$$
\eta^{ab}X_{0,ab}^{\mu}=0
$$
$$
\eta^{ab}\delta X^{\mu}_{,ab}=\delta.F^{\mu}(X_0,X_0',X_0'')
$$
and solve this to express $\delta X^{\mu}$ as the sum of a classical part (ie a particular solution to the inhomogeneous equation) and a quantum part (ie the general solution to the homogeneous equation). The quantum component is
$$
\delta X^{\mu}_q(\tau,\sigma)=-i\sum_{n\neq 0}(a^{\mu}(n)/n)exp(in(\tau-\sigma))
$$
where
$$
[a^{\mu}(n),a^{\nu}(m)]=n\eta^{\mu\nu}\delta[n+m]
$$
Note that in particular, $\square X_0^{\mu}(x)=0$ and hence the approximate corrected propagator equation becomes
$$
\square<T(\delta X^{\mu}(x).\delta X^{\nu}(x'))>
$$
$$
=\delta F^{\mu}(X_0,X_{0,a},X_{0,b})(x)X_0^{\nu}(x')
$$
$$
+\delta.F^{\mu}_{,1\rho}(X_0,X_0',X_0'')(x)<T(\delta X^{\rho}(x)\delta X^{\nu}(x'))>
$$
$$
+\delta.F^{\mu}_{,2\rho a}(X,X',X'')(x)<T(\delta X^{\rho}_{,a}(x).\delta X^{\nu}(x'))>
$$
$$
+\delta.F^{\mu}_{,3\rho ab}(X,X',X'')(x)<T(\delta X^{\rho}_{,ab}(x).\delta X^{\nu}(x'))>
$$
$$
-ig^{\mu\nu}(X_0(x))\delta^2(x-x')
$$
provided that we neglect the infinite term
$$
g_{\mu\nu,\alpha\beta}(X_0)<\delta X^{\alpha}(x)\delta X^{\beta}(x)>
$$
if we do not attach the perturbation tag $\delta$ to the $F$-term, ie, we do not regard 
$F$ as small, then the unperturbed Bosonic string $X_0$ satisfies
$$
\square X_0^{\mu}=F^{\mu}(X_0,X_0',X_0'')
$$
and then the corrected propagator $D^{\mu\nu}$ satisfies the approximate equation
$$
\square D^{\mu\nu}(x|x')=
$$
$$
+\delta.F^{\mu}_{,1\rho}(X_0,X_0',X_0'')(x)<T(\delta X^{\rho}(x)\delta X^{\nu}(x'))>_0
$$
$$
+\delta.F^{\mu}_{,2\rho a}(X,X',X'')(x)<T(\delta X^{\rho}_{,a}(x).\delta X^{\nu}(x'))>_0
$$
$$
+\delta.F^{\mu}_{,3\rho ab}(X,X',X'')(x)<T(\delta X^{\rho}_{,ab}(x).\delta X^{\nu}(x'))>_0
$$
$$
-i\eta^{\mu\nu}\delta^2(x-x')
$$
provided that we assume that the metric is a very small perturbation of that of Minkowski flat space-time. Here, $<.>_0$ denotes the unperturbed quantum average, ie, the quantum average in the absence of the external field $H_{\mu\nu}$ and also in the absence of metric perturbations of flat space-time. We write
$$
D^{\mu\nu}_0(x|x')=<T(\delta X^{\mu}(x).\delta X^{\nu}(x'))>_0
$$
Now observe that in the absence of such perturbations,
$$
\delta X^{\mu}(x)=-i\sum_{n\neq 0}(a^{\mu}(n)/n)exp(in(\tau-\sigma))
$$
so
$$
<T(\delta X^{\mu}(x).\delta X^{\nu}(x'))>_0=
$$
$$
-\sum_{n>0}(\theta(\tau-\tau')/n)exp(-in(\tau-\tau'-\sigma+\sigma'))
$$
$$
-\sum_{n>0}(\theta(\tau'-\tau)/n).exp(-in(\tau'-\tau-\sigma'+\sigma))
$$
$$
<T(\delta X^{\mu}_{,0}(x).\delta X^{\nu}(x'))>_0=
$$
$$
-\sum_{n>0}(\theta(\tau-\tau')/n)(-in)exp(-in(\tau-\tau'-\sigma+\sigma'))
$$
$$
-\sum_{n>0}(\theta(\tau'-\tau)/n).(in)exp(-in(\tau'-\tau-\sigma'+\sigma))
$$
$$
=i\sum_{n>0}(\theta(\tau-\tau')exp(-in(\tau-\tau'-\sigma+\sigma'))
$$
$$
-i\sum_{n>0}(\theta(\tau'-\tau)exp(-in(\tau'-\tau-\sigma'+\sigma))
$$
$$
<T(\delta X^{\mu}_{,1}(x).\delta X^{\nu}(x'))>_0=
$$
$$
-\sum_{n>0}(\theta(\tau-\tau')/n)(in)exp(-in(\tau-\tau'-\sigma+\sigma'))
$$
$$
-\sum_{n>0}(\theta(\tau'-\tau)/n).(-in)exp(-in(\tau'-\tau-\sigma'+\sigma))
$$
$$
=-i\sum_{n>0}(\theta(\tau-\tau')exp(-in(\tau-\tau'-\sigma+\sigma'))
$$
$$
+i\sum_{n>0}(\theta(\tau'-\tau)exp(-in(\tau'-\tau-\sigma'+\sigma))
$$
$$
=-<T(\delta X^{\mu}_{,0}(x).\delta X^{\nu}(x'))>_0
$$
Note that we have the alternate expression
$$
<T(\delta X^{\mu}_{,1}(x).\delta X^{\nu}(x'))>_0=
$$
$$
\partial_1D_0^{\mu\nu}(x|x')
$$
because differentiation w.r.t the spatial variable $\sigma$ commutes with the time ordering operation $T$. Further,
$$
<T(\delta X^{\mu}_{,00}(x).\delta X^{\nu}(x'))>_0
$$
$$
=<T(\delta X^{\mu}_{,11}(x).\delta X^{\nu}(x'))>_0
$$
$$
=\partial_1^2D_0^{\mu\nu}(x|x')
$$
$$
<T(\delta X^{\mu}_{,01}(x).\delta X^{\nu}(x'))>_0
$$
$$
=\partial_1<T(\delta X^{\mu}_{,0}(x).\delta X^{\nu}(x'))>_0
$$
$$
=\partial_1[i\sum_{n>0}(\theta(\tau-\tau')exp(-in(\tau-\tau'-\sigma+\sigma'))
$$
$$
-i\sum_{n>0}(\theta(\tau'-\tau)exp(-in(\tau'-\tau-\sigma'+\sigma))]=
$$
$$
i\sum_{n>0}(\theta(\tau-\tau')(in)exp(-in(\tau-\tau'-\sigma+\sigma'))
$$
$$
-i\sum_{n>0}(\theta(\tau'-\tau)(-in)exp(-in(\tau'-\tau-\sigma'+\sigma))=
$$
$$
-\sum_{n>0}(\theta(\tau-\tau')n.exp(-in(\tau-\tau'-\sigma+\sigma'))
$$
$$
-\sum_{n>0}(\theta(\tau'-\tau)n.exp(-in(\tau'-\tau-\sigma'+\sigma))
$$
\bigskip

\chapter{String theory and large deviations}

\section{Evaluating quantum string amplitudes using functional integration}

Assume that external lines are attached to a given string at space-time points $(\tau_r,\sigma_r), r=1,2,..., M$. Note that the spatial points $\sigma_r$ are fixed while the interaction times $\tau_r$ vary. Let $P_r(\sigma)$ denote the momentum of the $r^{th}$ external line at $\sigma$. At the point $\sigma=\sigma_r$ where it is attached, this momentum is specified as $P_r(\sigma_r)$. Thus noting that
$$
X(\tau,\sigma)=-i\sum_{n\neq 0}a(n)exp(in(\tau-\sigma))/n
$$
we have
$$
P_r(\sigma_r)=X_{,\tau}(\sigma_r)=\sum_{n\neq 0}a(n)exp(-in\sigma_r)
$$
This equation constrains the values of $a(n)$. We thus define the numbers $P_{rn}$ by
$$
P_r(\sigma_r)=\sum_nP_{rn}exp(-in\sigma_r)---(1)
$$
and the numbers $\{P_{rn}\}$ characterize the momentum of the external line attached at the point $\sigma_r$. Note that for reality of the momentum, we must have $P_{rn}=P_{r,-n}$ if we assume that the $P_{rn}'s$ are real and therefore
$$
P_r(\sigma_r)=2\sum_{n>0}P_{rn}cos(n\sigma_r)
$$
Note that $P_r(\sigma)$ is the momentum field corresponding to the $r^{th}$ external line and this is given to us. Thus, the $P_{rn}'s$ are determined by the equation
$$
P_r(\sigma)=2\sum_{n>0}P_{rn}cos(n\sigma)
$$
Equivalently, we fix the numbers $\{P_{rn}\}_{r,n}$ corresponding to the momenta carried by the external lines and then obtain the momentum $P_r(\sigma_r)$ of the line attached at $\sigma_r$ using (1).

The large deviation problem: Express the Hamiltonian density of the string having Lagrangian density $(1/2)\partial_{\alpha}X^{\mu}\partial^{\alpha}X_{\mu}$ in terms of
$X^{\mu},P_{\mu}=\partial_0X^{\mu}$. Show that it is given by
$$
H_0=(1/2)P^{\mu}P_{\mu}+(1/2)\partial_1X^{\mu}\partial_1X_{\mu}
$$
In the presence of a vector potential field $A_{\mu}(\sigma)$, the charged string field has the Hamiltonian density $H$ with $P_{\mu}$ replaced by $P_{\mu}+eA_{\mu}$. Write down the string field equations for this Hamiltonian and assuming that the vector potential is a weak Gaussian field, evaluate the rate function of the string field $X^{\mu}$.
\bigskip

LDP in general relativistic fluid dynamics.

\section{LDP for Poisson fields}

Small random perturbations in the metric field having a statistical model of the sum of a Gaussian and a Poisson field affect the fluid velocity field. Let
$X_1(x), x\in\Bbb R^n$ be a zero mean Gaussian random field with correlation
$R(x,y)=\Bbb E(X_1(x)X_1(y))$ and let $N(x),x\in\Bbb R^n$ be a Poisson random field with mean $\Bbb E(N(dx))=\lambda(x)dx$. Consider the family of random Poisson measure
$N_{\epsilon}(E)=\epsilon^aN(\epsilon^{-1}E)$ for any Borel subset $E$ of $\Bbb R^n$. We have
$$
\Bbb E[exp(\int_{\Bbb R^n}f(x)N(dx))]
$$
$$
=exp(\int_{\Bbb R^n}\lambda(x)(exp(f(x))-1)dx)
$$
Thus,
$$
M_{\epsilon}(f)=\Bbb Eexp(\int f(x)N_{\epsilon}(dx))
$$
$$
=exp(\int\epsilon^af(x)N(dx/\epsilon)))=exp(\int\epsilon^af(\epsilon x)N(dx))
$$
$$
=exp(\int\lambda(x)(exp(\epsilon^af(\epsilon x))-1)dx)
$$
Thus,
$$
log(M_{\epsilon}(\epsilon^{-a}f))=\int\lambda(x)(exp(f(\epsilon x))-1)dx
$$
$$
=\epsilon^{-n}\int\lambda(\epsilon^{-1}x)(exp(f(x))-1)dx
$$
Then,
$$
lim_{\epsilon\rightarrow 0}\epsilon^n.log(M_{\epsilon}(\epsilon^{-a}f))
$$
$$
=\lambda(\infty)\int(exp(f(x))-1)dx
$$
assuming that
$$
\lambda(\infty)=lim_{\epsilon\rightarrow 0}\lambda(x/\epsilon)
$$
exists. More generally, suppose we assume that this limit exists and depends only on the direction of the vector $x\in\Bbb R^n$ so that we can write
$$
\lambda(\infty,\hat x)=lim_{\epsilon\rightarrow 0}\lambda(x/\epsilon)
$$
then we get for the Gartner-Ellis limiting logarithmic moment generating function
$$
lim_{\epsilon\rightarrow 0}\epsilon^n.log(M_{\epsilon}(\epsilon^{-a}f))
$$
$$
=\int\lambda(\infty,\hat x)(exp(f(x))-1)dx=\bar\Lambda(f)
$$
\bigskip

\section{LDP for strings perturbed by mixture of Gaussian and Poisson noise}

Consider the string field equations
$$
\square X^{\mu}(\sigma)=f^{\mu}_k(\sigma)\sqrt{\epsilon}W^k(\tau)+g^{\mu}_k\epsilon dN_k(\tau)/d\tau
$$
where $W^k$ are independent white Gaussian noise processes and $N_k$ are independent Poisson processes. Assume that the rate of $N_k$ is $\lambda_k/\epsilon$. Calculate then the LDP rate functional of the string process $X^{\mu}$. Consider now a superstring with
Lagrangian density
$$
L=(1/2)\partial_{\alpha}X^{\mu}\partial^{\alpha}X_{\mu}+\bar\psi\rho^{\alpha}\partial_{\alpha}\psi
$$
Add to this an interaction Lagrangian
$$
\epsilon_{\alpha\beta}B_{\mu\nu}(X)X^{\mu}_{,\alpha}X^{\nu}_{,\beta}
$$
an another interaction Lagrangian
$$
J_{\alpha}(X)\bar\psi\rho^{\alpha}\psi
$$
Assume that $B_{\mu\nu}(X)$ and $J_{\alpha}(X)$ are small Gaussian fields. Calculate the joint rate function of the Bosonic and Fermionic string fields $X^{\mu},\psi$.
\bigskip

\chapter{ordinary, partial and stochastic differential equation models for biological systems,quantum gravity, Hawking radiation, quantum fluids,statistical signal processing, group representations, large deviations in statistical physics}

\section{A lecture on biological aspects of virus attacking the immune system}

We discuss here a mathematical model based on elementary fluid dynamics and quantum mechanics, how a virus attacks the immune system and how antibodies generated by the immune system of the body and the vaccination can wage a counter attack against this virus thereby saving the life. According to recent researches, a virus consists of a spherical body covered with dots/rods/globs of protein. When the virus enters into the body, as such it is harmless but becomes dangerous when it lodges itself in the vicinity of a cell in the lung where the protein blobs are released. The protein blobs thus penetrate into the cell fluid where the ribosome gives the command to replicate in the same way that normal cell replication occurs. Thus a single protein blob is replicated several times and hence the entire cell is filled with protein and thus gets destroyed completely. When this happens to every cell in the lung, breathing is hampered and death results. However, if our immune system is strong enough, then the blood generates antibodies the moment it recognizes that a stranger in the form of protein blobs has entered into the cell. These antibodies fight against the protein blobs and in fact the antibodies lodge themselves on the boundary of the cell body at precisely those points where the protein from the virus has penetrated thus preventing further protein blobs from entering into the cell. Whenever this fight between the antibodies and the virus goes on, the body develops fever because an increase in the body temperature disables the virus and enables the antibodies to fight against the proteins generated by the virus. If however the immune system is not powerful enough, sufficient concentration of antibodies is not
generated and the viral protein dominates over the antibodies in their fight finally resulting in death of the body. The Covid vaccine is prepared by generating a sample virus in the laboratory within a fluid like that present within the body and hence antibodies are produced within the fluid by this artifical process in the same way that the small-pox vaccine was invented by Edward Jenner by injecting the small-pox virus into a cow to generate the required antibodies. Now we discuss the mathematics of the virus, protein, cells and antibodies.

Assume that the virus particles have masses $m_1,..., m_N$ and charges $e_1,..., e_N$. The blood fluid has a velocity field ${\bf v}(t,{\bf r})$ whose oxygenated component circulates to all the organs of the body and whose de-oxygenated component is directed toward the lungs. This circulation is caused by the pumping of the heart. Assume that the heart pumps blood with a generation rate $g(t,{\bf r})$ having units of density per unit time or equivalently mass per unit volume per unit time. Assume also that the force of this pumping is represented by a force density field ${\bf f}(t,{\bf r})$. Also there may be present external electromagnetic fields ${\bf E}(t,{\bf r}),{\bf B}(t,{\bf r})$ within the blood field which interact with the current field within the blood caused by non-zero conductivity $\sigma$ of the blood. This current density is according to Ohm's law given by
$$
{\bf J}(t,{\bf r})=\sigma({\bf E}(t,{\bf r})+{\bf v}(t,{\bf r})\times{\bf B}(t,{\bf r}))
$$
The relevant MHD equations and the continuity equation are given by
$$
\rho(\partial_tv+(v,\nabla)v)=-\nabla p+\eta\nabla^2v+f+\sigma(E+v\times B)\times B
$$
$$
\partial_t\rho+div(\rho.v)=g
$$
Sometimes a part of the electromagnetic field within the blood is of internal origin, ie, generated by the motion of this conducting blood fluid. Thus, if we write $E_{ext},B_{ext} $ for the externally applied electromagnetic field and $E_{int},B_{int}$ for the internally generated field, then we have
$$
E_{int}(t,r)=-\nabla\phi_{int}-\partial_tA_{int}, B_{int}=\nabla\times A_{int}
$$
where
$$
A_{int}(t,r)=(\mu/4\pi)\int J(t-|r-r'|/c,r')d^3r'/|r-r'|,
$$
$$
\Phi_{int}(t,r)=-c^2\int_0^tdiv A_{int}dt
$$
The total electromagnetic field within the blood fluid is the sum of these two components:
$$
E=E_{ext}+E_{int}, B=B_{ext}+B_{int}
$$
We have control only over $E_{ext},B_{ext}$ and can use these control fields to manipulate the blood fluid velocity appropriately so that the virus does not enter into the lung region. Let ${\bf r}_k(t)$ denote the position of the $k^{th}$ virus particle. It satisfies the obvious differential equation
$$
d{\bf r}_k(t)/dt={\bf v}(t,{\bf r}_k(t))
$$
with initial conditions ${\bf r}_k(0)$ depending on where on the surface of the body the virus entered. If the virus is modeled as a quantum particle with nucleus at the classical position ${\bf r}_k(t)$ at time $t$, then the protein molecules that are stuck to this virus surface have relative positions $({\bf r}_{kj}, j=1,2,..., M\}$ and their joint quantum mechanical wave function $\psi_k(t,{\bf r}_{kj}, j=1,2,..., M)$ will satisfy the Schrodinger equation
$$
ih\partial_t\psi_k(t,{\bf r}_{kj},j=1,2,..., M)=\sum_{j=1}^M(-h^2/2m_{kj})\nabla_{r_{kj}}^2\psi_k(t,{\bf r}_{kj}, j=1,2,..., M)
$$
$$
+\sum_{j=1}^Me_kV_k(t,{\bf r}_{kj}-{\bf r}_k(t))\psi_k(t,{\bf r}_{kj}, j=1,2,..., M)
$$
where $V_k(t,{\bf r})$ is the binding electrostatic potential energy of the viral nucleus.
There may also be other vector potential and scalar potential terms in the above Schrodinger equation, these potentials coming from the prevalent electromagnetic field.
Some times it is more accurate to model the entire swarm of viruses as a quantum field rather than as individual quantum particles but the resulting dynamics becomes more complicated as it involves elaborate concepts from quantum field theory. Suppose we have solved for the joint wave functions $\psi_k$ of the protein blobs on the $k^{th}$ virus. Then the average position of the $l^{th}$ blob on the $k^{th}$ viral surface is given by
$$
<{\bf r}_{kl}>(t)=\int{\bf r}_{kl}\psi_k(t,{\bf r}_{kj}, j=1,2,..., M)\Pi_{j=1}^Md^3{\bf r}_{kj}, 1\leq k\leq M
$$
We consider the time at which these average positions of the $l^{th}$ protein blob on the $k^{th}$ viral surface hits the boundary $S=\partial V$ of a particular cell. This time is given by
$$
\tau(k,l)=inf\{t\geq 0:<{\bf r}_{kl}>(t)\in S\}
$$
At this time, replication of this protein blob due to cell ribosome action begins. Let $\lambda(n)dt$ denote the probability that in the time interval $[t,t+dt]$, $n$ protein blobs will be generated by the ribosome action on a single blob. Let $N_{kl}(t)$ denote the number of blobs due to the $(k,l)^{th}$ one present at time $t$. Then for
$$
P_{kln}(t)=Pr(N_{kl}(t)=n), n=1,2,...
$$
we have the Chapman-Kolmogorov Markov chain equations
$$
dP_{kln}(t)/dt=\sum_{1\leq r<n}P_{klr}(t)\lambda(kl,r,n)-P_{kln}(t)\sum_{1\leq r<n}P_{kln}(t)
$$
where $\lambda(kl,r,n)dt$ is the probability that in time $[t,t+dt]$, $n$ blobs will be
generated from $r$ blobs. Clearly
$$
\lambda(kl r,n)=r\lambda(n-r)
$$
because each protein blob generates blobs indepndently of the others and hence the probability of two blobs generating more than one each is zero. In another discrete time version of blob replication we can make use of the theory of branching processes.

Let $X(n)$ denote the number of blobs at time $n$. At time $n+1$, each blob splits into 
several similar blobs, each splitting being independent of the others. let $\xi(n,r)$ denote the number of blobs into which the $r^{th}$ blob splits. Then, we have the obvious relation
$$
X(n+1)=\sum_{r=1}^{X(n)}\xi(n,r)
$$
Hence, if
$$
F_n(z)=\Bbb E(z^{X(n)}
$$
is the moment generating function of $X(n)$ and if 
$$
\phi(z)=\Bbb E(z^{\xi(n,r)})
$$
is the moment generating function of the number of blobs into which each blob splits, then we get the recursion
$$
F_{n+1}(z)=\Bbb E(\phi(z)^{X(n)})=F_n(\phi(z))
$$
In order to mimick this dynamics in the continuous time situation, we assume that there are $X(t)$ blobs at time $t$. Each blob independently of the others, waits for a time $T$ 
that has a probability density $f(t)$ and then replicates into $N$ blobs. Usually, we assume $f(t)$ to be the exponential density. The aim is to calculate the moment generating function $F(t,z)=\Bbb E(z^{X(t)})$ of the number of blobs at time $t$. Assuming $f$ to be exponential with parameter $\lambda$, we find that $X(t+dt)$ will be the
\bigskip

The effect of quantum gravity on the motion of the virus globules: Since the size of the 
virus is on the nano/quantum scale, the effects of gravity on it motion will correspondingly be quantum mechanical effects. Consider the ADM action for quantum gravity
obtained as the integral
$$
S[q_{ab},N,N^a]=\int L(q_{ab},q_{ab}',N,N^a)d^4x
$$
where $q_{ab}$ are the spatial components of the metric tensor in the embedded 3-dimensional space at each time $t$ and $N,N^a$ appear in the Lagrangian but not their time derivatives. The $q_{ab}$ are six canonical position fields and the corresponding canonical momentum fields are
$$
P^{ab}=\partial L/\partial q_{ab}'
$$
The ADM Hamiltonian is
$$
H=\int(P^{ab}q_{ab}'-L)d^3x
$$
and these turn out to be fourth degree polynomials in the momenta but highly nonlinear in the position fields. Let $e^i_a$ denote a tetrad basis for the three dimensional embedded manifold. Thus,
$$
q_{ab}=e^i_ae^i_b
$$
The metric relative to this tetrad basis is therefore the identity $\delta_{ij}$ and the group of transformations of this tetrad that preserve the metric is therefore $SO(3)$ which can be regarded as $SU(2)$ using the adjoint map. Note that $SU(2)$ is in fact the covering group of $SO(3)$. The spin connection involves representing the spatial connection in the $SU(2)$-spin Lie algebra representation. The form of this connection can in terms of the spatial connection $\Gamma^a_{bc}$ can be derived from the fact that the corresponding covariant derivative $D_b$ should annihilate the tetrad $e^i_a$. Thus if $\Gamma^i_{jb}$ denotes the spin connection, then
$$
0=D_be^i_a=e^i_{a,b}-\Gamma^c_{ba}e^i_c+\Gamma^i_{jb}e^j_a=0
$$
or equivalently,
$$
\Gamma^i_{jb}=-e^a_j(e^i_{a,b}-\Gamma^c_{ba}e^i_c)=-e^a_je^i_{a:b}
$$
where a double dot denotes the usual spatial covariant derivative. Equivalently, we can lower the index $i$ using the identity $SO(3)$ metric and write
$$
\Gamma_{ijb}=-e^a_je_{ia:b}=e^a_{j:b}e_{ia}=e_{ja:b}e_i^a=-\Gamma_{jib}
$$
Thus, the spin connection $\Gamma_{ija}$ is antisymmetric w.r.t $(ij)$ and hence can be expressed as
$$
\Gamma_{ija}=\Gamma^k_a(\tau_k)_{ij}=\Gamma_{ka}(\tau_k)_{ij}
$$
where $\tau_i, i=1,2,3$ is the standard basis for the Lie algebra of $SO(3)$ which satisfy the usual angular momentum commutation relations:
$$
[\tau_i,\tau_j]=\epsilon(ijk)\tau_k
$$
Note that $(\tau_i)_{jk}=\epsilon(ijk)$ upto a proportionality constant. We observe that the spatial connection coefficients of the embedded 3-D manifold are
$$
\Gamma_{abc}=(1/2)(q_{ab,c}+q_{ac,b}-q_{bc,a})
$$
and the spatial curvature tensor is constructed out of these in the usual way. Our aim is to develop the calculus of general relativity in a form that resembles the calculus of the Yang-Mills non-Abelian gauge theories. This means that we must introduce a non-Abelian $SO(3)$ connection $\Gamma=\Gamma^i\tau_i$ so that the curvature tensor has the representation
$$
R=d\Gamma+\Gamma\wedge\Gamma=[d+\Gamma,d+\Gamma]
$$
where $\Gamma$ is an $SU(2)$-Lie algebra valued one form. Note that
$$
q_{ab,c}=(e^i_ae^i_b)_{,c}=e^i_{a,c}e^i_b+e^i_ae^i_{b,c}
$$
In fact, Cartan's equations of structure precisely formulate the expression for the curvature tensor in the language of non-Abelian gauge theory. We write
$$
\Gamma^k=\Gamma_a^kdx^a
$$
or equivalently,
$$
\Gamma=\Gamma^k\tau_k=\Gamma^k_a\tau_kdx^a
$$
and hence
$$
R=d\Gamma+[\Gamma,\Gamma]=\Gamma^k_{a,b}\tau_kdx^b\wedge dx^a+\Gamma^k_a\Gamma^l_b[\tau_k,\tau_l]dx^a\wedge dx^b
$$
$$
=(\Gamma^m_{a,b}-\Gamma^k_a\Gamma^l_b\epsilon(klm))\tau_m dx^b\wedge dx^a
$$
$$
=R^m_{ba}\tau_mdx^b\wedge dx^a
$$
where
$$
R^m_{ba}=(1/2)(\Gamma^m_{a,b}-\Gamma^m_{b,a})-\Gamma^k_a\Gamma^l_b\epsilon(klm)
$$
is the Riemann curvature tensor in the spin representation.

When the metric fluctuation components $\delta q_{ab},\delta N^a,\delta N$ are quantum fields is a coherent state of the gravitational field, we can write the corresponding fluctuations in the space-time components of the metric tensor along the following lines:
$$
g^{\mu\nu}=q^{\mu\nu}+N^2n^{\mu}n^{\nu}
$$
where
$$
q^{\mu\nu}=q_{ab}X^{\mu}_{,a}X^{\nu}_{,b}
$$
$$
T^{\mu}=X^{\mu}_{,0}=N^{\mu}+N.n^{\mu}=N^aX^{\mu}_{,a}+N.n^{\mu}
$$
where $N^a,N$ satisfy the orthogonal relations
$$
g_{\mu\nu}(T^{\mu}-N^aX^{\mu}_{,a})X^{\nu}_{,b}=0,
$$
$$
g_{\mu\nu}n^{\mu}n^{\nu}=1
$$
We can equivalently write
$$
g^{\mu\nu}=q_{ab}X^{\mu}_{,a}X^{\nu}_{,b}+N^2(X^{\mu}_{,0}-N^aX^{\mu}_{,a}).(X^{\nu}_{,0}-N^bX^{\nu}_{,b})
$$
When the metric $g^{\mu\nu}$ fluctuates, it results correspondingly in the fluctuation of
$q_{ab}$ and $N,N^a$ (these are totally ten in number corresponding to the fact that the original metric tensor $g^{\mu\nu}$ has ten components). The coordinate mapping functions
$X^{\mu}(x)$ are assumed to be fixed and non-fluctuating in this formalism. Thus, the fluctuations in the metric tensor $g^{\mu\nu}$ can be expressed in terms of the corresponding fluctuations in $q_{ab},N^a,N$ as
$$
\delta g^{\mu\nu}=\delta q_{ab}X^{\mu}_{,a}X^{\nu}_{,b}
$$
$$
+2N.\delta N.(X^{\mu}_{,0}-N^aX^{\mu}_{,a}).(X^{\nu}_{,0}-N^bX^{\nu}_{,b})
$$
$$
-N^2\delta N^aX^{\mu}_{,a}(X^{\nu}_{,0}-N^bX^{\nu}_{,b})
$$
$$
-N^2((X^{\mu}_{,0}-N^aX^{\mu}_{,a})X^{\nu}_{,b}\delta N^b
$$
Given the metric quantum fluctuations, the corresponding quantum fluctuations in the equation of motion of the virus particle will be
$$
d^2\delta X^{\mu}/d\tau^2+2\Gamma^{\mu}_{\alpha\beta}(x)\delta X^{\alpha}dX^{\alpha}/d\tau
$$
$$
\delta\Gamma^{\mu}_{\alpha\beta}(X)(dX^{\alpha}/d\tau).(dX^{\beta}/d\tau)
$$
$$
\Gamma^{\mu}_{\alpha\beta,\rho}(X)\delta X^{\rho}(dX^{\alpha}/d\tau).(dX^{\beta}/d\tau)=0
$$
The fluctuations in the particle position are quantum observables evolving in time. According to the Heisenberg picture, these observables evolve dynamically in time according to the above equation with the state of the gravitational field being prescribed. So solving the above equation for $\delta X^{\mu}(\tau)$ in terms of the metric fluctuation field $\delta g_{\mu\nu}(X)$, we can in principle evaluate the mean, correlation and higher order moments of the particle position field in the given state of the gravitational field. Once the mean and correlations of the virus position are known, we can calculate the probability dispersion of this position at the first time that the mean position of this virus first hits a given cell. This dispersion gives us an estimate of what range relative to its mean position will be affected by the virus. We could also give a classical stochastic description of the dynamics of the virus but since the virus size is also of the atomic dimensions, such a quantum probabilistic description is more accurate.
\bigskip

A cruder model for the virus motion would be to express its trajectory as the sum of a classical and a quantum trajectory in accordance with Newton-Heisenberg equation of motion with a force that is the sum of a classical component and a purely quantum fluctuating component:
$$
d^2/dt^2({\bf r}(t))=f_0({\bf r}(t))+\delta f({\bf r}(t))
$$
Solving this differential equation using first order perturbation theory,
$$
{\bf r}(t)={\bf r}_0(t)+\delta{\bf r}(t)
$$
with ${\bf r}_0(t)$ a purely classical trajectory and $\delta{\bf r}(t)$ a quantum, ie, operator valued trajectory, we get
$$
d^2{\bf r}_0(t)/dt^2=f_0({\bf r}_0(t)),
$$
$$
d^2\delta{\bf r}(t)/dt^2=f_0'({\bf r}_0(t))\delta{\bf r}(t)+\delta f({\bf r}_0(t))
$$
Here $\delta f({\bf r})$ is an operator valued field and $f_0({\bf r})$ is a classical c-number field. Expressing the first order perturbed trajectory in state variable form gives us
$$
d\delta{\bf r}(t)/dt=\delta{\bf v}(t), d\delta{\bf v}(t)/dt=f_0'({\bf r}_0(t))\delta{\bf r}(t)+\delta f({\bf r}_0(t))
$$
or equivalently
$$
\frac{d}{dt}\left(\begin{array}{cc}\delta{\bf r}(t)\\\delta{\bf v}(t)\end{array}\right)
$$
$$
=\left(\begin{array}{cc}{\bf 0}&{\bf I}\\{\bf f}_0'({\bf r}_0(t))&{\bf 0}\end{array}\right)\left(\begin{array}{cc}\delta{\bf r}(t)\\\delta{\bf v}(t)\end{array}\right)
$$
$$
+\left(\begin{array}{cc}{\bf 0}\\\delta f({\bf r}_0(t))\end{array}\right)
$$
Now we can also include damping terms in this model by allowing the force to be both position and velocity dependent:
$$
d^2{\bf r}_0(t)+\delta{\bf r}(t))/dt^2=f_0({\bf r}_0(t)+\delta{\bf r}(t),{\bf r}_0'(t)+\delta{\bf r}'(t))+\delta f({\bf r}_0(t)+\delta{\bf r}(t),{\bf r}_0'(t)+\delta{\bf r}'(t))
$$
which gives using first order perturbation theory,
$$
d^2{\bf r}_0(t)/dt^2=f_0({\bf r}_0(t),{\bf r}_0'(t)),
$$
$$
d^2\delta{\bf r}(t)/dt^2=f_{0,1}({\bf r}_0(t),{\bf r}_0'(t))\delta{\bf r}(t)+F_{0,2}({\bf r}_0(t),{\bf r}_0'(t))\delta{\bf r}'(t)+\delta f({\bf r}_0(t),{\bf r}_0'(t))
$$
After introducing the state vector as above, and solving the linearized equations using the usual state transition matrix obtained in terms of the Dyson series expansion, we get
$$
\delta{\bf r}(t)=\int_0^t\Phi(t,s)\delta f({\bf r}_0(s),{\bf r}_0'(s))ds
$$
where $\Phi(t,s)$ is a classical c-number function and $\delta f({\bf r},{\bf v})$ is a quantum operator valued function of phase space variables, ie, a quantum field on $\Bbb R^6$. 

MATLAB simulation studies:

[1] Simulation of branching processes in discrete time.

[2] Simulation of blood flow in an electromagnetic field.

Continuous time branching processes: Let $X(t)$ denote the number of particles at time $t$. Then $X(t+dt)$ will equal $X(t)+\xi(dt)$ where conditioned on $X(t)$, $\xi(dt)$ takes the value $r$ with probability $X(t)\lambda(r)dt$ for $r=1,2,...$ and the value $0$ with probability $1-X(t)\sum_r\lambda(r)dt$. This is because, each particle has the probability 
$\xi(r)dt$ of giving birth to $r$ offspring. This gives us the recursion
$$
\Bbb E(z^{X(t+dt)})=\sum_{r\geq 1}\Bbb E(z^{X(t)}z^rX(t))\lambda(r)+
$$
$$
+\Bbb E((1-\sum_rX(t)\lambda(r)dt)z^{X(t)})
$$
So writing
$$
F(t,z)=\Bbb E(z^{X(t)}
$$
we get
$$
F(t+dt,z)=\phi(z)z\partial F(t,z)/\partial z+F(t,z)-z(\partial F(t,z)/\partial z)\phi(1)
$$
where
$$
\phi(z)=\sum_r\lambda(r)z^r
$$
and hence we derive the following differential equation for the evolution of the generating function of the number of particles at time $t$:
$$
\partial F(t,z)/\partial t=z(\phi(z)-\phi(1))\partial F(t,z)/\partial z
$$

\section{More on quantum gravity}

Let
$$
E^a_i=\sqrt{det(q)}e^a_i, q=((q_{ab}))
$$
Then,
$$
det(E)=det((E^a_i))=det((E^i_a))^{-1}=det(q)^{3/2}det(e)^{-1}=det(q)
$$
Note that,
$$
q_{ab}=e^i_ae^i_b, det(q)=det(e)^2, \sqrt{det(q)}=det(e)=det((e^i_a))
$$
$$
D_b(E^a_i)=0
$$
follows from
$$
D_be^i_a=0
$$
and the expression
$$
E^a_i=det(e)e^a_i=\epsilon(a_1,a_2,a_3))e^1_{a_1}e^2_{a_2}e^3_{a_3}e^a_i
$$
Note also that defining
$$
F^i_a=\epsilon(aa_2a_3)\epsilon(ii_2i_3)e^{a_2}_{i_2}e^{a_3}_{i_3}
$$
we get
$$
e^a_iF^i_a=\epsilon(a_1a_2a_3)\epsilon(i_1i_2i_3)e^{a_1}_{i_1}e^{a_2}_{i_2}e^{a_3}_{i_3}
$$
$$
=6.det((e^a_i))=6.det(e)^{-1}
$$
The factor of $6$ comes because of the identity
$$
\epsilon(i_1i_2i_3)\epsilon(i_1i_2i_3)=6
$$
More generally,
$$
e^b_iF^i_a=\epsilon(aa_2a_3)\epsilon(i_1i_2i_3)e^b_{i_1}e^{a_2}_{i_2}e^{a_3}_{i_3}
$$
$$
=\epsilon(aa_2a_3)\epsilon(ba_2a_3)det(e)^{-1}=\delta^b_a.6.det(e)^{-1}
$$
Thus,
$$
F^i_a=6.det(e)^{-1}e^i_a
$$
Taking the inverse gives
$$
F^a_i=6^{-1}det(e)e^a_i=6^{-1}.E^a_i
$$
or equivalently,
$$
E^a_i=6.F^a_i
$$
We can derive another alternate expression for $F^a_i$ or equivalently for $E^a_i$. Let
$$
G^a_i=\epsilon(aa_2a_3)\epsilon(ii_2i_3)e^{i_2}_{a_2}e^{i_3}_{a_3}
$$
Then,
$$
e^i_bG^a_i=\epsilon(aa_2a_3)\epsilon(ba_2a_3)det(e)=6\delta^a_b.det(e)
$$
Thus
$$
G^a_i=6.e^a_idet(e)=6.E^a_i
$$
In particular,
$$
D_aE^a_i=0
$$
or equivalently,
$$
(\sqrt{q}e^a_i)_{,a}-\Gamma^j_{ia}E^a_j=0
$$
Note that this is the same as
$$
\sqrt{q}(e^a_{i:a}-\Gamma^j_{ia}e^a_j)=0
$$
which is the same as
$$
\sqrt{q}D_ae^i_a=0
$$
ie
$$
D_ae^i_a=0
$$
Substituting for the spin connection
$$
\Gamma^j_{ia}=\Gamma_{jia}=\Gamma^k_a(\tau_k)_{ji}=\Gamma^k_a\epsilon(kji)
$$
we get
$$
E^a_{i,a}-\Gamma^k_a\epsilon(kji)E^a_j=0---(1)
$$
In other words, the covariant derivative of $E^a_i$ w.r.t the spin connection $\Gamma^k_a$ vanishes. This result is of fundamental importance because it states that these position fields $E^a_i$ which are called the Ashtekar position fields, are covariantly flat w.r.t the spin connection. We wish to express the $SO(3)$ spin connection $\Gamma^k_a$ in terms of the Ashtekar canonical position fields $E^a_i$. To this end, we note that (1) can be expressed as
$$
\Gamma^k_a\epsilon(kji)=
$$

$$
\Gamma_{abc}=(1/2)((e^i_ae^i_b)_{,c}+(e^i_ae^i_c)_{,b}-(e^i_ce^i_b)_{,a})
$$

\bigskip

References

[1] Thomas Thiemann, "Modern Canonical Quantum General Relativity", Cambridge University Press.

[2] M.Green, J.Schwarz and E.Witten, "Superstring Theory", Cambridge University Press.

[3] Vijay Upreti, G.Sagar, H.Parthasarathy and V.Agarwal, Technical report, NSUT
on mathematical models for the Corona virus and their MATLAB simulation.

\section{Large deviation properties of virus combat with antibodies}

A virus is a small spherical blob around which the blood fluid moves. Let $n(t,r)$ denote the concentration in terms of number of virus blobs per unit volume and let $g(t,r)$ denote the net rate of virus generation per unit volume. The virus gradient current density is given by
$$
J_v(t,r)=-D\nabla n(t,r)
$$
and hence the conservation law for the virus number is given by
$$
div J_v+\partial_tn=g
$$
or equivalently,
$$
\partial_tn(t,r)-\nabla^2n(t,r)=g(t,r)
$$
Now let us take the drift current also into account caused by the presence of charge on the virus body interacting with the external electric field/ Let $v(t,r)$ denote the velocity of the fluid in which the viurs particles are immersed.
Then, if $E(t,r)$ is the electric field and $\mu$ the virus mobility, we have that the
total drift plus diffusion current density is given by
$$
J=n\mu E-D\nabla n
$$
and its conservation law is given by
$$
div J+\partial_tn=g
$$
or equivalently,
$$
\partial_tn-D\nabla^2n+\mu n div E+\mu E.\nabla n=g
$$
Let $q$ denote the charge carried by a single virus particle. Then the virus charge density is given by
$$
\rho=qn
$$
and by Gauss' law,
$$
div E=-\rho/\epsilon
$$
so our equation of continuity for the virus particles is given by
$$
\partial_tn-D\nabla^2n+(\mu q/\epsilon)n^2=g
$$
This is the pde satisfied by the virus number density. If $g(t,r)$ is the sum of a random Gaussian field and a random Poisson field both of small amplitude, then the problem is to calculate the rate functional of the field $n(t,r)$. Using this rate functional, we can
predict the probability of the rare event that the virus concentration in the fluid will exceed the critical concentration level defined by the solution to the above pde without the generation term by an amount more than a given threshold. Note that the mobility/drift term for the current is obtained as follows: let $v(t)$ denote the velocity of a given virus particle and $m$ its mass, $\gamma$ the viscous damping coefficient. Then, the equation of motion of the virus particle is given by
$$
mv'(t)=-\gamma v(t)+q E(t)
$$
and at equilibrium, $v'=0$ so that
$$
v(t)=qE(t)/\gamma=\mu E(t), \mu=q/\gamma
$$
Now let us consider the motion of the virus particle taking into account random fluctuating forces. Its equation of motion is
$$
mdv(t)=-\gamma v(t)+qE(t)+\sigma.dB(t)
$$
where $B$ is standard three dimensional Brownian motion. The solution to this stochastic differential equation is given by
$$
v(t)=\int exp(-\gamma(t-s)/m)(q/m)E(s)ds+\sigma\int_0^texp(-\gamma(t-s)/m)dB(s)
$$
The corresponding total drift current density is given by
$$
J_d(t,r)=nv=n(t,r)(\int exp(-\gamma(t-s)/m)(q/m)E(s,r)ds+\sigma\int_0^texp(-\gamma(t-s)/m)dB(s,r))
$$
where now $B(t,r)$ is a Gaussian random field, ie, a zero mean Gaussian field having correlations
$$
\Bbb E(B(t,r).B(s,r'))=min(t,s)R_B(r,r')
$$
Note that the pdf $p(t,v)$ of $v(t)$, ie, of a single virus particle satisfies the Fokker-Planck equation
$$
\partial_tp(t,v)=\gamma div(vp(t,v))-qE(t).\nabla_vp(t,v)+(\sigma^2/2)\partial^2p(t,v)
$$
$p(t,v)d^3v$ gives us the probability that the virus particle will have a velocity in the range $d^3v$ around $v$. $n(t,r)p(t,v)d^3rd^3v$ is the number of particles having position in the range $d^3r$ around $r$ and having a velocity in the range of $d^3v$ around $v$.
\bigskip

MATLAB simulations for virus dynamics based on a stochastic differential equation model: 

[a] Simulation of blood velocity flow:
The heart pumping force field per unit volume can be represented as
$$
f(t,r)=f_0(t)K(r)
$$
where $f_0(t)$ is a periodic function and $K(r)$ is a point spread function that is a smeared version of the Dirac $\delta$-function at the point $r_0$, the position of the heart. The equation of motion of a virus particle in this force field is given by
$$
dr(t)=v(t)dt, mdv(t)=-\gamma v(t)dt+f(t,r(t))dt+\sigma dB(t)--(1)
$$
The corresponding pdf is $p(t,r,v)$ in position space. If $n_0$ is the total number of virus particles in the fluid, then $n_0p(t,r,v)d^3rd^3v$ equals the total number of virus particles in the position range $d^3r$ and having velocity in the range $d^3v$ centred around $(r,v)$. It satisfies the Fokker-Planck equation
$$
\partial_tp(t,r,v)=-v.\nabla_rp(t,r,v)-f(t,r).\nabla_vp(t,r,v)++\gamma.div_v(v.p(t,r,v))+(\sigma^2/2)\nabla_v^2p(t,r,v)---(2)
$$
equations (1) and (2) can be discretized respectively in time and in space-time and simulated. For example, simulation of (1) will be based on the discrete time model
$$
r(t+\delta)=r(t)+\delta.v(t),
$$
$$
v(t+\delta)=v(t)-(\gamma\delta/m)v(t)+f(t,r(t))\delta+\sigma\sqrt{\delta}W(t)
$$
where $W(t), t=1,2,...$ are iid standard normal $\Bbb R^3$-valued random vectors. 
\bigskip

Simulation of blood flow in the body: The force exerted by the heart is localized around the position of the heart and as such this force density field can be represented as
$$
f(t,r)=f_0(t)K(r)
$$
where $f_0(t)$ is a periodic function of time and $K(r)$ is a smeared version of the Dirac 
$\delta$-function centred around the heart's position. The fluid-dynamical equations to be simulated are
$$
\rho(\partial_tv(t,r)+(v,\nabla)v(t,r))=-\nabla p+f+\eta\nabla^2v
$$
$$
div(\rho.v)+\partial_t\rho=g
$$
where $\rho(t,r)$ is the virus density and $g(t,r)$ is the rate of generation of virus mass per unit volume. Assuming an equation of state
$$
p=F(\rho)
$$
we get a set of four pde's for the four fields $\rho,v$ by replacing $\nabla p$ with
$F'(\rho)\nabla\rho$. Standard CFD (Computational fluid dynamics) methods can be used to discretize this dynamics and solve it in MATLAB.
\bigskip

The spread of virus from a body within a room: Let $C(t)$ denote the concentration of the virus at a certain point within the room. It increases due to the presence of a living organism carrying the virus and decreases due to the presence of ventilation, medicine etc. other effects. Thus, the differential equation for its dynamics can be expressed in the form
$$
dC(t)/dt=I(t)-\beta(X(t))C(t)
$$
where $I(t)$ denotes the rate of increase due to the external living organism and $X(t)$ denotes the other factors like medicine, ventilation etc. Formally, we can solve this equation to give
$$
C(t)=\Phi(t,0))C(0)+\int_0^t\Phi(t,s)I(s)ds
$$
where
$$
\Phi(t,s)=exp(-\int_s^t\beta(X(\tau))d\tau)
$$
This dynamical equation can be generalized in two directions, one, by including a spatial dependence of the virus concentration and describing the dynamics using a partial differential equation, and two, by including stochastic terms. Let $C(t,{\bf r})$  denote the virus concentration in the room at the spatial location ${\bf r}$ and at time $t$.
The virus diffuses from a region of higher concentration to a region of lower concentration in such a way that if $D$ denotes the diffusion coefficient, the virus diffusion current density, ie, number of virus particles flowing per unit area per unit time at $(t,{\bf r})$ is given by
$$
{\bf J}(t,{\bf r})=-D.\nabla C(t,{\bf r})
$$
Let $X(t,{\bf r})$ denote the external factors at the point ${\bf r}$ at time $t$. Its effect on the decrease in the virus concentration at the point ${\bf r}'$ at time $t$ can be described via an "influence function" $F({\bf r}',{\bf r}')$ which usually decreases with increase in the spatial distance $|{\bf r}'-{\bf r}|$. So we can to a good degree of approximation, describe the concentration dynamics as
$$
\partial_tC(t,{\bf r})=I(t,{\bf r})-div{\bf J}(t,{\bf r})-\int F({\bf r},{\bf r}')\beta(X(t,{\bf r}'))C(t,{\bf r}')d^3r'
$$
$$
=I(t,{\bf r})+D\nabla^2C(t,{\bf r})-\int F({\bf r},{\bf r}')\beta(X(t,{\bf r}')C(t,{\bf r}')d^3r'
$$
In the particular case, if 
$$
F({\bf r},{\bf r}')=\delta^3({\bf r}-{\bf r}')
$$
so that the external agents influence the decrease in virus only at the point where they are located, the above equation simplifies to
$$
\partial_tC(t,{\bf r})=I(t,{\bf r})+D\nabla^2C(t,{\bf r})-\beta(X(t,{\bf r}))C(t,{\bf r})
$$
This equation can be solved easily using Fourier transform methods. More generally, we can consider a situation when the influence of external agents located at different points ${\bf r}_1,...,{\bf r}_n$ takes place via nonlinear interactions so Volterra interaction terms are added to the dynamics to give
$$
\partial_tC(t,{\bf r})=
$$
$$
=I(t,{\bf r})+D\nabla^2C(t,{\bf r})-\sum_{n\geq 1}\int F_n({\bf r},{\bf r}_1,...,{\bf r}_n)\beta(X(t,{\bf r}_1),...,X(t,{\bf r}_n))C(t,{\bf r}_1)...C(t,{\bf r}_n)d^3r_1...d^3r_n
$$
Of course, this equation does not describe memory in the action of external agents. To take into account such memory effects, we further modify it to
$$
\partial_tC(t,{\bf r})=
$$
$$
=I(t,{\bf r})+D\nabla^2C(t,{\bf r})
$$
$$
-\sum_{n\geq 1}\int_{\tau_j\geq 0,j=1,2,...,n}F_n({\bf r},{\bf r}_1,...,{\bf r}_n)\beta(\tau_1,...,\tau_n, X(t-\tau_1,{\bf r}_1),...,X(t-\tau_n,{\bf r}_n))
$$
$$
.C(t-\tau_1,{\bf r}_1)...C(t-\tau_n,{\bf r}_n)d^3r_1...d^3r_nd\tau_1...d\tau_n
$$
We could also consider a model which takes the delay in the effect of external agents to be caused on account of the finite propagation velocity for the influence of virus at the point ${\bf r}'$ to take place at ${\bf r}$. Virus at
${\bf r}'$ has a concentration $C(t,{\bf r}')$ at time $t$ and its effect propagates at a speed of $v$ so that its effect at time $t$ at the point ${\bf r}$ will be described by a term $C(t-|{\bf r}-{\bf r}'|/v,{\bf r}')$. Our model then assumes the form
$$
\partial_tC(t,{\bf r})=I(t,{\bf r})+D\nabla^2C(t,{\bf r})
$$
$$
-\sum_{n\geq 1}\int F_n({\bf r},{\bf r}_1,...,.{\bf r}_n)\beta({\bf r}_1,...,{\bf r}_n,X(t-|{\bf r}-{\bf r}_1|/v,{\bf r}_1),...,X(t-|{\bf r}-{\bf r}_n|/v),{\bf r}_n)
$$
$$
.C(t-|{\bf r}-{\bf r}_1|/v,{\bf r}_1)..C(t-|{\bf r}-{\bf r}_n|/v,{\bf r}_n)d^3r_1..d^3r_n
$$

Remark: Consider the wave equation pde with source and nonlinear terms:
$$
\partial_t^2F(t,r)-c^2\nabla^2F(t,r)+\delta.\int H(\tau,r,F(t-\tau,r'))d\tau d^3r'=s(t,r)
$$
Solving this using first order perturbation theory gives
$$
F(t,r)=F_0(t,r)+\delta.F_1(t,r)+O(\delta^2)
$$
where
$$
\partial_t^2F_0(t,r)-c^2\nabla^2F_0(t,r)=s(t,r),
$$
$$
\partial_t^2F_1(t,r)-c^2\nabla^2F_1(t,r)=-\int H(\tau,r,F_0(t-\tau,r'))d\tau d^3r'
$$
The causal solution to this can be expressed in terms of retarded potentials:
$$
F_0(t,r)=-\int s(t-|{\bf r}-{\bf r}'|/c,{\bf r}')d^3r'/4\pi,
$$
$$
F_1(t,r)=(4\pi)^{-1}\int H(\tau,r',F_0(t-|{\bf r}-{\bf r}'|/c-\tau,{\bf r}''))d\tau d^3r'd^3r''
$$
It is clear from this expression therefore that delay terms coming from the finite speed of propagation of the virus effects can be described using second order partial differential equations in space-time like the wave equation with source terms.
\bigskip

Large deviation problems: Express the above pde as
$$
\partial_tC(t,r)=L(C(t,r))+s(t,r)+\epsilon.w(t,r)
$$
where $L$ is a nonlinear spatio-temporal integro-differential operator and $w$ is the noise field. Note that the operator $L$ depends upon the external medicine/ventilation etc. fields $X(t,r)$. So we can write $L=L(X)$. The problem is to design these external fields $X$ so that the probability of the concentration field exceeding a given threshold at some point in the room is a minimum. This probability can for small $\epsilon$ be described using the LDP rate function.
\bigskip

\section{Large deviations problems in quantum fluid dynamics}

The Navier-Stokes equation for an incompressible fluid can be expressed as
$$
\partial_tv(t,r)=L(v)(t,r)+f(t,r)-\rho^{-1}\nabla p(t,r)
$$
$$
div(v)=0
$$
where $L$ is the nonlinear Navier-Stokes vector operator:
$$
L(v)=-(v,\nabla)v+\eta\rho^{-1}\nabla^2v
$$
We can write
$$
v=curl\psi, div\psi=0, \Omega=curl v=-\nabla^2\psi
$$
and then get
$$
\partial_t\Omega+curl(\Omega\times v)=\eta\rho^{-1}\nabla^2\Omega+g(t,r), g=curl f
$$
Let $G$ denote the Green's function of the Laplacian. Then
$$
\partial_t\psi=-G.curl(\nabla^2\psi\times curl\psi)+\eta\rho^{-1}\nabla^2\psi-G.g
$$
Formally, we can express this as
$$
\partial_t\psi=F(\psi)+h
$$
where now $F$ is a nonlinear functional of the entire spatial vector field $\psi$ and
$h=-G.g=-G.curl f$ is the modified force field. Here $F(\psi)$ is a functional of $\psi$ at different spatial points but at the same time $t$ which is why after spatial discretization, it can be regarded as a nonlinear state variable system. Ideally speaking, we should therefore write
$$
\partial_t\psi(t,r)=F(r,\psi(t,r'),r'\in D)+h(t,r)
$$
where $D$ is the region of three dimensional space over which the fluid flows. Note that
$$
F(\psi)=-G.curl(\nabla^2\psi\times curl\psi)+\eta\rho^{-1}\nabla^2\psi
$$
Quantization: Introduce an auxiliary field $\phi(t,r)$ and define an action functional
$$
S[\psi,\phi]=\int\phi.partial_t\psi d^3rdt-\int\phi.F(\psi)d^3rdt-\int\phi.hd^3rdt
$$
Note that
$$
\int\phi.\partial_t\psi d^3rdt=(1/2)\int(\phi.\partial_t\psi-\psi.\partial_t\phi)d^3rdt
$$
$$
\int\phi.F(\psi)d^3rdt=-\int(G\phi).curl(\nabla^2\psi\times curl\psi)d^3rdt
$$
$$
=\int curl(G\phi).(\nabla^2\psi\times curl\psi)d^3rdt
$$
Quantization: The canonical momentum fields are
$$
P_{\psi}=\delta S/\delta\partial_t\psi=\phi/2,
$$
$$
P_{\phi}=\delta S/\delta\partial_t\phi=-\psi/2
$$
Then the Hamiltonian is given by
$$
H=\int(P_{\psi}.\partial_t\psi+P_{\phi}.\partial_t\phi)d^3r-S
$$
$$
=\int\phi.F(\psi)d^3rdt+\int\phi.hd^3rdt
$$
Surprisingly, no kinetic term, ie terms involving the canonical momenta appear here. The
Hamiltonian equations are then
$$
dP_{\psi}/dt=-\delta H/\delta\psi
$$
$$
d\psi/dt=\delta H/\delta P_{\psi}=0
$$
and likewise for $\phi$. Note that these equations imply that $\psi(t,r)=\psi(r)$ is time independent and hence $P_{\psi}$ varies linearly with time. 
\bigskip

Second quantization of the Schrodinger equation: Let
$$
L=i(\psi^*\partial_t\psi-\psi\partial_t\psi^*)-a(\nabla\psi^*)(\nabla\psi)
$$
be the Lagrangian density. Assume that $\psi$ and $\psi^*$ are independent canonical position fields. Then, the corresponding canonical momentum fields are
$$
P_{\psi}=\partial L/\partial\partial_t\psi=i\psi^*,
$$
$$
P_{\psi^*}=\partial L/\partial\partial_t\psi^*=-i\psi
$$
The Hamiltonian density corresponding to $L$ is then
$$
\mathcal H=P_{\psi}.\partial_t\psi+P_{\psi^*}.\partial_t\psi^*-L=
$$
$$
P_{\psi}\partial_t\psi+P_{\psi^*}\partial_t\psi^*-L=
$$
$$
a\nabla\psi^*.\nabla\psi
$$
and hence the second quantized Hamiltonian density of the free Schrodinger particle is
$$
H=\int\mathcal Hd^3x=a\int\nabla\psi^*.\nabla\psi.d^3x=-a\int\psi^*\nabla^2\psi.d^3x
$$
where of course, to get agreement with experiment,
$$
a=h^2/2m
$$
where $h$ is Planck's constant divided by $2\pi$. In the presence of an external potential, this Hamiltonian becomes
$$
H=\int(-a\psi^*\nabla^2\psi+V(t,r)\psi^*\psi)d^3r
$$
The canonical commutation relations are
$$
[\psi(t,r),P_{\psi}(t,r')]=ih\delta^3(r-r')
$$
or equivalently,
$$
[\psi(t,r),\psi^*(t,r')]=h\delta^3(r-r')
$$
We expand the wave operator field $\psi(t,r)$ in terms of the stationary state energy functions of the first quantized Schrodinger Hamiltonian operator
$$
-a\nabla^2+V
$$
to get
$$
\psi(t,r)=\sum_na(n)u_n(r)exp(-i\omega(n)t), \omega(n)=E(n)/h
$$
so that the above commutation relations give
$$
\sum_{n,m}[a(n),a(m)^*]u_n(r)u_m(r')^*exp(-i(\omega(n)-\omega(m))t)=h\delta^2(r-r')
$$
This implies that
$$
[a(n),a(m)^*]=\delta_{n,m}
$$
since
$$
\sum_nu_n(r)u_n(r')^*=\delta^3(r-r')
$$
Actually, since the electronic field is Fermionic, the above commutation relations should be replaced with anticommutation relations and hence all the above commutators should actually be anticommutators. This picture does not show the existence of the positron which is because the Schrodinger equation is a non-relativistic equation. If we use the Dirac relativistic wave equation, then positrons automatically appear.
\bigskip

Large deviations in quantum fluid dynamics: Let $\psi(t,r)$ be an operator valued function of time with $f$ and hence $g,h$ being classical random fields. Then solving 
$$
\partial_t\psi(t,r)=F(r,\psi(t,r'),r'\in D)+h(t,r)
$$
for $\psi(t,r)$, we obtain $\psi(t,r)$ in terms of its initial values $\psi(0,r)$ and the classical source $h$. Note that $\psi(0,r)$ is an operator valued function of the spatial coordinates. This solution can be obtained perturbatively. Hence, formally, we can write
$$
\psi(t,r)=Q(\psi(0,.),h(s,.),s\leq t)
$$
Assuming $h$ to be a "small" classical random field, we can ask about the statistics of
$\psi(t,r)$ in a given quantum state $\rho$, ie for things like the quantum moments
$$
Tr(\rho.\psi(t_1,r_1)...\psi(t_N,r_N))
$$
This will be a functional of $h(s,.), s\leq max(t_1,..,t_N)$ and hence this moment will be a classical random variable and we can ask about the large deviation properties of this moment after replacing $h$ by $\epsilon h$ with $\epsilon\rightarrow 0$.
\bigskip

\section{Lecture schedule for statistical signal processing}

[1] Revision of the basic concepts in linear algebra and probability theory.

[a] Probability spaces, random variables, probability distributions, characteristic functions, independence, convergence of random variables, emphasis on weak convergence of random variables in a metric space.

[2] Basic Bayesian detection and estimation theory.

[3] Various kinds of optimal estimators for parameters: Maximum likelihood estimator for random and non-random parameters, maximum aposteriori estimator for random parameters, nonlinear minimum mean square estimator as a conditional expectation, best linear minimum mean square estimator for random parameters, linear minimum unbiased minimum variance estimators of parameters in linear models, expanding the nonlinear mmse as a Taylor series. Emphasis on the orthogonality principle for optimal minimum linear and nonlinear mean square estimators of random parameters.

[4] Estimating parameters in linear and nonlinear time series models: AR, MA, ARMA models.

[5] Recursive estimation of parameters in linear models, the RLS algorithm.

[6] Estimation of random signals in noise: The orthogonality principle, the FIR Wiener filter, the optimal noncausal Wiener filter---derivation based on spectral factorization method of Wiener and on the innovations method of Kolmogorov. 

[7] Linear prediction theory. The infinite order Wiener predictor. Kolmogorov's formula for the minimum prediction error energy in terms of the power spectral density of the process. Finite order forward and backward linear prediction of stationary processes. The Levinson-Durbin algorithm and lattice filters. Significance of the reflection coefficients from the viewpoint of stability. Lattice filter realizations. Lattice to ladder filter conversion and vice-versa.

[8] An introduction to Brownian motion and stochastic differential equations. Application to real time nonlinear filtering theory. The Kalman-Bucy filter in continuous and discrete time. The Kushner-Kallianpur nonlinear filter and the extended Kalman filter. The disturbance observer in extended Kalman filtering. Applications of the extended Kalman filter to robotics.

[9] Mean and variance propagation in stochastic differential equations for nonlinear systems.

[10] Stochastic differential equations driven by Poisson fields.

[11] Estimating the frequencies of harmonic signals using time series methods and using high resolution eigensubspace estimators like the Pisarenko harmonic decomposition, MŪSIC and ESPRIT algorithm.

[12] Estimating bispectra and polyspectra of harmonic processes using multidimensional versions of the MUSIC and ESPRIT algorithm.

[13] An introduction to large deviation theory for evaluating the effects of low amplitude noise on the probabilities of deviation of the system from the stability zone.

[a] Large deviations for iid normal random variables.

[b] Cramer's theorem for empirical means of independent random variables.

[c] Sanov's theorem for empirical distributions of independent random variables.

[d] Large deviations problems in binary hypothesis testing. Evaluation of the asymptotic
miss probability rate given that the false alarm probability rate is arbitrarily close to zero using the optimal Neyman-Pearson test.

[e] The Gibbs conditioning principle for noninteracting and interacting particles:Given that the empirical distribution for iid random variables satisfies certain energy constraints corresponding to noninteracting and interacting potentials, evaluate the most probable value assumed by this empirical distribution using Sanov's relative entropy formula for the probability distribution of the empirical distribution. 

[f] Diffusion exit from a domain. Calculate the mean exit time of a diffusion process from a boundary given that the driving white Gaussian noise is weak.
\bigskip

[14] Large deviations for Poisson process driven stochastic differential equations.
Let $N(t),t\geq 0$ be a Poisson process with rate $\lambda$. Consider the scaled process
$N_{\epsilon}(t)=\epsilon.N(t/\epsilon)$. Then $N_{\epsilon}$ is a Poisson process whose jumps are in units of $\epsilon$ and whose rate is $\lambda/\epsilon$. Consider the stochastic differential equation
$$
dX(t)=F(X(t))dt+dN_{\epsilon}(t), t\geq 0
$$
The rate function of the process $N_{\epsilon},\epsilon\rightarrow 0$ is evaluated as follows:
$$
\Lambda(f)=\epsilon.log\Bbb E[exp(\epsilon^{-1}\int_0^Tf(t)dN_{\epsilon}(t))]
$$
$$
=\lambda.\int_0^T(exp(f(t))-1)dt
$$
so the corresponding rate function is
$$
I_N(X)=sup_f(\int_0^Tf(t)X'(t)-\lambda.\int_0^T(exp(f(t))-1)dt
$$
and this can be expressed in the form
$$
I_N(X)=\int_0^TJ(X'(t))dt
$$
where
$$
J(u)=sup_v(uv-\lambda.(exp(v)-1))
$$
and then the rate function of the solution $X(t)$ to the sde over the time interval $[0,T]$ is given by
$$
I_X(X)=\int_0^TJ(X'(t)-F(X(t)))dt
$$
More generally, consider a spatial Poisson field $N(t,d\xi)$ with a rate measure of 
$\lambda.dF(\xi)$, ie
$$
\Bbb E(N(t,E))=\lambda.tF(E)
$$
where $E$ takes values in the Borel sigma field of $\Bbb R^n$. $X(t)$ is assumed to satisfy the sde
$$
dX(t)=F(X(t))dt+\int g(X(t),u(t),\xi)N(dt,d\xi)
$$
where $u(t)$ is an input signal field. Now scale the Poisson in both time and amplitude to obtain the following sde satisfied by $X(t)$:
$$
dX_{\epsilon}(t)=F(X_{\epsilon}(t))dt+\epsilon\int g(X_{\epsilon}(t),u(t),\xi).N(dt/\epsilon, d\xi), t\geq 0
$$
We require to calculate the rate function of the process $X_{\epsilon}$ over the time interval $[0,T]$. To this end, we consider first a more general situation: Let $N(d\xi)$ be a spatial Poisson field on $\Bbb R^n$ which we write formally as $N(d\x_1,...d\xi_n)$.
We assume a constant rate field measure, ie,
$$
\Bbb E(N(d\xi_1,...,d\xi_n))=\lambda.d\xi_1...d\xi_n
$$
Consider a different scaling along the different axes:
$$
N_{\delta_1...\delta_n}(d\xi_1,...,d\xi_n)=\delta_1...\delta_nN(d\xi_1/\delta_1,...,d\xi_n/\delta_n)
$$
Then,
$$
M_{\delta_1...\delta_n}(f)=\Bbb Eexp(\int f(\xi_1,...,\xi_n)N_{\delta_1...\delta_n}(d\xi_1,...,d\xi_n)=
$$
$$
=exp(\lambda.\int(exp(\delta_1...\delta_n.f(\delta_1\xi_1,...\delta_n\xi_n))-1)d\xi_1...d\xi_n)
$$
$$
=exp(\lambda.(\delta_1...\delta_n)^{-1}\int(exp(\delta_1...\delta_n.f(\xi_1,...\xi_n))-1)d\xi_1...d\xi_n)
$$
from which we get for the scaled logarithmic moment generating function
$$
\delta_1...\delta_n.log(M_{\delta_1...\delta_n}((\delta_1...\delta_n)^{-1}f))
$$
$$
\lambda\int(exp(f(\xi_1,...\xi_n))-1)d\xi_1...d\xi_n
$$
so the rate function of this scaled Poisson field is given by
$$
I(\nu)=sup_f(\int f(\xi_1,...,\xi_n)d\nu(\xi_1,...,d\xi_n)-\lambda\int(exp(f(\xi_1,...,\xi_n))-1)d\xi_1...d\xi_n)
$$

\section{Basic concepts in group theory and group representation theory}

[1] Show that $SL(2,\Bbb C)$ and $SU(2,\Bbb C)$ are simply connected.

Hint: $SU(2,\Bbb C)$ is topologically isomorphic to $S^3$ via the representation
$$
g=\left(\begin{array}{cc}\alpha&\beta\\-\bar\beta&\bar\alpha\end{array}\right)
$$
where $\alpha,\beta\in\Bbb C$ are arbitrary subject to the only constraint $|\alpha|^2+|\beta|^2=1$. $S^3$ is simply connected, ie, any closed loop in $S^3$ can be deformed continuously to any single point. Note that $S^2$ is also simply connected but $S^2$ without the north pole is not simply connected because a loop around the north pole cannot be deformed into a point within this loop continuously without allowing the loop to intersect the north pole at some stage.

To see that $G=SL(2,\Bbb C)$ is simply connected, simply use the Iwasawa decomposition
$G=KAN$ with $K$ Maximally compact $=SU(2,\Bbb C)$ and $A$ Abelian diagonal, $N$ unipotent upper triangular and hence topologically, $G$ is isomorphic to $SU(2,\Bbb C)x\Bbb R^4$ whose fundamental group is the same as that of $SU(2,\Bbb C)$ which is zero since $SU(2,\Bbb C)$ is simply connected. 

[2] Show that the centre of the covering group of $G=SL(2,\Bbb R)$ is isomorphic to $\Bbb Z$ the additive group of integers. For again by the Iwasawa decomposition, $G=KAN$ where
now $K=SO(2)$ and hence, $G$ is topologically isomorphic to $SO(2)x\Bbb R^2$ whose fundamental group is that of $SO(2)$ which is $\Bbb Z$.

(Problem taken from V.S.Varadarajan, "Lie groups, Lie algebras and their representations", Springer).

Problem: Let $G$ be a real analytic group and let $\pi$ be a finite dimensional irreducible representation of $G$ in a vector space $V$. Let $B=\{v_1,..., v_n\}$ be a basis for $V$ and let $\pi$ denote the representation $\pi$ in the basis $B$, ie, we use the same notation for the representation and its matrix representation in the basis $B$. Let $F$ denote the vector space of all functions on $G$ spanned by $\pi_{ij}(g), 1\leq i,j\leq n$. Show that $dim F=n^2$, ie, the $\pi_{ij}$ are all $n^2$ linear independent functions on $G$. When $G$ is compact, this result follows from the Schur orthogonality relations. How does one do this problem for any real analytic group ? 

Problem: Deduce from the above result that the set of all finite linear combinations
$\sum_{g\in G}f(g)\pi(g)$ with $f$ a complex function on $G$ vanishing at all but a finite number of points equals the complete matrix algebra $M_n(\Bbb C)$. 

Let $A$ be the algebra of all finite linear combinations of the $\pi(g),g\in G$. This is the group algebra corresponding to the representation $\pi$ of $G$. Note that we are assuming that $\pi$ is an irreducible representation of $G$. Note that $A$ is a subalgebra of the full matrix algebra $M_n(\Bbb C)$. We wish to prove that $A=M_n(\Bbb C)$. Suppose
$T$ commutes with $A$, ie
$$
T\in M_n(\Bbb C), T\pi(g)=\pi(g)T\forall g\in G
$$
Then $\mathcal R(T)$ and $\mathcal N(T)$ are $\pi(G)-invariant$, ie, $A$ invariant subspaces. By irreducibility of $\pi$, it then follows that $\mathcal R(T)=V$ and $N(T)=0$, ie, $T$ is non-singular. Now let $c$ be an eigenvalue of $T$. Then $N(T-c)$ is a
$\pi$-invariant subspace and hence by irreducibility of $\pi$, 
$N(T-c)=\Bbb C^n$. Hence, $T=cI$. This shows that the centre of $A$ consists precisely of all scalar multiples of the identity and hence $A$ is the full matrix algebra $M_n(\Bbb C)$. Now we are in a position to prove that the functions $\pi_{ij}(g), 1\leq i,j\leq n$ form a linearly independent set. Suppose not. Then $\sum_{i,j}c(i,j)\pi_{ij}(g)=0\forall g\in G$ where some $c(i,j)$ is non-zero. This equation can be expressed as
$$
Tr(C\pi(g))=0\forall g\in G
$$
and hence since the set of linear combinations of $\pi(g), g\in G$ equals $M_n(C)$, it follows that
$$
Tr(CX)=0\forall X\in M_n(\Bbb C)
$$
from which we deduce that $C=0$. The claim is completely proved.

Note: This problem was taken from V.S.Varadarajan, "Lie groups, Lie algebras and their representations", Springer.
\bigskip

\section{Some remarks on quantum gravity using the ADM action}

Express
$$
K_{ab}=X^{\mu}_{,a}X^{\nu}_{,b}\nabla_{\mu}n_{\nu}
$$
in terms of $q_{ab},q_{ab,c},q_{ab,0}$. 
$$
\nabla_{\mu}n_{\nu}=n_{\nu,\mu}-\Gamma^{\sigma}_{\mu\nu}n_{\sigma}
$$
$$
X^{\mu}_{,a}X^{\nu}_{,b}n_{\nu,\mu}=X^{\nu}_{,b}n_{\nu,a}
$$
$$
=-n_{\nu}X^{\nu}_{,ab}
$$
since
$$
n_{\nu}X^{\nu}_{,b}=0
$$
$$
X^{\mu}_{,a}X^{\nu}_{,b}\Gamma^{\sigma}_{\mu\nu}n_{\sigma}
$$
$$
=X^{\mu}_{,a}X^{\nu}_{,b}\Gamma_{\sigma\mu\nu}n^{\sigma}
$$
$$
=(1/2)X^{\mu}_{,a}X^{\nu}_{,b}(q_{\sigma\mu,\nu}+q_{\sigma\nu,\mu}-q_{\mu\nu,\sigma})n^{\sigma}
$$
$$
X^{\mu}_{,a}X^{\nu}_{,b}q_{\sigma\mu,\nu}=
$$
$$
X^{\mu}_{,a}q_{\sigma\mu,b}=q_{\sigma a,b}-X^{\mu}_{,ab}q_{\sigma\mu}
$$
Thus,
$$
X^{\mu}_{,a}X^{\nu}_{,b}q_{\sigma\mu,\nu}n^{\sigma}=q_{\sigma a,b}n^{\sigma}
$$
$$
=q_{\sigma a,b}(X^{\sigma}_{,0}-N^cX^{\sigma}_{,c})
$$
$$
=q_{\sigma a,b}X^{\sigma}_{,0}-N^c(q_{ca,b}-q_{\sigma a}X^{\sigma}_{,bc})
$$
Likewise,
$$
X^{\mu}_{,a}X^{\nu}_{,b}q_{\sigma\nu,\mu}n^{\sigma}=q_{\sigma b,a}n^{\sigma}
$$
$$
X^{\mu}_{,a}X^{\nu}_{,b}q_{\mu\nu,\sigma}n^{\sigma}
$$
$$
=X^{\mu}_{,a}X^{\nu}_{,b}q_{\mu\nu,\sigma}(X^{\sigma}_{,0}-N^cX^{\sigma}_{,c})
$$
$$
=X^{\mu}_{,a}X^{\nu}_{,b}q_{\mu\nu,0}-N^cX^{\mu}_{,a}X^{\nu}_{,b}q_{\mu\nu,c}
$$
$$
=q_{ab,0}-q_{\mu\nu}(X^{\mu}_{,a}X^{\nu}_{,b})_{,0}-N^c(q_{ab,c}-q_{\mu\nu}(X^{\mu}_{,a}X^{\nu}_{,b})_{,c})
$$
$$
=q_{ab,0}-q_{a\nu}X^{\nu}_{,b0}-q_{\mu b}X^{\mu}_{,a0}-N^cq_{ab,c}+N^cq_{a\nu}X^{\nu}_{,bc}+N^cq_{\mu b}X^{\mu}_{,ac}
$$
Now
$$
q_{a\nu}X^{\nu}_{,b0}=q_{ab,0}-q_{a\nu,0}X^{\nu}_{,b}
$$
$$
q_{\mu b}X^{\mu}_{,a0}=q_{ab,0}-q_{\mu b,0}X^{\mu}_{,a}
$$
Alternately,
$$
q_{a\nu}X^{\nu}_{,b0}=q_{a0,b}-q_{a\nu,b}X^{\nu}_{,0}
$$
$$
q_{\mu b}X^{\mu}_{,a0}=q_{b0,a}-q_{\mu b,0}X^{\mu}_{,0}
$$
Now,
$$
g_{\mu\nu}(X^{\mu}_{,0}-N^aX^{\mu}_{,a})X^{\nu}_{,b}=0
$$
gives
$$
q_{0b}=N^aq_{ab}
$$
So we can substitute this expression for $q_{a0}$ in the above equations.
\bigskip

Hamilton-Jacobi equation and large deviations: 

\section{Diffusion exit problem}

Consider the diffusion exit problem in which the process $X(t)\in\Bbb R^n$ satisfies the sde
$$
dX(t)=f(X(t))dt+\sqrt{\epsilon}g(X(t))dB(t)
$$
The process starts at $x$ within a connected open set $G$ with boundary $\partial G$ and let $\tau(\epsilon)$ denote the first time at which the process hits the boundary. Then, we know (Reference:A.Dembo and O.Zeitouni, "Large deviations, Techniques and Applications") that
$$
lim_{\epsilon\rightarrow 0}\epsilon.log(\Bbb E(\tau(\epsilon))=V(x)
$$
where
$$
V(x)=inf(V(t,x,y):y\in\partial G, t\geq 0)
$$
where
$$
V(t,x,y)=inf\{\int_0^t|u(s)|^2/2: y=x+\int_0^tf(X(s))ds+\int_0^tg(X(s))u(s)ds\}
$$
Suppose $g(X(s))$ is an invertible square matrix. Then,
$$
V(T,x,y)=inf\{\int_0^T(X'(t)-f(X(t))^Tg(X(t))^Tg(X(t))(X'(t)-f(X(t))dt: X(0)=x,X(T)=y\}
$$
The problems of calculating $V(T,x,y)$ therefore amounts more generally to computing the extremum of the functional
$$
S[T,X]=\int_0^TL(X(t),X'(t))dt
$$
subject to the boundary conditions $X(0)=x,X(T)=y$. Denoting this infimum by $S(T,x,y)$, we know from basic classical mechanics that $S$ satisfies the Hamilton-Jacobi equation
$$
\partial_tS(t,x,y)+H(x,\partial_yS(t,x,y))=0
$$
where $H(x,p)$ the Hamiltonian is defined as the Legendre transform of the Lagrangian
$$
H(x,p)=inf_y(py-L(x,y))
$$
or equivalently,
$$
H(x,p)=py-L(x,y), p=\partial_yL(x,y)
$$
It is now clear that in the context of our diffusion exit problem,
$$
L(x,y)=(y-f(x))^Tg(x)^Tg(x)(y-f(x))
$$
and
$$
V(t,x,y)=S(t,x,y)
$$
Then
$$
V(x,y)=inf_{t\geq 0}V(t,x,y)=inf_{t\geq 0}S(t,x,y)=S(x,y)
$$
where $S(x,y)$ satisfies the stationary Hamilton-Jacobi equation
$$
H(x,\partial_y(S(x,y))=0
$$
We note that it is hard to compute the exact value of $\Bbb E(\tau(\epsilon))$ for $\epsilon>0$ because this involves computing the Green's function of the diffusion generator. However, computing the limiting value of $\epsilon.log\Bbb E(\tau(\epsilon))$
as $\epsilon\rightarrow 0$ is simpler as it involves only computing the stationary solution of the Hamilton-Jacobi equation. 
\bigskip

\section{Large deviations for the Boltzmann kinetic transport equation}

$$
f_{,t}(t,r,p)+(p,\nabla_r)f(t,r,p)+(F(t,r,p),\nabla_p)f(t,r,p)=
$$
$$
\int(w(p+q,q)f(t,r,p+q)-w(p,q)f(t,r,p))d^3q
$$
The rhs is interpreted as follows: $w(p+q,p)d^3q$ is the probability of a particle having momentum centred around $p+q$ in the momentum volume $d^3q$ to get scattered to the momentum $p$ after losing a momentum $q$ and likewise $w(p,q)d^3q$ is the probability of a particle having momentum $p$ to get scattered to a momentum centred around $q$ within a momentum volume of $d^3q$.
Making appropriate approximations gives us
$$
w(p+q,q)f(t,r,p+q)\approx w(p,q)f(t,r,p)+(q,\nabla_p)(w(p,q)f(t,r,p))
$$
$$
+(1/2)Tr(qq^T\nabla_p\nabla_p^T)w(p,q)f(t,r,p)
$$
It follows that
$$
\int(w(p+q,q)f(t,r,p+q)-w(p,q)f(t,r,p))d^3q
$$
$$
\approx(q,\nabla_p)(w(p,q)f(t,r,p))+(1/2)Tr(qq^T\nabla_p\nabla_p^T)(w(p,q)f(t,r,p))
$$
Now letting
$$
\int q.w(p,q)d^3q=A(p), \int qq^Tw(p,q)^3q=B(p)
$$
we can write down the approximated kinetic transport equation as
$$
\partial_tf(t,r,p)+(p,\nabla_r)f(t,r,p)+(F(t,r,p),\nabla_p)f(t,r,p)=
$$
$$
div_p(A(p)f(t,r,p))+(1/2)Tr(\nabla_p\nabla_p^TB(p)f(t,r,p))
$$
which is the Fokker-Planck equation that one encounters in diffusion processes described in the form of stochastic differential equations driven by Brownian motion.

The large deviation problem: If the scattering probability function $w(p,q)$ has small random fluctuations, the we can ask what would be the deviation in the resulting solution to the particle distribution function $f$ ? Of course, to answer this question, we must expand the solution $f$ in powers of a perturbation parameter tag $\delta$ that is attached the random vector and matrix valued functions $A(p),B(p)$. Specifically, assuming $w(p,q)$ to be of order $\delta$, we have that $A(p)$ and $B(p)$ are random functions respectively of orders $\delta$ and $\delta^2$. 
\bigskip

\section{Orbital integrals for $SL(2,\Bbb R)$ with applications to statistical image processing}
$$
G=SL(2,\Bbb R)
$$
Iwasawa decomposition of $G$ is
$$
G=KAN
$$
where
$$
K=\{u(\theta):0\leq\theta<2\pi\}, u(\theta)=\left(\begin{array}{cc}cos(\theta)&-sin(\theta)\\sin(\theta)&cos(\theta)\end{array}\right)
$$
$$
A=\{diag(a,a^{-1}):a\in\Bbb R-\{0\}\}
$$
$$
N=\{\left(\begin{array}{cc}1&n\\0&1\end{array}\right):n\in\Bbb R\}
$$
Define
$$
H=diag(1,-1), X=\left(\begin{array}{cc}0&1\\0&0\end{array}\right)
$$
$$
Y=\left(\begin{array}{cc}0&0\\1&0\end{array}\right)
$$
Then,
$$
[H,X]=2X, [H,Y]=-2Y, [X,Y]=H
$$
Consider the following element in the universal enveloping algebra of $G$:
$$
\Omega=4XY+H^2-2H
$$
Then,
$$
[H^2,X]=2(XH+HX)=2(2HX-2X)=4HX-4X, [XY,X]=-XH=-(HX-2X)=-HX+2X
$$
Thus,
$$
[\Omega,X]=0
$$
$$
[H^2,Y]=-2(HY+YH)=-2(2HY+2Y)=-4HY-4Y, [XY,Y]=HY
$$
Thus,
$$
[\Omega,Y]=0
$$
and hence
$$
[\Omega,H]=[\Omega,[X,Y]]=0
$$
Thus, $\Omega$ belongs to the centre of the universal enveloping algebra.

Now consider the orbital integral
$$
F_f(h)=\Delta(h)\int_{G/A}f(xhx^{-1})dx, h\in A
$$
Noting that
$$
G=KNA
$$
we can write
$$
G/A=KN
$$
and hence
$$
\int_{G/A}f(xhx^{-1})dx=\int_{K\times N}f(unhn^{-1}u^{-1})dudn
$$
$$
=\int_N\bar f(nhn^{-1})dn
$$
where
$$
\bar f(x)=\int_Kf(uxu^{-1})du
$$
Now let
$$
nhn^{-1}-h=M, n\in N,h\in A
$$
It is clear that $M$ is strictly upper triangular since $n$ is unipotent, ie, upper-triangular with ones on the diagonal. Then, for fixed $h$,
$$
dM=d(nhn^{-1})=|a-1/a|dn=\Delta(h)dn
$$
(Simply write
$$
n=\left(\begin{array}{cc}1&m\\0&1\end{array}\right), m\in\Bbb R
$$
and note that $dn=dm$ and
$$
n^{-1}=\left(\begin{array}{cc}1&-m\\0&1\end{array}\right)
$$
$$
h=diag(a,1/a)
$$
and observe that
$$
nhn^{-1}-h=\left(\begin{array}{cc}0&(1/a-a)m\\0&0\end{array}\right)
$$
)
Note that for $a=exp(t)$ and $h=diag[a,1/a]$, we have
$$
\Delta(h)=|e^t-e^{-t}|
$$
This gives
$$
\int_Nf(nhn^{-1})dn=\int_{\mathfrak n}f(M+h)(dn/dM)dM=\Delta(h)^{-1}\int_{\mathfrak n}f(M+h)dM
$$
where $\mathfrak n$ is the Lie algebra of $N$ and consists of all strictly upper triangular real matrices. Finally, using Weyl's integration formula,
$$
\int_Gf(x)dx=\int_Adh|\Delta(h)|^2\int_{G/A}f(xhx^{-1})dx=\int_{A\times\mathfrak n}|\Delta(h)|f(M+h)dhdM
$$
$$
=\int_{\Bbb R\times\Bbb R}|exp(t)-exp(-t)|f(a(t)+sX)dtds, a(t)=diag(exp(t),exp(-t))=exp(tH)
$$

Reference: V.S.Varadarajan, "Harmonic Analysis on Semisimple Lie Groups", Cambridge University Press.
\bigskip

\section{Loop quantum gravity (LQG)}

Here, first we introduce the $SO(3)$ connection coefficients
that realize the covariant derivatives of mixed spinors and tensors in such a way that the mixed covariant derivative of the tetrad becomes zero. In fact this equation defines the spinor connection coefficients in terms of the spatial Christoffel connection coefficients. We then introduce the quantities $K^j_a$ which are obtained by considering the spatial components $K_{ab}$ of the covariant derivative of the unit normal four vector to the three dimensional hypersurface at each time $t$ that has been embedded into $\Bbb R^4$. The Ashtekar momentum fields $E^a_j$ are obtained by scaling the tetrad $e^a_j$ with the square root of the determinant of the spatial metric $q_{ab}$ where $q_{ab}=e^j_ae^j_b$. The tetrad basis matrix $e^a_j$ or rather its inverse matrix $e^j_a$ is used to contract the $K_{ab}'s$ to obtain the $K^j_a$. The fact that that $K_{ab}'s$ are symmetric is manifested in the equation that $K^j_ae^j_b$ or equivalently $K^j_aE^j_b$ is antisymmetric in $(a,b)$ and this constraint is called the Gauss constraint. By Combining this with the fact that the spinor covariant derivative of $E^j_a$ or $E^a_j$ vanishes, we are able to obtain fundamental relation which states that the covariant derivative of $E^j_a$ or $E^a_j$ with respect to another connection called the Immirzi-Sen-Barbero-Ashtekar connection obtained by adding the the spinor connection $\Gamma^j_a$ with any scalar multiple of $K^j_a$ vanishes. This new connection is denoted by $A^j_a$. Further,
in accordance with the ADM action, the canonical position fields are the spatial metric coefficients $q_{ab}$ and the corresponding canonical momentum fields $P^{ab}$ are obtained in terms of the $K_{ab}$, namely the spatial components of the covariant derivative of the unit normal to the embedded three dimensional hypersurface.
Ashtekar proved that if we replace these canonical position fields by the $A^j_a$ and the canonical momentum fields by the $E^a_j$ then because the spin connection $\Gamma^j_a$ are homogeneous functions of the $E^a_j$, it follows that these new variables also satisfy the canonical commutation relations and hence can serve as canonical position and momentum fields. Moreover, he showed that the Hamiltonian constraint operator when expressed in term of these new canonical variables is just polynomial functional in these fields and hence all the renormalizability problems are overcome. Further the Gauss constraint in these new variables is simplified to the form that it states that the covariant derivative of the $E^a_j$ vanishes. This formalism means that the new connection $A^j_a$ behaves like a Yang-Mills connection for general relativity, the curvature of this connection being immediately related after appropriate contraction with the $E^a_j$ to the Einstein-Hilbert action. This means that one can discretize space into a graphical grid and replace the curvature terms in the Einstein-Hilbert action by parallel displacements of the connection $A^j_a$ around loops in the graph and the connection terms $A^j_a$ integrated over edges by $SO(3)$ holonomy group elements. In other words, the whole programme of quantizing the Einstein-Hilbert action using such discretized graphs becomes simply that of quantizing a function of $SO(3)$ holonomy group elements, one element associated with each edge of a graph and a holonomy group element corresponding to the curvature obtained by parallely transporting the connection around a loop.
\bigskip

\section{Group representations in relativistic image processing}

For $G=SL(2,\Bbb R)$, we've seen that
$$
\int_Gf(x)dx=\int\Delta(t)f(a(t)+sX)dtds, \Delta(t)=|exp(t)-exp(-t)|
$$
Now suppose that $f,g$ are two functions on $G$. Let $\chi$ be the character of a representation of $G$. Consider
$$
I(f,g)=\int_{G\times G}f(x)g(y)\chi(y^{-1}x)dxdy
$$
Let $u\in G$ and observe that
$$
I(fou^{-1},gou^{-1})=\int_{G\times G}f(u^{-1}x)g(u^{-1}y)\chi(y^{-1}x)dxdy
$$
$$
=\int_{G\times G}f(x)g(y)\chi(y^{-1}u^{-1}ux)dxdy=\int_{G\times G}f(x)g(y)\chi(y^{-1}x)dxdy
$$
$$
=I(f,g)
$$
Thus, $I(f,g)$ is an invariant function of two image fields defined on $G$. $I$ can be evaluated using the above integration formula. Note that this formula is true for any function $\chi$ on $G$, not necessarily a character. Now consider
$$
J(f,g)=\int_{G\times G}f(x)g(y)\chi(yx^{-1})dxdy
$$
We shall show that $J$ is invariant whenever $\chi$ is a class function an in particular, when it is a character. We have
$$
J(fou^{-1}, gou^{-1})=\int f(u^{-1}x).g(u^{-1}y)\chi(yx^{-1})dxdy
$$
$$
=\int f(u^{-1}x).g(u^{-1}y)\chi(yx^{-1})dxdy
$$
$$
=\int f(x)g(y)\chi(uyx^{-1}u^{-1})dxdy=J(f,g)
$$
because $\chi$ is a class function ie,
$$
\chi(uxu^{-1})=\chi(x)\forall u,x\in G
$$
\bigskip

Some other formulae regarding orbital integrals, Fourier transforms and Weyl's character formula. Let $G$ be a compact Lie group and consider Weyl's character formula for an irreducible representation of $G$:
$$
\chi=\frac{\sum_{s\in W}\epsilon(s)exp(s.(\lambda+\rho))}{\sum_{s\in W}\epsilon(s)exp(s.\rho)}
$$
where $\rho=(1/2)\sum_{\alpha\in\Delta_+}\alpha$ and $\lambda$ is a dominant integral weight. Note that the denominator can expressed as
$$
\Delta_1(exp(H))=\sum_{s\in W}\epsilon(s).exp(s.\rho(H))
$$
$$
=\Pi_{\alpha\in\Delta_+}(exp(\alpha(H)/2)-exp(-\alpha(H)/2))
$$
$$
=exp(\rho(H))\Pi_{\alpha\in\Delta_+}(1-exp(-\alpha(H)))=exp(\rho(H))\bar\Delta(exp(H))
$$
where
$$
\Delta(H)=\Pi_{\alpha\in\Delta_+}(exp(\alpha(H))-1)
$$
so that
$$
\bar\Delta(H)=\Pi_{\alpha\in\Delta_+}(exp(-\alpha(H))-1)=c.\Pi_{\alpha\in\Delta_+}(1-exp(-\alpha(H)))
$$
with $c$ a constant that equals $\pm 1$. Note also that $H$ varies over the Cartan-subalgebra $\mathfrak a$ and since $G$ is a compact group, it follows that $\alpha(H)$ is pure imaginary for all $\alpha\in\Delta$ and $H\in\mathfrak a$. Now, Weyl's integration formula gives for any function $f$ on $G$,
$$
\chi(f)=\int f(g)\chi(g)dg=
$$
$$
C.\int_{A\times G/A}|\Delta(h)|^2f(xhx^{-1})\chi(h)dhdx
$$
$$
=C.\int_A\bar\Delta(h)F_f(h)\chi(h)dh
$$
where $F_f$, the orbital integral of $f$, is defined by
$$
F_f(h)=\Delta(h)\int_{G/A}f(xhx^{-1})dx
$$
Substituting for $\chi(h)$ from Weyl's character formula, we get
$$
\chi(f)=C.\int_{\mathfrak a}F_f(exp(H)).\sum_{s\in W}\epsilon(s)exp(s.\lambda(H))dH
$$
$$
=C|W|\int_{\mathfrak a}F_f(exp(H))exp(\lambda(H))dH=C|W|\hat F_f(\lambda)
$$
where $\hat g$ denotes Fourier transform of a function $g$ defined on $\mathfrak a$.
This formula shows that the ordinary Euclidean Fourier transform of the orbital integral
on the Cartan subalgebra evaluated at a dominant integral weight equals the character of the corresponding irreducible representation when the group is compact.
\bigskip

Weyl's dimension formula:
$$
u(H)=\chi(H)\Delta(H)=\sum_{s\in W}\epsilon(s)\chi^s_{m_1...m_n}(H)
$$
where $\chi$ on the right is given by
$$
\chi_{m_1...m_n}(H)=exp(im_1\theta_1+...+im_n\theta_n)
$$
with $(\theta_1,...,\theta_n)$ denoting the standard toral coordinates of $H\in\mathfrak a$. Define the differential operator
$$
D=\Pi_{r>s}i^{-1}(\partial_{\theta_r}-\partial_{\theta_s})
$$
Then,
$$
(Du)(0)=\Pi_{r>s}(m_r-m_s)\sum_{s\in W}\chi^s(0)=|W|\Pi_{r>s}(m_r-m_s)
$$
on the one hand and on the other, using the fact that
$$
\Delta(H)=\sum_{s\in W}\epsilon(s).\chi^s_{n,n-1,...,1}(H)
$$
where
$$
\chi_{n-1,..., 0}(H)=exp(in\theta_1+...+i2\theta_2+i\theta_1)
$$
we get
$$
D\Delta(0)=|W|\Pi_{n\geq r>s\geq 1}(r-s)
$$
and hence
$$
(Du)(0)=\chi(0)(D\Delta(0))=d(\chi).|W|.\Pi_{r>s}(r-s)
$$
where
$$
d(\chi)=\chi(0)
$$
is the dimension of the irreducible representation. Equating these two expressions, we obtain Weyl's dimension formula
$$
d(\chi)=d(m_1,..., m_n)=\frac{\Pi_{r>s}(m_r-m_s)}{\Pi_{r>s}(r-s)}
$$
This argument works for the compact group $SU(n)$ where a dominant integral weight is parametrized by an $n$ tuple of non-negative integers $(m_1\geq m_2\geq ...\geq m_1\geq 0)$. In the general case of a compact semisimple Lie group $G$, we use Weyl's character in the form
$$
\chi(H)=\frac{\sum_{s\in W}\epsilon(s)exp(s.(\lambda+\rho)(H))}{\Delta(H)}
$$
$$
\Delta(H)=\sum_{s\in W}\epsilon(s).exp(s.\rho(H))=\Pi_{\alpha\in\Delta_+}(exp(\alpha(H)/2)-exp(-\alpha(H)/2))
$$
Then define
$$
D=\Pi_{\alpha\in\Delta_+}\partial(H_{\alpha})
$$
We get
$$
D\Delta(0)=|W|.\Pi_{\alpha\in\Delta_+}\rho(H_{\alpha})
$$
Thus, on the one hand,
$$
D(\chi\Delta)(0)=\chi(0)D\Delta(0)=d(\chi).|W|.\Pi_{\alpha\in\Delta_+}\rho(H_{\alpha})
$$
and on the other,
$$
D(\Delta.\chi)(0)=|W|\Pi_{\alpha\in\Delta_+}(\lambda+\rho)(H_{\alpha})
$$
Equating these two expressions gives us Weyl's dimension formula for the irreducible representation of any compact semisimple Lie group corresponding to the dominant integral weight $\lambda$ in terms of the roots:
$$
d(\chi)=d(\lambda)=\frac{\Pi_{\alpha\in\Delta_+}(\lambda+\rho)(H_{\alpha})}{\Pi_{\alpha\in\Delta_+}\rho(H_{\alpha})}
$$
$$
=\frac{\Pi_{\alpha\in\Delta_+}<\lambda+\rho,\alpha>}{\Pi_{\alpha\in\Delta_+}<\rho,\alpha>}
$$
\bigskip

\section{Approximate solution to the wave equation in the vicinity of the event horizon of a Schwarchild black-hole}

This calculation would enable us to expand the solution to the wave equation in the vicinity of a blackhole in terms of "normal modes" or eigenfunctions of the Helmholtz operator in Schwarzchild space-time. The coefficients in this expansion will be identified with the particle creation and annihilation operators which would therefore enable us to obtain a formula for the blackhole entropy assuming that it is in a thermal state. It would enable us to obtain the blackhole temperature at which it starts radiating particles in the Rindler space-time metric which is an approximate transformation of the Schwarzchild metric. By using the relationship between the time variable in the Rindler space-time and the time variable in the Schwarzchild space-time, we can thus obtain the characteristic frequencies of oscillation in the solution of the wave equation in Schwarzchild space-time in terms of the the gravitational constant $G$ and the mass $M$ of the blackhole. These characteristic frequencies can be used to calculate the critical Hawking temperature at which the blackhole radiates in terms of $G,M$ using the formula for the entropy of a Gaussian state.

Let
$$
g_{00}=\alpha(r)=1-2GM/c^2r, g_{11}=\alpha(r)^{-1}/c^2=c^{-2}(1-2GM/rc^2)^{-1},
$$
$$
g_{22}=-r^2/c^2, g_{33}=-r^2sin^2(\theta)/c^2
$$
The Schwarzchild metric is then
$$
d\tau^2=g_{\mu\nu}dx^{\mu}dx^{\nu}, x^0=t, x^1=r, x^2=\theta,x^3=\phi
$$
or equivalently,
$$
d\tau^2=\alpha(r)dt^2-\alpha(r)^{-1}dr^2/c^2-r^2d\theta^2/c^2-r^2sin^2(\theta)d\phi^2/c^2
$$
The Klein-Gordin wave equation in Schwarzchild space-time is
$$
(g^{\mu\nu}\psi_{,\nu}\sqrt{-g})_{,\mu}+\mu^2\sqrt{-g}\psi=0
$$
where
$$
\mu=mc/h
$$
with $h$ denoting Planck's constant divided by $2\pi$. Now,
$$
\sqrt{-g}=(g_{00}g_{11}g_{22}g_{33})^{1/2}=r^2sin(\theta)/c^3
$$
so the above wave equation expands to give
$$
\alpha(r)^{-1}r^2sin(\theta)\psi_{,tt}-((c^2/\alpha(r))r^2sin(\theta)\psi_{,r})_{,r}
$$
$$
-(c^2/r^2)r^2(sin(\theta)\psi_{,\theta}))_{,\theta}-(c^2/r^2sin^2(\theta))(r^2sin(\theta)\psi_{,\phi\phi}+\mu^2r^2sin(\theta)\psi=0
$$
This equation can be rearranged as
$$
(1/c^2)\partial_t^2\psi=(\alpha(r)/r^2)((r^2/\alpha(r))\psi_{,r})_{,r}
$$
$$
+(\alpha(r)/r^2)((sin(\theta))^{-1}(sin(\theta)\psi_{,\theta})_{,\theta}+sin(\theta)^{-2}\psi_{,\phi\phi})-\mu^2\alpha(r)\psi
$$
This wave equation can be expressed in quantum mechanical notation as
$$
(1/c^2)\partial_t^2\psi=(\alpha(r)/r^2)((r^2/\alpha(r))\psi_{,r})_{,r}
$$
$$
-(\alpha(r)/r^2)L^2\psi-\mu^2\alpha(r)\psi
$$
where
$$
L^2=-(\frac{1}{sin(\theta)}\frac{\partial}{\partial\theta}sin(\theta)\frac{\partial}{\partial\theta}+\frac{1}{sin^2(\theta)}\frac{\partial^2}{\partial\phi^2})
$$
is the square of the angular momentum operator in non-relativistic quantum mechanics. 
Thus, we obtain using separation of variables, the solution
$$
\psi(t,r,\theta,\phi)=\sum_{nlm}f_{nl}(t,r)Y_{lm}(\theta,\phi)
$$
where $f_{nl}(t,r)$ satisfies the radial wave equation
$$
(1/c^2)\partial_t^2f_{nl}(t,r)=(\alpha(r)/r^2)((r^2/\alpha(r))f_{nl_,r})_{,r}
$$
$$
-((\alpha(r)/r^2)l(l+1)+\mu^2\alpha(r))f_{nl}(t,r)
$$
where $l=0,1,2,...$. We write
$$
f_{nl}(t,r)=exp(i\omega t)g_{nl}(r)
$$
Then the above equation becomes
$$
(-\omega^2/c^2)g_{nl}(r)=(\alpha(r)/r^2)((r^2/\alpha(r))g_{nl}'(r))'
$$
$$
-((\alpha(r)/r^2)l(l+1)+\mu^2\alpha(r))g_{nl}(r)
$$
For each value of $l$, this defines an eigenvalue equation and on applying the boundary conditions, the possible frequencies $\omega$ assume discrete values say $\omega(n,l), n=1,2,...$. These frequenices will depend upon $GM$ and $\mu$ since $\alpha(r)$ depends on
$GM$. Let $g_{nl}(r)$ denote the corresponding set of orthonormal eigenfunctions w.r.t the volume measure 
$$
c^3\sqrt{-g}d^3x=r^2sin(\theta)dr.d\theta.d\phi
$$
The general solution can thus be expanded as
$$
\psi(t,r,\theta,\phi)=\sum_{nlm}[a(n,l,m)g_{nl}(r)Y_{lm}(\theta,\phi)exp(i\omega(nl)t)+a(n,l,m)^*\bar g_{nl}(r)\bar Y_{lm}(\theta,\phi)exp(-i\omega(nl)t)]
$$
where $n=1,2,..., l=0,1,2,..., m=-l,-l+1,..., l-1,l$. The Hamiltonian operator of the field can be expressed as
$$
H=\sum_{nlm}h\omega(nl)a(n,l,m)^*a(n,l,m)
$$
and the bosonic commutation relations are
$$
[a(nlm), a(n'l'm')^*]=h\omega(nl)\delta(nn')\delta(ll')\delta(mm')
$$
\bigskip

An alternate analysis of Hawking temperature.

The equation of motion of a photon, ie null radial geodesic is given by
$$
\alpha(r)dt^2-\alpha(r)^{-1}dr^2=0
$$
or
$$
dr/dt=\pm\alpha(r)=\pm(1-2m/r), m=GM/c^2
$$
The solution is
$$
\int rdr/(r-2m)=\pm t+c
$$
so that
$$
2m.r+2mln(r-2m)=\pm t+c
$$
The change in time $t$ when the particle goes from $2m+0$ to $2m-0$ is purely imaginary an is given by $\delta t=2m\pi i$ since $ln(r-2m)=ln|r-2m|+i.arg(r-2m)$ with $arg(r-2m)=0$ for $r>2m$ and $=\pi$ for $r<2m$. Thus, a photon wave oscillating at a frequency $\omega$
has its amplitude changed from $exp(i\omega t)$ to $exp(i\omega(t+2m\pi i))=exp(i\omega t-2m\pi\omega)$. The corresponding ratio of the probability when it goes from just outside to just inside the critical radius is obtained by taking the modulus square of the probability amplitude and is given by $P=exp(-4m\pi\omega)$. The energy of the photon within the blackhole is $h\omega$ (where $h$ is Planck's constant divided by $2\pi$) and hence by Gibbsian theory in equilibrium statistical mechanics, its temperature $T$ must be given by $P=exp(-h\omega/kT)$. Therefore
$$
h\omega/kT=4\pi m\omega
$$
or equivalently,
$$
T=4\pi mk/h=8\pi GMk/hc^2 
$$
This is known as the Hawking temperature and it gives us the critical temperature of the interior of the blackhole at photons can tunnel through the critical radius, ie the temperature at which the blackhole can radiate out photons.
\bigskip

\section{More on group theoretic image processing}

Problem: Compute the Haar measure on $SL(2,\Bbb R)$ in terms of the Iwasawa decomposition
$$
x=u(\theta)a(t)n(s)
$$
where
$$
u(\theta)=\left(\begin{array}{cc}cos(\theta)&sin(\theta)\\-sin(\theta)&cos(\theta)\end{array}\right)=exp(\theta(X-Y)),
$$
$$
a(t)=exp(tH)=diag(e^t,e^{-t}), n(s)=exp(sX)=\left(\begin{array}{cc}1&s\\0&1\end{array}\right)
$$

Solution: We first express the left invariant vector fields corresponding to $H,X,Y$ as linear combinations of $\partial/\partial t, \partial/\partial\theta, \partial/\partial s$ and then the reciprocial of the associated determinant gives the Haar density. Specifically, writing
$$
H^*=f_{11}(\theta,t,s)\partial/\partial\theta+f_{12}(\theta,t,s)\partial/\partial t+f_{13}(\theta,t,s)\partial/\partial s,
$$
$$
X^*=f_{21}(\theta,t,s)\partial/\partial\theta+f_{22}(\theta,t,s)\partial/\partial t+f_{23}(\theta,t,s)\partial/\partial s,
$$
$$
Y^*=f_{31}(\theta,t,s)\partial/\partial\theta+f_{32}(\theta,t,s)\partial/\partial t+f_{33}(\theta,t,s)\partial/\partial s,
$$
for the left invariant vector fields corresponding to $H,X,Y$ respectively, it would follow that the Haar measure on $SL(2,\Bbb R)$ is given by
$$
F(\theta,t,s)d\theta dt ds, F=|det((f_{ij}))|
$$
or more precisely, the Haar integral of a function $\phi(x)$ on $SL(2,\Bbb R)$ would be given by
$$
\int_{SL(2,\Bbb R)}\phi(x)dx=\int\phi(u(\theta)a(t)n(s))F(\theta,t,s)d\theta dt ds
$$
We have
$$
d/ds(u(\theta)a(t)n(s))=u(\theta)a(t)n(s)X
$$
so
$$
X^*=\partial/\partial s
$$
Then,
$$
d/dt(u(\theta)a(t)n(s))=u(\theta)a(t)Hn(s)=u(\theta)a(t)n(s)n(s)^{-1}Hn(s)
$$
Now,
$$
n(s)^{-1}Hn(s)=exp(-s.ad(X))(H)=H-s[X,H]=H-2sX
$$
So
$$
H^*-2sX^*=\partial/\partial t
$$
Finally,
$$
d/d\theta(u(\theta)a(t)n(s))=u(\theta)(X-Y)a(t)n(s)=u(\theta)a(t)n(s)(n(s)^{-1}a(t)^{-1}(X-Y)a(t)n(s))
$$
and
$$
a(t)^{-1}(X-Y)a(t)=exp(-t.ad(H))(X-Y)=exp(-2t)X-exp(2t)Y
$$
$$
n(s)^{-1}a(t)^{-1}(X-Y)a(t)n(s)=n(s)^{-1}(exp(-2t)X-exp(2t)Y)n(s)
$$
$$
=exp(-2t)X-exp(2t).exp(-s.ad(X))(Y)
$$
$$
exp(-s.ad(X))(Y)=Y-s[X,Y]+(s^2/2)[X,[X,Y]]=Y-sH-s^2X
$$
so
$$
exp(-2t)X^*-exp(2t)(Y^*-sH^*-s^2X^*)=\partial/\partial\theta
$$

Problem: Solve these three equations and thereby express $X^*,Y^*,H^*$ as linear combinations of $\partial/\partial s,\partial/\partial t,\partial/\partial\theta$.
\bigskip

Problem: Suppose $I(f,g)$ is the invariant functional for a pair of image fields $f,g$ defined on $SL(2,\Bbb R)$. If $f$ changes by a small random amount and so does $g$, then what is the rate functional of the change in $I(f,g)$ ? Suppose that for each irreducible character $\chi$ of $G$, we have constructed the invariant functional $I_{\chi}(f,g)$ for a pair of image fields $f,g$. Let $\hat G$ denote the set of all the irreducible characters of $G$. Consider the Plancherel formula for $G$:
$$
f(e)=\int_{\hat G}\chi(f)d\mu(\chi)
$$
where $\mu$ is the Plancherel measure on $\hat G$. In other words, we have
$$
\delta_e(g)=\int_{\hat G}\chi(g)d\mu(\chi)
$$
Then we can write
$$
I_{\chi}(f,g)=\int_{G\times G}f(x)g(y)\chi(yx^{-1})dxdy, \chi\in\hat G
$$
Let $\psi(g)$ be any class function on $G$. Then, by Plancherel's formula, we have
$$
\psi(g)=\psi*\delta_e(g)=\int_G\delta_e(h^{-1}g)\psi(h)dh=
$$
$$
\int\chi(h^{-1}g).\psi(h)d\mu(\chi)dh=\int_{\hat G}\psi*\chi(g).d\mu(\chi)
$$
and hence the invariant $I_{\psi}$ associated with a pair of image fields can be expressed as
$$
I_{\psi}(f,g)=\int f(x)g(y)\psi(yx^{-1})dxdy=
$$
$$
\int_{G\times G\times\hat G}\psi*\chi(yx^{-1})f(x)g(y)dxdy.d\mu(\chi)
$$
$$
=\int\psi(h)\chi(h^{-1}yx^{-1})f(x)g(y)dxdy.d\mu(\chi)dh
$$
$$
=\int_{G\times\hat G}\psi(h)I_{\chi}(f,g,h)dh.d\mu(\chi)
$$
where we define
$$
I_{\chi}(f,g,h)=\int_{G\times G}f(x)g(y)\chi(yx^{-1}h)dxdy, h\in G
$$
Note that $(f,g)\rightarrow I_{\chi}(f,g,h)$ is not an invariant. We define the Fourier transform of the class function $\psi$ by
$$
\hat\psi(\chi)=\int_G\psi(h)\chi(h^{-1})dh
$$
Then, we have
$$
\delta_e(g)=\int\chi(g)d\mu(\chi),
$$
For any function $f(g)$ on $G$, not necessarily a class function,
$$
f(g)=\int\delta_e(h^{-1}g)f(h)dh=\int\chi(h^{-1}g)f(h)dhd\mu(\chi)
$$
Formally, let $\pi_{\chi}$ denote an irreducible representation having character $\chi$. Then, defining
$$
\hat f(\pi_{\chi})=\int_Gf(g)\pi(g^{-1})dg=\int_Gf(g^{-1})\pi(g)dg
$$
(Note that we are assuming $G$ to be a unimodular group. In the case of $SL(2,\Bbb R)$, this anyway true), we have
$$
f(g)=\int Tr(\pi_{\chi}(h^{-1}g))f(h)dhd\mu(\chi)
$$
$$
=\int_{\hat G}Tr(\hat f(\pi_{\chi})\pi_{\chi}(g))d\mu(\chi)
$$
and hence,
$$
I_{\chi}(f,g,h)=\int_{G\times G}f(x)g(y)\chi(yx^{-1}h)dxdy
$$
$$
=\int f(x)g(y)Tr(\pi_{\chi}(y)\pi_{\chi}(x^{-1})\pi_{\chi}(h))dxdy
$$
$$
=Tr(\hat g_0(\pi_{\chi})\hat f(\pi_{\chi})\pi_{\chi}(h))
$$
where
$$
g_0(x)=g(x^{-1}), x\in G
$$
Then, for a class function $\psi$,
$$
I_{\psi}(f,g)=\int\psi(h)Tr(\hat g_0(\pi_{\chi})\hat f(\pi_{\chi})\pi_{\chi}(h))dhd\mu(\chi)
$$
but,
$$
\hat\psi(\chi)=\int\psi(g)Tr(\pi_{\chi}(g^{-1}))dg=
$$
$$
Tr(\hat\psi(\pi_{\chi}))
$$
We have,
$$
\int\psi(h)\pi_{\chi}(h)dh=\int\psi_0(h)\pi_{\chi}(h^{-1})dh
$$
$$
=\hat\psi_0(\pi_{\chi})
$$
We have, on the one hand
$$
\psi(x)=\int\psi(h)\chi(h^{-1}x)dhd\mu(\chi)
$$
and since $\psi$ is a class function, $\forall x,y\in G$,
$$
\psi(x)=\psi(yxy^{-1})=\int\psi(h)\chi(h^{-1}yxy^{-1})dh.d\mu(\chi)
$$
$$
=\int\psi(h)\chi(y^{-1}h^{-1}yx)dh.d\mu(\chi)
$$
$$
=\int\psi(h)\chi_1(h,x,a)d\mu(\chi)
$$
where
$$
\chi_1(h,x,a)=\int_Ga(y)\chi(y^{-1}h^{-1}yx)dy=\int_Ga(y)\chi(h^{-1}yxy^{-1})dy
$$
where $a(x)$ is any function on $G$ satisfying,
$$
\int_Ga(x)dx=1
$$

Remark: $SL(2,\Bbb R)$ under the adjoint action, acts as a subgroup of the Lorentz transformations in a plane, ie, in the $t-y-z$ space. Indeed, represent the space-time point $(t,x,y,z)$ by
$$
\Phi(t,x,y,z)=\left(\begin{array}{cc}t+z&x+iy\\x-iy&t-z\end{array}\right)
$$
Then define $(t',x',y',z')$ by
$$
\Phi(t',x',y',z')=A.\Phi(t,x,y,z)A^*
$$
where
$$
A\in SL(2,\Bbb R)
$$
Clearly, by taking determinants, we get
$$
t^{'2}-x^{'2}-y^{'^2}-z^{'2}=t^2-x^2-y^2-z^2
$$
and then choosing $A=u(\theta)=exp(\theta(X-Y)),a(v)=exp(vH),n(s)=exp(sX)$ successively gives us the results that 
$$
t'+z'=t+z.cos(2\theta)+x.sin(2\theta), t'-z'=t-z.cos(2\theta)-x.sin(2\theta)
$$
$$
x'+iy'=x.cos(2\theta)-z.sin(2\theta)
$$
which means that
$$
t'=t, x'=x.cos(2\theta)-z.sin(2\theta), z'=x.sin(2\theta)+z.cos(2\theta), y'=y
$$
for the case $A=u(\theta)$, which corresponds to a rotation of the $x-z$ plane around the $y$ axis by an angle $2\theta$. The case $A=a(v)$ gives
$$
x'+iy'=x+iy, t'+z'=e^{2v}(t+z), t'-z'=e^{-2v}(t-z)
$$
or equivalently,
$$
x'=x, y'=y, t'=t.cosh(2v)+z.sinh(2v), z'=t.sinh(2v)+z.cosh(2v)
$$
which corresponds to a Lorentz boost along the $z$ direction with a velocity of
$u=-tanh(2v)$. Finally, the case $A=n(s)$ gives
$$
t'=t+sx+s^2(t-z)/2, x'=x+s(t-z), z'=z+sx+s^2(t-z), y'=y
$$
This corresponds to a Lorentz boost along a definite direction in the xz plane. The direction and velocity of the boost is a function of $s$. To see what this one parameter group of transformations corresponding to $A=n(s)$ corresponds to, we first evaluate it for infinitesimal $s$ to get
$$
t'=t+sx, x'=x+s(t-z), z'=z+sx, y'=y
$$
This transformation can be viewed as a composite of two infinitesimal transformations, the first is
$$
x\rightarrow x-sz, z\rightarrow z+sx, t\rightarrow t
$$
This transformation corresponds to an infinitesimal rotation of the $xz$ plane around the $y$ axis by an angle $s$. The second is
$$
x\rightarrow x+st, t\rightarrow t+sx
$$
which corresponds to an infinitesimal Lorentz boost along the $x$ axis by a velocity of $s$. To see what the general non-infinitesimal case corresponds to, we assume that it corresponds to a boost with speed of $u$ along the unit vector $(n_x,0,n_z)$. Then the corresponding Lorentz transformation equations are
$$
(x',z')=\gamma(u)(xn_x+zn_z-ut)(n_x,n_z)+(x,z)-(xn_x+zn_z)(n_x,n_z)
$$
$$
t'=\gamma(u)(t-xn_x-zn_z)
$$
The first pair of equation is in component form,
$$
x'=\gamma(xn_x+zn_z-ut)n_x+x-(xn_x+zn_z)n_x,
$$
$$
z'=\gamma(xn_x+zn_z-ut)n_z+z-(xn_x+zn_z)n_z
$$
To get the appropriate matching, we must therefore have
$$
1+s^2/2=\gamma(u), s=-\gamma(u)n_x, -s^2/2=\gamma(u)n_z,
$$
which is impossible. Therefore, this transformation cannot correspond to a single Lorentz boost. However, it can be made to correspond to the composition of  a rotation in the xz plane followed by a Lorentz boost in the same plane.
Let $L_x(u)$ denote the Lorentz boost along the $x$ axis by a speed of $u$ and let 
$R_y(\alpha)$ denote a rotation around the $y$ axis by an angle $\alpha$ counterclockwise.
Then, we have already seen that the finite transformation corresponding to $A=n(s)$ is given by
$$
L(s)=lim_{n\rightarrow\infty}(L_x(-s/n)R_y(-s/n))^n=exp(-s(K_x+X_y))
$$
where $K_x$ is the infinitesimal generator of Lorentz boosts along the $x$ axis while
$X_y$ is the infinitesimal generator of rotations around the $y$ axis. Specifically,
$$
K_x=L_x'(0), X_y=R_y'(0)
$$
The characters of the discrete series of $SL(2,\Bbb R)$. Let $m$ be a nonpositive integer. Let $e_{m-1}$ be a highest weight vector of a representation $\pi_m$ of $\mathfrak sl(2,\Bbb R)$. Then
$$
\pi(X)e_{m-1}=0, \pi(H)e_{m-1}=(m-1)e_{m-1}
$$
Define
$$
\pi(Y)^re_{m-1}=e_{m-1-2r}, r=0,1,2,...
$$
Then
$$
\pi(H)e_{m-1-2r}=\pi(H)\pi(Y)^re_{m-1}=([\pi(H),\pi(Y)^r]+\pi(Y)^r\pi(H))e_{m-1}
$$
$$
=(m-1-2r)\pi(Y)^re_m=(m-1-2r)e_{m-1-2r}, r\geq 0
$$
Thus, considering the module $V$ of $SL(2,\Bbb R)$ defined by the span of $e_{m-2r}, r\geq 0$, we find that
$$
Tr(\pi(a(t))=Tr(\pi(exp(tH))=\sum_{r\geq 0}exp(t(m-1-2r))=\frac{exp((m-1)t)}{1-exp(-2t)}
$$
$$
=\frac{exp(mt)}{exp(t)-exp(-t)}
$$
Now,
$$
\pi(X)e_{m-1-2r}=\pi(X)\pi(Y)^re_{m-1}=[\pi(X),\pi(Y)^r]e_{m-1}
$$
$$
=(\pi(H)\pi(Y)^{r-1}+....+\pi(Y)^{r-1}\pi(H))e_{m-1}
$$
$$
=[(-2(r-1)\pi(Y)^{r-1}+\pi(Y)^{r-1}\pi(H)-2(r-2)\pi(Y)^{r-1}+\pi(Y)^{r-1}\pi(H)+...+\pi(Y)^{r-1}\pi(H)]e_{m-1}
$$
$$
=[-r(r-1)+r(m-1)]\pi(Y)^{r-1}e_{m-1}=r(m-r)e_{m+1-2r}
$$
Also note that
$$
\pi(Y)e_{m-1-2r}=e_{m-3-2r}, r\geq 0
$$

Problem: From these expressions, calculate $Tr(\pi(u(\theta))$ where
$u(\theta)=exp(\theta(X-Y))$. Show that it equals $exp(im\theta)/(exp(i\theta)-exp(-i\theta))$. In this way, we are able to determine all the discrete series characters for
$SL(2,\Bbb R)$.

\section{A problem in quantum information theory}

Let $K$ be a quantum operation, ie, a CPTP map. Its action on an operator $X$ is given in terms of the Choi-Kraus representation by
$$
K(X)=\sum_aE_aXE_a^*
$$
Define the following inner product on the space of operators:
$$
<X,Y>=Tr(XY^*)
$$
Then, calculate $K^*$ relative to this inner product in terms of the Choi-Kraus representation of $K$.

Answer:
$$
<K(X),Y>=Tr(K(X)Y^*)=Tr(\sum_aE_aXE_a^*Y^*)=\sum_aTr(XE_a^*Y^*E_a)
$$
$$
=Tr(X.(\sum_aE_a^*YE_a)^*)=Tr(X.K^*(Y)^*)=<X,K(^*(Y)>
$$
where
$$
K^*(X)=\sum_aE_a^*XE_a
$$

Problem: Let $A,B$ be matrices $A$ square and $B$ rectangular of the appropriate size so that
$$
X=\left(\begin{array}{cc}A&B\\B^*&I\end{array}\right)\geq 0
$$
Then deduce that
$$
A\geq BB^*
$$

Solution: For any two column vectors $u,v$ of appropriate size
$$
(u^*,v^*)x\left(\begin{array}{cc}u\\v\end{array}\right)\geq 0
$$
This gives
$$
u^*Au+u^*Bv+v^*B^*u+v^*v\geq 0
$$
Take
$$
v=Cu
$$
where $C$ is any matrix of appropriate size. Then, we get
$$
u^*(A+BC+C^*B^*+C^*C)u\geq 0
$$
this being true for all $u$, we can write
$$
A+BC+C^*B^*+C^*C\geq 0
$$
for all matrices $C$ of appropriate size. In particular, taking $C=-B^*$ gives us
$$
A-BB^*\geq 0
$$
\bigskip

\section{Hawking radiation}

Problem: For a spherically symmetric blackhole with a charge $Q$ at its centre, solve the static Einstein-Maxwell equations to obtain the metric. Hence, calculate the equation of motion of a radially moving photon in this metric and determine the complex time taken by the photon to travel from just outside the critical blackhole radius to just inside and denote this imaginary time by $i\Delta$. Note that $\Delta$ will be a function of the parameters $G,M,Q$ of the blackhole. Deduce that if $T$ is the Hawking temperature for quantum emission of photons for such a charged blackhole, then at frequency $\omega$,
$$
exp(-h\omega/kT)=exp(-\omega\Delta)
$$
and hence that the Hawking temperature is given by
$$
T=h/k\Delta
$$
Evaluate this quantity.
\bigskip

Let 
$$
D(\rho|\sigma)=Tr(\rho.(log(\rho)-log(\sigma)))
$$
the relative entropy between two states $\rho,\sigma$ defined on the same Hilbert space.
We know that it is non-negative and that if $K$ is a quantum operation, ie, a CPTP map, then
$$
D(\rho|sigma)\geq D(K(\rho)|K(\sigma))
$$
Consider three subsystems $A,B,C$ and let $\rho(ABC)$ be a joint state on these three subsystems. Let $\rho_m=I_B/d(B)$ the completely mixed state on $B$. $I_B$ is the identity operator on $\mathcal H_B$ and $d(B)=dim\mathcal H_B$. Let $K$ denote the quantum operation corresponding to partial trace on the $B$ system. Then consider the inequality
$$
D(\rho(ABC)|\rho_m\otimes\rho(AC))\geq D(K(\rho(ABC))|K(\rho_m\otimes\rho(AC)))
$$
$$
=D(\rho(AC)|\rho(AC))=0
$$
Evaluating gives us
$$
Tr(\rho(ABC)log(\rho(ABC))-Tr(\rho(ABC).(log(\rho(AC))-log(d(B)))\geq 0
$$
or equivalently,
$$
-H(ABC)+H(AC)+log(d(B))\geq 0
$$ 
Now, let $K$ denote partial tracing on the $C$-subsystem. Then,
$$
D(\rho(ABC)|\rho_m\otimes\rho(AC))\geq D(K(\rho(ABC))|K(\rho_m\otimes\rho(AC)))
$$
gives
$$
-H(ABC)+H(AC)+log(d(B))\geq D(\rho(AB)|I_m\otimes\rho(A))
$$
$$
=-H(AB)-Tr(\rho(AB)(log(\rho(A))-log(d(B)))
$$
$$
=-H(AB)+H(A)+d(B)
$$
giving
$$
H(AB)+H(AC)-H(A)-H(ABC)\geq 0
$$
\bigskip

Let the metric of space-time be given in spherical polar coordinates by
$$
d\tau^2=A(r)dt^2-B(r)dr^2-C(r)(d\theta^2+sin^2(\theta)d\phi^2)
$$
The equation of the radial null geodesic is given by
$$
A(r)-B(r)(dr/dt)^2=0
$$
or equivalently for an incoming radial geodesic,
$$
dr/dt=-\sqrt{A(r)/B(r)}=-F(r)
$$
say where
$$
F(r)=\sqrt{A(r)/B(r)}
$$
Let $F(r)$ have a zero at the critical radius $r_0$ so that we can write
$$
F(r)=G(r)(r-r_0)
$$
where $G(r_0)\neq 0$. Then, the time taken for the photon to go from $r_0+0$ to $r_0-0$ is given by
$$
\Delta=\int_{r_0-0}^{r_0+0}dr/((r-r_0)G(r))=G(r_0)^{-1}lim_{\epsilon\downarrow 0}(log(\epsilon)-log(-\epsilon))
$$
$$
=-i\pi.G(r_0)^{-1}
$$
so that the ratio change in the probability amplitude of a photon pulse of frequency $\omega$ is the quantity $exp(-\omega\pi.G(r_0)^{-1})$ and equating this to the Gibbsian probability $exp(-h\omega/kT)$ gives us the Hawking temperature as
$$
T=h.G(r_0)/\pi k
$$
\bigskip

\chapter{Research Topics on represenations,}

Using the Schrodinger equation to track the time varying probability density of a signal.
The Schrodinger equation naturally generates a family of pdf's as the modulus square of the evolving wave function and by controlling the time varying potential in which the Schrodinger evolution takes place, it is possible to track a time varying pdf, ie, encode this time varying pdf into the form the weights of a neural network. From the stored weights, we can synthesize the time varying pdf of the signal.

\section{Remarks on induced representations and Haar integrals}

Let $G$ be a compact group. Let $\chi$ be a character of an Abelian subgroup $N$ of $G$ in a semidirect product $G=HN$. Let $U=Ind^G_N\chi$ be the induced representation. Suppose a function $f$ on $G$ is orthogonal to all the matrix elements of $U$. Then we can express this condition as follows. First the representation space $V$ of $U$ consists of all functions
$h:G\rightarrow\Bbb C$ for which $h(xn)=\chi(n)^{-1}h(x)\forall x\in G,n\in N$ and $U$ acts on such a function to give $U(y)h(x)=h(y^{-1}x), x,y\in G$. The condition
$$
\int_Gf(x)U(x)dx=0
$$
gives
$$
\int_Gf(x)h(x^{-1}y)dx=0\forall y\in G\forall h\in V
$$
Now for $h\in V$, we have
$$
h(xn)=\chi(n)^{-1}h(x), x\in G, n\in N
$$
and hence for any representation $\pi$ of $G$,
$$
\int h(xn)\pi(x^{-1})dx=\chi(n)^{-1}\hat h(\pi)
$$
or
$$
\int h(x)\pi(n.x^{-1})dx=\chi(n)^{-1}\hat h(\pi)
$$
or
$$
\pi(n)\hat h(\pi)=\chi(n^{-1})\hat h(\pi), n\in N
$$
Define
$$
P_{\pi,N}=\int_N\pi(n)dn
$$
Assume without loss of generality that $\pi$ is a unitary representation of $G$. Then, $P_{\pi,N}$ is the orthogonal projection onto the space of all those vectors $w$ in the representation space of $\pi$ for which $\pi(n)w=w\forall n\in N$. Then the above equation gives on integrating w.r.t $n\in N$,
$$
P_{\pi,N}\hat h(\pi)=0, h\in V, \chi\neq 1
$$
and
$$
P_{\pi,N}\hat h(\pi)=\hat h(\pi), h\in V,\chi=1
$$
on using the Schur orthogonality relations for the characters of $N$. We therefore have
The equation
$$
\int_Gf(x)h(x^{-1}y)dx=0
$$
gives us
$$
\int_Gf(x)h(x^{-1}y)\pi(y^{-1})dxdy=0
$$
or
$$
\int f(x)h(z)\pi(z^{-1}x^{-1})dxdz=0
$$
or
$$
\hat h(\pi)\hat f(\pi)=0, h\in V
$$
It also implies
$$
\int f(x)h(x^{-1}y)\pi(y)dxdy=0
$$
or
$$
\int f(x)h(z)\pi(xz)dxdz=0
$$
or
$$
\hat f(\pi)^*\hat h(\pi)^*=0
$$
which is the same as the above. Suppose on the other hand that $f$ satisfies
$$
\int_Nf(xny)dn=0, x,y\in G---(1)
$$
Then, it follows that
$$
\int_{G\times N}f(xny)\pi(y^{-1})dydn=0
$$
which implies
$$
\int_{G\times N}f(z)\pi(z^{-1}xn)dzdn=0
$$
This implies
$$
\hat f(\pi)\pi(x)P_{\pi,N}=0, x\in G
$$
It also implies
$$
\int f(xny)\pi(x^{-1})dxdn=0
$$
which implies
$$
\int f(z)\pi(nyz^{-1})dzdn=0
$$
or equivalently,
$$
P_{\pi,N}\pi(y)\hat f(\pi)=0, y\in G
$$
In particular, (1) implies
$$
P_{\pi,N}\hat f(\pi)=\hat f(\pi)P_{\pi,N}=0
$$
for all representations $\pi$ of $G$. The condition (1) implies that
$P\pi(y)\hat f(\pi)=0$ where $P=P_{\pi,N}$ for all representations $\pi$ and hence for all
functions $h$ on $G$ it also implies that $P\hat h(\pi)\hat f(\pi)=0$. More precisely, it implies that for all representations $\pi$ and all functions $h$ on $G$,
$$
P_{\pi,N}\hat h(\pi)\hat f(\pi)=0---(2)
$$
Now suppose that $h_1$ belongs to the the representation space of $U=Ind^G_N\chi_0$ where 
$\chi_0$ is the identity character on $N$, ie, one dimensional unitary representation of $N$ equal to unity at all elements of $G$. Then, it is clear that $h_1$ can be expressed as
$$
h_1(x)=\int_Nh(xn)dn, x\in N
$$
for some function $h$ on $G$ and conversely if $h$ is an arbitrary function on $G$ for which the above integral exists, then $h_1$ will belong to the representation space of $U$. That is because $h_1$ as defined above satisfies
$$
h_1(xn)=h_1(x), x\in G,n\in N
$$
The condition that $f$ be orthogonal to the matrix elements of $U$ is that
$$
\int f(x)U(x^{-1})dx=0
$$
and this condition is equivalent to the condition that for all $h$ on $G$,
$$
\int_Gf(x)h_1(xy)dx=0, y\in G, h=\int_Nh(xn)dn
$$
and this condition is in turn equivalent to
$$
\int f(x)h_1(xy)\pi(y^{-1})dxdy=0
$$
or
$$
\int f(x)h_1(z)\pi(z^{-1}x)dxdz=0
$$
or equivalently,
$$
\hat h_1(\pi)\hat f(\pi)^*=0
$$
for all unitary represenations $\pi$ of $G$ and for all functions $h$ on $G$. Now,
$$
\hat h_1(\pi)=\int h(xn)\pi(x^{-1})dxdn=\int h(z)\pi(nz^{-1})dzdn=P.\hat h(\pi)
$$
and hence the condition is equivalent to
$$
P_{\pi,N}\hat h(\pi)\hat f(\pi)^*=0---(3)
$$
for all unitary representations $\pi$ of $G$ and for all functions $h$ on $G$. This condition is the same as (2) if we observe that $\hat f(\pi)$ can be replaced by $\hat f(\pi)^*$ because $f(x)$ being orthogonal to the matrix elements of $U(x)$ is equivalent to $f(x^{-1})$ being orthogonal to the same because for compact groups, a finite dimensional representation is equivalent to a unitary representation which means that
$U$ is an appropriate basis is a unitary representation. Thus, we have proved that a necessary and sufficient condition that a function $f$ be orthogonal to all the matrix elements of the representation of $G$ induced by the identity character $\chi_0$ on $N$ is that $f$ be a cusp form, ie, it should satisfies
$$
\int_Nf(xny)dn=0\forall x,y\in G
$$
Now suppose $\chi$ is an arbitary character of $N$ (ie a one dimensional unitary representation of $N$). Then let $U=Ind^G_N\chi$. The If $h_1$ belongs to the representation space of $U$, then $h_1(xn)=\chi(n)^{-1}h_1(x), x\in G,n\in N$. Now let $h(x)$ be any function on $G$. Define
$$
h_1(x)=\int_Nh(xn)\chi(n)dn
$$
Then,
$$
h_1(xn_0)=\int_Nh(xn_0n)\chi(n)dn=\int h(xn)\chi(n_0^{-1}n)dn
$$
$$
=\chi(n_0)^{-1}h_1(x), n_0\in N, x\in G
$$
It follows that $h_1$ belongs to the representation space of $U$. conversely if $h_1$ belongs to this representation space, then
$$
h_1(x)=\int_Nh_1(xn)\chi(n)dn
$$
Thus we have shown that $h_1$ belongs to the representation space of $U=Ind^G_N\chi$ iff
there exists a function $h$ on $G$ such that
$$
h_1(x)=\int h(xn)\chi(n)dn
$$
Note that this condition fails for non-compact groups since the integral $dn$ on $N$ cannot be normalized in general. Then, the condition for $f$ to be orthogonal to $U$ is that
$$
\int_Gf(x)U(x^{-1})dx=0---(4)
$$
and this condition is equivalent to
$$
\int f(x)h_1(xy)dx=0
$$
for all $h_1$ in the representation space of $U$ or equivalently,
$$
\int f(x)h(xyn)\chi(n)dxdn=0
$$
for all functions $h$ on $G$. This condition is equivalent to
$$
\int f(x)h(xyn)\chi(n)\pi(y^{-1})dxdydn=0
$$
for all unitary representations $\pi$ of $G$ which is in turn equivalent to
$$
\int_{G\times G\times N}f(x)h(z)\chi(n)\pi(nz^{-1}x)dxdzdn=0
$$
or equivalently,
$$
P_{\chi,\pi,N}.\hat h(\pi).\hat f(\pi)^*=0---(5)
$$
for all functions $h$ on $G$ and for all unitary representations $\pi$ of $G$ where
$$
P_{\chi,\pi,N}=P=\int_N\chi(n)\pi(n)dn
$$
Now consider the following condition on a function $f$ defined on $G$:
$$
\int_Nf(xny)\chi(n)dn=0, x,y\in G
$$
This condition is equivalent to
$$
\int_{G\times N}f(xny)\chi(n)\pi(y^{-1})dydn =0
$$
for all unitary representations $\pi$ of $G$ or equivalently,
$$
\int_{G\times N}f(z)\chi(n)\pi(z^{-1}xn)dzdn=0
$$
and this is equivalent to 
$$
\hat f(\pi)\pi(x)P=0, x\in G---(6)
$$ 
which is in turn equivalent to
$$
\hat f(\pi)\hat h(\pi)P=0---(7)
$$
for all functions $h$ on $G$ as follows by multiplying $(5)$ with $h(x^{-1})$ and integrating over $G$. This condition is the same as
$$
P_{\chi,\pi,N}\hat h(\pi)^*\hat f(\pi)^*=0
$$
or since $h(x)$ can be replaced by $h(x^{-1})$, the condition is equivalent to
$$
P_{\chi,\pi,N}\hat h(\pi)\hat f(\pi)^*=0---(8)
$$
for all functions $h$ on $G$ and all unitary representations $\pi$ of $G$. This condition is the same as (5). Thus, we have established that a necessary and sufficient condition for $f$ to be orthogonal to all the matrix elements of $U=Ind^G_N\chi$ is that $f$ should satisfy
$$
\int_Nf(xny)\chi(n)dn=0, x,y\in G
$$
In other words, $f$ should be a generalized cusp form corresponding to the character $\chi$ of $N$.
\bigskip

\section{On the dynamics of breathing in the presence of a mask}

Let $V$ denote the volume between the face and the mask. Oxygen passes from outside the mask to within which the person breathes in. While breathing out, the person ejects $CO_2$ along with of course some fraction of oxygen. Let $n_o(t)$ denote the number of oxygen molecules per unit volume
at time $t$ and $n_c(t)$ denote the number of carbon-dioxide molecules per unit volume at time $t$ in the region between the face and the mask. Let $n_r(t)$ denote the number of molecules of the other gases in the same region per unit volume. Let $\lambda_c(t)dt$ denote the probability of a $CO_2$ molecule escaping away the volume between the mask and the face in time $[t,t+dt]$. Then let $g_c(t)dt$ denote the number of $CO_2$ molecules being ejected from the nose and mouth of the person per unit time. Then, we have a simple conservation equation
$$
dn_c(t)/dt=-n_c(t)\lambda_c(t)+g_c(t)/V
$$
To this differential equation, we can add stochastic terms and express it as
$$
dn_c(t)/dt=-n_c(t)\lambda_c(t)+g_c(t)/V+w_c(t)
$$
We can ask the question of determining the statistics of the first time $\tau_c$ at which the $CO_2$ concentration $n_c(t)$ exceeds a certain threshold $N_c$.
Sometimes there can be a chemical reaction between the $CO_2$ trapped in the volume with some other gases. Likewise other gases within the volume can react and produce $CO_2$. If we take into account these chemical reactions, then there will be additional terms in the above equation describing the rate of change of concentration of $CO_2$. For example, consider a reaction of the form 
$$
a.CO_2+b.X\rightarrow c.Y+d.Z
$$
where $a,b,c,d$ are positive integers. $X$ denotes the concentration of the other gases, ie, gases other than $CO_2$ within the volume. At time $t=0$, there are $n_c(0)$ molecules of $CO_2$ and $n_X(0)$ molecules of $X$ and zero molecules of $Y$ and $Z$. Let $K$ denote the rate constant at fixed temperature $T$. Then, we have an equation of the form
$$
dn_c(t)/dt=-K.n_c(t)^an_X(t)^b---(1)
$$
according to Arrhenius' law of mass action. After time $t$, therefore, $n_c(0)-n_c(t)$ molecules of $CO_2$ have reacted with $(n_c(0)-n_c(t))b/a$ moles of $X$. Thus,
$$
n_X(t)=n_X(0)-(n_c(0)-n_c(t))b/a
$$
from which we deduce
$$
dn_X(t)/dt=(b/a)dn_c(t)/dt=(-Kb/a)n_c(t)^a.n_X(t)^b---(2)
$$
equations (1) and (2) have to be jointly solved for $n_c(t)$ and $n_X(t)$. This can be reduced to a single differential equation for $n_c(t)$:
$$
dn_c(t)/dt=-K.n_c(t)^a(n_X(0)-(n_c(0)-n_c(t))b/a)^b---(3)
$$
Likewise, we could also have a chemical reaction in which $CO_2$ is produced by a chemical reaction between two compounds $U,V$. We assume that such reaction is of the form
$$
pU+qV\rightarrow r.CO_2+s.M
$$
Let $n_U(0),n_V(0)$ denote the number of molecules of $U$ and $V$ at time $t=0$. Then,
the rate of at which their concentrations change with time is again given by Arrhenius' law of mass action
$$
dn_U(t)/dt=-K_1n_U(t)^pn_V(t)^q
$$
$$
n_V(t)=n_V(0)-q(n_U(0)-n_U(t))/p
$$
and then the number of $CO_2$ molecules generated after time $t$ due to this reaction will be given by
$$
n_c(t)=(r/p)(n_U(0)-n_U(t))
$$
and also
$$
n_V(t)=n_V(0)-(q/p)(n_U(0)-n_U(t))=n_V(0)-(q/r)n_c(t)
$$
and hence
$$
dn_c(t)/dt=(-r/p)dn_U(t)/dt=(r/p)K_1n_U(t)^p.n_V(t)^q
$$
$$
=(K_1r/p)(n_U(0)-pn_c(t)/r)^p(n_V(0)-(q/r)n_c(t))^q
$$
It follows that combining these two processes we can write
Now consider the combination of two such reactions. Writing down explicitly these reactions, we have
$$
aCO_2+bX\rightarrow c.Y+d.Z, p.U+q.V\rightarrow r.CO_2+s.Z
$$
At time $t=0$, we have initial conditions on $n_c(0),n_X(0), n_U(0),n_V(0)$. Then the rate at which $CO_2$ gets generated by the combination of these two reactions and the other non-chemical-reaction processes of diffusion exit etc. is given by
$$
dn_c(t)/dt=-Kn_c(t)^an_X(t)^b+(K_1r/p)n_U(t)^pn_V(t)^q+f(n_c(t),t)
$$
where
$$
dn_X(t)/dt=(-Kb/a)n_c(t)^an_X(t)^b, dn_U(t)/dt=-K_1n_U(t)^pn_V(t)^q,
$$
$$
dn_V(t)/dt=(-K_1q/p)n_U(t)^pn_V(t)^q
$$
Here, $f(n_c(t),t)$ denotes the rate at which $CO_2$ concentration changes due to the other effects described above. Of course, we could have more complex chemical reactions in of the sort
$$
a_0CO_2+a_1X_1+...+a_nX_n\rightarrow b_1Y_1+...+b_mY_m,
$$
for which the law of mass action would give
$$
dn_c(t)/dt=-Kn_c(t)^{a_0}n_{X_1}(t)^{a_1}...n_{X_n}(t)^{a_n},
$$
$$
n_{X_j}(t)=n_{X_j}(0)-(a_j/a_0)(n_c(0)-n_c(t)), j=1,2,..., n,
$$
$$
n_{Y_j}(t)=n_{Y_j}(0)+(b_j/a_0)(n_c(0)-n_c(t)), j=1,2,..., m
$$
\bigskip

\section{Plancherel formula for $SL(2,\Bbb R)$}

Let $G=KAN$ be the usual Iwasawa decomposition and let $B$ denote the compact/elliptic Cartan subgroup $B=\{exp(\theta(X-Y)), \theta\in[0,2\pi)\}$. Let $L$ denote the non-compact/hyperbolic Cartan subgroup $L=\{exp(tH), t\in\Bbb R\}$. Let $G_B$ denote the conjugacy class of $B$, ie, $G_B$ is the elliptic subgroup $gBg{-1}, g\in G$ and likewise $G_L$ will denote the conjugacy class of $L$ ie, the hyperbolic subgroup $gLg^{-1}, g\in G$. These two subgroups have only the identity element in common. Now let $\Theta_m$ denote the discrete series characters. We know that
$$
\Theta_m(u(\theta))=-sgn(m)\frac{exp(im\theta)}{exp(i\theta)-exp(-i\theta)}, m\in\Bbb Z, u(\theta)=exp(\theta(X-Y))
$$
$$
\Theta_m(a(t))=\frac{exp(mt)}{exp(t)-exp(-t)}, a(t)=exp(tH)
$$
Weyl's integration formula gives
$$
\Theta_m(f)=\int_G\Theta_m(x)f(x)dx=
$$
$$
\int_0^{2\pi}\Theta_m(u(\theta))F_{f,B}(theta)\Delta_B(\theta)d\theta
$$
$$
+\int_{\Bbb R}\Theta_m(a(t))F_{f,L}(t)\Delta_L(t)dt
$$
$$
=\int sgn(m)exp(im\theta)F_{f,B}(\theta)d\theta+\int_{\Bbb R}exp(-|mt|)F_{f,L}(t)dt
$$
$$
=sgn(m)\hat F_{f,B}(m)+\int_{\Bbb R}exp(-|mt|)F_{f,L}(t)dt
$$
where $F_{f,B}$ and $F_{f,L}$ are respectively the orbital integrals of $f$ corresponding to the elliptic and hyperbolic groups and if $a(\theta)$ is a function of $\theta$, then $\hat a(m)$ denotes its Fourier transform, ie, its Fourier series coefficients. Thus,
$$
F_{f,B}(\theta)=\Delta_B(\theta)\int_{G/B}f(xu(\theta)x^{-1})dx
$$
$$
F_{f,L}(a(t))=\Delta_L(t)\int_{G/L}f(xa(t)x^{-1})dx
$$
where
$$
\Delta(\theta)=(exp(i\theta)-exp(-i\theta)), \Delta(t)=exp(t)-exp(-t)
$$
Now
$$
\sum_{m odd}|m|\Theta_m(f)=\int_0^{2\pi}\Delta_L(t)F_{f,L}(t)\sum_{m odd}|m|\Theta_m(a(t))dt
$$
$$
+\int_{\Bbb R}\Delta_B(\theta)F_{f,B}(\theta).\sum_{m odd}|m|\Theta_m(u(\theta))d\theta
$$
Now,
$$
\sum_{m odd}|m|\Theta_m(a(t))=\sum_{m odd}|m|exp(-|mt|)/(exp(t)-exp(-t))
$$
$$
\sum_{m\geq 1, m odd}mexp(-mt)=\sum_{k\geq 0}(2k+1)exp(-(2k+1)t)
$$
$$
\sum_{m\geq 0}m.exp(-mt)=-d/dt(1-exp(-t))^{-1}=exp(-t)(1-exp(-t))^{-2},t\geq 0
$$
So
$$
\sum_{k\geq 0}(2k+1)exp(-(2k+1)t)=exp(-t)(2exp(-2t)(1-exp(-2t))^{-2}+(1-exp(-2t))^{-1})
$$
$$
=exp(-t)(1+exp(-2t))/(1-exp(-2t))^2=(exp(t)+exp(-t))/(exp(t)-exp(-t))^2
$$
Thus,
$$
\sum_{m odd}|m|exp(-|mt|)=2\sum_{m\geq 0}m.exp(-m|t|)=tanh(t)=w_1(t)
$$
say.
$$
\sum_{m odd}|m|\Theta_m(u(\theta))=\sum_{m odd}m.exp(im\theta)/(exp(i\theta)-exp(-i\theta))=w_2(\theta)
$$
say. Then, the Fourier transform of $F_{f,B}'(\theta)$ (ie Fourier series coefficients)
$$
\hat F_{f,B}'(m)=\int F_{f,B}'(\theta)exp(im\theta)d\theta=
$$
$$
-\int F_{f,B}(\theta)(im)exp(im\theta)d\theta=-im\hat F_{f,B}(m)
$$
Now, we've already seen that
$$
\Theta_m(f)=sgn(m)\hat F_{f,B}(m)+\int_{\Bbb R}exp(-|mt|)F_{f,L}(t)dt
$$
So
$$
-im\Theta_m(f)=sgn(m)(-im\hat F_{f,B}(m))-im\int exp(-|mt|)F_{f,L}(t)dt
$$
$$
=sgn(m)\hat F_{f,B}'(m)-im\int exp(-|mt|)F_{f,L}(t)dt
$$
Multiplying both sides by $sgn(m)$ gives us
$$
-i|m|\Theta_m(f)=\hat F_{f,B}'(m)-i|m|\int exp(-|mt|)F_{f,L}(t)dt
$$
Summing over odd $m$ gives
$$
-i\sum_{m odd}|m|\Theta_m(f)=\sum_{m odd}\hat F_{f,B}'(m)-i\int w_1(t)F_{f,L}(t)dt
$$
Now,
$$
\sum_m\hat F_{f,B}'(m)=F_{f,B}'(0)-i\int w_1(t)F_{f,L}(t)dt
$$
Now,
$$
F_{f,B}(\theta)=(exp(i\theta)-exp(-i\theta))\int_{G/B}f(xu(\theta)x^{-1})dx
$$
which gives on differentiating w.r.t $\theta$ at $\theta=0$
$$
F_{f,B}'(0)=2if(1)
$$
Thus, we get the celebrated Plancherel formula for $SL(2,\Bbb R)$,
$$
2f(1)=-\sum_m|m|\Theta_m(f)+\int w_1(t)F_{f,L}(t)dt
$$
Now we know that the Fourier transform 
$$
\hat F_{f,L}(\mu)=\int_{\Bbb R}F_{f,L}(t)exp(it\mu)dt
$$
equals $T_{i\mu}(f)$, the character of the Principal series corresponding to $i\mu, \mu\in\Bbb R$. Thus defining
$$
\hat w_1(\mu)=\int_{\Bbb R}w_1(t)exp(it\mu)dt
$$
the Plancherel formula can be expressed in explicit form:
$$
2f(1)=-\sum_m|m|\Theta_m(f)+\int_{\Bbb R}\hat w(\mu)^*.T_{i\mu}(f)d\mu
$$
This formula was generalized to all semisimple Lie groups and also to p-adic groups by Harish-Chandra.

\section{Some applications of the Plancherel formula for a Lie group to image processing} 

Let $G$ be a group and let its irreducible characters be $\chi, \chi\in\hat G$. let $d\mu(\chi)$ be its Plancherel measure:
$$
\delta_e(g)=\int\chi(g).d\mu(\chi)
$$
or equivalently,
$$
f(e)=\int_{\hat G}\chi(f)d\mu(\chi)
$$
for all $f$ in an appropriate Schwartz space. Consider an invariant for image pairs of the form
$$
I_{\chi}(f_1,f_2)=\int_{G\times G}f_1(x)\chi(xy^{-1})f_2(y)dxdy
$$
Then if $\psi$ is any class function on $\hat G$, we have
$$
\int_{\hat G}\chi(\psi).I_{\chi}(f_1,f_2)d\mu(\chi)=
$$
$$
\int f_1(x)\chi(\psi^0)\chi(xy^{-1})f_2(y)dxdyd\mu(\chi)
$$
$$
=\int f_1(x)\psi_1(xy^{-1})f_2(y)dxdy
$$
where
$$
\psi_1(x)=\int\chi(\psi^0)\chi(x)d\mu(\chi), \psi^0(x)=\psi(x^{-1})
$$
so that
$$
\chi(\psi^0)=\int\psi(x^{-1})\chi(x)dx
$$
Remark: let $\pi_{\chi}$ be an irreducible representation having character $\chi$. Then, we can write
$$
f(x)=\int Tr(\hat f(\pi_{\chi})\pi_{\chi}(x))d\mu(\chi)
$$
where
$$
\hat f(\pi_{\chi})=\int f(x)\pi_{\chi}(x^{-1}))dx
$$
Thus, in particular, if $f$ is a class function,
$$
f(x)=\int f(y)Tr(\pi_{\chi}(y^{-1})\pi_{\chi}(x))d\mu(\chi)dy
$$
$$
=\int f(y)\chi(y^{-1}x)d\mu(\chi)dy
$$
as follows from the Plancherel formula.

Problem: What is the relationship between $\psi_1(x)$ and $\psi(x)$ when $\psi$ is a class function ?

Note that if the group is compact, then if $\psi$ is a class function and $\pi$ is an irreducible unitary representation of dimension $d$, we have
$$
\hat\psi(\pi)=\int\psi(y)\pi(y)^*)dy=\int\psi(xyx^{-1})\pi(y^{-1})dydx
$$
$$
=\int\psi(y)\pi(x^{-1}y^{-1}x)dydx
$$
Now,
$$
\int\pi_{ab}(x^{-1}yx)dx=\int\pi_{ac}(x^{-1})\pi_{cd}(y)\pi_{db}(x)dx
$$
$$
=\pi_{cd}(y)\int\bar\pi_{ca}(x).\pi_{db}(x)dx=d^{-1}\pi_{cd}(y)\delta(c,d)\delta(a,b)
$$
$$
=d^{-1}\pi_{cc}(y)\delta(a,b)=d^{-1}\chi(y)\delta(a,b)
$$
where $\chi=Tr(\pi)$ is the character of $\pi$. This means that
$$
\hat\psi(\pi)=d^{-1}\chi(\psi)I
$$
Thus, for a class function on a compact group, we have
$$
\psi(g)=\sum_{\chi\in\hat G}d(\chi)^{-1}Tr(\chi(\psi^0)\pi(g))
$$
$$
=\sum_{\chi\in\hat G}d(\chi)^{-1}\chi(\psi^0)\chi(g)
$$
$$
=\sum_{\chi\in\hat G}d(\chi)^{-1}\chi(g)\int\psi(x)\bar\chi(x)dx
$$
This is the same as saying that
$$
\delta_g(x)=\sum_{\chi\in\hat G}d(\chi)^{-1}\chi(g)\bar\chi(x)
$$
Is this result true for an arbitrary semisimple Lie group when we replace $d(\chi)$ by the 
Plancherel measure/formal dimension on the space of irreducible characters ? In other words, is the result
$$
\delta_g(x)=\int_{\hat G}\chi(g)\bar\chi(x)d\mu(\chi)
$$
true ?
\bigskip

\chapter{Some problems in quantum information theory}

A problem in quantum information theory used in the proof of the coding theorem for Cq channels : Let $S,T$ be two non-negative operators with $S\leq 1$, ie, $0\leq S\leq 1, T\geq 0$. Then
$$
1-(S+T)^{-1/2}S(S+T)^{-1/2}\leq 2-2S+4T
$$
or equivalently,
$$
(S+T)^{-1/2}S(S+T)^{-1/2}\geq 2S-1-4T
$$
To prove this, we define the operators
$$
A=\sqrt T((T+S)^{-1/2}-1), B=\sqrt T
$$
Then, the equation
$$
(A-B)^*(A-B)\geq 0
$$
or equivalently,
$$
A^*A+B^*B\geq A^*B+B^*A
$$
gives
$$
((T+S)^{-1/2}-1)T+T((T+S)^{-1/2}-1)\leq T+((T+S)^{-1/2}-1)T((T+S)^{-1/2}-1)
$$
This equation can be rearranged go give
$$
2(T+S)^{-1/2}T+2T(T+S)^{-1/2}-4T\leq (T+S)^{-1/2}T.(T+S)^{-1/2}
$$
$$
=(T+S)^{-1/2}(T+S-S)(T+S)^{-1/2}=1-(T+S)^{-1/2}S(T+S)^{-1/2}
$$
A rearrangement of this equation gives
$$
4T+1+(S+T)^{-1/2}S(S+T)^{-1/2}\geq 2(S+T)^{-1/2}S(S+T)^{-1/2}+2(S+T)^{-1/2}T+2T(S+T)^{-1/2}
$$
So to prove the stated result, it suffices to prove that
$$
(S+T)^{-1/2}S(S+T)^{-1/2}+(S+T)^{-1/2}T+T(S+T)^{-1/2}\geq S
$$
\bigskip

Let $A,B$ be two states. The hypothesis testing problem amounts to choosing an operator
$0\leq T\leq 1$ such that
$$
C(T)=Tr(A(1-T))+Tr(BT)
$$
is a minimum. Now,
$$
C(T)=1+Tr((B-A)T)
$$
It is clear that this is a minimum iff
$$
T=\{B-A\leq 0\}=\{A-B\geq 0\}
$$
ie $T$ is the projection onto the space spanned by those eigenvectors of $B-A$ that have positive eigenvalues or equivalently the space spanned by those eigenvectors of $A-B$ that have non-negative eigenvalues. The minimum value of the cost function is then
$$
C(min)=Tr(A\{A<B\})+Tr(B\{A>B\})
$$
Note that the minimum is attained when $T$ is a projection. 

An inequality: Let $0\leq s\leq 1$. Then if $A,B$ are any two positive operators, we have
$$
Tr(A\{A<B\})\leq Tr(A^{1-s}B^s)
$$
To see this we note that
$$
Tr(A^{1-s}B^s)\leq Tr((B+(A-B)_+)^{1-s}B^s)\leq Tr(B+(A-B)_+)^{1-s}(B+(A-B)_+)^s)
$$
$$
=Tr(B+(A-B)_+)=Tr(B)+Tr((A-B)_+)=Tr(B)+Tr((A-B)\{A-B>0\})
$$
$$
=Tr(B\{A-B<0\})+Tr(A.\{A-B>0\})
$$
Other related inequalities are as follows:
$$
Tr(A^{1-s}B^s)\geq Tr((B-(B-A)_+)^{1-s}B^s)
$$
$$
\geq Tr((B-(B-A)_+)^{1-s}.(B-(B-A)_+)^s)=Tr((B-(B-A)_+)
$$
$$
=Tr(B)-Tr((B-A)_+)=Tr(B)-Tr((B-A)\{B>A\})
$$
$$
=Tr(B.\{A>B\})+Tr(A.\{B>A\})
$$
In particular, when $A,B$ are states, the minimum probability of discrimination error has the following upper bound:
$$
P(error)=Tr(A.\{B>A\})+Tr(B.\{A>B\})\leq min_{0\leq s\leq 1}Tr(A^{1-s}B^s)
$$
$$
=min_{0\leq s\leq 1}exp(\phi(s|A|B))
$$
where
$$
\phi(s|A|B)=log(Tr(A^{1-s}B^s))
$$
To obtain the minimum in the upper bound, we set the derivative w.r.t $s$ to zero to get
$$
d/ds(Tr(A^{1-s}B^s)=0
$$
or
$$
Tr(log(A)A^{1-s}B^s)=Tr(A^{1-s}B^slog(B))
$$
It is not clear how to solve this equation for $s$. Now define
$$
\psi_1(s)=log(Tr(A^{1-s}B^s))/s, \psi_2(s)=log(Tr(A^{1-s}B^s)/(1-s)
$$
Then an elementary application of Le'Hopital's rule gives us
$$
\psi_1(0+)=lim_{s\downarrow 0}\psi_1(s)=-D(A|B)=-Tr(A.(log(A)-log(B)),
$$
$$
\psi_2(1-)=lim_{s\uparrow 1}\psi_2(s)=-D(B|A)=-Tr(B.(log(B)-log(A))
$$
Consider now the problem of distinguishing between tensor product states $A^{\otimes n}$ and $B^{\otimes n}$. Let $P_n(e)$ denote the minimum error probability. Then, by the above,
$$
P_n(e)\leq min_{0\leq s\leq 1}exp(log Tr(A^{\otimes n(1-s)}B^{\otimes n s})
$$
Now,
$$
Tr(A^{\otimes n(1-s)}B^{\otimes ns})=(Tr(A^{1-s}B^s))^n
$$
Therefore, we can write
$$
limsup_{n\rightarrow\infty}n^{-1}log(P_n(e))\leq 
$$
$$
min_{0\leq s\leq 1}log Tr(A^{1-s}B^s)
$$
Define
$$
F(s)=Tr(A^{1-s}B^s)
$$
Then,
$$
F'(s)=-Tr(A^{1-s}log(A)B^s)+Tr(A^{1-s}B^slog(B))
$$
$$
F''(s)=Tr(A^{1-s}(log A)^2B^s)+Tr(A^{1-s}B^s(log(B))^2)
$$
$$
-Tr(A^{1-s}log(A)B^slog(B))+Tr(A^{1-s}.log(A).B^s.log(B))
$$
$$
=Tr(A^{1-s}(log A)^2B^s)+Tr(A^{1-s}B^s(log(B))^2)\geq 0
$$
Hence $F(s)$ is a convex function and hence $F'(s)$ is increasing. Further,
$$
F'(0)=-Tr(A.log(A))+Tr(A.log(B))=-D(A|B)\leq 0, F'(1)=-Tr(B.log(A))+Tr(B.log(B))=D(B|A)\geq 0
$$
Hence $F'(s)$ changes sign in $[0,1]$ and hence it must be zero at some $s\in[0,1]$, say
at $s_0$ and since $F$ is convex, it attains its minimum at $s_0$. 
\bigskip

Let $\rho,\sigma$ be two states and let 
$$
\rho=\sum_ip(i)|e_i><e_i|, \sigma=\sum_iq(i)|f_i><f_i|
$$
be their respective spectral decompositions. Define
$$
p(i)|<e_i|f_j>|^2=P(i,j), q(j)|<e_i|f_j>|^2=Q(i,j)
$$
Then $P,Q$ are both classical probability distributions. We have for $0\leq T\leq 1$,
$$
Tr(\rho(1-T))=\sum_{i,j}<e_i|\rho|f_j><f_j|1-T|e_i>
$$
$$
=\sum_{i,j}p(i)<e_i|f_j>(<f_j|e_i>-<f_j|T|e_i>)
$$
$$
=1-\sum_{i,j}p(i)<e_i|f_j>t(j,i)
$$
and likewise,
$$
Tr(\sigma T)=\sum_{i,j}<e_i|\sigma|f_j>t(j,i)=\sum_{i,j}q(j)<e_i|f_j>t(j,i)
$$
where
$$
t(j,i)=<f_j|T|e_i>
$$
So,
$$
Tr(\rho(1-T))+Tr(\sigma T)=1+\sum_{i,j}t(j,i)(q(j)-p(i))<e_i|f_j>t(j,i)
$$
Now,
$$
|\sum_{i,j}p(i)<e_i|f_j>t(j,i)|^2\leq\sum_{i,j}p(i)|e_i|f_j>|^2.\sum_{i,j}p(i)|t(j,i)|^2
$$
$$
=\sum_{i,j}p(i)|t(j,i)|^2
$$
Thus,
$$
Tr(\rho(1-T))\geq 1-\sum_{i,j}p(i)|t(j,i)|^2
$$
$$
Tr(\sigma T)=\sum_{i,j}q(j)<e_i|f_j>t(j,i)
$$
So
$$
Tr(\sigma T)^2\leq\sum_{i,j}q(j)|t(j,i)|^2
$$
Then, if $c$ is any positive real number,
$$
Tr(\rho(1-T))-c.Tr(\sigma T)\geq 1-\sum_{i,j}p(i)|t(j,i)|^2-c.\sum_{i,j}q(j)|t(j,i)|^2
$$
Suppose now that $T$ is a projection. Then,
$$
Tr(\rho(1-T))=\sum_ip(i)<e_i|1-T|e_i>=\sum_ip(i)<e_i|(1-T)^2|e_i>
$$
$$
=\sum_{i,j}p(i)|<e_i|1-T|f_j>|^2\geq_{i,j}a(i,j)|<e_i|1-T|f_j>|^2
$$
and likewise,
$$
Tr(\sigma T)=\sum_{i,j}q(j)|<f_j|T|e_i>|^2\geq\sum_{i,j}a(i,j)|<f_j|T|e_i>|^2
$$
where
$$
a(i,j)=min(p(i),q(j))
$$
Now,
$$
|<e_i|1-T|f_j>|^2+<e_i|T|f_j>|^2\geq (1/2)|<e_i|1-T|f_j>+<e_i|T|f_j>|^2
$$
$$
=(1/2)|<e_i|f_j>|^2
$$
Thus,
$$
Tr(\rho(1-T))+Tr(\sigma T)\geq (1/2)\sum_{i,j}a(i,j)|<e_i|f_j>|^2
$$
$$
=(1/2)\sum_{i,j}P(i,j)\chi_{p(i)\leq q(j)}+(1/2)\sum_{i,j}Q(i,j)\chi_{p(i)>q(j)}
$$
$$
=(1/2)\sum_{i,j}P(i,j)\chi_{P(i,j)\leq Q(i,j)}+(1/2)\sum_{i,j}Q(i,j)\chi_{P(i,j)>Q(i,j)}
$$
Now, for tensor product states, this result gives
$$
Tr(\rho^{\otimes n}(1-T_n))+Tr(\sigma^{\otimes n}T_n)\geq
$$
$$
(1/2)P^n(P^n\leq Q^n)+(1/2)Q^n(P^n>Q^n)
$$
for any measurement $T_n$ in the tensor product space, ie, $0\leq T_n\leq 1$. Now by the large deviation principle as $n\rightarrow\infty$, we have
$$
n^{-1}.log(P^n(P^n\geq Q^n))\approx -inf_{x\leq 0}I_1(x)
$$
where $I_1$ is the rate function under $P^n$ of the random family $n^{-1}log(P^n/Q^n)$.
Evaluating the moment generating function of this r.v gives us
$$
M_n(ns)=\Bbb E_{P^n}exp(s.log(P^n/Q^n))=\Bbb E_{P^n}(P^n/Q^n)^s
$$
$$
=(\Bbb E_P(P/Q)^s)^n=(\sum P^{1-s}Q^s)^n=M_s(P,Q)^n
$$
and hence the rate function is
$$
I_1(x)=sup_s(sx-log M_s(P,Q))=
$$
$$
sup_s(sx-log(\sum_{\omega}P(\omega)^{1-s}Q(\omega)^s))
$$
Then,
$$
inf_{x\leq 0}I_1(x)=sup_{s\geq 0}-log(\sum_{\omega}P(\omega)^{1-s}Q(\omega)^s))
$$
$$
=-inf_{s\geq 0}log(\sum_{\omega}P(\omega)^{1-s}Q(\omega)^s))
$$
This asymptotic result can also be expressed as
$$
liminf_{n\rightarrow\infty}inf_{0\leq T_n\leq 1}[n^{-1}log(Tr(\rho^{\otimes n}(1-T_n))+Tr(\sigma^{\otimes n}T_n))]
$$
$$
\geq -inf_{0\leq s\leq 1}[log(\sum P^{1-s}Q^s)]-inf_{0\leq s\leq 1}[log\sum Q^{1-s}P^s]
$$ 

Relationship between quantum relative entropy and classical relative entropy:
$$
Tr(\rho.log(\rho))=\sum_ip(i)log(p(i))
$$
$$
Tr(\rho.log(\sigma))=\sum_{i,j}p(i)log(q(j))<e_i|f_j>|^2
$$
$$
D(\rho|\sigma)=Tr(\rho.log(\rho)-\rho.log(\sigma))
$$
Therefore,
$$
D(\rho|\sigma)=\sum_{i,j}p(i)<e_i|f_j>|^2(log(p(i))+log(|<e_i|f_j>|^2))
$$
$$
-\sum_{i,j}p(i)<e_i|f_j>|^2(log(q(j))+log(|<e_i|f_j>|^2))
$$
(Note the cancellations)
$$
=\sum_{i,j}P(i,j)log(P(i,j))-\sum_{i,j}P(i,j)log(Q(i,j))
$$
$$
=D(P|Q)
$$
\bigskip

Let $X$ be a r.v. with moment generating function
$$
M(\lambda)=\Bbb E[exp(\lambda X)]
$$
Its logarithmic moment generating function is
$$
\Lambda(\lambda)=log(M(\lambda))
$$
Let $\mu=\Bbb E(X)$ be its mean. Then,
$$
\mu=\Lambda'(0)
$$
Further,
$$
\Lambda''(\lambda)\geq 0\forall\lambda
$$
ie $\Lambda$ is a convex function. Hence $\Lambda'(\lambda)$ is increasing for all $\lambda$. Now suppose $x\geq\mu$. Consider
$$
\psi(x,\lambda)=\lambda.x-\Lambda(\lambda)
$$
Since
$$
\partial_{\lambda}^2\psi(x,\lambda)=-\Lambda''(\lambda)<0
$$
it follows that $\partial_{\lambda}\psi(x,\lambda)$ is decreasing or equivalently that $\psi(x,\lambda)$ is concave in $\lambda$. Suppose $\lambda_0$ is such that
$$
\partial_{\lambda}\psi(x,\lambda_0)=0
$$
Then it follows that for a fixed value of $x$, $\psi(x,\lambda)$ attains its maximum at
$\lambda_0$. Now $\partial_{\lambda}\psi(x,\lambda_0)=0$ and $\partial_{\lambda}\psi(x,\lambda)$ is decreasing in $\lambda$. It follows that $\partial_{\lambda}\psi(x,\lambda)\leq 0$ for $\lambda\leq\lambda_0$. Hence, $\psi(x,\lambda)$ is decreasing for $\lambda\geq\lambda_0$. Also for $\lambda<\lambda_0$, $\partial_{\lambda}\psi(x,\lambda)\geq 0$
(since $\partial_{\lambda}\psi(x,\lambda_0)=0$ and $\partial_{\lambda}\psi(x,\lambda)$ is decreasing for all $\lambda$) which means that $\psi(x,\lambda)$ is increasing for $\lambda<\lambda_0$. Now
$$
\partial_{\lambda}\psi(x,0)=x-\mu\geq 0
$$
by hypothesis. Since $\partial_{\lambda}\psi(x,\lambda)$ is decreasing in $\lambda$, it follows that $\partial_{\lambda}\psi(x,\lambda)\geq 0$ for all $\lambda\leq 0$. This means that $\psi(x,\lambda)$ is increasing for all $\lambda\geq 0$. In particular,
$$
\psi(x,\lambda)\leq\psi(x,0)=x,\forall\lambda\leq 0
$$
Thus,
$$
I(x)=sup_{\lambda\in\Bbb R}\psi(x,\lambda)=sup_{\lambda\geq 0}\psi(x,\lambda)
$$
in the case when $x\geq\mu$. Consider now the case when $x\leq\mu$.
\bigskip

\section{Quantum information theory and quantum blackhole physics}

Let $\rho_1(0)$ be the state of the system of particles within the event horizon of a Schwarzchild blackhole and $\rho_2(0)$ the state outside. At time $t=0$, the state of the total system to the interior and the exterior of the blackhole is
$$
\rho(0)=\rho_1(0)\otimes\rho_2(0)
$$
Here, as a prototype example, we are considering the Klein-Gordon field of particles in the blackhole as well as outside it. Let $\chi(x)$ denote this KG field. Its action functional is given by
$$
S[\chi]=(1/2)\int(g^{\mu\nu}(x)\sqrt{-g(x)}\chi_{,\mu}(x)\chi_{,\nu}(x)-\mu^2\sqrt{-g(x)}\chi(x)^2)d^4x
$$
$$
=(1/2)\int[g^{\mu\nu}(x)\chi_{,\mu}(x)\chi_{,\nu}(x)-\mu^2\chi(x)^2]\sqrt{-g(x)}d^4x
$$
which in the case of the Schwarzchild blackhole reads
$$
S[\chi]=(1/2)\int[\alpha(r)^{-1}\chi_{,t}^2-\alpha(r)\chi_{,r}^2-r^{-2}\chi_{,\theta}^2-r^{-2}sin(\theta)^{-2}\chi_{,\phi}^2]r^2sin(\theta)dtdrd\theta d\phi=\int\mathcal Ld^4x
$$
The corresponding KG Hamiltonian can also be written down. The canonical position field is $\chi$ and the canonical momentum field is
$$
\pi=\delta S/\delta\chi_{,t}=\alpha(r)^{-1}\chi_{,t}
$$
so that the Hamiltonian density is
$$
\mathcal H=\pi.\chi_{,t}-\mathcal L=
$$
$$
(1/2)[\alpha(r)^{-1}\chi_{,t}^2+\alpha(r)\chi_{,r}^2+r^{-2}\chi_{,\theta}^2+r^{-2}sin(\theta)^{-2}\chi_{,\phi}^2]
$$
$$
=(1/2)[\alpha(r)\pi^2+\alpha(r)\chi_{,r}^2+r^{-2}\chi_{,\theta}^2+r^{-2}sin(\theta)^{-2}\chi_{,\phi}^2+\mu^2\chi^2]
$$
The KG field Hamiltonian is then
$$
H=\int\mathcal Hd^3x=\int\mathcal Hr^2sin(\theta)drd\theta d\phi
$$

The radial Hamiltonian: Assume that the angular oscillations of the KG field are as per the classical equations of motion of the field corresponding to an angular momentum square of $l(l+1)$. Then, we can write
$$
\chi(t,r,\theta,\phi)=\psi(t,r)Y_{lm}(\theta,\phi)
$$
an substituting this into the action functional, performing integration by parts using
$$
L^2Y_{lm}(\theta,\phi)=l(l+1)Y_{lm}(\theta,\phi),
$$
$$
L^2=-(sin(\theta)^{-1}\frac{\partial}{\partial\theta}sin(\theta)\frac{\partial}{\partial\theta}+\frac{1}{sin^2(\theta)}\frac{\partial^2}{\partial\phi^2}
$$
gives us the "radial action functional" as
$$
S_l[\phi]=(1/2)\int(\alpha(r)^{-1}\psi_{,t}^2-\alpha(r)\psi_{,r}^2-l(l+1)\psi^2/r^2-\mu^2\psi^2)r^2dtdr
$$
and the corresponding radial Hamiltonian as
$$
H_l(\psi,\pi)=(1/2)\int_0^{\infty}[\alpha(r)(\pi^2+\psi_{,r}^2)+(\mu^2+l(l+1)/r^2)\psi^2]r^2dr
$$
where $\psi=\psi(t,r),\pi=\pi(t,r)$, the canonical position and momentum fields are functions of only time and the radial coordinate. 
\bigskip

\section{Harmonic analysis on the Schwarz space of $G=SL(2,\Bbb R)$}

Let $f_{mn}(x,\mu)$ denote the matrix elements of the unitary principal series appearing in the Plancherel formula and let $\phi_{kmn}(x)$ the matrix elements of the discrete series appearing in the Plancherel formula. Then we have a splitting of the Schwarz space into the direct sum of the discrete series Schwarz space and the principal series Schwarz space. Thus, for any $f\in L^2(G)$, we have the Plancherel formula
$$
\int_G|f(x)|^2dx=\sum_{kmn}d(k)|<f|\phi_{kmn}>|^2+\int_{\Bbb R}|<f|f_{mn}(.,\mu)>|^2c(mn,\mu)d\mu
$$
where $d(k)'s$ are the formal dimensions of the discrete series while $c(mn,\mu)$ are determined from the wave packet theory for principal series, ie, from the asymptotic behaviour of the principal series matrix elements. The asymptotic behaviour of the principal and discrete matrix elements is obtained from the differential equations
$$
\Omega.f_{mn}(x,\mu)=-\mu^2f_{mn}(x,\mu)
$$
where $\Omega$ is the Casimir element for $SL(2,\Bbb R)$. We can express $x\in G$ as
$x=u(\theta_1)a(t)u(\theta_2)$ and $\Omega$ as a second order linear partial differential operator in $\partial/\partial t, \partial/\partial\theta_k, k=1,2$. Then, observing that
$$
f_{mn}(u(\theta_1)a(t)u(\theta_2),\mu)=exp(i(m\theta_1+n\theta_2))f_{mn}(a(t),\mu)
$$
we get on noting that the restriction of the Casimir operator to $\Bbb R.H$ is given by
$$
(\Omega f)(a(t))=\Delta_L(t)^{-1}\frac{d^2}{dt^2}\Delta_L(t)f(a(t))
$$
where
$$
\Delta_L(t)=e^t-e^{-t}
$$
as follows by applying both sides to the finite dimensional irreducible characters of $SL(2,\Bbb R)$ restricted to $\Bbb R.H$ on noting that the restriction of such a character is given by $(exp(mt)-exp(-mt))/(exp(t)-exp(-t))$, we get a differential equation of the form
$$
\frac{d^2}{dt^2}f_{mn}(a(t),\mu)-q_{mn}(t)f_{mn}(a(t),\mu)=-\mu^2f_{mn}(a(t),\mu)
$$
where $q_{mn}(t)$ is expressed in terms of the hyperbolic Sinh functions. From this equation, the asymptotic behaviour of the matrix elements $f_{mn}(a(t),\mu)$ as $|t|\rightarrow\infty$ may be determined. It should be noted that the discrete series modules can be realized within the principal series modules and hence the asymptotic behaviour of the matrix elements $f_{mn}(a(t),\mu)$ will determine that of the principal series when $\mu$ is non-integral and that of the discrete series when $\mu$ is integral. Specifically,
consider a principal series representation $\pi(x)$. Let $e_m$ denote its highest weight vector and define
$$
f_m(x)=<\pi(x)e_m|e_m>
$$
where now $x$ is regarded as an element of the Lie algebra $\mathfrak g$ of $G$. We have for any $a\in U(\mathfrak g)$, the universal enveloping algebra of $\mathfrak g$, the fact that 
$$
\pi(a).e_m=0\implies af_m(x)=f(x;a)=0
$$
This means that if we consider the $U(\mathfrak g)$-module $V=U(\mathfrak g)f_m$ and the map $T:\pi(U(\mathfrak g))e_m\rightarrow V$ defined by $T(\pi(a)e_m)=a.f_m$, then it is clear that $T$ is a well defined linear map with kernel
$$
K_T=\{\pi(a)e_m:a\in U(\mathfrak g), a.f_m=0\}
$$
Note that $\pi(a)e_m\in K_T$ implies $a.f_m=0$ implies $b.a.f_m=0$ implies $\pi(b)\pi(a)e_m=\pi(b.a)e_m\in K_T$ for any $a,b\in U(\mathfrak g)$. Thus, $K_T$ is $U(\mathfrak g)$ invariant.  Then, the $U(\mathfrak g)$-module $V$ is isomorphic to the quotient module
$\pi(U(\mathfrak g))e_m/K_T$. However, $\pi(U(\mathfrak g))e_m$ is an irreducible (discrete series) module. Hence, the submodule $K_T$ of it must be zero and hence, we obtain the fundamental conclusion that $V=U(\mathfrak g)f_m$ is an irreducible discrete series module. Note that $f_m$ is a principal series (reducible since $m$ is an integer)
matrix element and this means that we have realized the discrete series representations within the principal series. This also means that in order to study the asymptotic properties of the matrix elements of the discrete series, it suffices to study asymptotic properties of differential operators in $U(\mathfrak g)$ acting on the principal series matrix elements $f_m$.
\bigskip

\section{Monotonicity of quantum entropy under pinching}

Let $|\psi_1>,...,|\psi_n>$ be normalized pure states, not necessarily orthogonal. Let
$p(i), i=1,2,..., n$ be a probability distribution and let $|i>, i=1,2,..., n$ be an orthonormal set. Consider the pure state
$$
|\psi>=\sum_{i=1}^n\sqrt{p(i)}|\psi_i>\otimes|i>
$$
Note that this is a state because
$$
<\psi|\psi>=\sum_{i,j}\sqrt{p(i)p(j)}<\psi_j|\psi_i><j|i>=\sum_{i,j}\sqrt{p(i)p(j)}<\psi_j|\psi_i>\delta(j,i)=\sum_ip(i)=1
$$
We have
$$
\rho_1=Tr_2(|\psi><\psi|)=\sum_ip(i)|\psi_i><\psi_i|
$$
$$
\rho_2=Tr_1(|\psi><\psi|=\sum_i\sqrt{p(i)p(j)}<\psi_j|\psi_i>|i><j|
$$
Then by Schmidt's theorem,
$$
H(\rho_1)=H(\rho_2)\leq H(p)
$$
Now let $\sigma_1,\sigma_2$ be two mixed states and let $p>0, p+q=1$. Consider the spectral decompositions
$$
\sigma_1=\sum_ip(i)|e_i><e_i|, \sigma_2=\sum_iq(i)|f_i>f_i|
$$
Then,
$$
p\sigma_1+q\sigma_2=\sum_i(pp(i)|e_i><e_i|+qq(i)|f_i><f_i|)
$$
So by the above result,
$$
H(p\sigma_1+q\sigma_2)\leq H(\{pp(i),qq(i)\})
$$
$$
=-\sum_i(pp(i)log(pp(i))+qq(i)log(qq(i)))
$$
$$
=-p.log(p)-q.log(q)+pH(\{p(i)\})+qH(\{q_i\})=H(p,q)+p.H(\sigma_1)+q.H(\sigma_2)
$$
where
$$
H(p,q)=-p.log(p)-q.log(q)+pH(\{p(i)\})
$$
Now consider several mixed states $\sigma_1,...,\sigma_n$ and a probability distribution
$p(i), i=1,2,..., n$. Then 
$$
\sum_{i=1}^np(i)\sigma_i=p(1)\sigma_1+(1-p(1))\sum_{i=2}^np(i)/(1-p(1))\sigma_i
$$
So the same argument and induction gives
$$
H(\sum_{i=1}^np(i)\sigma_i)\leq H(p(1),1-p(1))+p(1)H(\sigma_1)+(1-p(1))\sum_{i=2}^n(p(i)/(1-p(1)))H(\sigma_i)
$$
$$
+H(p(i)/(1-p(1)), i=2,3,..., n)
$$
$$
=H(p(i), i=1,2,..., n)+\sum_{i=1}^np(i)H(\sigma_i)
$$
Remark: Let $\rho$ be a state and $\{P_i\}$ a PVM. Define the state
$$
\sigma=\sum_iP_i\rho.P_i=1-\sum_iP_i(1-\rho)P_i
$$
Then
$$
0\leq D(\rho|\sigma)=Tr(\rho.log(\rho))-Tr(\rho.log(\sigma))
$$
$$
=-H(\rho)-\sum_iTr(\rho.P_i.log(P_i\rho.P_i)P_i)=-H(\rho)-\sum_iTr(P_i\rho.P_i.log(P_i\rho.P_i)
$$
$$
=-H(\rho)-\sum_iTr(P_i\rho.P_i.log(\sum_jP_j\rho.P_j))=-H(\rho)+H(\sigma)
$$
Thus,
$$
H(\sigma)\geq H(\rho)
$$
which when stated in words, reads, pinching increases the entropy of a state.

Remark: We've made use of the following facts from matrix theory: 
$$
log(\sum_iP_i.\rho.P_i)=\sum_iP_i.log(\rho)P_i
$$
because the $P_i's$ satisfy $P_jP_i=0, j\neq i$. In fact, we write
$$
\sum_iP_i\rho.P_i=1-\sum_iP_i(1-\rho)P_i=1-A
$$
and use
$$
log(1-z)=-z-z^2/2-z^3/3-...
$$
combined with $P_jP_j=P_i\delta(i,j)$ to get
$$
log(\sum_iP_i\rho.P_i)=-A-A^2/2-A^3/3-...
$$
where
$$
A=\sum_iP_i(1-\rho)P_i, A^2=\sum_iP_i(1-\rho)^2P_i,..., A^n=-\sum_iP_i(1-\rho)^nP_i
$$
so that
$$
log(1-A)=\sum_iP_i.log(\rho)P_i
$$
A particular application of this result has been used above in the form
$$
H(\sum_i\sqrt{p(i)p(j)}<\psi_j|\psi_i>|i><j|)\leq H(\sum_{i,k}\sqrt{p(i)p(j)}<\psi_j|\psi_i><k|i><j|k>|k><k|)
$$
$$
=H(\sum_kp(k)|k><k|)=H(\{p(k)\})=-\sum_kp(k).log(p(k))
$$
\bigskip

\chapter{Lectures on Statistical Signal Processing}

[1] Large deviation principle applied to the least squares problem of estimating vector parameters from sequential iid  vector measurements in a linear model.
$$
X(n)=H(n)\theta+V(n), n=1,2,...
$$
$V(n)'s$ are iid and have the pdf $p_V$. Then $\theta$ is estimated based on the measurements $X(n), n=1,2,..., N$ using the maximum likelihood method as
$$
\hat\theta(N)=argmax_{\theta}N^{-1}\sum_{n=1}^Nlog(p_V(X(n)-H(n)\theta))
$$
Consider now the case when $H(n)$ is a matrix valued stationary process independent of $V(.)$. Then, the above maximum likelihood estimate is replaced by
$$
\hat\theta(N)=argmax_{\theta}\int[\Pi_{n=1}^Np_V(X(n)-H(n)\theta)]p_H(H(1),..., H(N))\Pi_{j=1}^N dH(j)
$$
In the special case when the $H(n)'s$ are iid, the above reduces to
$$
\hat\theta(N)=argmax_{\theta}N^{-1}\sum_{n=1}^Nlog(\int p_V(X(n)-H\theta)p_H(H)dH)
$$
The problem is to calculate the large deviation properties of $\hat\theta(N)$, namely the rate at which $P(|\hat\theta(N)-\theta|>\delta)$ converges to zero as $N\rightarrow\infty$.
\bigskip

\section{Large deviation analysis of the RLS algorithm}

Consider the problem of estimating
$\theta$ from the sequential linear model
$$
X(n)=H(n)\theta+V(n), n\geq 1
$$
sequentiall. The estimate based on data collected upto time $N$ is
$$
\theta(N)=[\sum_{n=1}^NH(n)^TH(n)]^{-1}[\sum_{n=1}^NH(n)^TX(n)]
$$
To cast this algorithm in recursive form, we define
$$
R(N)=\sum_{n=1}^NH(n)^TH(n), r(N)=\sum_{n=1}^NH(n)^TX(n)
$$
Then,
$$
R(N+1)=R(N)+H(N+1)^TH(N+1), r(N+1)=r(N)+H(N+1)^TX(N+1)
$$
Then,
$$
R(N+1)^{-1}=R(N)^{-1}-R(N)^{-1}H(N+1)^T(I+H(N+1)H(N+1)^T)^{-1}H(N+1)R(N)^{-1}
$$
so that
$$
\theta(N+1)=\theta(N)+K(N+1)(X(N+1)-H(N+1)\theta(N))
$$
where $K(N+1)$ is a Kalman gain matrix expressible as a function of $R(N),H(N+1)$.
Our aim is to do a large deviation analysis of this recursive algorithm. Define
$$
\delta\theta(N)=\theta(N)-\theta
$$
Then we have
$$
\delta\theta(N+1)=\delta\theta(N)+K(N+1)(X(N+1)-H(N+1)\theta-H(N+1)\delta\theta(N))
$$
$$
=\delta\theta(N)+K(N+1)(V(N+1)-H(N+1)\delta\theta(N))
$$
We assume that the noise process $V(.)$ is small and perhaps even non-Gaussian. This fact is emphasized by replacing $V(n)$ with $\epsilon.V(n)$. Then, the problem is to calculate the relationship between the LDP rate function of $\delta\theta(N)$ and $\delta\theta(N+1)$.
\bigskip

\section{Problem suggested by my colleague Prof. Dhananjay}

Large deviation analysis of magnet with time varying magnetic moment falling thorugh a coil. Assume that the coil is wound around a tube with $n_0$ turns per unit length extending from a height of $h$ to $h+b$. The tube is cylindrical and extends from $z=0$ to $z=d$. Taking relativity into account via the retarded potentials, if $m(t)$ denotes the magnetic moment of the magnet, then the vector potential generated by it is approximately obtained from the retarded potential formula as
$$
A(t,r)=\mu m(t-r/c)\times r/4\pi r^3
$$
where we assume that the magnet is located near the origin. This approximates to
$$
A(t,r)=\mu m(t)\times r/4\pi r^3-\mu m'(t)\times r/4\pi cr^2
$$
Now Assume that the magnet falls through the central axis with its axis always being oriented along the $-\hat z$ direction. Then, after time $t$, let its $z$ coordinate be
$\xi(t)$. We have the equation of motion
$$
M\xi''(t)=F(t)-Mg
$$
where $F(t)$ is the $z$ component of the force exerted by the magnetic field generated by the current induced in the coil upon the magnet. Now, the flux of the magnetic field generated by the magnet through the coil is given by
$$
\Phi(t)=n_0\int_{X_2+Y^2\leq R_T^2,h\leq z\leq h+b}B_z(t,X,Y,Z)dXdYdZ
$$
where
$$
B_z(t,X,Y,Z)=\hat z.B(t,X,Y,Z),
$$
$$
B(t,X,Y,Z)=curl A(t,{\bf R}-\xi(t)\hat z)
$$
where
$$
{\bf R}=(X,Y,Z)
$$
and where $R_T$ is the radius of the tube. The current through the coil is then
$$
I(t)=\Phi'(t)/R_0
$$
where $R_0$ is the coil resistance. The force of the magnetic field generated by this current upon the falling magnet can now be computed easily. To do this, we first compute the magnetic field $B_c(t,X,Y,Z)$ generated by the coil using the usual retarded potential formula
$$
B_c(t,X,Y,Z)=curl A_c(t,X,Y,Z)
$$
$$
A_c(t,X,Y,Z)=(\mu/4\pi)\int_0^{2n_0\pi}(I(t-|{\bf R}-R_T(\hat x.cos(\phi)+\hat y.sin(\phi))-\hat z(b/2n_0\pi)\phi|/c).
$$
$$
.(-\hat x.sin(\phi)+\hat y.cos(\phi))/|{\bf R}-R_T(\hat x.cos(\phi)+\hat y.sin(\phi))-\hat z(b/2n_0\pi)\phi|)R_Td\phi
$$
Note that $A_c(t,X,Y,Z)$ is a function of $\xi(t)$ and also of $m(t)$ and $m'(t)$. It follows that $I(t)$ is a function of $\xi(t),\xi'(t),m(t),m'(t),m''(t)$.  Then we calculate the interaction energy between the magnetic fields generated by the magnet and by the coil as
$$
U_{int}(\xi(t),\xi'(t),t)=(2\mu)^{-1}\int(B_c(t,X,Y,Z),B(t,X,Y,Z))dXdYdZ
$$
and then we get for the force exerted by the coil's magnetic field upon the falling magnet as
$$
F(t)=-\partial U_{int}(\xi(t),\xi'(t),t)/\partial\xi(t)
$$

Remark: Ideally since the interaction energy between the two magnetic fields appears in the Lagrangian $\int(E^2/c^2-B^2)d^3R/2\mu$, we should write down the Euler-Lagrange equations using the Lagrangian
$$
L(t,\xi'(t),\xi'(t))=M\xi'(t)^2/2-U_{int}(\xi(t),\xi'(t),t)-Mg\xi(t)
$$
in the form
$$
\frac{d}{dt}\partial L/\partial\xi'-\partial L/\partial\xi=0
$$
which yields
$$
M\xi''(t)-\frac{d}{dt}\partial U/\partial\xi'+\partial U_{int}/\partial\xi+Mg=0
$$
\bigskip

\section{Fidelity between two quantum states}

Let $\rho,\sigma$ be two quantum states in the same Hilbert space $\mathcal H_1$. Let $P(\rho),P(\sigma)$ denote respectively the set of all purifications of $\rho$ and $\sigma$ w.r.t the same reference Hilbert space $\mathcal H_2$. Thus, if $|u>\in\mathcal H_1\otimes\mathcal H_2$ is a purification of $\rho$, ie $|u>\in\mathcal P(\rho)$, then
$$
\rho=Tr_2|u><u|
$$
and likewise for $\sigma$. Then, the fidelity between $\rho$ and $\sigma$ is defined as
$$
F(\rho,\sigma)=sup_{|u>\in P(\rho), |v>\in P(\sigma)}|<u|v>|
$$

Theorem: 
$$
F(\rho,\sigma)=Tr(|\sqrt{\rho}.\sqrt{\sigma}|)
$$
Proof: Let $|u>\in P(\rho)$. Then, we can write
$$
|u>=\sum_i|u_i\otimes w_i>
$$
where $\{|w_i>\}$ is an onb for $\mathcal H_2$ and
$$
\rho=\sum_i|u_i><u_i|
$$
Likewise if $|v>$ is a purification of $\sigma$, we can write
$$
|v>=\sum_i|v_i\otimes w_i>
$$
where
$$
\sigma=\sum_i|v_i><v_i|
$$
But then
$$
<u|v>=\sum_i<u_i|v_i>
$$
Define the matrices
$$
X=\sum_i|u_i><w_i|, Y=\sum_i|v_i><w_i|
$$
Then we have
$$
\rho=XX^*, \sigma=YY^*
$$
Hence, by the polar decomposition theorem, we can write
$$
X=\sqrt{\rho}U, Y=\sqrt{\sigma}V
$$
where $U,V$ are unitary matrices. Then,
$$
<u|v>=\sum_i<u_i|v_i>=Tr\sum_i|v_i><u_i|= Tr(YX^*)=Tr(\sqrt{\sigma}VU^*\sqrt{\rho})
$$
$$
=Tr(\sqrt{\rho}\sqrt{\sigma}VU^*)=Tr(\sqrt{\rho}.\sqrt{\sigma}W)
$$
where $W=VU^*$ is a unitary matrix. It is then clear from the singular value decomposition of $\sqrt{\rho}.\sqrt{\sigma}$ that the maximum of $|<u|v>|$ which equals the maximum of
$|Tr(\sqrt{\rho}.\sqrt{\sigma}W)|$ as $W$ varies over all unitary matrices is simply the sum of the singular values of $\sqrt{\rho}.\sqrt{\sigma}$, ie,
$$
F(\rho,\sigma)=Tr|\sqrt{\rho}\sqrt{\sigma}|
$$
$$
=Tr[(\sqrt{\rho}\sigma.\sqrt{\rho})^{1/2}]=Tr[(\sqrt{\sigma}.\rho.\sqrt{\sigma})^{1/2}]
$$
\bigskip

\section{Stinespring's representation of a TPCP map, ie, of a quantum operation}

Let $K$ be a quantum operation. Thus, for any state $\rho$,
$$
K(\rho)=\sum_{a=1}^NE_a\rho.E_a^*, \sum_{a=1}^NE_a^*E_a=1
$$
We claim that $K$ can be represented as
$$
K(\rho)=Tr_2(U(\rho\otimes |u><u|)U^*)
$$
where $\rho$ acts in $\mathcal H_1$, $|u>\in\mathcal H_2$ and $U$ is a unitary operator in $\mathcal H_1\otimes\mathcal H_2$. To see this, we define
$$
V^*=[E_1^*,E_2^*,..., E_N^*]
$$
Then,
$$
V^*V=I_p
$$
In other words, the columns of $V$ are orthonormal. We assume that
$$
\rho\in\Bbb C^{p\times p}
$$
so
$$
V\in\Bbb C^{Np\times p}
$$
Note that
$$
V\rho V^*=((E_a\rho E_b^*))_{1\leq a,b\leq N}
$$
as an $Np\times Np$ block structured matrix, each block being of size $p\times p$. Now define
$$
|u>=[1,0,...,0]^T\in\Bbb C^N
$$
Then,
$$
|u><u|\otimes\rho=\left(\begin{array}{cc}\rho&0\\0&0\end{array}\right)\in\Bbb C^{Np\times Np}
$$
Since the columns of $V$ are orthonormal, we can add columns to it to make it a square $Np\times Np$ unitary matrix $U^*$:
$$
U=[V,W]\in\Bbb C^{Np\times Np}, U^*U=I_{Np}
$$
Then,
$$
U(|u><u|\otimes\rho)U^*=[V,W]\left(\begin{array}{cc}\rho&0\\0&0\end{array}\right)\left(\begin{array}{cc}V^*\\W^*\end{array}\right)
$$
$$
=V\rho V^*
$$
from which it follows that
$$
Tr_1(U(|u><u|\otimes\rho)U^*)=Tr_1(V\rho V^*)=\sum_aE_a\rho E_a^*
$$

Remark: Consider the block structured matrix
$$
{\bf X}=(({\bf B}_{ij}))_{1\leq i,j\leq N}
$$
where each ${\bf B}_{ij}$ is a $p\times p$ matrix. Then
$$
Tr_1({\bf X})=\sum_i{\bf B}_{ii}\in\Bbb C^{p\times p}
$$
and
$$
Tr_2({\bf X})=((Tr({\bf B}_{ij})))_{1\leq i,j\leq N}\in\Bbb C^{N\times N}
$$
\bigskip

\section{RLS lattice algorithm, statistical performance analysis}

The RLS lattice algorithm involves updating the forward and backward prediction parameters simultaneously in time and order. In order to carry out a statistical performance analysis, we first simply
look at the block processing problem involving estimating the forward prediction filter parameter vector ${\bf a}_{N,p}$ of order $p$ based on data collected upto time $N$.
Let
$$
{\bf x}(n)=[x(n),x(n-1),..., x(0)]^T, z^{-r}{\bf x}(n)=[x(n-r),x(n-r-1),..., x(0),0,...,0]^T, r=1,2,...
$$
where these vectors are all of size $N+1$. Form the data matrix of order $p$ at time $N$:
$$
{\bf X}_{N,p}=[z^{-1}{\bf x}(n),..., z^{-p}{\bf x}(n)]\in\Bbb R^{N+1\times p}
$$
Then the forward prediction filter coefficient estimate is
$$
{\bf a}_{N,p}=({\bf X}_{N,p}^T{\bf X}_{N,p})^{-1}{\bf X}_{N,p}^T{\bf x}(N)
$$
If $x(n),n\in\Bbb Z$ is a zero mean stationary Gaussian process with autocorrelation
$R(\tau)=\Bbb E(x(n+\tau)x(n))$, then the problem is to calculate the statistics of
${\bf a}_{N,p}$. To this end, we observe that
$$
N^{-1}(z^{-i}{\bf x}(N))^T(z^{-j}{\bf x}(N))=
$$
$$
N^{-1}\sum_{n=max(i,j)}^Nx(n-i)x(n-j)\approx R(j-i)
$$
So we write
$$
N^{-1}(z^{-i}{\bf x}(N))^T(z^{-j}{\bf x}(N))=R(j-i)+\delta R_N(j,i)
$$
where now $\delta R_N(j,i)$ is a random variable that is a quadratic function of the Gaussian process $\{x(n)\}$ plus a constant scalar. Equivalently,
$$
N^{-1}{\bf X}_{N,p}^T{\bf X}_{N,p}={\bf R}_p+\delta{\bf R}_{N,p}
$$
where
$$
{\bf R}_p=((R(j-i)))_{1\leq i,j\leq p}
$$
is the statistical correlation matrix of size $p$ and $\delta{\bf R}_{N,p}$ is a random matrix. Likewise,
$$
N^{-1}(z^{-i}{\bf x}(N))^T{\bf x}(N)=N^{-1}\sum_{n=i}^Nx(n-i)x(n)=R(i)+\delta R_N(0,i)
$$
Thus,
$$
N^{-1}{\bf X}_{N,p}^T{\bf x}(n)={\bf r}(1:p)+\delta{\bf r}_N(0,1:p)
$$
where
$$
{\bf r}(1:p)=[R(1),..., R(p)]^T, \delta{\bf r}_N(1:p)=[\delta R_N(0,1),...,\delta R_N(0,p)]^T
$$
\bigskip

\section{Approximate computation of the matrix elements of the principal and discrete series for $G=SL(2,\Bbb R)$}

These matrix elements $f_{mn}(x,\mu)$ for $\mu$ non-integral are the irreducible principal series matrix elements and $f_{mm}(x,m)=f_m(x)$ for $k$ integral generate the Discrete series $D_m$-modules when acted upon by $U(\mathfrak g)$, the universal enveloping algebra of $G$.  These matrix elements satisfy the equations
$$
f_{mn}(u(\theta_1)xu(\theta_2),\mu)=exp(i(m\theta_1+n\theta_2))f_{mn}(x,\mu),
$$
$$
\Omega f_{mn}(x,\mu)=-\mu^2f_{mn}(x,\mu)
$$
the second of which, in view of the first and the expression of the Casimir operator $\Omega$ in terms of $t,\theta_1,\theta_2$ in the singular value decomposition $x=u(\theta_1)a(t)u(\theta_2)$ simplifies to an equation of the form
$$
d^2f_{mn}(t,\mu)/dt^2-q_{mn}(t)f_{mn}(t,\mu)+\mu^2f_{mn}(t,\mu)=0
$$
where $f_{mn}(t,\mu)=f_{mn}(a(t),\mu)$. This determines the matrix elements of the principal series. For the discrete series matrix elements, we observe that these are
$$
\phi_{kmn}(x)=<\pi_k(x)e_m|e_n>
$$
where $\pi_k$ is the reducible principal series for integral $k$. Replacing $k$ by $-k$ where $k=1,2,...$, the discrete series matrix elements are
$$
\phi_{-kmn}(x)=<\pi_{-k}(x)e_m|e_n>, m,n=-k-1,-k-2,...
$$
with
$$
\pi(X)e_{-k-1}=0
$$
Note that we can using the method of lowering operators, write
$$
\pi_{-k}(Y)^re_{-k-1}=c(k,r)e_{-k-1-r}, r=0,1,2,...
$$
where $c(k,r)$ are some real constants. Thus, writing
$$
f_{k+1}(x)=<\pi_{-k}(x)e_{-k-1}|e_{-k-1}>, x\in G,
$$
we have for the discrete series matrix elements,
$$
\phi_{-k,-k-1-r,-k-1-s}(x)=<\pi_{-k}(xY^r)e_{-k-1}|\pi_{-k}(Y^s)e_{-k-1}>
$$
$$
=<\pi_{-k}(Y^{*s}xY^r)e_{-k-1}|e_{-k-1}>=f_{k+1}(Y^{*s};x;Y^r)
$$
In this way, all the discrete series matrix elements can be determined by applying differential operators on $f_{k+1}(x)$.

Now given an image field $h(x)$ on $SL(2,\Bbb R)$, in order to extract invariant features, or to estimate the group transformation element parameter from a noisy version of it, we must first evaluate its matrix elements w.r.t the discrete and principal series, ie,
$$
\int_Gh(x)\phi_{-k,-k-1-r,-k-1-s}(x)dx, \int_Gh(x)f_{mn}(x,\mu)dx
$$
Now,
$$
\int_Gh(x)\phi_{-k,-k-1-r,-k-1-s}(x)dx=\int_Gh(x)f_{k+1}(Y^{*s};x;Y^r)dx
$$
$$
=\int_Gh(Y^s;x;Y^{*r})f_{k+1}(x)dx
$$
\bigskip

\section{Proof of the converse part of the Cq coding theorem}

Let $\phi(i),i=1,2,..., N$ be a code.
Note that the size of the code is $N$. Let $s\leq 0$ and define
$$
p_s=argmax_pI_s(p,W)
$$
where
$$
I_s(p,W)=min_{\sigma}D_s(p\otimes W|p\otimes\sigma)
$$
where
$$
p\otimes W=diag[p(x)W(x), x\in A], p\otimes\sigma=diag[p(x)sigma,x\in A]
$$
Note that
$$
D_s(p\otimes W|p\otimes\sigma]=log(Tr((p\otimes W)^{1-s}(p\otimes\sigma)^s)/(-s)
$$
$$
=(1/-s)\log sum_xTr(p(x)W(x))^{1-s}(p(x)\sigma)^s)
$$
$$
=
(1/-s)log\sum_xp(x)Tr(W(x)^{1-s}\sigma^s)
$$
so that
$$
lim_{s\rightarrow 0}D_s(p\otimes W|p\otimes\sigma)=
$$
$$
\sum_xp(x)Tr(W(x)log(W(x))-Tr(W_p.log(\sigma))
$$
where
$$
W_p=\sum_xp(x)W_x
$$
Note that this result can also be expressed as
$$
lim_{s\rightarrow 0}D_s(p\otimes W|p\otimes\sigma)=\sum_xp(x)D(W(x)|\sigma)
$$
We define
$$
f(t)=Tr(tW_x^{1-s}+(1-t)\sum_yp_s(y)W_y^{1-s})^{1/(1-s)}, t\geq 0
$$
Note that
$$
I_s(p,W)=(Tr(\sum_yp(y)W_y^{1-s})^{1/(1-s)})^{1-s}
$$
and since by definition of $p_s$, we have
$$
I_s(p_s,W)\geq I_s(p,W)\forall p,
$$
it follows that
$$
f(t)\leq f(0)\forall t\geq 0
$$
Thus
$$
f'(0)\leq 0
$$
This gives
$$
Tr[(W_x^{1-s}-\sum_yp_s(y)W_y^{1-s}).(\sum_yp_s(y)W_y^{1-s})^{s/(1-s)}]\leq 0
$$
or equivalently,
$$
Tr(W_x^{1-s}.(\sum_yp_s(y)W_y^{1-s})^{s/(1-s)})\leq Tr(\sum_yp_s(y)W_y^{1-s})^{1/(1-s)}
$$
Now define the state
$$
\sigma_s=(\sum_yp_s(y)W_y^{1-s})^{1/(1-s)}/Tr(\sum_yp_s(y)W_y^{1-s})^{1/(1-s)}
$$
Then we can express the above inequality as
$$
Tr(W_x^{1-s}\sigma_s^s)\leq [Tr(\sum_yp_s(y)W_y^{1-s})^{1/(1-s)}]^{1-s}
$$
Remark: Recall that
$$
D_s(p\otimes W|p\otimes\sigma]=(1/-s)log\sum_xp(x)Tr(W_x^{1-s}\sigma^s)
$$
$$
=(1/-s)log(Tr(A\sigma^s)
$$
where
$$
A=\sum_xp(x)W_x^{1-s}
$$
Note that we are all throughout assuming $s\leq 0$. Application of the reverse Holder inequality gives for all states $\sigma$
$$
Tr(A\sigma^s)\geq (Tr(A^{1/(1-s)}))^{1-s}.(Tr(\sigma))^s=(Tr(A^{1/(1-s)}))^{1-s}
$$
with equality iff $\sigma$ is proportional to $A^{1/(1-s)}$, ie, iff
$$
\sigma=A^{1/(1-s)}/Tr(A^{1/(1-s)})
$$
Thus,
$$
min_{\sigma}D_s(p\otimes W|p\otimes\sigma)=
$$
$$
(-1/s)[Tr(\sum_xp(x)W_x^{1-s})^{1/(1-s)}]^{1-s}
$$
The maximum of this over all $p$ is attained when $p=p_s$ and we denote the corresponding value of $\sigma$ by $\sigma_s$. Thus,
$$
\sigma_s=\frac{(\sum_xp_s(x)W_x^{1-s})^{1/(1-s)}}{Tr(\sum_xp_s(x)W_x^{1-s})^{1/(1-s)}}
$$
Now let $\phi(i), i=1,2,..., N$ be a code with $\phi(i)\in A^n$. Let $Y_i,i=1,2,..., N$ be detection operators, ie, $Y_i\geq 0, \sum_iY_i=1$. Let $\epsilon(\phi)$ be the error probability corresponding to this code and detection operators. Then
$$
1-\epsilon(\phi)=N^{-1}\sum_{i=1}^NTr(W_{\phi(i)}Y_i)
$$
$$
=N^{-1}Tr(S(\phi)T)
$$
where
$$
S(\phi)=N^{-1}diag[W_{\phi(i)}, i=1,2,..., N], T=diag[Y_i, i=1,2,..., N]
$$
Also define for any state $\sigma$, the state
$$
S(\sigma)=N^{-1}diag[\sigma_n,...,\sigma_n]=N^{-1}.I_N\otimes\sigma_n
$$
We observe that
$$
Tr(S(\sigma_n)T)==N^{-1}\sum_iTr(\sigma_nY_i)=Tr(\sigma.N^{-1}.\sum_{i=1}^NY_i)
$$
$$
=N^{-1}Tr(\sigma_n)=1/N
$$
where $\sigma_n=\sigma^{\otimes n}.$ Then for $s\leq 0$ we have by monotonicity of quantum Renyi entropy, (The relative entropy between the pinched states cannot exceed the relative entropy between the original states)
$$
N^{-s}(1-\epsilon(\phi))^{1-s}=(Tr(S(\phi)T))^{1-s}(Tr(S(\sigma_{ns})T))^s
$$
$$
\leq (Tr(S(\phi)T))^{1-s}(Tr(S(\sigma_{ns})T))^s+(Tr(S(\phi)(1-T)))^{1-s}(Tr(S(\sigma_{ns})(1-T)))^s
$$
$$
\leq Tr(S(\phi)^{1-s}S(\sigma_{ns})^s)
$$
Note that the notation $\sigma_{ns}=\sigma_s^{\otimes n}$ is being used. Now, we have
$$
(Tr(S(\phi)^{1-s}S(\sigma_{ns})^s))=(N^{-1}\sum_iTr(W_{\phi(i)}^{1-s}\sigma_{ns}^s))
$$
$$
log(Tr(W_{\phi(i)}^{1-s}\sigma_{ns}^s)
$$
$$
=\sum_{l=1}^nlog(Tr(W_{\phi_l(i)}^{1-s}\sigma_s^s))
$$
$$
=n.n^{-1}\sum_{l=1}^nlog(Tr(W_{\phi_l(i)}^{1-s}\sigma_s^s))
$$
$$
\leq n.log(n^{-1}.\sum_{l=1}^nTr(W_{\phi_l(i)}^{1-s}\sigma_s^s))
$$
$$
\leq n.log(Tr(\sum_yp_s(y)W_y^{1-s})^{1/(1-s)}]^{1-s})
$$
$$
=n(1-s).log(Tr(\sum_xp_s(x)W_x^{1-s})^{1/(1-s)})
$$
Thus,
$$
Tr(S(\phi)^{1-s}S(\sigma_{ns})^s)\leq (Tr(\sum_xp_s(x)W_x^{1-s})^{1/(1-s)})^{n(1-s)}
$$
and hence
$$
N^{-s}(1-\epsilon(\phi))^{1-s}\leq (Tr(\sum_xp_s(x)W_x^{1-s})^{1/(1-s)})^{n(1-s)}
$$
or equivalently,
$$
1-\epsilon(\phi)\leq exp((s/(1-s))log(N)+n.log(Tr(\sum_xp_s(x)W_x^{1-s})^{1/(1-s)}))
$$
$$
=exp((s/(1-s))log(N)+n.log(Tr(\sum_xp_s(x)W_x^{1-s})^{1/(1-s)}))
$$
$$
=exp(ns((1-s)log(N)/n+s^{-1}log(Tr(\sum_xp_s(x)W_x^{1-s})^{1/(1-s)}))---(a)
$$
where $s\leq 0$. Now, for any probability distribution $p(x)$ on $A$, we have
$$
d/ds Tr[(\sum_xp(x)W_x^{1-s})^{1/(1-s)}]=
$$
$$
(1/(1-s)^2)Tr[(\sum_xp(x)W_x^{1-s})^{1/(1-s)}.log(\sum_xp(x)W_x^{1-s})]
$$
$$
-(1/(1-s))[Tr(\sum_xp(x)W_x^{1-s}log(W_x))].[Tr(\sum_xp(x)W_x^{1-s})^{s/(1-s)}]
$$
and as $s\rightarrow 0$, this converges to
$$
d/ds Tr[(\sum_xp(x)W_x^{1-s})^{1/(1-s)}]|_{s=0}=
$$
$$
\sum_xp(x)Tr(W_x.(log(W_p)-log(W_x)))=-\sum_xp(x)D(W_x|W_p)
$$
$$
=-(H(W_p)-\sum_p(x)H(W_x))=-(H(Y)-H(Y|X))=-I_p(X,Y)
$$
where
$$
W_p=\sum_xp(x)W_x
$$
Write $N=N(n)$ and define
$$
R=limsup_nlog(N(n))/n, \phi=\phi_n
$$
Also define
$$
I_s(X,Y)=-(1/s(1-s))log(Tr(\sum_xp_s(x)W_x^{1-s})^{1/(1-s)}))
$$
Then, we've just seen that
$$
lim_{s\rightarrow 0}I_s(X,Y)=I_0(X,Y)\leq I(X,Y)
$$
where $I_0(X,Y)=I_{p_0}(X,Y)$ and $I(X,Y)=sup_pI_p(X,Y)$. and
$$
1-\epsilon(\phi_n)\leq exp((ns/(1-s))(log(N(n))/n-I_s(X,Y)))
$$
It thus follows that if
$$
R>I(X,Y)
$$
then by taking $s<0$ and $s$ sufficiently close to zero we would have
$$
R>I_s(X,Y)
$$
and then we would have
$$
limsup_n(1-\epsilon(\phi_n))\leq limsup_nexp((ns/(1-s))(R-I_s(X,Y))=0
$$
or equivalently,
$$
liminf_n\epsilon(\phi_n)=1
$$
This proves the converse of the Cq coding theorem, namely, that if the rate of information transmission via any coding scheme is more that the Cq capacity $I(X,Y)$, then the asymptotic probability of error in decoding using any sequence of decoders will always converge to unity.
\bigskip

\section{Proof of the direct part of the Cq coding theorem}

Our proof is based on Shannon's random coding method. Consider the same scenario as above.
Define
$$
\pi_i=\{W_{\phi(i)}\geq 2NW_{p_n}\}, i=1,2,..., N
$$
where $N=N(n)$, $\phi(i)\in A^n$ with $\phi(i), i=1,2,..., N$ being iid with probability distribution $p_n=p^{\otimes n}$ on $A^n$. Define the detection operators
$$
Y(\phi(i))=(\sum_{j=1}^N\pi)^{-1/2}.\pi_i.(\sum_{j=1}^N\pi_j)^{-1/2}, i=1,2,..., N
$$
Then $0\leq Y_i\leq 1$ and $\sum_{i=1}^NY_i=1$. Write
$$
S=\pi_i, T=\sum_{j\neq i}\pi_j
$$
Then,
$$
Y(\phi(i))=(S+T)^{-1/2}S(S+T)^{-1/2}, 0\leq S\leq 1, T\geq 0
$$
and hence we can use the inequality
$$
1-(S+T)^{-1/2}S(S+T)^{-1/2}\leq 2-2S+4T
$$
Then,
$$
1-Y(\phi(i))=\leq 2-2\pi_i+4\sum_{j:j\neq i}\pi_j
$$
and hence
$$
\epsilon(\phi)=N^{-1}\sum_{i=1}^NTr(W_{\phi(i)}(1-Y(\phi(i))))
$$
$$
\leq N^{-1}\sum_{i=1}^NTr(W(\phi(i))(2-2\pi_i)+4N^{-1}\sum_{i\neq j}Tr(W(\phi(i))\pi_j)
$$
Note that since for $i\neq j$, $\phi(i)$ and $\phi(j)$ are independent random variables in $A^n$, it follows that $W(\phi(i))$ and $\pi_j$ are also independent operator valued random variables. Thus,
$$
\Bbb E(\epsilon(\phi))\leq (2/N)\sum_{i=1}^N\Bbb E(Tr(W(\phi(i))\{W(\phi(i))\leq 2NW_{p_n}\}))
$$
$$
+(4/N)\sum_{i\neq j, i,j=1,2,..., N}Tr(\Bbb E(W(\phi(i))).\Bbb E\{W(\phi(j))>2NW_{p_n}\})
$$
$$
=2\sum_{x_n\in A^n}p_n(x_n)Tr(W(x_n)\{W(x_n)\leq 2N.W(p_n)\})
$$
$$
+4(N-1)\sum_{x_n,y_n\in A^n}p_n(x_n)p_n(y_n)Tr(W(x_n)\{W(y_n)>2NW(p_n)\})
$$
$$
=
(2/N)\sum_{x_n}p_n(x_n)Tr(W(x_n)\{W(x_n)\leq 2N.W(p_n)\})
$$
$$
+4(N-1)\sum_{y_n\in A^n}p_n(y_n)Tr(W(p_n)\{W(y_n)>2NW(p_n)\})
$$
$$
\leq 2\sum_{x_n}p_n(x_n)Tr(W(x_n)^{1-s}(2NW(p_n))^s)
$$
$$
+4N\sum_{y_n}p_n(y_n)Tr(W(y_n)^{1-s}W(p_n)^s(2N)^{s-1})
$$
for $0\leq s\leq 1$. This gives
$$
\Bbb E(\epsilon(\phi_n))\leq
$$
$$
2^{s+2}N^s\sum_{x_n\in A^n}p_n(x_n)Tr(W(x_n)^{1-s}W(p_n)^s)
$$
$$
=2^{s+2}N^s[\sum_{x\in A}p(x)Tr(W(x)^{1-s}W(p)^s)]^n
$$
$$
=2^{s+2}.exp(sn(log(N(n)/n+s^{-1}.log(\sum_xp(x)Tr(W(x)^{1-s}W(p)^s))))
$$
\bigskip

\section{Induced representations}

Let $G$ be a group and $H$ a subgroup. Let $\sigma$ be a unitary representation of $H$ in a Hilbert space $V$. Let $\mathcal F$ be the space of all $f:G\rightarrow V$ for which
$$
f(gh)=\sigma(h)^{-1}f)(g), h\in H, g\in G
$$
Then define
$$
U(g)f(x)=\rho(g,g^{-1}xH)f(g^{-1}x), g,x\in G, f\in\mathcal F
$$
where 
$$
\rho(g,x)=(d\mu(xH)/d\mu(gxH))^{1/2}, g,x\in G
$$
with $\mu$ a measure on $G/H$. $U$ is called the representation of $G$ induced by the representation $\sigma$ of $H$. We define
$$
\parallel f\parallel=\int_G|f(g)|^2d\mu(g)
$$
Then, unitarity of $U(g)$ is easy to see:
$$
\int_{G/H}\parallel U(g)f(x)\parallel^2d\mu(x)=
$$
$$
\int_{G/H}(\rho(g,g^{-1}xH))^2|f(g^{-1}x)|^2d\mu(xH)=
$$
$$
\int (d\mu(g^{-1}xH)/d\mu(xH))|f(g^{-1}x)|^2d\mu(xH)=\int_{G/H}|f(g^{-1}x)|^2d\mu(g^{-1}xH)
$$
$$
=\int_{G/H}|f(x)|^2d\mu(xH)
$$
proving the claim. Note that
$$
(\rho(g_1,g_2xH)\rho(g_2,xH))^2=(d\mu(g_2xH)/d\mu(g_1g_2xH))(d\mu(xH)/d\mu(g_2xH))=
$$
$$
d\mu(xH)/d\mu(g_1g_2xH)=\rho(g_1g_2,xH)^2
$$
ie
$$
\rho(g_1g_2,xH)=\rho(g_1,g_2xH)\rho(g_2,xH), g_1,g_2,x\in G
$$
ie, $\rho$ is a scalar cocycle. 

Remark: $|f(x)|^2$ can be regarded a function on $G/H$  since it assumes the same value on
$xH$ for any $x\in G$. Indeed for $x\in G,h\in H$, $|f(xh)|^2=|\sigma(h)^{-1}f(x)|^2=|f(x|^2$ since $\sigma(h)$ is a unitary operator.
\bigskip

There is yet another way to describe the induced representation, ie, in a way that yields a unitary representation of $G$ that is equivalent to $U$. The method is as follows: Let $X=G/H$ and let $\mathcal F_1$ be the space of all square integrable functions $f:X\rightarrow V$ w.r.t the measure $\mu_X$ induced by $\mu$ on $X$. Let $\gamma$ be any cocycle of $(G,X)$ with values in the space of unitary operators on $V$, ie, 
$$
\gamma:G\times X\rightarrow \mathcal U(V)
$$
satisfies
$$
\gamma(g_1g_2,x)=\gamma(g_1,g_2x)\gamma(g_2,x), g_1,g_2\in G, x\in X
$$
Then define a representation $U_1$ of $G$ in $\mathcal F_1$ by
$$
U_1(g)f(x)=\rho(g,g^{-1}x)\gamma(g,g^{-1}x)f(g^{-1}x), g\in G, x\in X, f\in\mathcal F_1
$$
Let $s:X\rightarrow G$ be a section of $X=G/H$ such that $s(H)=e$. This means that
for any $x\in X, s(x)H=x$, ie, $s(x)$ belong to the coset $x$. Define
$$
b(g,x)=s(gx)^{-1}gs(x)\in G, g\in G, x\in X
$$
Then,
$$
b(g,x)H=s(gx)^{-1}gx=H
$$
because
$$
gs(x)H=gx, s(gx)H=gx
$$
Now observe that
$$
b(g_1,g_2x)b(g_2,x)=s(g_1g_2x)^{-1}g_1s(g_2x)s(g_2x)^{-1}g_2s(x)=s(g_1g_2x)^{-1}g_1g_2s(x)
$$
$$
=b(g_1g_2,x), g_1,g_2\in G, x\in X
$$
implying that $\gamma=\sigma(b)$ is a $(G,X,V)$ cocycle, ie,
$$
\gamma(g_1g_2,x)=\sigma(b(g_1g_2,x))=\sigma(b(g_1,g_2x)b(g_2,x))
$$
$$
=\sigma(b(g_1,g_2x))\sigma(b(g_2,x))=\gamma(g_1,g_2x)\gamma(g_2,x)
$$
with $\gamma$ assuming values in the set of unitary operators in $V$.
Now let
$$
f(g)=\gamma(g,H)^{-1}f_1(gH)=T(f_1)(g)\in V, f_1\in\mathcal F_1, g\in G
$$
Then
$$
U(g)T(f_1)(g_1)=U(g)f(g_1)=\rho(g,g^{-1}g_1H)f(g^{-1}g_1)
$$
$$
=\rho(g,g^{-1}g_1H)\gamma(g^{-1}g_1,H)^{-1}f_1(g^{-1}g_1H)
$$
$$
=\rho(g,g^{-1}g_1H)(\gamma(g^{-1},g_1H)\gamma(g_1,H))^{-1}f_1(g^{-1}g_1H)
$$
On the other hand,
$$
T(U_1(g)f_1)(g_1)=\gamma(g_1,H)^{-1}U_1(g)f_1(g_1H)
$$
$$
=\gamma(g_1,H)^{-1}\rho(g,g^{-1}g_1H)\gamma(g,g^{-1}g_1H)f_1(g^{-1}g_1H)
$$
$$
=\rho(g,g^{-1}g_1H)\gamma(g_1,H)^{-1}\gamma(g,g^{-1}g_1H)f_1(g^{-1}g_1H)
$$
Now,
$$
\gamma(g_1,H)=\gamma(gg^{-1}g_1,H)=\gamma(g,g^{-1}g_1H)\gamma(g^{-1}g_1,H)
$$
or equivalently,
$$
\gamma(g_1,H)^{-1}\gamma(g,g^{-1}g_1H)=\gamma(g^{-1}g_1,H)^{-1}
$$
Thus,
$$
T(U_1(g)f_1)(g_1)=\rho(g,g^{-1}g_1H)\gamma(g^{-1}g_1,H)^{-1}f_1(g^{-1}g_1H)
$$
$$
=\rho(g,g^{-1}g_1H)T(f_1)(g^{-1}g_1)=U(g)T(f_1)(g_1)
$$
In other words,
$$
TU_1(g)=U(g)T
$$
Thus, $U_1$ and $U$ are equivalent unitary representations.
\bigskip

\section{Estimating non-uniform transmission line parameters and line voltage and current using the extended Kalman filter}

Abstract: The distributed parameters of a non-uniform transmission line are assumed to depend on a finite set of unknown parameters, for example their spatial Fourier series coefficients. The line equations are setup taking random voltage and current line loading into account with the random loading being assumed to be white Gaussian noise in the time and spatial variables. By passing over to the spatial Fourier series domain, these line stochastic differential equations are expressed as a truncated sequence of coupled stochastic differential equations for the spatial Fourier series components of the line voltage and current. The coefficients that appear in these stochastic differential equations are the spatial Fourier series coefficients of the distributed line parameters and these coefficients in turn are linear functions of the unknown parameters. By separating out these complex stochastic differential equations into their real and imaginary part, the line equations become a finite set of sde's for the extended state defined to be the real and imaginary components of the line voltage and current as well as the parameters on which the distributed parameters depend. The measurement model is assumed to be white Gaussian noise corrupted versions of the line voltage and current sampled at a discrete set of spatial points on the line and this measurement model can be expressed as a linear transformation of the state variables, ie, of the spatial Fourier series components of the line voltage and current plus a white measurement noise vector. The extended Kalman filter equations are derived for this state and measurement model and our real time line voltage and current and distributed parameter estimates show excellent agreement with the true values based on MATLAB simulations.
\bigskip

\section{Multipole expansion of the magnetic field produced by a time varying current distribution}

Let $J(t,r)$ be the current density. The magnetic vector potential generated by it is given by
$$
A(t,r)=(\mu/4\pi)\int J(t-|r-r'|/c,r')d^3r'/|r-r'|
$$
$$
=(\mu/4\pi)\sum_{n\geq 0}(1/c^nn!)\int(\partial_t^nJ(t,r'))|r-r'|^{n-1}d^3r'
$$
For $|r|$ greater than the source dimension and with $\theta$ denoting the angle between the vectors $r$ and $r'$, we have the expansion
$$
|r-r'|^{\alpha}=(r^2+r^{'2}-2r.r')^{\alpha/2}
$$
$$
=r^{\alpha}(1+(r'/r)^2-2r.r'/r^2)^{\alpha/2}=r^{\alpha}(1+(r'/r)^2-2(r'/r)cos(\theta))^{\alpha/2}
$$
$$
=r^{\alpha}\sum_{m\geq 0}P_{m,\alpha}(cos(\theta))(r'/r)^m
$$
where the generating function of $P_{m,\alpha}(x), m=0,1,2,...$ is given by
$(1+x^2-2x.cos(\theta))^{\alpha/2}$. Then,
$$
A(t,r)=(\mu/4\pi)\sum_{n,m\geq 0}(1/c^nn!)r^{n-1-m}\int\partial_t^nJ(t,r')r{'m}P_{n-1}(r.r'/rr')d^3r'
$$
A useful approximation to this exact expansion is obtained by taking
$$
|r-r'|^{\alpha}\approx (r^2-2r.r')^{\alpha/2}\approx r^{\alpha}(1-\alpha r.r'/r^2)
$$
$$
=r^{\alpha}-\alpha.r^{\alpha-2}r.r'
$$
Further,
$$
\partial_t^nJ(t,r')(r.r')=(r'\times\partial_t^nJ(t,r'))\times r+(\partial_t^nJ(t,r').r)r'
$$
In the special case when the source is a wire carrying a time varying current $I(t)$, we can write
$$
J(t,r')d^3r'=I(t)dr'
$$
and then the above equation gets replaced by
\newpage

\chapter{Some aspects of stochastic processes}

{\bf End semester exam, M.Tech Ist semester, Stochastic processes and queueing theory\\Attempt any five questions, each question carries ten marks}
\bigskip

[1] In a single server queue, let the arrivals take place at the times $S_n=\sum_{k=1}^nX_k, n\geq 1$, let the service time of the $n^{th}$ customer be $Y_n$ and let his waiting time in the queue be $W_n$. Derive the recursive relation
$$
W_{n+1}=max(0,W_n+Y_n-X_{n+1}), n\geq 1
$$
where $W_1=0$. If the $X_n's$ are iid and the $Y_n's$ are also iid and independent of $\{X_n\}$, then define $T_n=Y_n-X_{n+1}$ and prove that $W_{n+1}$ has the same distribution
as that of $max_{0\leq k\leq n}F_k$ where $F_n=\sum_{k=1}^nT_k, n\geq 1$ and $F_0=0$. When
$X_n,Y_n$ are respectively exponentially distributed with parameters $\lambda, \mu$ respectively explain the sequence of steps based on ladder epochs and ladder heights by which you will calculate the distribution of the equilibrium waiting time
$W_{\infty}=lim_{n\rightarrow\infty}W_n$ in the case when $\mu>\lambda$. What happens when
$\mu\leq\lambda$ ?
\bigskip

Solution: The $(n+1)^{th}$ customer arrives at the time $S_{n+1}$. The previous, ie, the $n^{th}$ customer departs at time $S_n+W_n+Y_n$. If this time of his departure is smaller than $S_{n+1}$, (ie, $S_n+W_n+Y_n-S_{n+1}=W_n+Y_n-X_{n+1}<0$), then the $(n+1)^{th}$ customer finds the queue empty on arrival and hence
his waiting time $W_{n+1}=0$. If on the other hand, the time of departure $S_n+W_n+Y_n$ of the $n^{th}$ customer is greater than $S_{n+1}$, the arrival time of the $(n+1)^{th}$ customer, then the $n+1^{th}$ customer has to wait for a time 
$$
W_{n+1}=(S_n+W_n+Y_n)-S_{n+1}=W_n+Y_n-X_{n+1}
$$
before his service commences. Therefore, we have the recursion relation
$$
W_{n+1}=max(0,W_n+Y_n-X_{n+1}), n\geq 0
$$
Note that $W_1=0$. This can be expressed as
$$
W_{n+1}=max(0,W_n+T_n), T_n-Y_n-X_{n+1}
$$
By iteration, we find
$$
W_2=max(0,T_1), W_3=max(0,T_2+max(0,T_1))=max(0,max(T_2,T_1+T_2))=max(0,T_2,T_1+T_2)
$$
$$
W_4=max(0,W_3+T_3)=max(0,T_3+max(0,max(T_2,T_1+T_2))=max(0,max(T_3,max(T_2+T_3,T_1+T_2+T_3))))
$$
$$
=max(0,T_3,T_2+T_3,T_1+T_2+T_3)
$$
Continuing, we find by induction that
$$
W_{n+1}=max(0,\sum_{k=r}^nT_k, r=1,2,..., n)
$$
Now the $T_k's$ are iid. Therefore, $W_{n+1}$ has the same distribution as
$$
max(0,\sum_{k=r}^nT_{n+1-k}, r=1,2,..., n)
$$
$$
=max(0,\sum_{k=1}^rT_k, r=1,2,..., n)
$$
In particular, the equilibrium distribution of the waiting time, ie of $W_{\infty}=lim_{n\rightarrow\infty}W_n$ is the same as that of
$$
max(0,\sum_{k=1}^{\infty}T_k)
$$
provided that this limiting r.v. exists, ie, the provided that the infinite series converges w.p.1. Let $F_n=\sum_{k=1}^nT_k, n\geq 1, F_0=0$

Define $\tau_1$ denote the first time $k\geq 1$ at which $F_k>F_0=0$. For $n\geq 1$,
let $\tau_{n+1}$ denote the first time $k\geq 1$ at which $F_k>F_{\tau_n}$, ie, the $n+1^{th}$ time $k$ at which $F_k$ exceeds all of $F_j,j=0,1,..., k-1$. Then, it is clear that $W_{\infty}$ has the same distribution as that of $lim F_{\tau_n}$, provided that
$\tau_n$ exists for all $n$. In the general case, we have
$$
P(W_{\infty}\leq x)=lim_{n\rightarrow\infty}P(F_{\tau_n}\leq x)
$$
where if for a given $\omega$, $\tau_N(\omega)$ exists for some positive integer $N=N(\omega)$ but $\tau_{N+1}(\omega)$ does not exist, then we set $\tau_k(\omega)=\tau_N(\omega)$ for all $k>N$. It is also clear using stopping time theory, that the
r.v's $U_{n+1}=F_{\tau_{n+1}}-F_{\tau_n}, n=1,2,...$ are iid and non-negative and we can write 
$$
W_{\infty}=\sum_{n=0}^{\infty}U_n, U_0=0
$$
Thus, the distribution of $W_{\infty}$ is formally the limit of $F_U^{*n}$ where
$F_U$ is the distribution of $U_1=F_{\tau_1}$ with $\tau_1=min(k\geq 1, F_k>0)$.
Note also that for $x>0$,
$$
1-F_U(x)=P(U>x)=\sum_{n=0}^{\infty}P(max_{k\leq n}F_k\leq x<F_{n+1})
$$
\bigskip

[2] Consider a priority queue with two kinds of customers labeled $C_1,C_2$ and a single
server. Customers of type $C_1$ arrive at the rate $\lambda_1$ and customers of type $C_2$ arrive at the rate $\lambda_2$. A customer of type $C_1$ is served at the rate $\mu_1$ while a customer of type $C_2$ is served at the rate $\mu_2$. If a customer of type $C_1$ is present in the queue, then his service takes immediate priority over a customer of type $C_2$, otherwise the queue runs on a first come first served basis. Let $X_1(t)$ denote the number of type $C_1$ customers and $X_2(t)$ the number of type $C_2$ customers in the queue at time $t$. Set up the Chapman-Kolmogorov differential equations for
$$
P(t,n_1,n_2)=Pr(X_1(t)=n_1,X_2(t)=n_2), n_1,n_2=0,1,...
$$
and also set up the equilibrium equations for
$$
P(n_1,n_2)=lim_{t\rightarrow\infty}P(t,n_1,n_2)
$$
\bigskip

Solution: Let $\theta(n)=1 if n\geq 0$ and $\theta(n)=0$ if $n<0$ and let $\delta(n,m)=1$ if $n=m$ else $\delta(n,m)=0$.
$$
P(t+dt,n_1,n_2)=P(t,n_1-1,n_2)\lambda_1dt.\theta(n_1-1)+P(t,n_1,n_2-1)\lambda_2dt.\theta(n_2-1)+
$$
$$
P(t,n_1+1,n_2)\mu_1dt+P(t,n_1,n_2+1)\mu_2.dt.\delta(n_1,0)
$$
$$
+P(t,n_1,n_2)(1-(\lambda_1+\lambda_2)dt-\mu_1.dt.\theta(n_1-1)-\mu_2dt.\theta(n_2-1).\delta(n_1,0))
$$
Essentially, the term $\delta(n_1,0)$ has been inserted to guarantee that the $C_2$ type of customer is currently being served only when there is no $C_1$ type of customer in the queue and the term $\theta(n_2-1)$ term has been inserted to guarantee that the $C_2$ type of customer can depart from the queue only if there is at least one such in the queue.
The product $\delta(n_1,0)\theta(n_2-1)$ term corresponds to the situation in which there is no $C_1$ type of customer and at least one $C_2$ type of customer which means that a $C_2$ type of customer is currently being served and hence will depart immediately after his service is completed. From these equations follow the Chapman-Kolmogorov equations:
$$
\partial_tP(t,n_1,n_2)=
$$
$$
P(t,n_1-1,n_2)\lambda_1\theta(n_1-1)+P(t,n_1,n_2-1)\lambda_2.\theta(n_2-1)+
$$
$$
P(t,n_1+1,n_2)\mu_1+P(t,n_1,n_2+1)\mu_2.\delta(n_1,0)
$$
$$
-P(t,n_1,n_2)(\lambda_1+\lambda_2)-\mu_1.\theta(n_1-1)-\mu_2.\theta(n_2-1).\delta(n_1,0))
$$
The equilibrium equations are obtained by setting the rhs to zero, ie $\partial_tP(n_1,n_2)=0$.
\bigskip

[3] Write short notes on any three of the following:

[a] Construction of Brownian motion on $[0,1]$ from the Haar wavelet functions including the proof of almost sure continuity of the sample paths.

Solution: Define the Haar wavelet functions as

$H_{n,k}(t)=2^{(n-1)/2}$ for $(k-1)/2^n\leq t\leq k/2^n$, $H_{n,k}(t)=-2^{(n-1)/2}$ for $k/2^n<t\leq (k+1)/2^n$ whenever $k$ is in the set $I(n)=\{k:k odd, k=0,1,..., 2^{n-1}\}$ for n=1,2,... and put $H_0(t)=1$. Here $t\in[0,1]$. It is then easily verified that the functions $H_{nk}, k\in I(n), n\geq 1, H_{00}$ form an orthonormal basis for $L^2[0,1]$. 
We set $I(0)=\{0\}$.

Define the Schauder functions
$$
S_{nk}(t)=\int_0^tH_{nk}(s)ds
$$
Then, from basic properties of orthonormal bases, we have
$$
\sum_{nk}H_{nk}(t)H_{nk}(s)=\delta(t-s)
$$
and hence
$$
\sum_{nk}S_{nk}(t)S_{nk}(s)=min(t,s)
$$
Let $\xi(n,k), k\in I(n), n\geq 1,\xi(0,0)$ be iid $N(0,1)$ random variables and define the random functions
$$
B_N(t)=\sum_{n=0}^N\sum_{k\in I(n)}\xi(n,k)S_{nk}(t), N\geq 1
$$
Then since $S_{nk}(t)$ are continuous functions, $B_N(t)$ is also continuous for all $N$.
Further, the $S_{nk}'s$ for a fixed $n$ are non-overlapping functions and
$$
max_t|S_{nk}(t)|=2^{-(n+1)/2}
$$
Thus,
$$
max_t\sum_{k\in I(n)}|S_{nk}(t)|=2^{-(n+1)/2}
$$
Let 
$$
b(n)=max(|\xi(nk)|:k\in I(n)\}
$$
Then, by the union bound
$$
P(b(n)>n)\leq 2^n.P(|\xi(nk)|>n)=2^{n+1}\int_n^{\infty}exp(-x^2/2)dx/\sqrt{2\pi}
$$
$$
\leq K.2^{n+1}n.exp(-n^/2)
$$
Thus,
$$
\sum_{n\geq 1}P(b(n)>n)<\infty
$$
which means that by the Borel-Cantelli lemma, for a.e.$\omega$, there is a finite $ N(\omega)$ such that for all $n>N(\omega)$, $b(n,\omega)<n$. This means that for a.e.$\omega$,
$$
\sum_{n=1}^N\sum_{k\in I(n)}|\xi(n,k,\omega)||S_{nk}(t)|
$$
$$
\leq\sum_{n=1}^Nn.2^{-(n+1)/2}
$$
which converges as $N\rightarrow\infty$. This result implies that the sequence of continuous processes $B_N(t), t\in[0,1]$ converges uniformly to a continuous process $B(t), t\in[0,1]$ a.e. Note that continuity of each term and uniform convergence imply continuity of the limit. That $B(t), t\in[0,1]$ is a Gaussian process with zero mean and covariance $min(t,s)$ as can be seen from the joint characteristic function of the time samples of $B_N(.)$. Alternately, $B(.)$ being an infinite linear superposition of Gaussian r.v's must be a Gaussian process, its mean zero is obvious since $\xi(n,k)$ have mean zero and the covariance follows from
$$
\Bbb E(B_N(t)B_N(s))=\sum_{n,m=0}^N\sum_{k,r\in I(n)}\Bbb E(\xi(nk)\xi(mr))S_{nk}(t)S_{mr}(s)
$$
$$
=\sum_{n,m=0}^N\sum_{k,r\in I(n)}\delta(n,m)\delta(k,r)S_{nk}(t)S_{mr}(s)
$$
$$
=\sum_{n=0}^N\sum_{k\in I(n)}S_{nk}(t)S_{nk}(s)\rightarrow\sum_{n=0}^{\infty}\sum_{k\in I(n)}S_{nk}(t)S_{nk}(s)=min(t,s)
$$
\bigskip

[b] $L^2$-construction of Brownian motion from the Karhunen-Loeve expansion.

The autocorrelation of $B(t)$ is
$$
R(t,s)=\Bbb E(B(t)B(s))=min(t,s)
$$
Its eigenfunctions $\phi(t)$ satisfy
$$
\int_0^1R(t,s)\phi(s)ds=\lambda\phi(t), t\in[0,1]
$$
Thus,
$$
\int_0^tsphi(s)ds+t\int_t^1\phi(s)ds=\lambda\phi(t)
$$
Differentiating,
$$
t\phi(t)+\int_t^1\phi(s)-t\phi(t)=\lambda\phi'(t)
$$
Again differentiating
$$
-\phi(t)=\lambda\phi''(t)
$$
Let $\lambda=1/\omega^2$. Then, the solution is
$$
\phi(t)=A.cos(\omega t)+B.sin(\omega t)
$$
$$
\int_0^ts\phi(s)ds=A((t/\omega)sin(\omega t)+(cos(\omega t)-1)/\omega^2)+
$$
$$
B(-t.cos(\omega t)/\omega+sin(\omega t)/\omega^2)
$$
$$
\int_t^1\phi(s)ds=A(sin(\omega)-sin(\omega t))/\omega+(B/\omega)(cos(\omega t)-\cos(\omega))
$$
Substituting all these expressions into the original integral equation gives
$$
A((t/\omega)sin(\omega t)+(cos(\omega t)-1)/\omega^2)+
$$
$$
B(-t.cos(\omega t)/\omega+sin(\omega t)/\omega^2)
$$
$$
+tA(sin(\omega)-sin(\omega t))/\omega+(Bt/\omega)(cos(\omega t)-\cos(\omega))
$$
$$
=(A.cos(\omega t)+B.sin(\omega t))/\omega^2
$$
On making cancellations and comparing coefficients,
$$
A.sin(\omega)=0, B.cos(\omega)=0, A/\omega^2=0,
$$
Therefore,
$$
A=0, \omega=(n+1/2)\pi, n\in\Bbb Z
$$
and for normalization
$$
\int_0^1\phi^2(t)dt=1
$$
we get
$$
B^2\int_0^1sin^2(\omega t)dt=1
$$
or
$$
(B^2/2)(1-sin(2\omega)/2\omega)=1
$$
or
$$
B=\sqrt 2
$$
Thus, the KL expansion for Brownian motion is given by
$$
B(t)=\sum_{n\in\Bbb Z}\sqrt{\lambda_n}\xi(n)\phi_n(t)
$$
$$
=\sum_{n\in\Bbb Z}((n+1/2)\pi)^{-2}\xi(n).\sqrt 2sin((n+1/2)\pi t)
$$
where $\xi(n), n=1,2,...$ are iid $N(0,1)$ r.v's.
\bigskip

[c] Reflection principle for Brownian motion with application to the evaluation of the distribution of the first hitting time of the process at a given level.

[d] The Borel-Cantelli lemmas with application to proving almost sure convergence of a sequence of random variables.

[e] The Feynman-Kac formula for solving Schrodinger's equation using Brownian motion with complex time.

[f] Solving Poisson's equation $\nabla^2u(x)=s(x)$ in an open connected subset of $\Bbb R^n$ with the Dirichlet boundary conditions using  Brownian motion stop times.
\bigskip

[4] A bulb in a bulb holder is changed when it gets fused. Let $X_n$ denote the time for which the $n^{th}$ bulb lasts. If $n$ is even, then $X_n$ has the exponential density $\lambda.exp(-\lambda.x)u(x)$ while if $n$ is odd then $X_n$ has the exponential density $\mu.exp(-\mu.x)u(x)$. Calculate (a) the probability distribution of the number of bulbs $N(t)=max(n:S_n\leq t)$ changed in $[0,t]$ where $S_n=X_1+...+X_n, n\geq 1$ and $S_0=0$. Define 
$$
\delta(t)=t-S_{N(t)},\gamma(t)=S_{N(t)+1}-t
$$
Pictorially illustrate and explain the meaning of the r.v's $\delta(t),\gamma(t)$. What happens to the distributions of these r.v's as $t\rightarrow\infty$ ?
\bigskip

Solution: Let $n\geq 1$.
$$
P(N(t)\geq n)=P(S_n\leq t)=P(\sum_{1\leq k\leq n,k even}X_k+\sum_{1\leq k\leq n, n odd}X_k\leq t)
$$
Specifically,
$$
P(N(t)\geq 2n)=P(\sum_{k=1}^nX_{2k}+\sum_{k=1}^nX_{2k-1}\leq t)
$$
$$
=F^{*n}*G^{*n}(t)
$$
where 
$$
F(t)=1-exp(-\lambda t), G(t)=1-exp(-\mu t)
$$
Note that with $f(t)=F'(t), g(t)=G'(t)$, we have
$$
f^{n*}(t)=\lambda^nt^{n-1}exp(-\lambda t)/(n-1)!,
$$
$$
g^{n*}(t)=\mu^nt^{n-1}exp(-\mu t)/(n-1)!
$$
Equivalently, in terms of Laplace transforms,
$$
L(f(t))=\lambda/(\lambda+s), L(g(t))=\mu/(\mu+s)
$$
so 
$$
L(f^{*n}(t))=\lambda^n/(\lambda+s)^n, L(g^{*n}(t))=\mu^n/(\mu+s)^n
$$
and hence
$$
f^{*n}(t)*g^{*n}(t)=L^{-1}((\lambda\mu)^2/(\lambda+s)^n(\mu+s)^n)=\psi_n(t)
$$
say. Then,
$$
P(N(t)\geq 2n)=\int_0^t\psi_n(x)dx
$$
Likewise,
$$
P(N(t)\geq 2n-1)=P(S_{2n-1}\leq t)=\int_0^t\phi_n(x)dx
$$
where
$$
\phi_n(t)=g^{n*}(t)*f^{*(n-1)}(t)=L^{-1}(\lambda^{n-1}\mu^n/(\lambda+s)^{n-1}(\mu+s)^n)
$$
$\delta(t)=t-S_{N(t)}$ is the time taken from the last change of the bulb to the current time. $\gamma(t)=S_{N(t)+1}-t$ is the time taken from the current time upto the next change of the bulb. Thus, for $0\leq x\leq t$,
$$
P(\delta(t)\leq x)=P(S_{N(t)}>t-x)=\sum_{n\geq 1}P(S_n>t-x, N(t)=n)
$$
$$
=\sum_{n\geq 1}P(t-x<S_n\leq t<S_{n+1})=\sum_{n\geq 1}P(t-x<S_n<t,X_{n+1}>t-S_n)
$$
$$
=\sum_{n\geq 1}\int_{t-x}^tP(X_{n+1}>t-s)P(S_n\in ds)
$$
Likewise,
$$
P(\gamma(t)\leq x)=P(S_{N(t)+1}\leq t+x)=\sum_{n\geq 0}P(S_{n+1}\leq t+x, S_n\leq t<S_{n+1})
$$
$$
=\sum_{n\geq 0}P(S_n\leq t<S_{n+1}\leq t+x)=\sum_{n\geq 0}\int_0^tP(S_n\in ds)P(t-s<X_{n+1}<t+x)
$$
\bigskip

[5] [a] Let $\{X(t):t\in\Bbb R\}$ be a real stochastic process with autocorrelation
$\Bbb E(X(t_1).X(t_2))=R(t_1,t_2)$. Assume that there exists a finite positive real number
$T$ such that $R(t_1+T,t_2+T)=R(t_1,t_2)\forall t_1,t_2\in\Bbb R$. Prove that $X(t)$ is a mean square periodic process, ie,
$$
\Bbb E[(X(t+T)-X(t))^2]=0\forall t\in\Bbb R
$$
Hence show that $X(t)$ has a mean square Fourier series:
$$
lim_{N\rightarrow\infty}\Bbb E[|X(t)-\sum_{n=-N}^Nc(n)exp(jn\omega_0t)|^2]=0
$$
where $\omega_0=2\pi/T$ and
$$
c(n)=T^{-1}\int_0^TX(t)exp(-jn\omega_0t)dt
$$

[b] Define a cyclostationary and wide sense cyclostationary process with an example. Give an example of a wide sense cyclostationary process that is not cyclostationary.
\bigskip

Solution: A process $X(t),t\in\Bbb R$ is said to be cyclostationary if its finite dimensional probability distribution functions are invariant under shifts by integer multiples of some fixed positive $T$, ie, if
$$
P(X(t_1)\leq x_1,..., X(t_N)\leq x_N)=P(X(t_1+T)\leq x_1,..., X(t_N+T)\leq x_N)
$$
for all $x_1,...,x_N,t_1,..., t_N, N\geq 1$. For example, the process
$$
X(t)=\sum_nc(n)exp(jn\omega t)
$$
where $c(n)$ are any r.v's is cyclostationary with period $2\pi/\omega$. In fact, this process is a periodic process. A wide sense cyclostationary process is a process
whose mean and autocorrelation function are periodic with period $T$ for some fixed $T>0$, ie, if 
$$
\mu(t)=\Bbb E(X(t)), R(t,s)=\Bbb E(X(t)X(s))
$$
then
$$
\mu(t+T)=\mu(t), R(t+T,s+T)=R(t,s)\forall t,s\in\Bbb R
$$
An example of a WSS process that is not stationary is
$$
X(t)=cos(\omega_1t+\phi_1)+cos(\omega_2t+\phi_2)+cos(\omega_3t+\phi_1+\phi_2)
$$
where $\omega_1+\omega_2\neq\omega_3$ and $\phi_1,\phi_2$ are iid r.v's uniformly distributed over $[0,2\pi)$. This process has second moments that are invariant under time shifts but its third moments are not invariant under time shifts. Likewise, an example
of a WS-cyclostationary process that is not cyclostationary is the process $X(t)+Y(t)$ with $Y(t)$ having the same structure as $X(t)$ but independent of it with frequencies
$\omega_4,\omega_5,\omega_6$ (and phases $\phi_4,\phi_5,\phi_4+\phi_5$ with $\phi_1,\phi_2,\phi_4,\phi_5$ being independent and uniform over $[0,2\pi)$) such that
$(\omega_1+\omega_2-\omega_3)/(\omega_4+\omega_5-\omega_6)$ is irrational. This process is also in fact WSS. To see why it is not cyclostationary, we compute its third moments:
$$
\Bbb E(X(t)X(t+t_1)X(t+t_2))=\Bbb E(cos(\omega_3t+\phi_1+\phi_2).cos(\omega_1(t+t_1)+\phi_1).cos(\omega_2(t+t_2)+\phi_2))
$$
$$
+\Bbb E(cos(\omega_3t+\phi_1+\phi_2).cos(\omega_1(t+t_2)+\phi_1).cos(\omega_2(t+t_1)+\phi_2))
$$
$$
=(1/4)(cos((\omega_1+\omega_2-\omega_3)t+\omega_1t_1+\omega_2t_2)
$$
$$
+cos((\omega_1+\omega_2-\omega_3)t+\omega_2t_1+\omega_1t_2))
$$
and likewise
$$
\Bbb E(Y(t)Y(t+t_1)Y(t+t_2))=
$$
$$
=(1/4)(cos((\omega_4+\omega_5-\omega_6)t+\omega_4t_1+\omega_5t_2)
$$
$$
+cos((\omega_4+\omega_5-\omega_6)t+\omega_5t_1+\omega_4t_2))
$$
Thus, if 
$$
Z(t)=X(t)+Y(t)
$$
then
$$
\Bbb E(Z(t)Z(t+t_1)Z(t+t_2))=\Bbb E(X(t)X(t+t_1)X(t+t_2))+\Bbb E(Y(t)Y(t+t_1)Y(t+t_2))
$$
$$
=(1/4)(cos((\omega_1+\omega_2-\omega_3)t+\omega_1t_1+\omega_2t_2)
$$
$$
+cos((\omega_1+\omega_2-\omega_3)t+\omega_2t_1+\omega_1t_2))
$$
$$
+(1/4)(cos((\omega_4+\omega_5-\omega_6)t+\omega_4t_1+\omega_5t_2)
$$
$$
+cos((\omega_4+\omega_5-\omega_6)t+\omega_5t_1+\omega_4t_2))
$$
which is clearly not periodic in $t$ since $(\omega_1+\omega_2-\omega_3)/(\omega_4+\omega_5-\omega_6)$ is irrational.
\bigskip

[6] Explain the construction of the Lebesgue measure as a countably additive non-negative set function on the Borel subsets of $\Bbb R$ whose value on an interval coincides with the interval's length.
\bigskip

Solution: We first define the Lebesgue measure $\mu$ on the field $\mathcal F$ of finite disjoint unions of intervals of the form $(a,b]$ in $(0,1]$ as the sum of lengths. Then, we define the outer measure $\mu^*$ on the class of all subsets of $(0,1]$ as
$$
\mu^*(E)=inf\{\sum_n\mu(E_n): E_n\in\mathcal F, E\subset\bigcup_nE_n\}
$$
We prove countable subadditivity of $\mu^*$. Then, we say that $E$ is a measurable subset of $(0,1]$ if
$$
\mu^*(EA)+\mu^*(E^cA)=\mu^*(A)\forall A\subset (0,1]
$$
Then, if $\mathcal M$ denotes the class of all measurable subsets, we prove that $\mathcal M$ is a field and that $\mu^*$ is finitely additive on $\mathcal M$.
We then prove that $\mathcal M$ is a $\sigma$ field and that $\mu^*$ is countably additive on $\mathcal M$. Finally, we prove that $\mathcal M$ contains the field $\mathcal F$ and hence the $\sigma$-field generated by $\mathcal F$, namely, the Borel $\sigma$-field on $(0,1]$.
\bigskip

[7] [a] What is a compact subset of $\Bbb R^n$ ? Give an example and characterize it means of (a) the Heine-Borel property and (b) the Bolzano-Weierstrass property. Prove that the countable intersection of a sequence of decreasing non-empty compact subsets of $\Bbb R^n$ is non-empty.

A compact subset $E$ of $\Bbb R^n$ is a closed and bounded subset. By bounded, we mean that 
$$
sup\{|x|:x\in E\}<\infty
$$
By closed, we mean that if $x_n\in E$ is a sequence in $E$ such that $lim x_n=x\in\Bbb R^n $ exists, then $x\in E$. For example,
$$
E=\{x\in\Bbb R^n:|x|\leq 1\}
$$
It can be shown that $E$ is compact iff every open cover of it has finite subcover, ie,
if $O_{\alpha},\alpha\in I$ are open sets in $\Bbb R^n$ such that
$$
E\subset \bigcup_{\alpha\in I}O_{\alpha}
$$
then there is a finite subset
$$
F=\{\alpha_1,...,\alpha_N\}\subset I
$$
such that
$$
E\subset\bigcup_{k=1}^NO_{\alpha_k}
$$
\bigskip

(b) What is a regular measure on $(\Bbb R^n,\mathcal B(\Bbb R^n))?$.  Show that every probability measure on $(\Bbb R^n,\mathcal B(\Bbb R^n))$ is regular.

Solution: A measure $\mu$ on $(\Bbb R^n,\mathcal B(\Bbb R^n))$ is said to be regular if given any $B\in\mathcal B(\Bbb R^n)$ and $\epsilon>0$, there exists an open set $F$ and a closed set $C$ in $\Bbb R^n$ such that $C\subset B\subset F$ and $\mu(F-C)<\epsilon$. Let us now show that any probability measure (and hence any finite measure) on $(\Bbb R^n,\mathcal B(\Bbb R^n))$ is regular. Let $P$ be a probability measure on the above space.
Define $\mathcal C$ to be the family of all sets $B\in\mathcal B(\Bbb R^n)$ such that $\forall\epsilon>0$, there exists a closed set $C$ and an open set $F$ such that
$C\subset B\subset F$ and $P(F-C)<\epsilon$. We prove that $\mathcal C$ is a  $\sigma$-field and then since it contains all closed sets (If $C$ is a closed set and
$F_n=\{x\in\Bbb R^n:d(x,C)<1/n\}$, then $F_n$ is open, $C\subset C\subset F_n$ and
$P(F_n-C)\rightarrow 0$ because $F_n\downarrow C$), it will follow that $\mathcal C=\mathcal B(\Bbb R^n)$ which will complete the proof of the result. 
Let $B\in\mathcal C$ and let $\epsilon>0$. Then, there exists a closed $C$ and an open $F$
such that $C\subset B\subset F$ and $P(F-C)<\epsilon$. Then $F^c$ is closed, $C^c$ is open, $F^c\subset B^c\subset C^c$ and $P(C^c-F^c)=P(F-C)<\epsilon$, thereby proving that $B^c\in\mathcal C$, ie, $\mathcal C$ is closed under complementation. Now, let $B_n\in\mathcal C, n=1,2,...$ and let $\epsilon>0$. Then, for each $n$ we can choose a closed $C_n$ and an open $F_n$ such that $C_n\subset B_n\subset F_n$ with $P(F_n-C_n)<\epsilon/2^n$. Then, $\bigcup_{n=1}^NC_n$ is closed for any finite integer $N$, $\bigcup_{n=1}^{\infty}F_n$ is open and for $B=\bigcup_nB_n$, we have that
$$
\bigcup_{n=1}^NC_n\subset B\subset\bigcup_nF_n, N\geq 1
$$
and further
$$
P(\bigcup_nF_n-\bigcup_{n=1}^NC_n)\leq P((\bigcup_{n=1}^N(F_n-C_n))\bigcup\bigcup_{n>N}F_n)
$$
$$
\leq\sum_{n=1}^NP(F_n-C_n)+\sum_{n>N}P(F_n)\leq\sum_{n=1}^N\epsilon/2^n+\sum_{n>N}P(F_n)
$$
which converges as $N\rightarrow\infty$ to $\epsilon$ and hence for sufficiently large $N$, it must be smaller than $2\epsilon$ proving thereby that $B\in\mathcal C$, ie, $\mathcal C$ is closed under countable unions. This completes the proof that $\mathcal C$ is a $\sigma$-field and hence the proof of the result.
\bigskip

(c) What is a consistent family of finite dimensional probability distributions on $(\Bbb R^{\Bbb Z_+},\mathcal B(\Bbb R^{\Bbb Z_+})) ?$

Solution: Such a family is a family of probability distributions $F(x_1,...,x_N,n_1,...,n_N)$ on $\Bbb R^N$ for each set of positive integers $n_1,...,n_N$and $N=1,2,...$ satisfying

(a) 
$$
F(x_{\sigma 1},..., x_{\sigma N},n_{\sigma 1},..., n_{\sigma N})=F(x_1,...,x_N,n_1,...,n_N)
$$
for all permutations $\sigma$ of $(1,2,..., N)$ and

(b) 
$$
lim_{x_N\rightarrow\infty}F(x_1,...,x_N,n_1,...,n_N)=F(x_1,...,x_{N-1},n_1,...,n_{N-1})
$$

(d) Prove Kolmogorov's consistency theorem, that given any consistent family of finite dimensional probability distributions on $(\Bbb R^{\Bbb Z_+},\mathcal B(\Bbb R^{\Bbb Z_+}))$, there exists a probability space and a real valued stochastic process defined on this space whose finite dimensional distributions coincide with the given ones.

Let $F(x_1,...,x_N;n_1,...,n_N), n_1,...,n_N\geq 0, x_1,...,x_N\in\Bbb R, N\geq 1$ be a consistent family of probability distributions. Thus $F(x_1,...,x_N;1,2,...,N)=F_N(x_1,...,x_N)$ is a probability distribution on $\Bbb R^N$ with the property
that
$$
lim_{x_N\rightarrow\infty}F_N(x_1,...,x_N)=F_{N-1}(x_1,...,x_{N-1})
$$
The aim is to construct a (countably additive) probability measure $P$ on $(\Bbb R^{\Bbb Z_+},\mathcal B(\Bbb R^{\Bbb Z_+})$ such that
$$
P(B\times\Bbb R^{\Bbb Z_+})=P_N(B), B\in\mathcal B(\Bbb R^N)
$$
where
$$
P_N(B)=\int_BF_N(x_1,...,x_N)dx_1...dx_N
$$
Define
$$
Q(B\times\Bbb R^{Z_+})=P_N(B), B\in\mathcal B(\Bbb R^N)
$$
Then by the consistency of $F_N, N\geq 1$, $Q$ is a well defined finitely additive probability measure on the field
$$
\mathcal F=\bigcup_{N\geq 1}(\mathcal B(\Bbb R^N)\times\Bbb R^{\Bbb Z_+})
$$
Note that 
$$
\mathcal B_N=\mathcal B(\Bbb R^N)\times\Bbb R^{\Bbb Z_+}
$$
is a field on $\Bbb R^{\Bbb Z_+}$ for every $N\geq 1$ and that
$$
\mathcal B_N\subset\mathcal B_{N+1},N\geq 1
$$
It suffices to prove that $Q$ is countably additive on $\mathcal F$, or equivalently, if $\tilde B_N\in\mathcal\mathcal B_N$ is a sequence such that $\tilde B_N\downarrow\phi$, then $Q(\tilde B_N)\downarrow 0$. Equivalently, it suffices to show that if
$B_N\in\mathcal B(\Bbb R^N)$ is a sequence such that
$$
\tilde B_N=B_N\times\Bbb R^{\Bbb Z_+}\downarrow\phi
$$
then
$$
P_N(B_N)\downarrow 0
$$
Suppose that this is false. Then, without loss of generality, we may assume that there is a $\delta>0$ such that $P_N(B_N)\geq\delta\forall N\geq 1$.
By regularity of probability measures on $\Bbb R^N$ for $N$ finite, we can find for each $N\geq 1$, a compact set $K_N\in\mathcal B(\Bbb R^N)$ such that $K_N\subset B_N$ and
$P_N(B_N-K_N)<\delta/2^N$. Now consider
$$
\tilde K_N=(K_1\times\Bbb R^{N-1})\cap(K_2\cap\Bbb R^{N-2})\cap...\cap(K_{N-1}\times\Bbb R)\times K_N
$$
Then 
$$
\tilde K_N\subset\Bbb K_N\subset\Bbb R^N
$$
and $\tilde K_N$ being a closed subset of the compact set $K_N$ is also compact. $\tilde K_N$ is closed because each $K_j,j=1,2,..., N$ is compact and hence closed. 
Define
$$
\hat K_N=\tilde K_N\times\Bbb R^{\Bbb Z_+}, N\geq 1
$$
Then, it is clear that $\hat K_N$ is decreasing and that
$$
\hat K_N\subset\tilde B_N=B_N\times\Bbb R^{\Bbb Z_+}
$$
Further,
$$
Q(\hat K_N)=Q(\tilde B_N)-Q(\tilde B_N-\hat K_N)
$$
Now,
$$
Q(\tilde B_N)=P_N(B_N)\geq\delta
$$
and
$$
Q(\tilde B_N-\hat K_N)=P_N(B_N-\bigcap_{k=1}^N(K_k\times\Bbb R^{N-k}))
$$
$$
=P_N(\bigcup_{k=1}^N(B_N-(K_k\times\Bbb R^{N-k}))
$$
$$
\leq P_N(\bigcup_{k=1}^N(B_k-K_k)\times\Bbb R^{N-k})
$$
$$
\leq\sum_{k=1}^NP_k(B_k-K_k)=\sum_{k=1}^N\delta/2^k\leq\delta/2
$$
Thus,
$$
Q(\hat K_N)\geq\delta/2
$$
and hence $\hat K_N$ is non-empty for each $N$. Hence, so is $\tilde K_N$ for each $N$.
Now $\hat K_N\subset\tilde B_N\downarrow\phi$. Hence $\hat K_N\downarrow
\phi$ and we shall obtain a contradiction to this as follows: Since $\tilde K_N$ is non-empty, we may choose 
$$
(x_{N1},...,x_{NN})\in\tilde K_N, N=1,2,...
$$
Then,
$$
x_{N1}\in K_1, (x_{N1},x_{N2})\in K_2,..., (x_{N1},...,x_{NN})\in K_N
$$
Since $K_1$ is compact, $x_{N1},N\geq 1$ has a convergent subsequence say
$\{x_{N_1,1}\}$. Now, $(x_{N_1,1},x_{N_1,2})\in K_2$ and since $K_2$ is compact,
this has a convergent subsequence say $(x_{N_2,1},x_{N_2,2})$. Then, $\{x_{N_2,1}\}$ is a subsequence of $\{x_{N_1,1}\}$ and hence also converges. Continuing this way, we finally obtain a convergent sequence $(y_{n,1},..., y_{n,N}), n=1,2,...$ in $K_N$  such that
$(y_{n,1},...y_{n,j})\in K_j, j=1,2,..., n$ and of course $(y_{n,1},..., y_{n,j}), n\geq 1$ converges. Let 
$$
y_j=lim_ny_{n,j}, j=1,2,...
$$
Then, it is clear that since $K_j$ is closed,
$$
(y_1,..., y_j)\in K_j, j=1,2,...
$$
Hence
$$
(y_1,y_2,...y_N)\in\tilde K_N, N\geq 1
$$
and hence,
$$
(y_1,y_2,...)\in\hat K_N, N\geq 1
$$
which means that
$$
(y_1,y_2,...)\in\bigcap_{N\geq 1}\hat K_N
$$
which contradicts the fact that the rhs is an empty set. This proves the Kolmogorov consistency theorem.
\newpage

\section{Problem suggested by Prof.Dhananjay}

When the magnet falls within the tube of radius $R_0$ with the coil having $n_0$ turns per unit length and extending from $z=h$ to $z=h+a$ and with the tube extending from $z=0$ to
$z=d$, we write down the equations of motion of the magnet. We usually may assume that at time $t$, the magnetic moment vector of the magnet is ${\bf m}$ and that its centre does not deviate from the central axis of the tube. This may be safely assumed provided that the tube's radius is small. However suppose we do not make this assumption so that we aim a the most general possible non-relativistic situation in which the magnetic fields are strong which may cause the position of the CM of the magnet to deviate from the central axis of the cylinder. At time $t$, assume that the CM of the magnet is located at
${\bf r}_0(t)=(\xi_1(t),\xi_2(t),\xi_3(t))$. We require a differential equation for $\xi(t)$. The magnet at time $t$ is assumed to have experienced a rotation ${\bf R}(t)$ (a $3\times 3$ rotation matrix). Thus, it moment at time $t$ is given by
$$
{\bf m}(t)=m_0{\bf R}(t)\hat z
$$
assuming that at time $t=0$, the moment pointed along the positive $z$ direction. The magnetic field produced by the magnet at time $t$ is given by
$$
{\bf B}_m(t,{\bf r})=curl (\mu/4\pi)({\bf m}(t)\times({\bf r}-{\bf r}_0(t))/|{\bf r}-{\bf r}_0(t)|^3)
$$
and the flux of this magnetic field through the coil is
$$
\Phi(t)=n_0\int_h^{h+a}dz\int_{x^2+y^2\leq R_0^2}B_{mz}(t,x,y,z)dxdy
$$
The angular momentum vector of the magnet is proportional to its magnetic moment. So by the torque equation, we have
$$
d{\bf m}(t)/dt=\gamma.{\bf m}(t)\times{\bf B}_c(t,{\bf r}_0(t))---(1)
$$
where ${\bf B}_c(t,{\bf r})$ is the magnetic field produced by the coil. The coil current is
$$
I(t)=\Phi'(t)/R_L
$$
where $R_L$ is the coil resistance. Then, the magnetic field produced by the coil is
$$
{\bf B}_c(t,{\bf r})=(\mu/4\pi)I(t)\int_0^{2n_0\pi}[(-\hat x.sin(\phi)+\hat y.cos(\phi))\times({\bf r}-R_0(cos(\phi)\hat x+sin(\phi)\hat y)-(a/2\pi n_0)\phi\hat z]
$$
$$
/|{\bf r}-R_0(cos(\phi)\hat x+sin(\phi)\hat y)-(a/2\pi n_0)\phi\hat z|^3R_0d\phi
$$
This expression can be used in (1). Note that $I(t)$ depends on $\Phi(t)$ which in turn is a function of ${\bf m}(t)$. Since in fact $I(t)$ is proportional to $\Phi'(t)$, it follows that $I(t)$ is function of ${\bf m}(t)$ and ${\bf m}'(t)$. Note that ${\bf B}_c(t, {\bf r})$ is proportional to $I(t)$ and is therefore also a function of ${\bf m}(t)$ and ${\bf m}'(t)$. However, we also note that ${\bf B}_m(t,{\bf r})$ is also a function of ${\bf r}_0(t)$. Hence, $\Phi(t)$ is also a function of ${\bf r}_0(t)$ and  hence $I(t)$ is also a function of ${\bf r}_0(t)$ and ${\bf r}_0'(t)$. Thus, ${\bf B}_c(t,{\bf r})$ is a function of ${\bf r}_0(t),{\bf r}_0'(t),{\bf m}(t),{\bf m}'(t)$, ie, we can express it as
$$
{\bf B}_c(t,{\bf r}|{\bf r}_0(t),{\bf r}_0'(t),{\bf m}(t),{\bf m}'(t))
$$
Thus, we write (1) as
$$
d{\bf m}(t)/dt=\gamma.{\bf m}(t)\times{\bf B}_c(t,{\bf r}_0(t)|{\bf r}_0(t),{\bf r}_0'(t),{\bf m}(t),{\bf m}'(t))---(2)
$$
Next, the force exerted on the magnet by the magnetic field produced by the coil is given by
$$
{\bf F}(t)=\nabla_{{\bf r}_0(t)}({\bf m}(t).{\bf B}_c((t,{\bf r}_0(t)|{\bf r}_0(t),{\bf r}_0'(t),{\bf m}(t),{\bf m}'(t))
$$
which means that the rectilinear equation of motion of the magnet must be given by
$$
M{\bf r}_0''(t)={\bf F}(t)-Mg\hat z
$$
However to be more precise, the term $V=-{\bf m}(t).{\bf B}_c(t,{\bf r}_0(t)|{\bf r}_0(t),{\bf r}_0'(t),{\bf m}(t),{\bf m}'(t))$ represents a potential energy of interaction between the magnet and the reaction field produced by the coil. So its recilinear equation of motion should be given by
$$
d/dt\partial L/\partial{\bf r}_0'(t)-\partial L/\partial{\bf r}_0(t)=0
$$
where $L$, the Lagrangian is given by
$$
L=M|{\bf r}_0'(t)|^2/2-V-mgz_0(t)
$$
Thus the rectilinear equation of motion of the coil is given by
$$
M{\bf r}_0''(t)+d/dt(\nabla_{{\bf r}_0'(t)}{\bf m}(t).{\bf B}_c(t,{\bf r}_0(t)|{\bf r}_0(t),{\bf r}_0'(t),{\bf m}(t),{\bf m}'(t)))
$$
$$
-\nabla_{{\bf r}_0(t)}({\bf m}(t).{\bf B}_c(t,{\bf r}_0(t)|{\bf r}_0(t),{\bf r}_0'(t),{\bf m}(t),{\bf m}'(t)))+mg\hat z={\bf 0}---(3)
$$
(2) and (3) constitute the complete set of six coupled ordinary differential equations for the six components of ${\bf m}(t),{\bf r}_0(t)$.

\section{Converse of the Shannon coding theorem} Let $A$ be an alphabet. Let $u_1,..., u_M\in\Bbb A^n$ with
$$
P_{u_i}=P
$$
where for $u\in A^n$, $P_u(x)=N(x|u)/n, x\in A$ with $N(x|u)$ denoting the number of times that $x$ has occurred in $u$. Let $\nu_x(y), x\in A,y\in B$ be the discrete memory channel transition probability from the alphabet $A$ to the alphabet $B$. Define
$$
Q(y)=\sum_xP(x)\nu_x(y), y\in B
$$
let $D_1,..., D_M$ be disjoint sets in $B^n$ such that $\nu_{u_k}(D_k)>1-\epsilon, k=1,2,..., n$, where $\nu_u(v)=\Pi_{k=1}^n\nu_{u(k)}(v(k))$ for $u=(u(1),..., u(n)), v=(v(1),..., v(n))$. Let 
$$
E_k=D_k\cap T(n,\nu_{u_k},\delta), k=1,2,..., M
$$
where $T(n,f,\delta)$ is the set of $\delta$-Bernoulli typical sequences of length $n$  having probability distribution $f$. For $u\in A^n$ with $P_u=P$, define
$T(n,\nu_u,\delta)$ to consist of all $v\in B^n$ for which
$$
v_x\in T(N(x|u),\nu_x,\delta)\forall x\in A
$$
Then we claim that if $v\in T(n,\nu_u,\delta)$, then
$$
v\in T(n,Q,\delta\sqrt a)
$$
The notation used is as follows. For $x\in A$, and $u\in A^n,v\in B^n$, define
$$
v_x=(v(k):u(k)=x, k=1,2,..., n)
$$
Note that $v_x\in B^{N(x|u)}$. To prove this claim, we note that for $y\in B$,
$$
|N(y|v)-nQ(y)|=|\sum_xN(y|v_x)-n\sum_xP(x)\nu_x(y)|
$$
$$
=|\sum_xN(y|v_x)-\sum_xN(x|u)\nu_x(y)|\leq\sum_x|N(y|v_x)-N(x|u)\nu_x(y)|
$$
$$
\leq\sum_x\delta\sqrt{N(x|u)\nu_x(y)(1-\nu_x(y))}
$$
$$
\leq\delta\sqrt a.\sqrt{\sum_xN(x|u)\nu_x(y)(1-\nu_x(y))}
$$
$$
=\delta\sqrt{an}.\sqrt{\sum_xP(x)\nu_x(y)(1-\nu_x(y))}
$$
Now,
$$
\sum_xP(x)\nu_x(y)=Q(y),
$$
$$
\sum_xP(x)\nu_x(y)^2\geq(\sum_xP(x)\nu_x(y))^2=Q(y)^2
$$
and therefore,
$$
|N(y|v)-nQ(y)|\leq \delta\sqrt{an}\sqrt{Q(y)(1-Q(y)}
$$
proving the claim. For $v\in E_k$, we have $v\in T(n,\nu_{u_k},\delta)$ and hence by the result just proved, $v\in T(n,Q,\delta\sqrt a)$. 
\bigskip

Now let $v\in T(n,Q,\delta)$.

It follows that since
$$
Q^n(v)=\Pi_{y\in B}Q(y)^{N(y|v)}
$$
and since
$$
|N(y|v)-nQ(y)|\leq\delta\sqrt{nQ(y)(1-Q(y))}, y\in B
$$
we have
$$
nQ(y)-\delta\sqrt{nQ(y)(1-Q(y))}\leq N(y|v)\leq nQ(y)+\delta\sqrt{nQ(y)(1-Q(y))}, y\in B
$$
and therefore,
$$
\Pi_yQ(y)^{nQ(y)+\delta\sqrt{nQ(y)(1-Q(y))}}\leq Q_n(v)\leq \Pi_yQ(y)^{nQ(y)-\delta\sqrt{nQ(y)(1-Q(y))}}
$$
Thus summing over all $v\in T(n,Q,\delta)$, we get
$$
2^{-nH(Q)-c\sqrt n}\mu(T(n,Q,\delta))\leq Q_n(T(n,Q,\delta)\leq 2^{-nH(Q)+c\sqrt n}\mu(T(n,Q,\delta))
$$
where
$$
c=-\delta\sum_y\sqrt{Q(y)(1-Q(y))}log_2(Q(y))
$$
and where $\mu(E)$ denotes the cardinality of a set $E$. In particular,
$$
\mu(T(n,Q,\delta))\leq 2^{nH(Q)+c.sqrt n}
$$
It follows that 
$$
\sum_{k=1}^M\mu(E_k)=\mu(\bigcup_kE_k)\leq\mu(T(n,Q,\delta\sqrt a))\leq 2^{nH(Q)+c\sqrt {an}}
$$
On the other hand,
$$
\nu_{u_k}(D_k)>1-\epsilon
$$
and by the Chebyshev inequality combined with the union bound,
$$
\nu_{x,N(x|u)}(T(N(x|u),\nu_x,\delta))\geq 1-b/\delta^2
$$
and hence
$$
\nu_u(T(n,\nu_u,\delta))\geq 1-ab/\delta^2
$$
for any $u\in A^n$. In particular,
$$
\nu_{u_k}(E_k)\geq 1-ab/\delta^2-\epsilon, k=1,2,..., M
$$
Further, by the above argument, we have for $v\in T(n,\nu_u,\delta)$,
$$
\nu_{x,N(x|u)}(v_x)\leq 2^{N(x|u)H(\nu_x)+c\sqrt n}
$$
and hence,
$$
\nu_u(v)\leq 2^{-\sum_xN(x|u)H(\nu_x)-c\sqrt n}
$$
$$
=2^{-nI(P,\nu)+c\sqrt n}
$$
where
$$
I(P,\nu)=\sum_xP(x)H(\nu_x), P=P_u
$$
From this we easily deduce that
$$
\nu_{u_k}(E_k)\leq 2^{-nI(P,\nu)+c\sqrt n}\mu(E_k), k=1,2,..., M
$$
Thus,
$$
(1-ab/\delta^2-\epsilon)2^{nI(P,\nu)-c\sqrt n}\leq\mu(E_k), k=1,2,..., M
$$
so that
$$
M(1-ab/\delta^2-\epsilon)2^{nI(P,\nu)-c\sqrt n}\leq\sum_k\mu(E_k)\leq 2^{nH(Q)+c\sqrt n}
$$
or equivalently,
$$
log(M)/n\leq H(Q)-I(P,\nu)+O(1/\sqrt n)
$$
which proves the converse of Shannon's coding theorem.
\bigskip

\chapter{Problems in electromagnetics and gravitation}

The magnetic field produced by a magnet of given magnetic dipole moment in a background gravitational field taking space-time curvature into account. The Maxwell equations in curved space-time are
$$
(F^{\mu\nu}\sqrt{-g})_{,\nu}=J^{\mu}\sqrt{-g}
$$
or equivalently
$$
(g^{\mu\alpha}g^{\nu\beta}\sqrt{-g}F_{\alpha\beta})_{,\nu}=J^{\mu}\sqrt{-g}
$$
Writing
$$
g_{\mu\nu}(x)=\eta_{\mu\nu}+h_{\mu\nu}(x)
$$
where $\eta$ is the Minkowski metric and $h$ is a small perturbation, we have approximately
$$
g^{\mu\nu}=\eta_{\mu\nu}-h^{\mu\nu}, h^{\mu\nu}=\eta_{\mu\alpha}\eta_{\nu\beta}h_{\alpha\beta}
$$
and
$$
\sqrt{-g}=1-h/2, h=\eta_{\mu\nu}h_{\mu\nu}
$$
We have
$$
F_{\mu\nu}=A_{\nu,\mu}-A_{\mu,\nu}
$$
and writing
$$
A_{\mu}=A_{\mu}^{(0)}+A_{\mu}^{(1)}
$$
where $A{\mu}^{(0)}$ is the zeroth order solution, ie, without gravity and $A_{\mu}^{(1)}$ the first order correction to the solution, ie, it is a linear functional of $h_{\mu\nu}$, we obtain using first order perturbation theory
$$
F_{\mu\nu}=F_{\mu\nu}^{(0)}+F_{\mu\nu}^{(1)}
$$
where
$$
F_{\mu\nu}^{(k)}=A_{\nu,\mu}^{(k)}-A_{\mu,\nu}^{(k)}, k=0,1
$$
$$
g^{\mu\alpha}g^{\nu\beta}\sqrt{-g}=
$$
$$
(\eta_{\mu\alpha}-h^{\mu\alpha})(\eta_{\nu\beta}-h^{\nu\beta})(1-h/2)
$$
$$
=\eta_{\mu\alpha}\eta_{\nu\beta}-f_{\mu\nu\alpha\beta}
$$
where
$$
f_{\mu\nu\alpha\beta}=\eta_{\mu\alpha}\eta_{\nu\beta}h/2+\eta_{\mu\alpha}h^{\nu\beta}
$$
$$
+\eta_{\nu\beta}h^{\mu\alpha}
$$
Then
$$
g^{\mu\alpha}g^{\nu\beta}\sqrt{-g}F_{\alpha\beta}=
$$
$$
\eta_{\mu\alpha}\eta_{\nu\beta}F_{\alpha\beta}^{(0)}
$$
$$
+\eta_{\mu\alpha}\eta_{\nu\beta}F_{\alpha\beta}^{(1)}-f_{\mu\nu\alpha\beta}F_{\alpha\beta}^{(0)}
$$
$$
=F^{0\mu\nu}+F^{1\mu\nu}
$$
where
$$
F^{0\mu\nu}=\eta_{\mu\alpha}\eta_{\nu\beta}F_{\alpha\beta}^{(0)},
$$
$$
F^{1\mu\nu}=\eta_{\mu\alpha}\eta_{\nu\beta}F_{\alpha\beta}^{(1)}-f_{\mu\nu\alpha\beta}F_{\alpha\beta}^{(0)}
$$
Also
$$
J^{\mu}\sqrt{-g}=J^{0\mu}+J^{1\mu}, J^{0\mu}=J^{\mu}, J^{1\mu}=-J^{\mu}h/2
$$
The gauge condition
$$
(A^{\mu}\sqrt{-g})_{,\mu}=0
$$
becomes
$$
0=(g^{\mu\nu}\sqrt{-g}A_{\nu})_{,\mu}=
$$
$$
[(\eta_{\mu\nu}-h^{\mu\nu})(1-h/2)A_{\nu}]_{,\mu}
$$
$$
=[(\eta_{\mu\nu}-k_{\mu\nu})(A_{\nu}^{(0)}+A_{\nu}^{(1)})]_{,\mu}
$$
where
$$
k_{\mu\nu}=h^{\mu\nu}-h\eta_{\mu\nu}/2
$$
and hence equating coefficients of zeroth and first orders of smallness on both sides gives
$$
\eta_{\mu\nu}A_{\nu,\mu}^{(0)}=0,
$$
$$
(k_{\mu\nu}A_{\nu}^{(0)})_{,\mu}=\eta_{\mu\nu}A_{\nu,\mu}^{(1)}
$$
\bigskip

\chapter{Some more aspects of quantum information theory} 

\section{An expression for state detection error}

Let $\rho,\sigma$ be two states and let their spectral decompositions be
$$
\rho=\sum_a|e_a>p(a)<e_a|, \sigma=\sum_a|f_a>q(a)<f_a|
$$
so that $<e_a|e_b>=\delta(a,b), <f_a|f_b>=\delta(a,b)$. The relative entropy between these two states is
$$
D(\rho|\sigma)=Tr(\rho.(log(\rho)-log(\sigma)))
$$
$$
=\sum_ap(a)log(p(a))-\sum_{a,b}p(a)log(q_b)|<e_a|f_b>|^2
$$
$$
=\sum_{a,b}(p(a)|<e_a|f_b>|^2(log(p(a)|<e_a|f_b>|^2)-log(q_b|<e_a|f_b>|^2))
$$
$$
=\sum_{a,b}(P(a,b)log(P(a,b))-log(Q(a,b)))=D(P|Q)
$$
where $P,Q$ are two bivariate probability distributions defined by
$$
P(a,b)=p(a)|<e_a|f_b>|^2, Q(a,b)=q(b)|<e_a|f_b>|^2
$$
Now let $0\leq T\leq 1$, ie, $T$ is a detection operator. Consider the detection error probabiity
$$
2P(\epsilon)=Tr(\rho(1-T))+Tr(\sigma T)
$$
$$
=\sum_a(p(a)<e_a|1-T|e_a>-q(a)<f_a|T|f_a>)
$$
$$
=\sum_ap(a)(1-<e_a|T|e_a>)-\sum_{a,b}q(a)|<f_a|e_b>|^2<e_b|T|f_a>
$$
Suppose $T$ is a projection, ie, $T^2=T^*=T$. Then,
$$
Tr(\rho(1-T))=\sum_ap(a)<e_a|1-T)|e_a>=\sum_ap(a)<e_a|(1-T)^2|e_a>
$$
$$
=\sum_{a,b}p(a)|<e_a|1-T|f_b>|^2
$$
Likewise,
$$
Tr(\sigma T)=\sum_{a,b}q(b)<e_a|T|f_b>|^2
$$
Hence,
$$
2P(\epsilon)=\sum_{a,b}(p(a)|<e_a|1-T|f_b>|^2+q(b)<e_a|T|f_b>|^2)
$$
$$
\geq\sum_{a,b}min(p(a),q(b))(|<e_a|1-T|f_b>|^2+q(b)<e_a|T|f_b>|^2)
$$
$$
\geq\sum_{a,b}(1/2)min(p(a),q(b))|<e_a|f_b>|^2
$$
where we have used
$$
2(|z|^2+|w|^2)\geq|z+w|^2
$$
for any two complex numbers $z,w$. Note that
$$
\sum_{a,b}(1/2)min(p(a),q(b))|<e_a|f_b>|^2=
$$
$$
(1/2)min_{0\leq t(a,b)\leq 1, a,b=1,2,..., n}\sum_{a,b}[p(a)|<e_a|f_b>|^2(1-t(a,b))+q(b)|<e_a|f_b>|^2t(a,b))]
$$
because if $u,v$ are positive real numbers, then as $t$ varies over $[0,1]$,
$tu+(1-t)v$  attains is minimum value of $v$ when $t=0$ if $u\geq v$ and $u$ when $t=1$ if
$u<v$. In other words, 
$$
min_{t\in[0,1]}(tu+(1-t)v)=min(u,v)
$$
In other words we have proved that
$$
2P(\epsilon)\geq (1/2)min_{0\leq t(a,b)\leq 1, a,b=1,2,..., n}\sum_{a,b}[P(a,b)(1-t(a,b))+Q(a,b)t(a,b)]
$$
The quantity on the rhs is the minimum error probability for the classical hypothesis testing problem between the two bivariate probability distributions $P,Q$. 
\bigskip

\section{Operator convex functions and operator monotone functions}

[1] Let $f$ be operator convex. Then, if $K$ is a contraction matrix, we have
$$
f(K^*XK)\leq K^*f(X)K
$$
for all Hermitian matrices $X$. By contraction we mean that $K^*K\leq I$. To see this
define the following positive (ie non-negative) matrices
$$
L=\sqrt{1-K^*K}, M=\sqrt{1-KK^*}
$$
Then by Taylor expansion and the identity
$$
K(K^*K)^m=(KK^*)^mK, m=1,2,...,
$$
we have
$$
KL=MK, LK^*=K^*M
$$
Define the matrix
$$
U=\left(\begin{array}{cc}K&-M\\L&K^*\end{array}\right)
$$
and observe that since $L^*=L,M^*=M$ and
$$
K^*K+L^2=I, -K^*M+LK^*=0, -MK+KL=0, M^2+KK^*=I
$$
it follows that $U$ is a unitary matrix. Likewise define the unitary matrix
$$
V=\left(\begin{array}{cc}K&M\\L&-K^*\end{array}\right)
$$
Also define for two Hermitian matrices $A,B$
$$
T=\left(\begin{array}{cc}A&0\\0&B\end{array}\right),
$$
Then, we obtain the identity
$$
(1/2)(U^*TU+V^*TV)=\left(\begin{array}{cc}K^*AK+LBL&0\\0&KBK^*+MAM\end{array}\right)
$$
Using operator convexity of $f$, we have
$$
f((1/2)(U^*TU+V^*TV))\leq (1/2)(f(U^*TU)+f(V^*TV))=(1/2)(U^*f(T)U+V^*f(T)V)
$$
which in view of the above identity, results in
$$
f(K^*AK+LBL)\leq K^*f(A)K+Lf(B)L
$$
and
$$
f(KBK^*+MAM)\leq Kf(B)K^*+Mf(A)M
$$
In particular, taking $B=0$, the first gives
$$
f(K^*AK)\leq K^*f(A)K+f(0)L^2
$$
and choosing $A=0$, the second gives
$$
f(KBK^*)\leq Kf(B)K^*+f(0)M^2
$$
In particular, if in addition to being operator convex, $f$ satisfies $f(0)\leq 0$, then
we get
$$
f(K^*AK)\leq K^*f(A)K, f(KBK^*)\leq Kf(B)K^*
$$
for any contraction $K$. Now we show that if $f$ is operator convex and $f(0)\leq 0$, then
for operators $K_1,..., K_p$ that satisfy
$$
\sum_{j=1}^pK_j^*K_j\leq I
$$
we have
$$
f(\sum_{j=1}^pK_j^*A_jK_j)\leq\sum_{j=1}^pK_j^*f(A_j)K_j
$$
In fact, defining
$$
K=\left(\begin{array}{ccccc}K_1&0&...&0\\K_2&0&...&0\\..&..&...&..\\K_p&0&...&0\end{array}\right)
$$
we observe that
$$
K^*K=\left(\begin{array}{cc}\sum_{j=1}^pK_j^*K_j&0\\0&0\end{array}\right)\leq I
$$
ie $K$ is a contraction and hence by the above result
$$
f(K^*TK)\leq K^*f(T)K
$$
Taking
$$
T=diag[A_1,A_2,..., A_p]
$$
we get
$$
K^*TK=\left(\begin{array}{cc}\sum_{j=1}^pK_j^*A_jK_j&0\\0&0\end{array}\right)
$$
and hence,
$$
\left(\begin{array}{cc}f(\sum_{j=1}^pK_j^*A_jK_j)&0\\0&f(0)\end{array}\right)
$$
$$
\leq\left(\begin{array}{cc}\sum_{j=1}^pK_j^*f(A_j)K_j&0\\0&0\end{array}\right)
$$
which results in the desired inequality. Note that from the above discussion, we also have
$$
f(\sum_jK_jB_jK_j^*)\leq\sum_jK_jf(B_j)K_j^*
$$
\bigskip

\section{A selection of matrix inequalities}

[1] If $A,B$ are square matrices the eigenvalues of $AB$ are the same as those of $BA$. 
If $A$ is non-singular, then this is obvious since
$$
A^{-1}(AB)A=BA
$$
Likewise if $B$ is non-singular then also it is obvious since
$$
B^{-1}(BA)B=AB
$$
If $A$ is singular, we can choose a sequence $c_n\rightarrow 0$ so that $A+c_nI$ is non-singular for every $n$ and then
$$
(A+c_nI)^{-1}(A+c_nI)B(A+c_nI)=B(A+c_nI)\forall n
$$
Thus, $(A+c_nI)B$ has the same eigenvalues as those of $B(A+c_nI)$ for each $n$. Taking the limit $n\rightarrow\infty$, the result follows since the eigenvalues of $(A+cI)B$ are continuous functions of $c$.

[2] Let $A$ be any square diagonable matrix. Let $\lambda_1,...,\lambda_n$ be its eigenvalues. We have for any unit vector $u$,
$$
|<u,Au>|^2\leq <u,A^*Au>
$$
Let $e_1,..., e_n$ be a basis of unit vectors of $A$ corresponding to the eigenvalues $\lambda_1,..., \lambda_n$ respectively. Assume that
$$
|\lambda_1>\geq|\lambda_2|\geq...\geq|\lambda_n|
$$
Let $M$ be any $k$ dimensional subspace and consider the $n-k+1$ dimensional subspace $W=span\{e_k, e_{k+1}..., e_n\}$. Then $dim(M\cap W)\geq 1$ and hence we can choose a unit vector $u\in M\cap W$. Then writing $u=c_ke_k+...+c_ne_n$, it follows that 
$$
|<u,Au>|=|c_k\lambda_ke_k+...+c_n\lambda_ne_n|\leq|\lambda_k|.(|c_k|+...+|c_n|)
$$
Suppose we Gram-Schmidt orthonormalize $e_k,e_{k+1},..., e_n$ in that order. The resulting vectors are
$$
w_k=e_k, w_{k+1}\in span(e_k,e_{k+1}),..., w_n\in span(e_k,..., e_n)
$$
Specifically,
$$
w_{k+1}=r(k+1,k)e_k+r(k+1,k+1)e_{k+1},..., w_n=r(n,k)e_k+...+r(n,n)e_n
$$
Then we can write
$$
u=d_kw_k+...+d_mw_n, |d_k|^2+...+|d_n|^2=1, <w_r,w_s>=\delta(r,s), r,s=k,k+1,..., n
$$
and we find that
$$
Aw_k=\lambda_kw_k, Aw_{k+1}=r(k+1,k)\lambda_ke_k+r(k+1,k+1)\lambda_{k+1}e_{k+1},
$$
$$
Aw_n=r(n,k)\lambda_ke_k+...+r(n,n)\lambda_ne_n
$$
Diagonalize $A$ so that
$$
A=\sum_{k=1}^n\lambda_ke_kf_k^*, f_k^*e_j=\delta(k,j)
$$
Then,
$$
<f_k,Ae_k>=\lambda_k
$$
Thus,
$$
|\lambda_k|\leq \parallel f_k\parallel.\parallel Ae_k\parallel
$$

Let $A^{\otimes_a k}$ denote the $k$-fold tensor product of $A$ with itself acting on the k fold antisymmetric tensor product of $\Bbb C^n$. Then, the eigenvalues of $A^{\otimes_an}$ are $\lambda_{i_1}...\lambda_{i_k}$ with $n\geq i_1>i_2>...>i_k\geq 1$ and likewise its singular values are $s_{i_1}...s_{i_k}, n\geq i_1>i_2>...>i_k\geq 1$ where $s_1\geq s_2\geq...\geq s_n\geq 0$ are the singular values of $A$. Now if $T$ is any operator whose eigenvalue having maximum magnitude is $\mu_1$ and whose maximum singular value is $s_1$, then we have
$$
|\mu_1|=spr(T)=lim_n\parallel T^n\parallel^{1/n}\leq\parallel T\parallel=s_1
$$
Applying this result to $T=A^{\otimes_ak}$ gives us
$$
\Pi_{i=1}^k|\lambda_i|\leq\Pi_{i=1}^ks_i, k=1,2,..., n
$$
\bigskip

[3] If $A,B$ are such that $AB$ is normal, then
$$
\parallel AB\parallel\leq\parallel BA\parallel
$$
In fact since $AB$ is normal, its spectral radius coincides with its spectral norm and further, since $AB$ and $BA$ have the same eigenvalues, the spectral radii of $AB$ and $BA$ are the same and finally since the spectral radius of any matrix is always smaller than its spectral norm ($\parallel T^n\parallel^{1/n}\leq\parallel T\parallel$), the result follows immediately:
$$
\parallel AB\parallel=spr(AB)=spr(BA)\leq\parallel BA\parallel
$$
\bigskip

[1] Show that if $S,T\geq 0$ and $S\leq 1$, then
$$
1-(S+T)^{-1/2}S(S+T)^{-1/2}\leq 2-2S+4T
$$

[2] Show that if $S_k,T_k,R_k, k=1,2$ are positive operators in a Hilbert space such that
(a) $[S_1,S_2]=[T_1,T_2]=[[R_1,R_2]=0$ and $S_k+T_k\leq R_k, k=1,2,$, then for any
$0\leq s\leq 1$, we have Lieb's inequality:
$$
R_1^{1-s}R_2^s\geq S_1^{1-s}S_2^s+T_1^{1-s}T_2^s
$$
First we show that this inequality holds for $s=1/2$, then we show that if it holds for
$s=\mu,\nu$, then it also holds for $s=(\mu+\nu)/2$ and that will complete the proof.
$$
|<y|S_1^{1/2}S_2^{1/2}+T_1^{1/2}T_2^{1/2}|x>|^2
$$
$$
\leq(|<S_1^{1/2}y,S_2^{1/2}x>|+|<T_1^{1/2}y,T_2^{1/2}x>|)^2
$$
$$
\leq[\parallel S_1^{1/2}y\parallel.\parallel S_2^{1/2}x\parallel+\parallel T_1^{1/2}y\parallel.\parallel T_2^{1/2}x\parallel]^2
$$
$$
\leq[\parallel S_1^{1/2}y\parallel^2+\parallel T_1^{1/2}y\parallel^2]
$$
$$
\times[\parallel S_2^{1/2}x\parallel^2+\parallel T_2^{1/2}x\parallel^2]
$$
$$
=<y,(S_1+T_1)y><x,(S_2+T_2)x>\leq <y,R_1y><x,R_2x>
$$
Then, replacing $y$ by $R_1^{-1/4}R_2^{1/4}x$ and $x$ by $R_2^{-1/4}R_1^{1/4}x$ in this inequality gives us (recall that $R_1$ and $R_2$ are assumed to commute)
$$
<x|R_1^{-1/4}R_2^{1/4}(S_1^{1/2}S_2^{1/2}+T_1^{1/2}T_2^{1/2})R_2^{-1/4}R_1^{1/4}|x>
$$
$$
\leq <x,R_1^{1/2}R_2^{1/2}x>
$$
Now define the matrices
$$
B=R_1^{-1/4}R_2^{1/4}(S_1^{1/2}S_2^{1/2}+T_1^{1/2}T_2^{1/2}, A=R_2^{-1/4}R_1^{1/4}
$$
Then it is clear that
$$
AB=S_1^{1/2}S_2^{1/2}+T_1^{1/2}T_2^{1/2}
$$
is a Hermitian positive matrix because $[S_1,S_2]=[T_1,T_2]=0$ and we also have the result
that the eigenvalues of $AB$ are the same as those of $BA$. Further, from the above inequality, the $k^{th}$ largest eigenvalue of $BA$ is $\leq$ the $k^{th}$ largest eigenvalue of $R_1^{1/2}R_2^{1/2}$. It follows therefore that the $k^{th}$ largest eigenvalue of the Hermitian matrix $AB$ is $\leq$ the $k^{th}$ largest eigenvalue of the Hermitian matrix $R_1^{1/2}R_2^{1/2}$. However, from this result, we cannot deduce
the desired inequality. To do so, we note that writing
$$
S_1^{1/2}S_2^{1/2}+T_1^{1/2}T_2^{1/2}
$$
we have proved that
$$
|<y,Cx>|^2\leq<y,R_1y><x,R_2x>
$$
This implies
$$
|<R_1^{-1/2}y,CR_2^{-1/2}x>|\leq\parallel y\parallel.\parallel x\parallel
$$
for all $x,y$ and hence
$$
|<y,R_1^{-1/2}CR_2^{-1/2}x>|/\parallel y\parallel.\parallel x\parallel\leq 1
$$
for all nonzero $x,y$. Taking supremum over $x,y$ yields
$$
\parallel R_1^{-1/2}CR_2^{-1/2}\parallel\leq 1
$$
and by normality of
$$
R_1^{-1/4}R_2^{-1/4}CR_1^{-1/4}R_2^{-1/4}=
$$
$$
(R_1^{1/4}R_2^{-1/4})(R_1^{-1/2}CR_1^{-1/4}R_2^{-1/4})=PQ
$$
where
$$
P=(R_1^{1/4}R_2^{-1/4}), Q=(R_1^{-1/2}CR_1^{-1/4}R_2^{-1/4})
$$
we get
$$
\parallel PQ\parallel\leq\parallel QP\parallel=
$$
$$
\parallel R_1^{-1/2}CR_2^{-1/2}\parallel\leq 1
$$
which implies since $PQ$ is positive that
$$
PQ\leq I
$$
or equivalently
$$
(R_1^{-1/4}R_2^{-1/4}CR_1^{-1/4}R_2^{-1/4})\leq I
$$
from which we deduce that (Note that $R_1$ and $R_2$ commute)
$$
C\leq R_1^{1/2}R_2^{1/2}
$$
proving the claim. 

Now suppose it is valid for some $s=\mu,\nu$. Then
we have
$$
R_1^{1-\mu}R_2^{\mu}\geq S_1^{1-\mu}S_2^{\mu}+T_1^{1-\mu}T_2^{\mu},
$$
and
$$
R_1^{1-\nu}R_2^{\nu}\geq S_1^{1-\nu}S_2^{\nu}+T_1^{1-\nu}T_2^{\nu},
$$
Then, replacing $R_1$ by $R_1^{1-\mu}R_2^{\mu}$, $R_2$ by $R_1^{1-\nu}R_2^{\nu}$,
$S_1$ by $S_1^{1-\mu}S_2^{\mu}$, $S_2$ by $S_1^{1-\nu}S_2^{\nu}$, $T_1$ by
$T_1^{1-\mu}T_2^{\mu}$ and $T_2$ by $T_1^{1-\nu}T_2^{\nu}$ and noting that these replacements satisfy all the required conditions of the theorem, gives us
$$
R_1^{1-(\mu+\nu)/2}R_2^{(\mu+\nu)/2}\geq S_1^{1-(\mu+\nu)/2}S_2^{(\mu+\nu)/2}+T_1^{1-(\mu+\nu)/2}T_2^{(\mu+\nu)/2},
$$
ie the result also holds for $s=(\mu+\nu)/2$. Since it holds for $s=0,1/2,1$ we deduce by induction that it holds for all dyadic rationals in $[0,1]$, ie, for all $s$ of the form
$k/2^n, k=0,1,..., 2^n, n=0,1,...$. Now every real number in $[0,1]$ is a limit of dyadic rationals and hence by continuity of the inequality, the theorem has been proved.

An application example:

Remark: Let $T$ be a Hermitian matrix of size $n\times n$. Let $M$ be an $n-k+1$-dimensional subspace and let $\lambda_1\geq...\geq\lambda_n$ be the eigenvalues of $T$ arranged in decreasing order. Then,
$$
sup_{x\in M}<x,Tx>\geq\lambda_k
$$
and thus
$$
inf_{dim N=n-k+1}sup_{x\in N}<x,Tx>=\lambda_k
$$
Likewise, if $M$ is a $k$ dimensional subspace, then
$$
inf_{x\in M}<x,Tx>\leq\lambda_k
$$
and hence
$$
sup_{dim N=k}inf_{x\in N}<x,Tx>=\lambda_k
$$
Thus,
$$
\lambda_k(AB)=\lambda_k(BA)\leq\lambda_k(C), k=1,2,..., n
$$
where
$$
C=R_1^{1/2}R_2^{1/2}
$$
\bigskip

Let $A,B$ be two Hermitian matrices. Choose a subspace $M$ of dimension $n-k+1$ so that
$$
\lambda_k(A)^{\downarrow}=sup_{x\in M}<x,Ax>
$$
Then,
$$
\lambda_k(A-B)^{\downarrow}=inf_{dim N=n-k+1}sup_{x\in N}<x,(A-B)x>
$$
$$
\leq sup_{x\in M}<x,(A-B)x>=sup_{x\in M}(<x,Ax>-<x,Bx>)
$$
$$
\leq sup_{x\in M}<x,Ax>-inf_{x\in M}<x,Bx>
$$
$$
=\lambda_k(A)^{\downarrow}-inf_{x\in M}<x,Bx>
$$
$$
\lambda_k(A-B)^{\downarrow}\leq\lambda_k(A)^{\downarrow}-inf_{x\in M}<x,Bx>
$$
$$
\leq\lambda_k(A)^{\downarrow}-\lambda_n(B)^{\downarrow}
$$
Now let $M$ be any $n-k+1$ dimensional subspace such that
$$
inf_{x\in M}<x,Bx>=\lambda_{n-k+1}(B)^{\downarrow}=\lambda_k(B)^{\uparrow}
$$
Then,
$$
\lambda_k(A-B)^{\downarrow}\leq sup_{x\in M}<x,(A-B)x>\leq sup_{x\in M}<x,Ax>-inf_{x\in M}<x,Bx>
$$
$$
\leq\lambda_k(A)^{\downarrow}-\lambda_k(B)^{\uparrow}
$$
This is a stronger result than the previous one.

\section{Matrix Holder Inequality}

Now let $A,B$ be two matrices. Consider $(AB)^{\otimes_ak}$. its maximum singular value square is
$$
\parallel(AB)^{\otimes_ak}\parallel^2
$$
which is the maximum eigenvalue of $((AB)^*(AB))^{\otimes_ak}$. This maximum eigenvalue equals the product of the first $k$ largest eigenvalues of $(AB)^*(AB)$, ie, the product of the squares of the first $k$ largest singular values of $AB$. Thus, we can write
$$
\parallel(AB)^{\otimes_ak}\parallel=\Pi_{j=1}^ks_j(AB)^{\downarrow}
$$
On the other hand,
$$
\parallel(AB)^{\otimes_ak}\parallel=\parallel A^{\otimes_ak}.\parallel B^{\otimes_ak}\parallel
$$
$$
\leq\parallel A^{\otimes_ak}\parallel.\parallel B^{\otimes_ak}\parallel
$$
Now, in the same way, $\leq\parallel A^{\otimes_ak}\parallel$ equals $\Pi_{j=1}^k s_j(A)^{\downarrow}$ and likewise for $B$. Thus we obtain the fundamental inequality
$$
\Pi_{j=1}^ks_j(AB)^{\downarrow}\leq\Pi_{j=1}^ks_j(A)^{\downarrow}s_j(B)^{\downarrow}, k=1,2,..., n
$$
and hence we easily deduce that
$$
s(AB)\leq s(A).s(B)
$$
which is actually a shorthand notation for
$$
\sum_{j=1}^ks_j(AB)^{\downarrow}\leq\sum_{j=1}^ks_j(A)^{\downarrow}.s_j(B)^{\downarrow}, k=1,2,..., n
$$
In particular, taking $k=n$ gives us for $p,q\geq 1$ and $1/p+1/q=1$,
$$
Tr(|AB|))=\sum_{j=1}^ns_j(AB)\leq(\sum_js_j(A)^p)^{1/p}.(\sum_js_j(B)^q)^{1/q}
$$
Now, 
$$
Tr(|A|^{1/p})=\sum_js_j(A)^{1/p}
$$
and likewise for $B$. Therefore, we have proved the Matrix Holder inequality
$$
Tr(|AB|)\leq (Tr(|A|^p))^{1/p}.(Tr(|B|^q))^{1/q}
$$
for any two matrices $A,B$ such that $AB$ exists and is a square matrix. In fact, the method of our proof implies immediately that if $\psi(.)$ is any unitarily invariant matrix norm, then
$$
\psi(AB)\leq\psi(|A|^p).^{1/p}.\psi(|B|^q)^{1/q}
$$
This is seen as follows. Since $\psi$ is a unitarily invariant matrix norm, 
$$
\psi(A)=\Phi(s(A))
$$
where $\Phi()$ is a convex mapping from $\Bbb R_+^n$ into $\Bbb R_+$ that satisfies the triangle inequality and in fact all properties of a symmetric vector norm and also since
$$
a^p/p+b^q/q\geq ab, a,b>0
$$
we get for $t>0$,
$$
\Phi(x.y)=\phi(tx.y/t)\leq\phi(t^px^p)/p+\phi(y^q/t^q)/q
$$
$$
\leq t^p\phi(x^p)/p+\phi(y^q)/qt^q
$$
Now $ut^p+v/t^q$ is minimized when 
$$
put^{p-1}-qv/t^{q+1}=0
$$
or equivalently, when
$$
t^{p+q}=qv/pu
$$
Substituting this value of $t$ immediately gives
$$
\Phi(x.y)\leq\Phi(x^p)^{1/p}.\Phi(y^q)^{1/q}
$$
and hence since
$$
s(A.B)\leq s(A).s(B)
$$
(in the sense of majorization), it follows that for two matrices $A,B$,
$$
\psi(A.B)=\Phi(s(A.B))\leq\Phi(s(A).s(B))\leq\Phi(s(A)^p)^{1/p}.\Phi(s(B)^q)^{1/q}
$$
$$
=\psi(|A|^p)^{1/p}.\Phi(|B|^q)^{1/q}
$$
Note that the eigenvalues (or equivalently the singular values) of $|A|^p$ are same as $s(A)^p$. 

Remark: We are assuming that if $x,y$ are positive vectors such that $x<y$ in the sense of majorization, then
$$
\Phi(x)\leq\Phi(y)
$$
This result follows since if $a<b$ are two positive real numbers, then
$$
a=tb+(1-t)0
$$
for $t=a/b\in [0,1]$ and by using the convexity of $\Phi$ and the fact that $\Phi(0)=0$.
If $x<y$, then $\sum_{i=1}^kx_i\leq\sum_{i=1}^ky_i, k=1,2,..., n$ and hence writing
$\xi_k=\sum_{i=1}^kx_i, \eta_k=\sum_{i=1}^ky_i, k=1,2,..., n$, we get by regarding
$\Phi(x)$ as a function of $\xi_k, k=1,2,..., n$ and noting that convexity of $\Phi$ in $x$ implies convexity of $\Phi$ in $\xi$ and the above result that
$$
\Phi(\xi)=\Phi(\xi_1,...,\xi_n)\leq\Phi(\eta_1,\xi_2,...,\xi_n)\leq
$$
$$
\Phi(\eta_1,\eta_2,\xi_3,...,\xi_n)\leq...\leq\Phi(\eta_1,...,\eta_n)=\Phi(\eta)
$$
\bigskip

\chapter{Lecture Plan for Statistical Signal Processing}

Course Instructor:Harish Parthasarathy.

[1] Revision of basic probability theory: Probability spaces, random variables, expectation, variance, covariance, characteristic function, convergence of random variables, conditional expectation. (four lectures)

[2] Revision of basic random process theory:Kolmogorov's existence theorem, stationary stochastic processes, autocorrelation and power spectral density, higher order moments and higher order spectra, Brownian motion and Poisson processes, an introduction to stochastic differential equations. (four lectures)

[3] Linear estimation theory: Estimation of parameters in linear models and sequential linear models using least squares and weighted least squares methods, estimation of parameters in linear models with coloured Gaussian measurement noise using the maximum likelihood method, estimation of parameters in time series models (AR,MA,ARMA) using least squares method with forgetting factor, Estimation of parameters in time series models using the minimum mean square method when signal correlations are known. Linear estimation theory as a computation of an orthogonal projection of a Hilbert space onto a subspace, the orthogonality principle and normal equations. (four lectures)

[4] Estimation of non-random parameters using the Maximum likelihood method, estimation of random parameters using the maximum aposteriori method, basic statistical hypothesis testing, the Cramer-Rao lower bound. (four lectures)

[5] non-linear estimation theory: Optimal nonlinear minimum mean square estimation as a conditional expectation, applications to non-Gaussian models. (four lectures)

[6] Finite order linear prediction theory: The Yule-Walker equations for estimating the AR parameters. Fast order recursive prediction using the Levinson-Durbin algorithm, reflection coefficients and lattice filters, stability in terms of reflection coefficients. (four lectures)

[7] The recursive least squares algorithm for parameter estimation linear models, simultaneous time and order recursive prediction of a time series using the RLS-Lattice filter. (four lectures)

[8] The non-causal Wiener filter, causal Wiener filtering using Wiener's spectral factorization method and using Kolmogorov's innovations method, Kolmogorov's formula for the mean square infinite order prediction error. (four lectures)

[9] An introduction to stochastic nonlinear filtering: The Kushner-Kallianpur filter, the Extended Kalman filter approximation in continuous and discrete time, the Kalman-Bucy filter for linear state variable stochastic models. (four lectures)

[10] Power spectrum estimation using the periodogram and the windowed periodogram, mean and variance of the periodogram estimate, high resolution power spectrum estimation using eigensubspace methods like Pisarenko harmonic decomposition, MUSIC and ESPRIT algorithms.
Estimation of the power spectrum using time series models. (four lectures)

[11] Some new topics: 

[a] Quantum detection and estimation theory, Quantum Shannon coding theory. 
(four lectures)

[b] Quantum filtering theory (four lectures)

Total: Forty eight lectures.
\bigskip

[1] Large deviations for the empirical distribution of a stationary Markov process.

[2] Proof of the Gartner-Ellis theorem.

[3] Large deviations in the iid hypothesis testing problem.

[4] Large deviations for the empirical distribution of Brownian motion process.

Let $X_n,n\geq 0$ be a Markov process in discrete time. Consider
$$
M_N(f)=\Bbb Eexp(\sum_{n=0}^Nf(X_n))
$$
Then, given $X_0=x$
$$
M_N(f)=(\pi_f)^N(1)
$$
where $\pi$ is the transition probability operator kernel:
$$
\pi(g)(x)=\int\pi(x,dy)g(y)
$$
and
$$
\pi_f(g)(x)=\int exp(f(x))\pi(x,dy)g(y)
$$
ie
$$
\pi_f(x,dy)=exp(f(x))\pi(x,dy)
$$
If we assume that the state space $E$ is discrete, then $\pi(x,y)$ is a stochastic matrix and $\pi_f(x,y)=exp(f(x))\pi(x,y)$. Let $D_f$ denote the diagonal matrix $diag[f(x):x\in E]$. Then
$$
\pi_f=exp(D_f)\pi
$$
and
$$
M_N(f)=(exp(D_f)\pi)^N(1)
$$
The Gartner-Ellis limiting logarithmic moment generating function is then
$$
\Lambda(f)=lim_{N\rightarrow\infty}N^{-1}.log((exp(D_f)\pi)^N(1))=log(\lambda_m(f))
$$
where $\lambda_m(f)$ is the maximum eigenvalue of the matrix $exp(D_f)\pi$.
Now let
$$
I(\mu)=sup_{u>0}\sum_x\mu(x)log((\pi u)(x))/u(x))
$$
where $\mu$ is a probability distribution on $E$. We wish to show that
$$
I(\mu)=sup_f(\sum_xf(x)\mu(x)-\lambda_m(f))
$$
$$
=sup_f(<f,\mu>-\lambda_m(f))
$$
\newpage

\section{End-Semester Exam, duration:four hours, Pattern Recognition}

[1] [a] Consider $N$ iid random vectors ${\bf X}_n, n=1,2,..., N$ where
${\bf X}_n$ has the probability density $\sum_{k=1}^K\pi(k)N({\bf x}|\mu_k,\Sigma_k)$, ie,
a Gaussian mixture density. Derive the optimal ML equations for estimating the parameters
$\theta=\{\pi(k),\mu_k,\sigma_k):k=1,2,..., K\}$ from the measurements ${\bf X}=\{{\bf X}_n:n=1,2,..., N\}$ by maximizing
$$
ln(p({\bf X}|\theta))=\sum_{n=1}^Nln(\sum_{k=1}^K\pi(k)N({\bf X}_n|\mu_k,\Sigma_k))
$$
\bigskip

[b] By introducing latent random vectors ${\bf z}(n)=\{z(n,k):k=1,2,..., K\}, n=1,2,..., N$ which are iid with $P(z(n,k)=1)=\pi(k)$ such that for each $n$, exactly one of the $z(n,k)'s$ equals one and the others are zero, cast the ML parameter estimation equation in the recursive EM form, namely as
$$
\theta_{m+1}=argmax_{\theta}Q(\theta,\theta_m)
$$
where
$$
Q(\theta,\theta_m)=\sum_{\bf z}ln(p({\bf X,Z}|\theta)).p({\bf Z}|{\bf X},\theta_m)
$$
with ${\bf Z}=\{{\bf z}(n):n=1,2,..., N\}$. Specifically, show that
$$
Q(\theta,\theta_m)=\sum_{n,k}\gamma_{nk}({\bf X},\theta_m)ln(\pi(k)N({\bf X}_n|\mu(k),\Sigma_k)))p({\bf Z}|{\bf X},\theta_m)
$$
where
$$
\gamma_{n,k}({\bf X},\theta_m)=\Bbb E(z(n,k)|{\bf X},\theta_m)
$$
$$
=\sum_{\bf Z}z(n,k)p({\bf Z}|{\bf X},\theta_m)=\sum_{z(n,k)}z(n,k)p(z(n,k)|{\bf X},\theta_m)
$$
Derive a formula for $p(z(n,k)|{\bf X},\theta_m)$. Derive the optimal equations for $\theta_{m+1}$ by maximizing $Q(\theta,\theta_m)$.
\bigskip

[2] Let ${\bf X}_n, n=1,2,..., N$ be vectors in $\Bbb R^N$. Choose $K<<N$ and vectors
$\{\mu(k):k=1,2,..., K\}$ and a partition 
$$
\mathcal P=E_1\cup E_2\cup...\cup E_K
$$
of $\{1,2,..., N\}$ into $K$ disjoint non-empty subsets such that
$$
\mathcal E=\sum_{k=1}^K\sum_{n\in E_k}\parallel{\bf X}_n-\mu(k)\parallel^2
$$
is a minimum. Interpret this "Clustering principle" from the viewpoint of data compression and image segmentation. Explain how you would obtain an iterative solution to this clustering problem by defining
$$
r(n,k)=1, n\in E_k, r(n,k)=0, n\notin E_k
$$
and expressing
$$
\mathcal E=\sum_{n,k}r(n,k)\parallel{\bf X}_n-\mu(k)\parallel^2
$$
Relate this clustering problem to the problem of calculating the $\mu(k)'s$ and the $\pi(k)'s$ in the Gaussian mixture model. Specifically, compare the equations for calculating $r(n,k),\mu(k)$ in this clustering problem to the equations for calculating the same in the GMM. Explain how the GMM model is a "smoothened version" of the clustering model by remarking that the event ${\bf X}_n\in E_k$ is to be interpreted in the GMM context by saying that ${\bf X}_n$ has the $N({\bf x}|\mu(k),\Sigma_k)$ distribution and that $\sum_{n=1}^Nr(n,k)/N=card(E_k)/N$ corresponds to the probability $\pi(k)$ of the 
$k^{th}$ component in the GMM occurring. Derive the explicit value that $r(n,k)$ corresponds to in the GMM by showing that the optimal $\pi(k)$ in the GMM satisfies
$$
\pi(k)=\sum_{n=1}^N\frac{\pi(k)N({\bf X}_n|\mu_k,\Sigma_k)}{\sum_{m=1}^K\pi(m)N({\bf X}_n|\mu(m),\Sigma_m)}
$$
Compare this with the quantity $\pi(k)=\sum_{n=1}^N(r(n,k)/N), \sum_{n,k}r(n,k)=N$ in the clustering problem.
\bigskip

[3] Let $\{Z(n):n=1,2,..., N\}$ be a Markov chain in discrete state space and let $\pi(Z(n),Z(n+1)|\theta)$ be its one step transition probability distribution. Let $\pi_0(Z(0))$ denote its initial distribution. Let ${\bf X}=({\bf X}(n):n=0,1,..., N)$ be the HMM emission random variables with the conditional probability density $p(X(n)|Z(n),\theta)$
defined. 

[a] Show that the EM algorithm for estimating $\theta$ recursively amounts to maximizing
$Q(\theta,\theta_0)$ where
$$
Q(\theta,\theta_0)=
$$
$$
\sum_{n=0}^N\Bbb E(ln(p(X(n)|Z(n),\theta))|{\bf X},\theta_0)
$$
$$
+\sum_{n=0}^{N-1}\Bbb E(ln(\pi(Z(n),Z(n+1)|\theta_0))|{\bf X},\theta_0)
$$

[b] Show that for evaluating the conditional expectation in [a], we require the conditional probabilities $p(Z(n)|{\bf X},\theta_0)$ and $p(Z(n),Z(n+1)|{\bf X},\theta_0)$. Explain how you would evaluate these conditional probabilities recursively by first proving the relations
$$
p({\bf X}|Z(n),\theta_0)=p(X(N),X(N-1),..., X(n)|Z(n),\theta_0).p(X(n-1),X(n-2),...,X(0)|Z(n),\theta_0)
$$
and
$$
p({\bf X}|Z(n+1),Z(n),\theta_0)=
$$
$$
p(X(N),..., X(n+1)|Z(n+1,\theta_0).p(X(n-1),...,X(0)|Z(n),\theta_0)
$$
\bigskip

4 [a] How would you approximate (based on matching only first and second order statistics) an $N$-dimensional random vector ${\bf X}$ by a $p<<N$ dimensional random vector ${\bf S}$ using the linear model
$$
{\bf X}\approx {\bf WS}+\mu+{\bf V}=\hat{\bf X}+{\bf V}, \hat{\bf X}={\bf WS}+\mu  
$$
where ${\bf S,V}$ are zero mean mutually uncorrelated random vectors with covariances
$\Sigma_S, \Sigma_V$ respectively and ${\bf W}$ is an $N\times p$ matrix to be determined. To solve this problem, you must assume that $\Bbb E({\bf X})$ is well approximated by $\mu$ and $Cov({\bf X})$ is well approximated by ${\bf W}\Sigma_S{\bf W}^T+\Sigma_V$.
\bigskip

[b] Now consider the maximum likelihood version of the above PCA in which the data is modeled by an iid sequence:
$$
{\bf X}_n={\bf W}{\bf S}_n+\mu+{\bf V}_n, n=1,2,..., N
$$
where the ${\bf S}_n's$ are iid $N({\bf 0},\Sigma_S)$ random vectors, the ${\bf V}_n's$ are iid $N({\bf 0},\Sigma_v)$ random vectors with the two sequences being mutually independent. Calculate the equations that determine the optimal choice of ${\bf W},\mu$ that would maximize $p({\bf X}_n,n=1,2,...,N|{\bf W},\mu)$. Also by assuming the ${\bf S}_n's$ to be latent random vectors, derive the EM algorithm recursive equations for estimating ${\bf W},\mu$ from the ${\bf X}_n,n=1,2,..., N$. 
\bigskip

Solution:
$$
p(\{{\bf X}_n\}_{n=1}^N,\{{\bf S}_n\}_{n=1}^N|{\bf W},\mu)
$$
$$
=p(\{{\bf X}_n\}_{n=1}^N|\{{\bf S}_n\},{\bf W},\mu)p(\{{\bf S}_n\}_{n=1}^N)
$$
$$
=\Pi_{n=1}^N[f_{\bf V}({\bf X}_n-{\bf WS}_n-\mu)f_{\bf S}({\bf S}_n)]
$$
So in the EM algorithm, the function to be maximized is
$$
Q(\theta,\theta_0)
$$
where
$$
\theta=({\bf W},\mu), \theta_0=({\bf W}_0,\mu_0)
$$
and
$$
Q(\theta,\theta_0)=\sum_{n=1}^N\Bbb E[log(f_{\bf V}({\bf X}_n-{\bf WS}_n-\mu))|\{{\bf X}_m\}_{m=1}^N,{\bf W}_0,\mu_0)]
$$
$$
+\sum_{n=1}^Nlog(f_{\bf S}({\bf S}_n))|\{{\bf X}_m\}_{m=1}^N,{\bf W}_0,\mu_0)
$$
$$
=
\sum_{n=1}^N\Bbb E[log(f_{\bf V}({\bf X}_n-{\bf WS}_n-\mu))|{\bf X}_n,{\bf W}_0,\mu_0)]
$$
$$
+\sum_{n=1}^Nlog(f_{\bf S}({\bf S}_n))|{\bf X}_n,{\bf W}_0,\mu_0)
$$
Note that only the first sum in this expression must be maximized since the second one does not depend on ${\bf W},\mu$. However, if $\Sigma_S, \Sigma_v$ also have to be estimated then the parameter vector would be
$$
\theta=({\bf W},\mu,\Sigma_S,\Sigma_v)
$$
and then the second term would also count. Note further, that
$$
p(\{{\bf S}_n\}_{n=1}^N|\{{\bf X}_n\}_{n=1}^N,{\bf W},\mu)=
$$
$$
[\Pi_{n=1}^Nf_{\bf V}({\bf X}_n-{\bf WS}_n-\mu)f_{\bf S}({\bf S}_n)]/\Pi_{n=1}^Nf_{\bf X}({\bf X}_n|{\bf W},\mu)
$$
with $f_{\bf X}$ being $N(\mu,{\bf W}\Sigma_s{\bf W}^T+\Sigma_v)$
\bigskip

[c] If $\Sigma_V=\sigma^2{\bf I}$, then derive the EM recursions for estimating ${\bf W},\mu,\Sigma_S,\sigma^2$ and compare with the exact maximum likelihood equations. Explain why this process is called principle component analysis and explain how much data compression we have achieved by this process.
\bigskip

[d] Consider now the following data matrix version of PCA. ${\bf X}$ is a given  $N\times M$ real matrix. Show that it has a singular value decomposition (SVD)
$$
{\bf X}={\bf U}\Sigma{\bf V}
$$
where ${\bf U}$ and ${\bf V}$ are respectively $N\times N$ and $M\times M$ real orthogonal matrices and $\Sigma$ is a diagonal $N\times M$ matrix with positive diagonal entries either $N$ or $M$ in number, depending on whether $N$ or $M$ is smaller. Let $r<min(N,M)$ and then using the SVD, construct an $N\times M$ matrix ${\bf Y}$ of rank $r$ so that $\parallel{\bf X}-{\bf Y}\parallel_F^2$ is a minimum where $\parallel Z\parallel _F=\sqrt{Tr({\bf Z}^T{\bf Z})}$ is the Frobenius norm of the real matrix ${\bf Z}$. 
\newpage

[5] Consider the binary hypothesis testing problem in which under the hypothesis $H_1$, each sample of the iid sequence $X(n), n\geq 1$ has the probability density $p_1(X)$ while under $H_0$, each sample of this iid sequence has the probability density $p_0(X)$. Show that the Neyman-Pearson criterion of decision based on the observations $X(n), n=1,2,..., N$ leads to the test of deciding $H_1$ in the case when $N^{-1}\sum_{n=1}^NL(X(n))\geq\eta$ and deciding $H_0$ in the case when $N^{-1}\sum_{n=1}^NL(X(n))<\eta$ where
$$
L(X)=ln(p_1(X)/p_0(X))
$$
and the threshold $\eta(N)$ is such that 
$$
0<\alpha=\int_{Z_1}\Pi_{n=1}^N(p_0(X(n))dX(n))=P(N^{-1}\sum_{n=1}^NL(X(n))>\eta|H_0)
$$
Here, $Z_1$ is the region where $H_1$ is decided. Estimate this probability by the LDP and
calculate using the LDP the value of $\eta$ for given $\alpha$. Hence use this threshold $\eta$ to calculate the other error probability 
$$
\beta(N)=\int_{Z_1^c}\Pi_{n=1}^N(p_1(X(n))dX(n))=P(N^{-1}\sum_{n=1}^NL(X(n))<\eta|H_1)
$$
and show that for any $\alpha>0$ however small, $\beta(N)$ if $\eta=\eta(N)$ is obtained as above for the given $\alpha$, then $\beta(N)$ converges to zero at the rate of $D(p_0|p_1)$, ie,
$$
N^{-1}log(\beta(N))\rightarrow D(p_0|p_1), N\rightarrow\infty
$$
where
$$
D(p_0|p_1)=\int p_0(X).log(p_0(X)/p_1(X)).dX
$$
\bigskip

[6] This problem deals with non-parametric density estimation. Let $p_1(X), p_0(X)$ be two different probability densities of a random vector $X\in\Bbb R^n$. To estimate $p_1$ at a given point $X$, we take samples from this distribution and draw a sphere of minimum radius $R_1$ with centre at $X$, for which the first sample point falls within this sphere. Let $V(R_1)$ denote the volume of such a sphere. Then, estimate $p_1(X)$ as $1/V(R_1)$. Likewise do so for estimating $p_0(X)$. Let the minimum radius for this be $R_0$. Justify using the law of large numbers that such a non-parametric density estimate is reliable. Show that given a sample point $X$, in order to test whether $X$ came from $p_1$ or from $p_0$, we use the ML criterion, namely decide $p_1$ if $\hat p_1(X)\geq \hat p_0(X)$ and decide $p_0$ otherwise, where $\hat p$ denotes the estimate of $p$ obtained using the above scheme. Show that this MLE method is the same as the minimum distance method, in the sense that if the first sample point that we decide $p_1$ if $R_1(X)<R_0(X)$ and decide $p_0$ otherwise where $R_1(X)$ is the distance of the first point taken from the $p_1$ sample from $X$ and likewise $R_0(X)$ is the distance of the first point taken from the $p_0$ sample from $X$.
\bigskip

[7] Let a group $G$ act on a set $\mathcal M$ and let $f_0,f_1:\mathcal M\rightarrow\Bbb R$ be two image fields defined on the set $\mathcal M$ such that
$$
f_1(x)=f_0(g_0^{-1}x)+w(x), x\in\mathcal M
$$
where $g_0\in G$ and $w$ is a zero mean Gaussian noise field. Choose a point $x_0\in\mathcal M$ and replace the above model by
$$
\tilde f_1(g)=\tilde f_0(g_0^{-1}g)+w(g), g\in G
$$
where
$$
\tilde f_1(g)=f_1(gx_0),\tilde f_0(g)=f_0(gx_0)
$$
Let $\pi$ be a unitary representation of $G$ in a the vector space $\Bbb C^p$. Define
the Fourier transform of a function $F$ on $G$ evaluated at $\pi$ by
$$
\mathcal F(F)(\pi)=\int_GF(g)\pi(g)dg
$$
where $dg$ is the left invariant Haar measure on $G$. Show that the above image model on $G$ gives on taking Fourier transforms
$$
\mathcal F(\tilde f_1)(\pi)=\pi(g_0)\mathcal F(\tilde f_0)(\pi)+\mathcal F(w)(\pi)
$$
Let $H$ be the subgroup of $G$ consisting of all $h\in G$ for which $hx_0=x_0$. Show that
$\tilde f_k(gh)=\tilde f_k(g), k=1,2, g\in g, h\in H$. Hence, deduce that if
$$
P_H=\int_H\pi(h)dh
$$
where $dh$ denotes the Haar measure on $H$, then $P_H$ is the orthogonal projection onto the set of all vectors $v\in\Bbb C^p$ for which $\pi(h)v=v\forall h\in H$ provided that $H$ is unimodular, ie, its left invariant Haar measure coincides with its right invariant Haar measure. Deduce further that 
$$
\mathcal F(\tilde f_k)(\pi)=\tilde F(\tilde f_k)(\pi)P_H, k=0,1
$$
Hence show that the above model for the image Fourier transforms is equivalent to
$$
\mathcal F(\tilde f_1)(\pi)P_H=\pi(g_0)\mathcal F(\tilde f_0)(\pi)P_H+\mathcal F(w)(\pi)P_H
$$
From this model, explain how by varying the representation $\pi$, we can estimate the transformation $g_0$ by minimizing
$$
\mathcal E(g_0)=\sum_kW(k)\parallel\mathcal F(\tilde f_1)(\pi_k)P_H-\pi_k(g_0)\mathcal F(\tilde f_0)(\pi_k)P_H\parallel^2
$$
where $\parallel.\parallel$ is the Frobenius norm and $W(k), k\geq 1$ are positive weights. By starting with an initial guess estimate $g_{00}$ for $g_0$ and writing
$g_0=(1+\delta X).g_{00}$ where $\delta X$ is small and belongs to the Lie algebra of $G$, derive a set of linear equations for $\delta X$ and solve them. For this, you can use the approximate identity 
$$
\pi((1+\delta X)g_{00})=\pi(exp(\delta X)g_{00})=exp(d\pi(\delta X))\pi(g_{00})=(1+d\pi(\delta X))\pi(g_{00})
$$
where $d\pi$ is the differential of $\pi$, ie, it is the representation of the Lie algebra of $G$ corresponding to the representation $\pi$ of $G$. Note that $d\pi$ is a linear map and hence if $\{L_1,..., L_N\}$ denotes a basis for the Lie algebra of $G$ then
$$
\delta X=\sum_{m=1}^Nx(k)L_k, d\pi(\delta X)=\sum_{m=1}^Nx(k)d\pi(L_k)
$$
The problem of estimating $\delta X$ is then equivalent to estimating $x(k), k=1,2,..., N$. Also explain how you would approximately evaluate the mean and covariance of the random variables $x(k), k=1,2,..., N$ in terms of the noise correlations.
\bigskip

Estimating the EEG parameters from the speech parameters using the EKF. The EEG model is
$$
X_E(t+1)=A(\theta)X_E(t)+W_E(t+1)\in\Bbb R^N
$$
where
$$
A(\theta)=A_0+\sum_{k=1}^p\theta(k)A_k
$$
$$
\theta=(\theta(k):k=1,2,..., p)^T\in\Bbb R^p
$$
is the EEG parameter vector to be estimated. The speech model on the other hand is
$$
X_S(t)=\sum_{k=1}^mrs(k)X_S(t-k)+r_s(m+1)+W_S(t)\in\Bbb R
$$
The speech parameter vector is
$$
{\bf r}_s=(r_s(k);k=1,2,..., m+1)^T\in\Bbb R^{m+1}
$$
We assume that the speech and EEG parameters are correlated, ie, there exists an affine relationship connecting the two of the form
$$
{\bf r}_s={\bf C}\theta+{\bf d}
$$
The regression matrices ${\bf C,d}$ are assumed to be fixed, ie, they do not vary from patient to patient. These matrices can be estimated using the following procedure: Choose a bunch of $K$ patients numbered $1,2,..., K$ having various kinds of brain diseases. For each patient, collect his EEG and speech data and apply the EKF separately to these data to using the above models and noisy measurement models of the form
$$
{\bf Z}_E(t)={\bf X}_E(t)+{\bf V}_E(t), Z_S(t)=X_S(t)+V_S(t)
$$
Denote the parameter estimates for the $k^{th}$ patient by $\hat\theta_k,\hat rs_k$ for $k=1,2,..., , p$. Then the regression matrices ${\bf C,d}$ are estimated by minimizing
$$
E({\bf C,d})=\sum_{k=1}^K\parallel\hat rs_k-{\bf C}\hat\theta_k-{\bf d}\parallel^2
$$
For convenience of notation, we write
$$
\theta_k=\hat\theta_k, rs_k=\hat rs_k
$$
and then the optimal equations for the regression matrices obtained by setting the gradient of $E({\bf C,d})$ to zero are given by
$$
\sum_{k=1}^K(rs_k-{\bf C}\theta_k-{\bf d})\theta_k^T=0,
$$
$$
\sum_{k=1}^K(rs_k-{\bf C}\theta_k-{\bf d})=0
$$
which can be expressed as
$$
{\bf C}{\bf R}_{\theta}+{\bf d}\mu_{\theta}^T={\bf R}_{s\theta},
$$
$$
{\bf C}\mu_{\theta}+{\bf d}=\mu_s
$$
where
$$
{\bf R}_{\theta}=K^{-1}.\sum_{k=1}^K\theta_k.\theta_k^T\in\Bbb R^{p\times p}
$$
$$
\mu_{\theta}=K^{-1}\sum_{k=1}^K\theta_k\in\Bbb R^{p\times 1}
$$
$$
{\bf R}_{s\theta}=K^{-1}\sum_{k=1}^Krs_k\theta_k^T\in\in\Bbb R^{(m+1)\times(m+1)},
$$
$$
\mu_s=K^{-1}\sum_{k=1}^Krs_k\in\Bbb R^{(m+1)\times 1}
$$
Eliminating ${\bf d}$ from these equations gives
$$
{\bf CR}_{\theta}+(\mu_s-{\bf C}\mu_{\theta})\mu_{\theta}^T={\bf R}_{s\theta}
$$
giving
$$
{\bf C}=({\bf R}_{s\theta}-\mu_s\mu_{\theta}^T)({\bf R}_{\theta}-\mu_{\theta}\mu_{\theta}^T)^{-1}
$$
and
$$
\mu_s-{\bf C}\mu_{\theta}
$$
This solves the problem of determining the regression coefficient matrices. Now given a fresh patient, we wish to estimate his EEG parameters from only his speech data. To this end, we substitute into the speech model the expression for the speech parameters in terms of the EEG parameters using our regression scheme: First transform the speech model into a state variable model by introducing state variables:
$$
\xi_1(t)=X_S(t-1),..., \xi_m(t)=X_S(t-m)
$$
and
$$
\xi(t)=[\xi_1(t),..., \xi_m(t)]^T
$$
The speech model then becomes in state variable form
$$
\xi(t+1)={\bf B}(r_s)\xi(t)+u.rs_{m+1}+bW_s(t)
$$
where
$$
{\bf B}(rs)=\left(\begin{array}{ccccc}rs1&r2&rs3&...&rsm\\1&0&0&...&0\\..&..&..&...&..\\0&0&0..&1&0\end{array}\right)
$$
We can write
$$
{\bf B}(rs)=\sum_{k=1}^mrs_k{\bf B}_k
$$
and hence our speech model becomes
$$
\xi(t+1)=(\sum_{k=1}^mrs_k{\bf B}_k)\xi(t)+u.rs_{m+1}+bW_s(t)
$$
Now
$$
rs_k=({\bf C}\theta)_k+d_k, k=1,2,..., m+1
$$
Writing
$$
F_1(\xi,\theta)=\sum_{k=1}^mrs_k{\bf B}_k\xi+u.rs_{m+1}
$$
$$
=\sum_{k=1}^m(({\bf C}\theta)_k+d_k){\bf B}_k\xi+u({\bf C}\theta)_{m+1}+d_{m+1})
$$
we find that
$$
\partial F_1/\partial\xi=[\sum_{k=1}^m((C\theta)_k+d_k){\bf B}_k]
$$
$$
\partial F_1/\partial\theta=[\sum_{k=1}^mC_{k1}{\bf B}_k\xi+C_{m+1,1}{\bf u},\sum_{k=1}^mC_{k2}{\bf B}_k\xi+C_{m+1,2}{\bf u},...,\sum_{k=1}^mC_{kp}{\bf B}_k\xi+C_{m+1,p}{\bf u}]
$$
\bigskip

\section{Quantum Stein's theorem}

We wish to discriminate between two states $\rho,\sigma$ using a sequence of iid measurements, ie, measurements taken when the states become $\rho^{\otimes n}$ and $\sigma^{\otimes n}, n=1,2,...$. Let $0\leq T_n\leq 1$ be the detection operators. If the state is $\rho^{\otimes n}$, then the probability of a wrong decision is
$P_1(n)=Tr(\rho^{\otimes n}(1-T_n))$ while if the state is $\sigma^{\otimes n}$, the probability of a wrong decision is $P_2(n)=Tr(\sigma^{\otimes n}T_n)$. The theorem that we prove is the following: If the measurement operators $\{T_n\}$ are such that $P_2(n)\rightarrow 0$, then $P_1(n)$ cannot go to zero at a rate faster than $exp(-nD(\sigma|\rho))$, ie,
$$
liminf_{n\rightarrow}n^{-1}.log(P_1(n))\geq -D(\sigma|\rho)
$$
and conversely, for any $\delta>0$, there there exists a sequence $\{T_n\}$ of detection operators such that $P_2(n)\rightarrow 0$ and simultaneously
$$
limsup_nn^{-1}.log(P_1(n))\leq -D(\sigma|\rho)+\delta
$$
To prove this, we first use the monotonicity of the Renyi entropy to deduce that
for any detection operator $T_n$,
$$
(Tr(\rho^{\otimes n}(1-T_n)))^sTr(\sigma^{\otimes n}(1-T_n))^{1-s}
$$
$$
\leq Tr(\rho^{\otimes ns)}\sigma^{\otimes n(1-s)})
$$
for any $s\leq 0$. Therefore, if
$$
Tr(\sigma^{\otimes n}T_n)\rightarrow 0
$$
then
$$
Tr(\sigma^{\otimes n}(1-T_n))\rightarrow 1
$$
and we get from the above on taking logarithms,
$$
limsup n^{-1}.log(Tr(\rho^{\otimes n}(1-T_n))\geq log(Tr(\rho^s\sigma^{1-s}))/s
$$
or equivalently writing
$$
\alpha=limsup n^{-1}.log(P_1(n))
$$
we get
$$
\alpha\geq s^{-1}Tr(\rho^s\sigma^{1-s})\forall s\leq 0
$$
Taking $lim s\uparrow 0$ in this expression gives us
$$
\alpha\geq -D(\sigma|\rho)=-Tr(\sigma(log(\sigma)-log(\rho)))
$$
This proves the first part. For the second part, we choose $T_n$ in accordance with the optimal Neyman-Pearson test. Specifically, choose 
$$
T_n=\{exp(nR)\rho^{\otimes n}\geq\sigma^{\otimes n}\}
$$
where the real number $R$ will be selected appropriately.
\newpage

\section{Question paper, short test, Statistical signal processing, M.Tech}

[1] Consider the AR(p) process
$$
x[n]=-\sum_{k=1}^pa[k]x[n-k]+w[n]
$$
where $w[n]$ is white noise with autocorrelation
$$
\Bbb E(w[n]w[n-k]]=\sigma_w^2\delta[n-kj
$$
Assume that
$$
A(z)=1+\sum_{k=1}^pa[k]z^{-k}=\Pi_{m=1}^r(1-z_mz^{-1})\Pi_{m=r+1}^{r+l}(1-z_mz^{-1})(1-\bar z_mz^{-1})
$$
where $p=r+2l$ and $z_1,..., z_{r+l}$ are all within the unit circle with $z_1,..., z_r$ being real and $z_{r+1},..., z_{r+l}$ being complex with non-zero imaginary parts.
Assume that the measured process is
$$
y[n]=x[n]+v[n]
$$
where $v[.]$ is white, independent of $w[.]$ and with autocorrelation $\sigma_v^2\delta[n]$. Assume that
$$
\sigma_w^2+\sigma_v^2A(z)A(z^{-1})=B(z)B(z^{-1})
$$
with
$$
B(z)=K\Pi_{m=1}^p(1-p_mz^{-1})
$$
where $p_1,..., p_p$ are complex numbers within the unit circle. Find the value of $K$ in terms of $\sigma_w^2,\sigma_v^2, a[1],..., a[p]$. Determine the transfer function of the optimal non-causal and causal Wiener filters for estimating the process $x[.]$ based on $y[.]$ and cast the causal Wiener filter in time recursive form. Finally, determine the mean square estimation errors for both the filters.
\bigskip

[2] Consider the state model
$$
{\bf X}[n+1]={\bf A}{\bf X}[n]+{\bf b}w[n+1]
$$
and an associated measurement model
$$
{\bf z}[n]={\bf Cx}[n]+{\bf v}[n]
$$
Given an AR(2)process
$$
x[n]=-a[1]x[n-1]-a[2]x[n-2]+w[n]
$$
and a measurement model
$$
z[n]=x[n]+v[n]
$$
with $w[.]$ and $v[.]$ being white and mutually uncorrelated processes with variances of $\sigma_w^2$ and $\sigma_v^2$ respectively. Cast this time series model in the above state variable form with ${\bf A}$ being a $2\times 2$ matrix, ${\bf X}[n]$ a $2\times 1$ state vector, ${\bf b}$ a $2\times 1$ vector and ${\bf C}$ a $1\times 2$ matrix. Write down explicitly the Kalman filter equations for this model and prove that in the steady state, ie, as $n\rightarrow\infty$, the Kalman filter converges to the optimal causal Wiener filter. For proving this, you may assume an appropriate factorization of $\sigma_v^2A(z)A(z^{-1})+\sigma_w^2$ into a minimum phase and a non-minimum phase part where $A(z)=1+a[1]z^{-1}+a[2]z^{-2}$ is assumed to be minimum phase.
Note that the steady state Kalman filter is obtained by setting the state estimation error conditional covariance matrices for prediction and filtering, namely ${\bf P}[n+1|n]$, ${\bf P}[n|n]$ as well as the Kalman gain $K[n]$ equal to constant matrices and thus express the Kalman filter in steady state form as
$$
\hat {\bf X}[n+1|n+1]={\bf D}\hat{\bf X}[n|n]+{\bf f}z[n+1]
$$
where ${\bf D,f}$ are functions of the steady state values of ${\bf P}[n|n]$ and ${\bf P}[n+1|n]$ which satisfy the steady state Riccati equation. 
\bigskip

[3] Let ${\bf X}_k,k=1,2,..., N$ and $\theta$ be random vectors defined on the same probability space with the conditional distribution of ${\bf X}_1,...,{\bf X}_N$ given $\theta$ being that of iid $N(\theta,{\bf R}_1)$ random vectors. Assume further that the distribution of $\theta$ is $N(\mu,{\bf R}_2)$. Calculate (a) $p({\bf X}_1,...,{\bf X}_N|\theta)$,(b) $p(\theta|{\bf X}_1,...,{\bf X}_N)$, (3) the ML estimate of $\theta$ given
${\bf X}_1,...,{\bf X}_N$ and (4) the MAP estimate of $\theta$ given ${\bf X}_1,...,{\bf X}_N$. Also evaluate
$$
Cov(\hat\theta_{ML}-\theta), Cov(\theta_{MAP}-\theta)
$$
Which one has a larger trace ? Explain why this must be so.
\bigskip

[4] Let ${\bf X}_1,...,{\bf X}_N$ be iid $N(\mu, {\bf R})$. Evaluate the Fisher information matrix elements 
$$
-\Bbb E[\frac{\partial^2}{\partial\mu_a\partial\mu_b}ln(p({\bf X}_1,...,{\bf X}_N|\mu,{\bf R})]
$$
$$
-\Bbb E[\frac{\partial^2}{\partial\mu_a\partial R_{bc}}ln(p({\bf X}_1,...,{\bf X}_N|\mu,{\bf R})]
$$
$$
-\Bbb E[\frac{\partial^2}{\partial R_{ab}\partial R_{cd}}ln(p({\bf X}_1,...,{\bf X}_N|\mu,{\bf R})]
$$
Taking the special case when the ${\bf X}_k's$ are 2-dimensional random vectors, evaluate the Cramer-Rao lower bound for $Var(\hat\mu_a), a=1,2$ and $Var(\hat R_{ab}), a,b=1,2$ where $\hat\mu$ and $\hat{\bf R}$ are respectively unbiased estimates of $\mu$
and ${\bf R}$.
\bigskip

[5] Let $\rho(\theta)$ denote a mixed quantum state in $\Bbb C^N$ dependent upon a classical parameter vector $\theta=(\theta_1,...\theta_p)^T$. Let $X,Y$ be two observables with eigendecompositions
$$
X=\sum_{a=1}^N|e_a>x(a)<e_a|, Y=\sum_{a=1}^N|f_a>y(a)<f_a|, <e_a|e_b>=<f_a|f_b>=\delta_{ab}
$$
First measure $X$ and note its outcome $x(a)$. Then allowing for state collapse, allow the state of the system to evolve from this collapsed state $\rho_c$ to the new state
$$
T(\rho_c)=\sum_{k=1}^mE_k\rho_cE_k^*, \sum_{k=1}^mE_k^*E_k=I
$$
Then measure  $Y$ in this evolved state and note the outcome $y(b)$. Calculate the joint probability distribution $P(x(a),y(b)|\theta)$. Explain by linearizing about some $\theta_0$, how you will calculate the maximum likelihood estimate of $\theta$ as a function of the measurements $x(a),y(b)$. Also explain how you will calculate the CRLB for $\theta$. Finally, explain how you will select the observables $X,Y$ to be measured in such a way that the CRLB is minimum when $\theta$ is a scalar parameter.
\bigskip

[6] For calculating time recursive updates of parameters in a linear AR model, we require to apply the matrix inversion lemma to a matrix of the form
$$
(\lambda.{\bf R}+{\bf bb}^T)^{-1}
$$
while for calculating order recursive updates, we require to use the projection operator update formula in the form
$$
{\bf P}_{span(,\xi)}={\bf P}_W+{\bf P}_{{\bf P}_W^{\perp}\xi}
$$
and also invert a block structured matrices of the form
$$
\left(\begin{array}{cc}{\bf A}&{\bf b}\\{\bf b}^T&c\end{array}\right)
$$
and
$$
\left(\begin{array}{cc}c&{\bf b}^T\\{\bf b}&{\bf A}\end{array}\right)
$$
in terms of ${\bf A}^{-1}$. Explain these statements in the light of the recurisive least squares lattice algorithm for simultaneous time and order recursive updates of the prediction filter parameters.
\bigskip

\section{Question paper on statistical signal processing, long test}

\subsection{Atom excited by a quantum field}

[1] Let the state at time $t=0$ of an atom be $\rho_A(0)$ and that of the field be
the coherent state $|\phi(u)><\phi(u)|$ so that the state of both the atom and the field at time $t=0$ is 
$$
\rho(0)=\rho_A(0)\otimes|\phi(u)><\phi(u)|
$$
Let the Hamiltonian of the atom and field in interaction be
$$
H(t)=H_A+\sum_k\omega(k)a(k)^*a(k)+\delta.V(t)
$$
where $H_A$ acts in the atom Hilbert space, $H_F=\sum_k\omega(k)a(k)^*a(k)$ acts in the field
Boson Fock space and the interaction Hamiltonian $V(t)$ acts in the tensor product of both the spaces. Assume that $V(t)$ has the form
$$
V(t)=\sum_k(X_k(t)\otimes a_k+X_k(t)^*\otimes a(k)^*)
$$
where $X_k(t)$ acts in the atom Hilbert space. Note that writing
$$
exp(itH_A)X_k(t)exp(-itH_A)=Y_k(t), 
$$
$$
exp(itH_F)a_k.exp(-itH_F)=a_k.exp(-i\omega(k)t)=a_k(t)
$$
we have on defining the unperturbed (ie, in the absence of interaction) system plus field Hamiltonian
$$
H_0=H_A+H_F=H_A+\sum_k\omega(k)a(k)^*a(k)
$$
that
$$
W(t)=exp(itH_0)V(t).exp(-itH_0)=\sum_k(Y_k(t)\otimes a_k(t)+Y_k(t)^*\otimes a_k(t)^*)
$$
show that upto $O(\delta^2)$ terms, the total system $\otimes$ field evolution operator
$U(t)$ can be expressed as
$$
U(t)=U_0(t).S(t)
$$
where
$$
U_0(t)=exp(-itH_0)=exp(-itH_A).exp(-itH_F)
$$
is the unperturbed evolution operator and
$$
S(t)=1-i\delta\int_0^tW(s)ds-\delta^2\int_{0<s_2<s_1<t}W(s_1)W(s_2)ds_1ds_2
$$
$$
=1-i\delta S_1(t)-\delta^2S_2(t)
$$
Show that with this approximation, the state of the system and bath at time $T$ is given by
$$
\rho(T)=U_0(T)(1-i\delta S_1(T)-\delta^2S_2(T))\rho(0).(1+i\delta S_1(T)-\delta^2S_2(T)^*)U_0(T)^*
$$
Evaluate using this expression, the atomic state at time $T$, as
$$
\rho_A(T)=Tr_2\rho(T)
$$
Use the fact that $\rho(0)=\rho_A(0)\otimes|\phi(u)><\phi(u)|$ and that
$$
Tr(a(k)|\phi(u)><\phi(u)|)=<\phi(u)|a(k)|\phi(u)>=u(k),
$$
$$
Tr(a(k)^*|\phi(u)><\phi(u)|)=<\phi(u)|a(k)^*|\phi(u)>=\bar u(k),
$$
$$
Tr(a(k)a(m)|\phi(u)><\phi(u)|)=<\phi(u)|a(k)a(m)|\phi(u)>=u(k)u(m),
$$
$$
Tr(a(k)a(m)^*|\phi(u)><\phi(u)|)=<\phi(u)|a(k)a(m)^*|\phi(u)>
$$
$$
=<\phi(u)|[a(k),a(m)^*]|\phi(u)>+<\phi(u)|a(m)^*a(k)|\phi(u)>
$$
$$
=\delta(k,m)+u(k)\bar u(m)
$$
and
$$
Tr(a(k)^*a(m)|\phi(u)><\phi(u)|)=<\phi(u)|a(k)^*a(m)|\phi(u)>=
$$
$$
=\bar u(k)u(m)
$$
\bigskip

Remark: More generally, we can expand any general initial state of the system (atom) and bath (field) as
$$
\rho(0)=\int F_A(u,\bar u)\otimes |\phi(u)><\phi(u)|dud\bar u
$$
$$
=\int F_A(u,\bar u)exp(-|u|^2)\otimes |e(u)><e(u)|du.d\bar u
$$
In this case also, we can calculate assuming that $\rho(0)$ is the initial state, final state
at time $T$ using the formula
$$
\rho(T)=U_0(T)(1-i\delta S_1(T)-\delta^2S_2(T))\rho(0).(1+i\delta S_1(T)-\delta^2S_2(T)^*)U_0(T)^*
$$
by evaluating this using the fact that if $Y_1,Y_2$ are system operators, then
$$
(Y_1\otimes a_k)(\int F(u,\bar u)\otimes|\phi(u)><\phi(u)|dud\bar u)(Y_2\otimes a_m)
$$
$$
=\int Y_1F(u,\bar u)Y_2exp(-|u|^2)\otimes a_k|e(u)><e(u)|a_m dud\bar u---(1)
$$
Now
$$
a_k|e(u)>=u(k)|e(u)>
$$
$$
a_k^*|e(u)>=\partial|e(u)>/\partial u(k)
$$
so (1) evaluates to after using integration by parts,
$$
\int Y_1F_1(u,\bar u)Y_2u(k)exp(-|u|^2)\otimes(\partial/\partial\bar u(m))(|e(u)><e(u)|)du.d\bar u
$$
$$
=\int(-\partial/\partial\bar u(m))(Y_1F(u,\bar u)Y_2).exp(-|u|^2u(k)))\otimes|e(u)><e(u)|dud\bar u
$$
$$
=\int(-\partial/\partial\bar u(m)+u(m))(Y_1F(u,\bar u)Y_2u(k))|\phi(u)><\phi(u)|du.d\bar u
$$
Likewise,
$$
(Y_1\otimes a_k^*)(\int F(u,\bar u)\otimes|\phi(u)><\phi(u)|dud\bar u)(Y_2\otimes a_m)
$$
$$
=\int Y_1F(u,\bar u)Y_2\otimes a_k^*|\phi(u)><\phi(u)|a_m dud\bar u
$$
$$
=\int Y_1F(u,\bar u)Y_2exp(-|u|^2)\otimes(\partial/\partial u(k)).(\partial/\partial\bar u(m))(|e(u)><e(u)|)du.d\bar u
$$
$$
=\int(\partial^2/\partial u(k)\partial\bar u(m))(Y_1F(u,\bar u)Y_2.exp(-|u|^2))\otimes|e(u)><e(u)|du.d\bar u
$$
$$
=\int(exp(|u|^2)(\partial^2/\partial u(k)\partial\bar u(m))(Y_1F(u,\bar u)Y_2.exp(-|u|^2)))\otimes|\phi(u)><\phi(u)|du.d\bar u
$$
$$
=\int (\partial/\partial u(k)+\bar u(k))(\partial/\partial\bar u(m)-u(m))(Y_1F(u,\bar u)Y_2)\otimes|\phi(u)<\phi(u)|dud\bar u
$$
and likewise terms like
$$
(Y_1\otimes a_k^*)(\int F(u,\bar u)\otimes|\phi(u)><\phi(u)|dud\bar u)(Y_2\otimes a_m^*)
$$
and
$$
(Y_1\otimes a_k)(\int F(u,\bar u)\otimes|\phi(u)><\phi(u)|dud\bar u)(Y_2\otimes a_m^*)
$$
We leave it as an exercise to the reader to evaluate these terms and hence derive upto the second order Dyson series approximation for the evolution operator the approximate final state of the atom and field and then by partial tracing over the field variables, to derive the approximate final atomic state under this interaction of atom and field (ie, system and bath).
\bigskip

[2] If the state of the system is given by $\rho=\rho_0+\delta.\rho_1$, then evaluate the 
Von-Neumann entropy of $\rho$ as a power series in $\delta$.

Hint:
\bigskip

\subsection{Estimating the power spectral density of a stationary Gaussian process using an AR model:Performance analysis of the algorithm}

The process $x[n]$ has mean zero and autocorrelation $R(\tau)$. It's $p^{th}$ order prediction error process is given by
$$
e[n]=x[n]+\sum_{k=1}^pa[k]x[n-k]
$$
where the $a[k]'s$ are chosen to minimize $\Bbb Ee[n]^2$. The optimal normal equations are
$$
\Bbb E(e[n]x[n-m])=0, m=1,2,..., p
$$
and these yield the Yule-Walker equations
$$
\sum_{k=1}^pR[m-k]a[k]=-R[m], m=1,2,..., p
$$
or equivalently in matrix notation,
$$
{\bf a}=-{\bf R}_p^{-1}{\bf r}_p
$$
where
$$
{\bf R}_p=((R[m-k]))_{1\leq m,k\leq p}, {\bf r}_p=((R[m]))_{m=1}^p
$$
The minimum mean square prediction error is then
$$
\sigma^2=\Bbb E(e[n]^2=\Bbb E(e[n]x[n])=R[0]+\sum_{k=1}^pa[k]R[k]=R[0]+{\bf a}^T{\bf r}_p
$$
$$
=R[0]-{\bf r}_p^T{\bf R}_p^{-1}{\bf r}_p
$$
If the order of the AR model $p$ is large, then $e[n]$ is nearly a white process, in fact in the limit $p\rightarrow\infty$, $e[n]$ becomes white, ie, $\Bbb E(e[n]x[n-m])$ becomes zero for all $m\geq 1$ and hence so does $\Bbb E(e[n]e[n-m])$. The ideal known statistics AR spectral estimate is given by
$$
S_{AR,p}(\omega)=S(\omega)=\frac{\sigma^2}{|A(\omega)|^2},
$$
$$
A(\omega)=1+{\bf a}^T{\bf e}(\omega),
$$
$$
{\bf e}(\omega)=((exp(-j\omega m)))_{m=1}^p
$$
Since in practice, we do not have the exact correlations available with us, we estimate the AR parameters using the least squares method, ie, by minimizing
$$
\sum_{n=1}^N(x[n]+{\bf a}^T{\bf x}_n)^2
$$
where
$$
{\bf x}_n=((x(n-m)))_{m=1}^p
$$
The finite data estimate of ${\bf R}_p$ is
$$
\hat R_p=N^{-1}\sum_{n=1}^N{\bf x}_n{\bf }_n^T
$$
and that of ${\bf r}_p$ is
$$
\hat r_p=N^{-1}\sum_{n=1}^Nx[n]{\bf x}_n
$$
We write $R$ for $R_p$ and $\hat R=R+\delta R$ for $\hat R_p$. Likewise, we write
$r$ for $r_p$ and $r+\delta r$ for $\hat r_p$ for notational convenience. Note that $\hat R$ and $\hat r$ are unbiased estimates of $R$ and $r$ respectively and hence $\delta R$ and $\delta r$ have zero mean. Now the least squares estimate $\hat a=a+\delta a $ of $a$ is given by
$$
\hat a=a+\delta a=-\hat R^{-1}\hat r=-(R+\delta R)^{-1}(r+\delta r)
$$
$$
=-(R^{-1}-R^{-1}\delta RR^{-1}+R^{-1}\delta R.R^{-1}\delta R.R^{-1})(r+\delta r)
$$
$$
=a-R^{-1}\delta r-R^{-1}\delta Ra+R^{-1}\delta R.R^{-1}\delta r+R^{-1}\delta R.R^{-1}\delta R.a
$$
upto quadratic orders in the data correlation estimate errors. Thus, we can write
$$
\delta a=-R^{-1}\delta r-R^{-1}\delta Ra+R^{-1}\delta R.R^{-1}\delta r+R^{-1}\delta R.R^{-1}\delta R.a
$$
Likewise, the estimate of the prediction error energy $\sigma^2$ is given by
$$
\hat\sigma^2=\sigma^2+\delta\sigma^2=\hat R[0]+\hat a^T\hat r
$$
and hence
$$
\delta\sigma^2=\delta R[0]+a^T\delta r+\delta a^Tr+\delta a^T\delta r
$$
$$
=\delta R[0]+a^T\delta r+(-R^{-1}\delta r-R^{-1}\delta Ra+R^{-1}\delta R.R^{-1}\delta r+R^{-1}\delta R.R^{-1}\delta r)^T\delta r
$$
$$
+(-R^{-1}\delta r-R^{-1}\delta Ra)^T\delta r
$$
again upto quadratic orders in the data correlation estimate errors. In terms of the fourth moments of the process $x[n]$, we can now evaluate the bias and mean square fluctuations of $\hat a$ and $\hat\sigma^2$ w.r.t their known statistics values $a,\sigma^2$ respectively. First we observe that the bias of $\hat a$ is
$$
\Bbb E(\delta a)=\Bbb E(-R^{-1}\delta r-R^{-1}\delta Ra+R^{-1}\delta R.R^{-1}\delta r+R^{-1}\delta R.R^{-1}\delta R.a)
$$
$$
=\Bbb E(R^{-1}\delta R.R^{-1}\delta r)+\Bbb E(R^{-1}\delta R.R^{-1}\delta R.a)
$$
Specifically, for the $i^{th}$ component of this, we find that with $P=R^{-1}$, and using Einstein's summation convention over repeated indices,
$$
\Bbb E(\delta a_i)=
$$
$$
\Bbb E(P_{ij}\delta R_{jk}P_{km}\delta r_m)+\Bbb E(P_{ij}\delta R_{jk}P_{km}\delta R_{ml}a_l)
$$
$$
=P_{ij}P_{km}\Bbb E(\delta R_{jk}\delta r_m)+P_{ij}P_{km}a_l\Bbb E(\delta R_{jk}\delta R_{ml})
$$
Now,
$$
\Bbb E(\delta R_{jk}.\delta r_m)=N^{-2}\sum_{n,l=1}^N\Bbb E[(x[n-k]x[n-j]-R[j-k])(x[l]x[l-m]-R[m])]
$$
$$
=N^{-2}\sum_{n,l=1}^N(\Bbb E(x[n-k]x[n-j]x[l]x[l-m])-R[j-k]R[m])
$$
upto fourth degree terms in the data. Note that here we have not yet made any assumption that $x[n]$ is a Gaussian process, although we are assuming that the process is stationary at least upto fourth order. Writing
$$
\Bbb E(x[n_1]x[n_2]x[n_3]x[n_4])=C(n_2-n_1,n_3-n_1,n_4-n_1)
$$
we get
$$
\Bbb E(\delta R_{jk}.\delta r_m)=
$$
$$
N^{-2}\sum_{n,l=1}^NC(k-j,l-n+k,l-n+k-m)
$$
\bigskip

\subsection{Statistical performance analysis of the MUSIC and ESPRIT algorithms}

Let 
$$
\hat R=R+\delta R
$$
be an estimate of the signal correlation matrix $R$. Let $v_1,...,v_p$ denote the signal eigenvectors and $v_{p+1},..., v_N$ the noise eigenvectors. The signal subspace eigen-matrix is
$$
V_S=[v_1,..., v_p]\in\Bbb C^{N\times p}
$$
and the noise subspace eigen-matrix is
$$
V_N=[v_{p+1},..., v_N]
$$
so that
$$
V=[V_S|V_N]
$$
is an $N\times N$ unitary matrix. The signal correlation matrix model is
$$
R=EDE^*+\sigma^2I
$$
$\sigma^2$ is estimated as $\hat\sigma^2$ which is the average of the $N-p$ smallest eigenvalues of $R$. In this expression,
$$
E=[e(\omega_1),..., e(\omega_p)]
$$
is the signal manifold matrix where
$$
e(\omega)=[1,exp(j\omega),..., exp(j(N-1)\omega)]^T
$$
is the steering vector along the frequency $\omega$. Note that since $N>p$, $E$ has full column rank $p$. The signal eigenvalues of $R$ are
$$
\lambda_i=\mu_i+\sigma^2, i=1,2,..., p
$$
while the noise eigenvalues are
$$
\lambda_i=\sigma^2, i=p+1,..., N
$$
Thus,
$$
Rv_i=\lambda_iv_i, i=1,2,..., N
$$
The known statistics exact MUSIC spectral estimator is
$$
P(\omega)=(1-N^{-1}\parallel V_S^*e(\omega)\parallel^2)^{-1}
$$
which is infinite when and only when $\omega\in\{\omega_1,...,\omega_p\}$. Alternately, the zeros of the function
$$
Q(\omega)=1/P(\omega)=1-N^{-1}\parallel V_S^*e(\omega)\parallel^2
$$
are precisely $\{\omega_1,...,\omega_p\}$ because, the $V_N^*e(\omega)=0$ iff
$e(\omega)$ belongs to the signal subspace $span\{e(\omega_i):1\leq i\leq p\}$ iff
$\{e(\omega),e(\omega_i),i=1,2,..., p\}$ is a linearly dependent set iff
$\omega\in\{\omega_1,...,\omega_p\}$, the last statement being a consequence of the fact that $N\geq p+1$ and the Van-Der-Monde determinant theorem. Now, owing to finite data effects, $v_i$ gets perturbed to $v_i+\delta v_i$ for each $i=1,2,..., p$ so that correspondingly, $V_S$ gets perturbed to $\hat V_S=V_S+\delta V_S$. Then $Q$ gets perturbed to (upto first orders of smallness, ie, upto $O(\delta R)$
$$
Q(\omega)+\delta Q(\omega)=1-N^{-1}\parallel(V_S+\delta V_S)^*e(\omega)\parallel^2
$$
so that
$$
\delta Q(\omega)=(-2/N)Re(e(\omega)^*\delta V_SV_S^*e(\omega))
$$
Let $\omega$ be one of the zeros of $Q(\omega)$, ie, one of the frequencies of the signal.
Then, owing to finite data effects, it gets perturbed to $\omega+\delta\omega$ where
$$
Q'(\omega)\delta\omega+\delta Q(\omega)=0
$$
ie,
$$
\delta\omega=-\delta Q(\omega)/Q'(\omega)
$$
and
$$
Q'(\omega)=(-2/N)Re(e(\omega)^*P_Se'(\omega))=(-2/N)Re<e(\omega)|P_S|e'(\omega)>
$$
where 
$$
P_S=V_SV_S^*
$$
is the orthogonal projection onto the signal subspace.  Note that by first order perturbation theory
$$
|\delta v_i>=\sum_{j\neq i}\frac{|v_j><v_j|\delta R|v_i>}{\lambda_i-\lambda_j}, i=1,2,..., N
$$
and
$$
\delta\lambda_i=<v_i|\delta R|v_i>
$$
So
$$
\delta V_S=column(\sum_{j\neq i}|v_j><v_j|\delta R|v_i>/(\lambda_i-\lambda_j):i=1,2,..., p]
$$
$$
=[Q_1\delta R|v_1>,...,Q_p\delta R|v_p>]
$$
where
$$
Q_i=\sum_{j\neq i}|v_j><v_j|/(\lambda_i-\lambda_j)
$$
We can express this as
$$
\delta V_S=[Q_1,..., Q_p](I_p\otimes\delta R.V_S)
$$
and hence
$$
\delta Q(\omega)=(-2/N)Re(<e(\omega)|\delta V_SV_S^*|e(\omega))>)
$$
$$
=(-2/N)Re(<e(\omega)S(I_p\otimes\delta R).V_SV_S^*|e(\omega)>)
$$
$$
=(-2/N)Re(<e(\omega)|S(I_p\otimes\delta R)P_S|e(\omega)>)
$$
where as before $P_S=V_SV_S^*$ is the orthogonal projection onto the signal subspace and
$$
S=[Q_1,Q_2,..., Q_p]
$$
The evaluation of $\Bbb E((\delta\omega)^2)$ can therefore be easily done once we evaluate
$\Bbb E(\delta R(i,j)\delta R(k,m))$ and this is what we now propose to do.
\newpage

Statistical performance analysis of the ESPRIT algorithm. Here, we have two correlation matrices $R,R_1$ having the structures
$$
R=EDE^*+\sigma^2I, R_1=ED\Phi^*E^*+\sigma^2Z
$$
where $E$ has full column rank, $D$ is a square, non-singular positive definite matrix and
$\Phi$ is a diagonal unitary matrix. The structure of $Z$ is known. We have to estimate the diagonal entries of $\Phi$ which give us the signal frequencies. Let as before
$V_S$ denote the matrix formed from the signal eigenvectors of $R$ and $V_N$ that formed from the noise eigenvectors of $R$. $V=[V_S|V_N]$ is unitary. When we form finite data estimates, $R$ is replaced by $\hat R=R+\delta R$ and $R_1$ by $\hat R_1=R_1+\delta R_1$.
The estimate of $\sigma^2$ is $\hat\sigma^2=\sigma^2+\delta\sigma^2$ obtained by averaging the $N-p$ smallest eigenvalues of $\hat R$. The estimate of $R_S=EDE^*$ is then
$$
\hat R_S=R_S+\delta R_S=\hat R-\hat\sigma^2I=R-\sigma^2I+\delta R-\delta\sigma^2
$$
Thus,
$$
\delta R_S=\delta R-\delta\sigma^2I
$$
Likewise, the estimate of $R_{S1}=ED\Phi^*E^*$ is 
$$
\hat R_{S1}=R_{S1}+\delta R_{S1}=\hat R_1-\hat\sigma^2Z=R_1-\sigma^2Z+\delta R_1-\delta\sigma^2Z
$$
so that
$$
\delta R_{S1}=\delta R_1-\delta\sigma^2Z
$$
The rank reducing numbers of the matrix pencil 
$$
R_S-\gamma R_{S1}=ED(I-\gamma\Phi^*)E^*
$$
are clearly the diagonal entries of $\Phi$. Since $R(E)=R(V_S)$, it follows that these rank reducing numbers are the zeroes of the polynomial equation
$$
det(V_S^*(R_S-\gamma R_{S1})V_S)=0
$$
or equivalently of
$$
det(M-\gamma V_S^*R_{S1}V_S)=0
$$
where
$$
M=V_S^*R_SV_S=diag[\mu_1,...,\mu_p]
$$
with
$$
\mu_k=\lambda_k-\sigma^2, kj=1,2,..., p
$$
being the nonzero eigenvalues of $R_S$. When finite data effects are taken into consideration, these become the rank reducing numbers of
$$
det(M+\delta M-\hat\gamma(V_S+\delta V_S)^*(R_{S1}+\delta R_{S1})(V_S+\delta V_S))=0
$$
or equivalently, upto first order perturbation theory, the rank reducing number
$\gamma$ gets perturbed to $\hat\gamma=\gamma+\delta\gamma$, where
$$
det(M-\gamma X+\delta M-\gamma.\delta X-\delta\gamma X)=0
$$
where
$$
X=V_S^*R_{S1}V_S, \delta X=\delta V_S^*R_{S1}V_S+V_S^*R_{S1}\delta V_S+V_S^*\delta R_{S1}V_S
$$
Let $\xi$ be a right eigenvector of $M-\gamma X$ and $\eta^*$ a left eigenvector of the same corresponding to the zero eigenvalue. Note that $M,X,\delta M,\delta X$ are all Hermitian but $M-\gamma X$ is not because $\gamma=exp(j\omega_i)$ is generally not real.
Then, let $\xi+\delta\xi$ be a right eigenvector of 
$$
M-\gamma X+\delta M-\gamma\delta X-\delta\gamma X
$$
corresponding to the zero eigenvalue. In terms of first order perturbation theory,
$$
0=(M-\gamma X+\delta M-\gamma\delta X-\delta\gamma X)(\xi+\delta\xi)
$$
$$
=(M-\gamma X)\delta\xi+(\delta M-\gamma\delta X)\xi-\delta\gamma X\xi
$$
Premultiplying by $\eta^*$ gives us
$$
\delta\gamma\eta^*X\xi=\eta^*(\delta M-\gamma\delta X)\xi
$$
or equivalently,
$$
\delta\gamma=\frac{<\eta|\delta M-\gamma\delta X\|\xi>}{<\eta|X|\xi>}
$$
which is the desired formula.
\bigskip

\section{Quantum neural networks for estimating EEG pdf from speech pdf based on Schrodinger's equation}

{\bf Reference}:Paper by Vijay Upreti, Harish Parthasarathy, Vijayant Agarwal and Guguloth Sagar.

Quantum neural networks for estimating the marginal pdf of one signal from the marginal pdf of another signal. The idea is that the essential characteristics of either speech or EEG signals are contained in their empirical pdf's, ie, the proportion of time that the signal amplitude spends within any Borel subset of the real line. This hypothesis leads one to believe that by estimating the pdf of the EEG data from that of speech data alone, we can firstly do away with the elaborate apparatus required for measuring EEG signals and secondly we can extract from the EEG pdf, the essential characteristics of the EEG signal parameters which determine the condition of a diseased brain. It is a fact that since the Hamiltonian operator $H(t)$ in the time dependent Schrodinger equation is Hermitian, the resulting evolution operator $U(t,s), t>s$ is unitary causing thereby the wave function modulus square to integrate to unity at any time $t$ if at time $t=0$ it does integrate to unity. Thus, the Schrodinger equation naturally generates a family of pdf's. Further, based on the assumption that the speech pdf and the EEG pdf for any patient are correlated in the same way (a particular kind of brain disease causes a certain distortion in the EEG signal and a correlated distortion in the speech process with the correlation being patient independent), we include a potential term in Schrodinger's equation involving speech pdf and hence while tracking a given EEG pdf owing to the adaptation algorithm of the QNN, the QNN also ensures that the output pdf coming from Schrodinger's equation will also maintain a certain degree of correlation with the speech pdf. The basic idea behind the quantum neural network is to modulate the potential by a "weight process" with the weight increasing whenever the desired pdf exceeds the pdf generated by Schrodinger's equation using the modulus of the wave function squared. A hieuristic proof of why the pdf generated by the Schrodinger equation should increase when the weight process and hence th modulated potential increases is provided at the end of the paper.
\bigskip

First consider the following model for estimating the pdf $p(t,x)$
of one signal via Schrodinger's wave equation which naturally generates a continuous family of pdf's via the modulus square of the wave function:
$$
ih\partial_t\psi(t,x)=(-h^2/2)\partial_x^2\psi(t,x)+W(t,x)V(x)\psi(t,x)
$$
The weight process $W(t,x)$ is generated in such a way that $|\psi(t,x)|^2$ tracks the given pdf $p(t,x)$. This adaptation scheme is
$$
\partial_tW(t,x)=-\beta W(t,x)+\alpha(p(t,x)-|\psi(t,x)|^2)
$$
where $\alpha,\beta$ are positive constants. This latter equation guarantees that if
$W(t,x)$ is very large and positive (negative), then the term $-\beta W(t,x)$ is very large and negative (positive) causing $W$ to decrease (increase). In other words,
$-\beta.W$ is a threshold term that prevents the weight process from getting too large in magnitude which could cause the break-down of the quantum neural network. We are assuming that $V(x)$ is a non-negative potential. Thus, increase in $W$ will cause an increase in
$WV$ which as we shall see by an argument given below, will cause a corresponding increase in $|\psi|^2$ and likewise a decrease in $W$ will cause a corresponding decrease in $|\psi|^2$. Now we add another potential term $p_0(t,x)V_0(x)$ to the above Schrodinger dynamics where $p_0$ is the pdf of the input signal. This term is added because we have the understanding that the output pdf $p$ is in some way correlated with the input pdf. Then the dynamics becomes
$$
ih\partial_t\psi(t,x)=(-h^2/2)\partial_x^2\psi(t,x)+W(t,x)V(x)\psi(t,x)+p_0(t,x)V_0(x)\psi(t,x)
$$

Statistical performance analysis of the QNN: Suppose $p_0(t,x)$ fluctuates to $p_0(t,x)+\delta p_0(t,x)$ and $p(t,x)$ to $p(t,x)+\delta p(t,x)$. Then the corresponding QNN output wave function $\psi(t,x)$ will fluctuate to $\psi(t,x)+\delta\psi(t,x)$ where $\delta\psi$ satisfies the linearized Schrodinger equation and likewise the weight process $W(t,x)$ will fluctuate to $W(t,x)+\delta W(t,x)$. The linearized dynamics for these fluctuations are
$$
ih\partial_t\delta\psi(t,x)=(-h^2/2)\partial_x^2\delta\psi(t,x)+V(x)(W\delta\psi+\psi\delta W)+V_0(x)(p_0\delta\psi+\psi\delta p_0)
$$
$$
\partial_t\delta W=-\beta\delta W+\alpha(\delta p-\bar\psi.\delta\psi-\psi.\delta\bar\psi)
$$
Writing
$$
\psi(t,x)=A(t,x).exp(iS(t,x)/h)
$$
we get
$$
\delta\psi=(\delta A+iA\delta S/h)exp(iS/h)
$$
so that with prime denoting spatial derivative,
$$
\delta\psi'=(\delta A'+iA'\delta S/h+iA\delta S'/h+iS'\delta A/h-AS'\delta S/h^2)exp(iS/h)
$$
$$
\delta\psi''=(\delta A''+iA''\delta S/h+2iA'\delta S'/h+iA\delta S''/h+2iS'\delta A'/h+iS''\delta A/h-AS'\delta S'/h^2
$$
$$
-A'S'\delta S/h^2-AS''\delta S/h^2-AS'\delta S'/h^2-S^{'2}\delta A/h^2-iAS^{'2}\delta S/h^3).exp(iS/h)
$$
Thus we obtain the following differential equation for the wave function fluctuations caused by random fluctuations in the EEG and speech pdf's:
$$
ih\partial_t\delta A-\partial_t(A\delta S)-\delta A.\partial_tS-A\delta S\partial_tS/h
$$
$$
=(-h^2/2)\delta A''-ihA''\delta S/2-ihA'\delta S'-ihA.\delta S''/2-ihS'\delta A'
$$
$$
-ihS''\delta A/2+AS'\delta S'/2+A'S'\delta S/2+AS''\delta S/2+AS'\delta S'/2+S^{'2}\delta A/2+iAS^{'2}\delta S/2h
$$
$$
+V_0(x)p_0(t,x)(\delta A+iA\delta S/h)+V_0(x)A\delta p_0(t,x)
$$
$$
+V(x)W(t,x)(\delta A+iA\delta S/h)+V(x)A\delta W)
$$
On equating real and imaginary parts in this equation, we get two pde's for the two functions $A$ and $S$. However, it is easier to derive these equations by first deriving the unperturbed pde's for $A,S$ and then forming the perturbations. The unperturbed equations are (See Appendix A.1)
$$
h\partial_tA=-h\partial_xA.\partial_xS-(1/2)A\partial_x^2S
$$
$$
-A\partial_tS=-((h^2/2)\partial_x^2A-(1/2)A(\partial_xS)^2)+WVA+p_0V_0A
$$
$$
\partial_tW(t,x)=-\beta(W(t,x)-W_0(x))+\alpha(p(t,x)-A^2)
$$

Forming the variation of this equation around a fixed operating point $p_0(x),p(x)=|\psi(t,x)|^2$ and $W(x)$ with $\psi(t,x)=A(x).exp(-iEt+iS_0(x))$ gives us 
$$
h\partial_t\delta A=-h\delta A'.S_0'-hA'\delta S'-(1/2)S_0''\delta A-(1/2)A\delta S'',
$$
$$
-A\partial_t\delta S+E\delta A=(-h^2/2)\delta A''-(1/2)S_0^{'2}\delta A-AS_0'\delta S''
$$
$$
+VW\delta A+VA\delta W+V_0A\delta p_0+V_0p_0\delta A,
$$
$$
\partial_t\delta W=-\beta\delta W+\alpha.\delta p-2\alpha A\delta A
$$
\bigskip

{\bf Appendix}

A.1

Proof that increase in potential causes increase in probability density determined as modulus square of the wave function.

Consider Schrodinger's equation in the form
$$
ih\partial_t\psi(t,x)=(-h^2/2)\partial_x^2\psi(t,x)+V(x)\psi(t,x)
$$
We write
$$
\psi(t,x)=A(t,x).exp(iS(t,x)/h)
$$
where $A,S$ are real functions with $A$ non-negative. Thus, $|\psi|^2=A^2$. Substituting this expression into Schrodinger's equation and equating real and imaginary parts gives
$$
h\partial_tA=-h\partial_xA.\partial_xS-(1/2)A\partial_x^2S)
$$
$$
-A\partial_tS=-((h^2/2)\partial_x^2A-(1/2)A(\partial_xS)^2)+VA
$$
We write
$$
S(t,x)=-Et+S_0(x)
$$
so that $E$ is identified with the total energy of the quantum particle. Then,
the second equation gives on neglecting $O(h^2)$ terms,
$$
(\partial_xS_0)=\pm\sqrt{2(E-V(x))}
$$
and then substituting this into the first equation with $\partial_tA=0$ gives
$$
S_0''/S_0'=-2hA'/A
$$
or
$$
ln(S_0')=-2h.ln(A)+C
$$
so
$$
A=K_1(S_0')^{-1/2}=K_2(E-V(x))^{-1/4}
$$
which shows that as long as $V(x)<E$, $A(x)$ and hence $A(x)^2=|\psi(x)|^2$ will increase
with increasing $V(x)$ confirming the fact that by causing the driving potential to increase when $p_E-|\psi|^2$ is large positive, we can ensure that $|\psi|^2$ will increase and viceversa thereby confirming the a validity of our adaptive algorithm for tracking the EEG pdf.

A.2 Calculating the empirical distribution of a stationary stochastic process.

Let $x(t)$ be a stationary stochastic process. The empirical density estimate of
$N$ samples of this process spaced $\tau_k,k=1,2,..., n-1$ seconds apart from the first sample, ie, an estimate of the joint density of $x(t), x(t+\tau_k), k=1,2,..., N-1$ is given bylog(
$$
\hat p_T(x_1,...,x_N;\tau_1,...,\tau_{N-1})=T^{-1}\int_0^T\Pi_{k=1}^N\delta(x(t+\tau_k)-x_k)dt
$$
where $\tau_0=0$. The Gartner-Ellis theorem in the theory of large deviations provides a reliability estimate of how this empirical density performs. Specifically, if $f$ is a function of $N$ variables, then the moment generating functional of $\hat p_T(.,\tau_1,...,\tau_{N-1})$ is given by
$$
\Bbb E[exp(\int\hat p_T(x_1,...,x_N;\tau_1,...,\tau_{N-1})f(x_1,...,x_N)dx_1...dx_N)]
$$
$$
=\Bbb E[exp(T^{-1}\int_0^Tf(x(t),x(t+\tau_1),...,x(t+\tau_{N-1}))dt)]
$$
$$
=M_T(f|\tau_1,...,\tau_{N-1})
$$
say. Then, the rate functional for the family of random fields $\hat p_T(.,\tau_1,...\tau_{N-1}), T\rightarrow\infty$ is given by
$$
I(p)=sup_f(\int f(x_1,...,x_N)p(x_1,...,x_N)dx_1...dx_N-\Lambda(f))
$$
where
$$
\Lambda(f)=lim_{T\rightarrow\infty}T^{-1}log(M_T(T.f))
$$
assuming that this limit exists. It follows that as $T\rightarrow\infty$, the asymptotic probability of $\hat p_T$ taking values in a set $E$ is given by
$$
T^{-1}log(Pr(\hat p_T(.,\tau_1,...,\tau_{N-1})\in E)\approx -inf(I(p):p\in E)
$$
or equivalently,
$$
Pr(\hat p_T(.,\tau_1,...,\tau_{N-1})=p)\approx exp(-TI(p)), T\rightarrow\infty
$$
\bigskip

Appendix A.3

We can also use the electroweak theory or more specially, the Dirac-Yang-Mills non-Abelian gauge theory for matter and gauge fields to track a given pdf. Such an algorithm enables us to make use of elementary particles in a particle accelerator to simulate probability densities for learning about the statistics of signals.

Let $A_{\mu}^a(x)\tau_a$ be a non-Abelian gauge field assuming values in a Lie algebra $\mathfrak g$ which is a Lie sub-algebra of $\mathfrak u(N)$. Here, $\tau_a, a=1,2,..., N$ are Hermitian generators of $\mathfrak g$. Denote their structure constants by $\{C(abc)\}$, ie,
$$
[\tau_a,\tau_b]=iC(abc)\tau_c
$$
Then the gauge covariant derivative is
$$
\nabla_{\mu}=\partial_{\mu}+ie A_{\mu}^a\tau_a
$$
and the corresponding Yang-Mills antisymmetric gauge field tensor, also called the curvature of the gauge covariant derivative is given by
$$
ieF_{\mu\nu}=[\nabla_{mu},\nabla_{\nu}]=
$$
$$
=ie(A_{\nu,\mu}^a-A_{\mu,\nu}^a)-eC(abc)A_{\mu}^bA_{\nu}^c)\tau_a
$$
so that
$$
F_{\mu\nu}=F_{\mu\nu}^a\tau_a
$$
where
$$
F_{\mu\nu}^a=A_{\nu,\mu}^a-A_{\mu,\nu}^a)-eC(abc)A_{\mu}^bA_{\nu}^c
$$
The action functional for the matter field $\psi(x)$ having $4N$ components and the gauge potentials $A_{\mu}^a(x)$ is then given by (S.Weinberg, The quantum theory of fields, vol.II)
$$
S[A,\psi]=\int Ld^4x,
$$
$$
L=\bar\psi(\gamma^{\mu}(i\partial_{\mu}+eA_{\mu}^a\tau_a)-m)\psi-(1/4)F_{\mu\nu}^aF^{a\mu\nu}
$$
where
$$
\bar\psi=\psi^*\gamma^0
$$
The equations of motion obtained by setting the variation of $S$ w.r.t the matter wave function $\psi$ and the gauge potentials $A$ are given by
$$
[\gamma^{\mu}(i\partial_{\mu}+eA_{\mu}^a\tau_a)-m]\psi=0,
$$
$$
D_{\nu}F^{\mu\nu}=J^{\mu}
$$
where $D_{\mu}$ is the gauge covariant derivative in the adjoint representation, ie,
$$
[\partial_{\nu}+ieA_{\nu},F^{\mu\nu}]=J^{\mu}
$$
or equivalently in terms of components,
$$
\partial_{\nu}F^{a\mu\nu}-eC(abc)A_{\nu}^bF^{c\mu\nu}=J^{a\mu}
$$
with $J^{a\mu}$ being the Dirac-Yang-Mills current:
$$
J^{a\mu}=-e\bar\psi(\gamma^{\mu}\otimes\tau_a)\psi
$$
it is clear that $\psi$ satisfies a Dirac-Yang-mills matter wave equation and it can be expressed in the form of Dirac's equation with a self-adjoint Hamiltonian owing to which
$J^{a\mu}$ is a conserved current and in particular, the $|\psi|^2$ integrates over space to give a constant at any time which can be set to unity by the initial condition. In other words, $|\psi(x)|^2$ is a probability density at any time $t$ where $x=(t,{\bf r})$.
Now it is easy to make the potential in this wave equation adaptive, ie, depend on the error $p(x)-|\psi(x)|^2$. Simply add an interaction term $W(x)(p(x)-|\psi(x)|^2)V(x)$ to
to the Hamiltonian. This is more easily accomplished using the electroweak theory where there is in addition to the Yang-Mills gauge potential terms a photon four potential
$B_{\mu}$ with $B_0(x)=W(x)(p(x)-|\psi(x)|^2)V(x)$. This results in the following wave equation
$$
[\gamma^{\mu}(i\partial_{\mu}+eA_{\mu}^a\tau_a+eB_{\mu})-m]\psi=0
$$
The problem with this formalism is that a photon Lagrangian term $(-1/4)(B_{\nu,\mu}-B_{\mu,\nu})(B^{\nu,\mu}-B^{\mu,\nu})$ also has to be added to the total Lagrangian which results in the Maxwell field equation
$$
F^{ph\mu\nu}_{,\nu}=-e\bar\psi\gamma^{\mu}\psi
$$
where
$$
F^{ph}_{\mu\nu}=B_{\nu,\mu}-B_{\mu,\nu}
$$
is the electromagnetic field. In order not to violate these Maxwell equations, we must add another external charge source term to the rhs so that the Maxwell equation become
$$
F^{ph\mu\nu}_{,\nu}=-e\bar\psi\gamma^{\mu}\psi+\rho(x)\delta^{\mu}_0
$$
$\rho(x)$ is an external charge source term that is given by
$$
-e\psi^*\psi+\rho(x)=-\nabla^2(W(x)(p(x)-|\psi(x)|^2)V(x))=-\nabla^2B_0
$$
so that
$$
B_0=W(p-|\psi|^2)V
$$
The change in the Lagrangian produced by this source term is $-\rho(x)B_0(x)$. This control charge term can be adjusted for appropriate adaptation. Thus in effect our Lagrangian acquires two new terms, the first is
$$
e\bar\psi\gamma^0B_0\psi=e\psi^*\psi.B_0
$$
and the second is
$$
-\rho.B_0
$$
where
$$
\rho=-\nabla^2(W(p-\psi^*\psi))+e\psi^*\psi
$$
The dynamics of the wave function $\psi$ then acquires two new terms, the first coming
the variation w.r.t $\bar\psi$ of the term $e\psi^*\psi.B_0=e\bar\psi\gamma^0\psi.B_0$, this term is $eB_0\gamma^0\psi$, the second is the term coming from the variation w.r.t $\bar\psi$ of the term $-\rho.B_0=B_0\nabla^2(W(p-\bar\psi\gamma^0\psi))$. This variation gives
$$
(\nabla^2B_0)(W\gamma^0\psi)
$$
The resulting equations of motion for the wave function get modified to
$$
[\gamma^{\mu}(i\partial_{\mu}+eA^a_{\mu}\tau_a+eB_{\mu})-m+\nabla^2B_0.W\gamma^0]\psi=0
$$
This equation is to be supplemented with the Maxwell and Yang-Mills gauge field equations and of course the weight update equation for $W(x)$.
\bigskip

\section{Statistical Signal Processing, M.Tech, Long Test\author{Max marks:50}}

[1] [a] Prove that the quantum relative entropy is convex, ie, if $(\rho_i,\sigma_i), i=1,2$ are pairs of mixed states in $\Bbb C^d$, then
$$
D(p_1\rho_1+p_2\rho_2|p_1\sigma_1+p_2\sigma_2)\leq p_1D(\rho_1|\sigma_1)+p_2D(\rho_2|\sigma_2)
$$
for $p_1,p_2\geq 0, p_1+p_2=1$.

hint: First prove using the following steps that if $K$ is a quantum operation, ie, TPCP map, then for two states $\rho,\sigma$, we have
$$
D(K(\rho)|K(\sigma))\leq D(\rho|\sigma)---(a)
$$
Then define
$$
\rho=\left(\begin{array}{cc}p_1\rho_1&0\\0&p_2\rho_2\end{array}\right),
$$
$$
\sigma=\left(\begin{array}{cc}p_1\sigma_1&0\\0&p_2\sigma_2\end{array}\right),
$$
and $K$ to be the operation given by
$$
K(\rho)=E_1\rho.E_1^*+E_2\rho.E_2^*
$$
where
$$
E_1=[I|0], E_2=[0|I]\in\Bbb C^{d\times 2d}
$$
Note that 
$$
E_1^*E_1+E_2^*E_2=I_{2d}
$$
To prove (a), make use of the result on the asymptotic error rate in hypothesis testing on tensor product states: Given that the first kind of error, ie, probability of detecting $\rho^{\otimes n}$ given that the true state is $\sigma^{\otimes n}$ goes to zero, the minimum (Neyman-Pearson test) probability of error of the second kind goes to zero at the rate of $-D(\rho|\sigma)$ (this result is known as quantum Stein's lemma). Now consider any sequence of tests $T_n$, ie $0\leq T_n\leq I$ on $(\Bbb C^d)^{\otimes n}$. Then consider using this test for testing between the pair of states $K^{\otimes n}(\rho^{\otimes n})$ and $K^{\otimes n}(\sigma^{\otimes n})$. Observe that the optimum performance for this sequential hypothesis testing problem is the same as that of the dual problem of testing $\rho^{\otimes n}$ versus $\sigma^{\otimes n}$ using the dual test $K^{*\otimes n}(T_n)$. This is because
$$
Tr(K^{\otimes n}(\rho^{\otimes n})T_n)=Tr(\rho^{\otimes n}K^{*\otimes n}(T_n))
$$
etc. It follows that the optimum Neyman-Pearson error probability of the first kind for 
testing $\rho^{\otimes n}$ versus $\sigma^{\otimes n}$ will be much smaller than that of
testing $K^{\otimes n}(\rho^{\otimes n})$ versus $K^{\otimes n}(\sigma^{\otimes n})$ since
$K^{*\otimes n}(T_n)$ varies only over a subset of tests as $T_n$ varies over all tests in
$(\Bbb C^d)^{\otimes n}$. Hence going to the limit gives the result
$$
exp(-nD(\rho|\sigma))\leq exp(-nD(K(\rho)|K(\sigma))
$$
asymptotically as $n\rightarrow\infty$ which actually means that
$$
D(\rho|\sigma)\geq D(K(\rho)|K(\sigma))
$$
\bigskip

[b] Prove using [a] that the Von-Neumann entropy $H(\rho)=-Tr(\rho.log(\rho))$ is a concave function, ie if $\rho, \sigma$ are two states, then
$$
H(p_1\rho+p_2\sigma)\geq p_1H(\rho)+p_2H(\sigma), p_1,p_2\geq 0, p_1+p_2=1
$$

hint: Let $K$ be any TPCP map. Let 
$$
\rho=\left(\begin{array}{cc}p_1\rho_1&0\\0&p_2\rho_2\end{array}\right)\in\Bbb C^{2d\times 2d},
$$
and
$$
\sigma= I_{2d}/2d
$$
where $\rho_k\in\Bbb C^{d\times d}, k=1,2$. Then choose
$$
E_1=[I|0], E_2=[0|I]\in\Bbb C^{d\times 2d}
$$
and
$$
K(X)=E_1XE_1^*+E_2XE_2^*\in\Bbb C^{d\times d}, X\in\Bbb C^{2d\times 2d}
$$
Note that
$$
E_1^*E_1+E_2^*E_2=I_d
$$
Show that
$$
K(\sigma)=K(I_{2d}/2d)=I_d/d
$$
Then we get using [a],
$$
D(K(\rho)|K(\sigma))\leq D(\rho|sigma)
$$
Show that this yields
$$
H(p_1\rho_1+p_2\rho_2)+log(d)\geq p_1H(\rho_1)+p_2H(\rho_2)-H(p)+log(2d)
$$
where
$$
H(p)=-p_1log(p_1)-p_2log(p_2)
$$
Show that
$$
H(p)\leq log(2)
$$
and hence deduce the result.
\bigskip

[2] [a] Let ${\bf X}[n], n\geq 0$ be a $M$-vector valued AR process, ie, it satisfies the first order vector difference equation
$$
{\bf X}[n]={\bf A}{\bf X}[n-1]+{\bf W}[n], n\geq 1
$$
Assume that ${\bf W}[n], N\geq 1$ is an iid $N({\bf 0},{\bf Q})$ process independent of
${\bf X}[0]$ and that ${\bf X}[0]$ is a Gaussian random vector. Calculate the value of
$$
{\bf R}[0]=\Bbb E({\bf X}[0]{\bf X}[0]^T)
$$
in terms of ${\bf A}$ and ${\bf Q}$ so that ${\bf X}[n], n\geq 0$ is wide-sense stationary in the sense that
$$
\Bbb E({\bf X}[n]{\bf X}[n-m]^T)={\bf R}[m]
$$
does not depend on $n$. In such a case, evaluate the ACF ${\bf R}[m]$ of $\{{\bf X}[n]\}$
in terms of ${\bf A}$ and ${\bf Q}$. Assume that ${\bf A}$ is diagonable with all eigenvalues being strictly inside the unit circle. Express your answer in terms of the eigenvalues and eigen-projections of ${\bf A}$.
\bigskip

[b] Assuming ${\bf X}[0]$ known, evaluate the maximum likelihood estimator of ${\bf A}$
given the data $\{{\bf X}[n]:1\leq n\leq N\}$. Also evaluate the approximate mean and covariance of the estimates of the matrix elements of ${\bf A}$ upto quadratic orders
in the correlations $\{{\bf R}[m]\}$ based on the data ${\bf X}[n], 1\leq n\leq N$ and upto $O(1/N)$. In other words, calculate using perturbation theory, $\Bbb E(\delta A_{ij})$ and $\Bbb E(\delta A_{ij}\delta A_{km})$ upto $O(1/N)$ where
$$
\hat A-{\bf A}=\delta{\bf A}
$$
where $\hat A$ is the maximum likelihood estimator of ${\bf A}$. 

hint: Write
$$
N^{-1}\sum_{n=1}^N{\bf X}[n-i]{\bf X}[n-j]^T-{\bf R}[j-i]=\delta R(j,i)
$$
and then evaluate $\delta{\bf A}$ upto quadratic orders in $\delta R(j,i)$. Finally, use 
the fact that since ${\bf X}[n], n\geq 0$ is a Gaussian process, 
$$
\Bbb E(X_a[n_1]X_b[n_2]X_c[n_3]X_d[n_4])=R_{ab}[n_1-n_2]R_{cd}[n_3-n_4]+R_{ac}[n_1-n_3]R_{bd}[n_2-n_4]+R_{ad}[n_1-n_4]R_{bc}[n_2-n_3]
$$
where
$$
((R_{ab}(\tau)))_{1\leq a,b\leq M}={\bf R}(\tau)=\Bbb E({\bf X}[n]{\bf X}[n-\tau]^T)
$$
\bigskip

[c] Calculate the Fisher information matrix for the parameters $\{A_{ij}\}$ and hence the corresponding Cramer-Rao lower bound.

[3] Let ${\bf X}[n]$ be a stationary $M$-vector valued zero mean Gaussian process with ACF 
${\bf R}(\tau)=\Bbb E({\bf X}[n]{\bf X}[n-\tau]^T)\in\Bbb R^{M\times M}$. Define its power spectral density matrix by
$$
{\bf S}(\omega)=\sum_{\tau=-\infty}^{\infty}{\bf R}(\tau)exp(-j\omega\tau)\in\Bbb C^{M\times M}
$$
Prove that ${\bf S}(\omega)$ is a positive definite matrix. Define its windowed periodogram estimate by
$$
\hat S_N(\omega)=\hat X_N(\omega)(\hat X_N(\omega))^*
$$
where
$$
\hat X_N(\omega)=N^{-1/2}\sum_{n=0}^{N-1}{\bf W}(n){\bf X}(n)exp(-j\omega n)
$$
where ${\bf W}(n)$ is an $M\times M$ matrix valued window function.
Calculate
$$
\Bbb E\hat S_N(\omega), Cov((S_N(\omega_1))_{ab},(S_N(\omega_2))_{cd})
$$
and discuss the limits of these as $N\rightarrow\infty$. Express these means and covariances in terms of 
$$
\hat W(\omega)=\sum_{\tau=-\infty}^{\infty}W(\tau)exp(-j\omega\tau)
$$
\bigskip

[4] This problem deals with causal Wiener filtering of vector valued stationary processes.
Let ${\bf X}(n),{\bf S}(n),n\in\Bbb Z$ be two jointly WSS processes with correlations
$$
{\bf R}_{XX}(\tau)=\Bbb E({\bf X}(n){\bf X}(n-\tau)^T), {\bf R}_{SX}(\tau)=\Bbb E({\bf S}(n){\bf X}(n-\tau)^T)
$$
Let
$$
{\bf S}_{XX}(z)=\sum_{\tau}{\bf R}_{XX}(\tau)z^{-\tau}
$$
Prove that
$$
{\bf R}_{XX}(-\tau)={\bf R}_{XX}(\tau)^T
$$
and hence
$$
{\bf S}_{XX}(z^{-1})={\bf S}_{XX}(z)^T
$$
Assume that this power spectral density matrix can be factorized as
$$
{\bf S}_{XX}(z)={\bf L}(z){\bf L}(z^{-1})^T
$$
where ${\bf L}(z)$ and ${\bf G}(z)={\bf L}(z)^{-1}$ admit expansions of the form
$$
{\bf L}(z)=\sum_{n\geq 0}{\bf l}[n]z^{-n}, |z|\geq 1
$$
$$
{\bf G}(z)=\sum_{n\geq 0}{\bf g}[n]z^{-n}, |z|\geq 1
$$
with
$$
\sum_{n\geq 0}\parallel{\bf l}[n]\parallel<\infty,
$$
$$
\sum_{n\geq 0}\parallel{\bf g}[n]\parallel<\infty
$$
Then suppose
$$
{\bf H}(z)=\sum_{n\geq 0}{\bf h}[n]z^{-n}
$$
is the optimum causal matrix Wiener filter that takes ${\bf X}[.]$ as input and outputs an estimate $\hat S[.]$ of ${\bf S}[.]$ in the sense that
$$
\hat S[n]=\sum_{k\geq 0}{\bf h}[k]{\bf X}[n-k]
$$
with $\{{\bf h}[k]\}$ chosen so that
$$
\Bbb E[\parallel{\bf S}[n]-\hat S[n]\parallel^2]
$$
is a minimum, then prove that ${\bf h}[.]$ satisfies the matrix Wiener-Hopf equation
$$
\sum_{k\geq 0}{\bf h}[k]{\bf R}_{XX}[m-k]={\bf R}_{SX}[m], m\geq 0
$$
Show that this equation is equivalent to
$$
[{\bf H}(z){\bf S}_{XX}(z)-{\bf S}_{SX}(z)]_+=0
$$
Prove that the solution to this equation is given by
$$
{\bf H}(z)=[{\bf S}_{SX}(z){\bf L}(z^{-1})^{-T}]_+{\bf L}(z)^{-1}
$$
$$
=[{\bf S}_{SX}(z){\bf G}(z^{-1})^T]_+{\bf G}(z)
$$
where
$$
[{\bf S}_{SX}(z){\bf G}(z^{-1})^T]_+
$$
$$
=[\sum {\bf R}_{SX}[n]{\bf g}[m]^Tz^{-n+m}]_+=
$$
$$
\sum_{n\geq 0}z^{-n}\sum_{m\geq 0}{\bf R}_{SX}[n+m]{\bf g}[m]^T
$$
Prove that this is a valid solution for a stable Causal matrix Wiener filter, it is sufficient that
$$
\sum_{n\geq 0}\parallel{\bf R}_{SX}[n]\parallel<\infty
$$
\bigskip

[5] Write short notes on any two of the following:

[a] Recursive least squares lattice filter for prediction with time and order updates of the prediction filter coefficients.

[b] Kolmogorov's formula for the entropy rate of a stationary Gaussian process in discrete time in terms of the power spectral density and its relationship to the infinite order one step prediction error variance.

[c] The Cq Shannon coding theorem and its converse and a derivation of the classical 
Shannon noisy coding theorem and its converse from it.

[d] Cramer-Rao lower bound for vector parameters and vector observations.

[e] The optimum detector in a binary quantum hypothesis testing problem between two mixed states.
\bigskip

\section{Monotonocity of quantum relative Renyi entropy}

Let $A,B$ be positive definite matrices and let $0\leq t\leq 1$, then prove that
$$
(tA+(1-t)B)^{-1}\leq t.A^{-1}+(1-t).B^{-1}---(1)
$$
ie, the map $A\rightarrow A^{-1}$ is operator convex on the space of positive matrices.

Solution: Proving (1) is equivalent to proving that
$$
A^{1/2}(tA+(1-t)B)^{-1}.A^{1/2}\leq tI+(1-t)A^{1/2}B^{-1}A^{1/2}
$$
which is the same as proving that
$$
(tI+(1-t)X)^{-1}\leq tI+(1-t)X^{-1}---(2)
$$
where
$$
X=A^{-1/2}BA^{-1/2}
$$
It is enough to prove (2) for any positive matrix $X$. By the spectral theorem for positive matrices, to prove (2), it is enough to prove
$$
(t+(1-t)x)^{-1}\leq t+(1-t)/x---(3)
$$
for all positive real numbers $x$. Proving (3) is equivalent to proving that
$$
(t+(1-t)x)(tx+1-t)\geq x
$$
for all $x>0$, or equivalently,
$$
(t^2+(1-t)^2)x+t(1-t)x^2+t(1-t)\geq x
$$
or equivalently, adding and subtracting $2t(1-t)x$ on both sides to complete the square,
$$
-2t(1-t)x+t(1-t)x^2+t(1-t)\geq 0
$$
or equivalently,
$$
1+x^2-2x\geq 0
$$
which is true. 
\bigskip

Problem: For $p\in(0,1)$, show that $x\rightarrow -x^p$, $x\rightarrow x^{p-1}$ and $x\rightarrow x^{p+1}$ are operator convex on the space of positive matrices. 

Solution: Consider the integral
$$
I(a,x)=\int_0^{\infty}u^{a-1}du/(u+x)
$$
It is clear that this integral is convergent when $x\geq 0$ and $0<a<1$. Now
$$
I(a,x)=x^{-1}\int_0^{\infty}u^{a-1}(1+u/x)^{-1}du
$$
$$
=x^{a-1}\int_0^{\infty}y^{a-1}(1+y)^{-1}dy
$$
on making the change of variables $u=xy$. Again make the change of variables
$$
1-1/(1+y)=y/(1+y)=z, y=1/(1-z)-1=z/(1-z)
$$
Then
$$
I(a,x)=x^{a-1}\int_0^1z^{a-1}(1-z)^{-a+1}(1-z)(1-z)^{-2}dz=x^{a-1}\int_0^1z^{a-1}(1-z)^{-a}dz
$$
$$
=x^{a-1}\beta(a,1-a)=x^{1-a}\Gamma(a)\Gamma(1-a)=x^{a-1}\pi/sin(\pi a)
$$
Thus,
$$
x^{a-1}=(sin(\pi a)/\pi)I(a,x)=(sin(\pi a)/\pi)\int_0^{\infty}u^{a-1}du/(u+x)
$$
and since for $u>0$, as we just saw, $x\rightarrow (u+x)^{-1}$ is operator convex, the proof that $f(x)=x^{a-1}$ is operator convex when $a\in(0,1)$. Now,
Now,
$$
u^{a-1}x/(u+x)=u^{a-1}(1-u/(u+x))=u^{a-1}-u^a/(u+x)
$$
Integrating from $u=0$ to $u=\infty$ and making use of the previous result gives
$$
x^a=(sin(\pi a)/\pi)\int_0^{\infty}(u^{a-1}-u^a/(u+x))du
$$
from which it immediately becomes clear that $x\rightarrow -x^a$ is operator convex on the space of positive matrices. Now consider $x^{a+1}$. We have
$$
u^{a-1}x^2/(u+x)=u^{a-1}((u+x)^2-u^2-2ux)/(u+x)=
$$
$$
u^{a-1}(u+x)-u^{a+1}/(u+x)-2u^ax/(u+x)=u^{a-1}(u+x)-u^{a+1}/(u+x)-2u^a(1-u/(u+x))
$$
and integration thus gives
$$
x^{a+1}=(sin(\pi a)/\pi)\int_0^{\infty}(u^{a-1}x+u^{a+1}/(u+x)-u^a)du
$$
from which it becomes immediately clear that $x^{a+1}$ is operator convex. 
\bigskip

Monotonicity of Renyi entropy: Let $s\in\Bbb R$. For two positive matrices $A,B$, consider the function
$$
D_s(A|B)=log(Tr(A^{1-s}B^s))
$$
Our aim is to show that for $s\in(0,1)$ and a TPCP map $K$, we have for two states $A,B$,
$$
D_s(K(A)|K(B))\geq D_s(A|B)
$$
while for $s\in(-1,0)$, we have
$$
D_s(K(A)|K(B))\leq D_s(A|B)
$$
For $a\in(0,1)$,  and $x>0$ we've seen that
$$
x^{a+1}=sin(\pi a)/\pi)\int_0^{\infty}(u^{a-1}x+u^{a+1}/(u+x)-u^a)du
$$
and hence for $s\in(-1,0)$, replacing $a$ by $-s$ gives
$$
x^{1-s}=-(sin(\pi s)/\pi)\int_0^{\infty}(u^{a-1}x+u^{a+1}/(u+x)-u^a)du
$$
Let $s\in(-1,0)$. Then the above formula shows that $x^{1-s}$ is operator convex.
Hence, if $A,B$ are any two density matrices and $K$ a TPCP map, then
$$
K(A)^{1-s}\leq K(A^{1-s})
$$
whence
$$
B^{s/2}K(A)^{1-s}B^{s/2}\leq B^{s/2}K(A^{1-s})B^{s/2}
$$
for any positive $B$ and taking taking trace gives
$$
Tr(K(A)^{1-s}B^s)\leq Tr(K(A^{1-s})B^s)
$$
Replacing $B$ by $K(B)$ here gives
$$
Tr(K(A)^{1-s}K(B)^s)\leq Tr(K(A^{1-s})K(B)^s)
$$
Further since $s\in (-1,0)$ it follows that $x^s$ is operator convex. Hence
$$
K(B)^s\leq K(B^s)
$$
and so
$$
A^{1/2}K(B)^sA^{1/2}\leq A^{1/2}K(B^s)A^{1/2}
$$
for any positive $A$ and replacing $A$ by $K(A^{1-s})$ and then taking trace gives
$$
Tr(K(A^{1-s})K(B)^s)\leq Tr(K(A^{1-s})K(B^s))
$$

More generally, we have the following:

If $X$ is any matrix, and $s\in(0,1)$ and $A,B$ positive, and $K$ any TPCP map
$$
Tr(X^*K(A)^{1-s}XK(B)^s)\geq Tr(X^*K(A)^{1-s}XK(B^s))
$$
(since $K(B)^s\geq K(B^s)$ because $B\rightarrow B^s$ is operator concave and since $K(A)^{1-s}X$ is positive)
$$
=Tr(K(A)^{1-s}XK(B^s)X^*)\geq Tr(K(A^{1-s})XK(B^s)X^*)
$$
(since $K(A)^{1-s}\geq K(A^{1-s})$ because $A\rightarrow A^{1-s}$ is operator concave and since $XK(B^s)X^*$ is positive)
$$
=Tr(X^*K(A^{1-s})XK(B^s))
$$
Since $X$ is an arbitary complex square matrix, this equation immediately can be written as
$$
K(A)^{1-s}\otimes K(B)^s\geq K(A^{1-s})\otimes K(B^s)
$$
for all positive $A,B$ and $s\in(0,1)$. We can express this equation as
$$
K(A)^{1-s}\otimes K(B)^s\geq (K\otimes K)(A^{1-s}\otimes B^s)
$$
and hence if $I$ denotes the identity matrix of the same size as $A,B$, then
$$
Tr(K(A)^{1-s}K(B)^s)\geq Tr((A^{1-s}\otimes B^s)(K\otimes K)(Vec(I).Vec(I)^*))
$$
Note that if $\{e_k:k=1,2,...,n\}$ is an onb for $\Bbb C^n$, the Hilbert space on which
$A,B$ act, then
$$
I=\sum_k|e_k><e_k|, Vec(I)=\sum_k|e_k\otimes e_k>=|\xi>
$$
$$
Vec(I).Vec(I)^*=|\xi><\xi|=\sum_{ij}E_{ij}\otimes E_{ij}
$$
where
$$
E_{ij}=|e_i><e_j|
$$
Thus, for $s\in(0,1)$,
$$
Tr(K(A)^{1-s}K(B)^s)\geq \sum_{i,j}Tr(K_{ij}A^{1-s}K_{ji}B^s)
$$
where
$$
K_{ij}=K(E_{ij})
$$
Likewise, if $s\in(0,1)$ so that $s-1\in(-1,0), 2-s\in(1,2)$, then both the maps
$A\rightarrow A^{s-1}$ and $A\rightarrow A^{2-s}$ are operator convex and hence for any
TPCP map $K$ , $K(A)^{s-1}\leq K(A)^{s-1}, K(A)^{2-s}\leq K(A^{2-s})$, from which we immediately deduce that
$$
Tr(X^*K(A)^{2-s}XK(B)^{s-1})\leq Tr(X^*K(A^{2-s})XK(B^{s-1}))
$$
or equivalently,
$$
K(A)^{2-s}\otimes K(B)^{s-1}\leq K(A^{2-s})\otimes K(B^{s-1})
$$
which result in the inequality
$$
Tr(K(A)^{2-s}K(B)^{s-1})\leq\sum_{i,j}Tr(K_{ij}A^{2-s}K_{ji}B^{s-1})
$$

Remark: Lieb's inequality states that for $s\in(0,1)$, the map
$$
(A,B)\rightarrow A^{1-t}\otimes B^t=f_t(A,B)
$$
is joint operator concave, ie, if $A_1,A_2,B_1,B_2$ are positive and $p,q>0,p+q=1$ then with
$$
A=pA_1+qA_2, B=pB_1+qB_2
$$
we have
$$
f_t(A,B)\geq pf_t(A_1,B_1)+qf_t(A_2,B_2)
$$
From this, we get the result that
$$
(A,B)\rightarrow Tr(X^*A^{1-t}XB^t)=H_t(A,B|X)
$$
is jointly concave. Now let $K$ denote the pinching
$$
K(A)=(A+UAU^*)/2
$$
where $U$ is a unitary matrix. Then, we get
$$
H_t(K(A),K(B)|X)\geq (1/2)H_t(A,B|X)+(1/2)H_t(UAU^*,UBU^*|X)
$$
and replacing $X$ by $I$ in this inequality gives
$$
Tr(K(A)^{1-t}K(B)^t)\geq Tr(A^{1-t}B^t)
$$
which proves the monotonicity of the Renyi entropy when $t\in(0,1)$ under pinching operations $K$ of the above kind. Now, this inequality will also hold for any pinching $K$ since any pinching can be expressed as a composition of such elementary pinchings of the above kind. We now also wish to show the reverse inequality when $t\in(-1,0)$, ie show that if $t\in(0,1)$, then
$$
Tr(K(A)^{2-t}K(B)^{t-1})\leq Tr(A^{2-t}B^{t-1})
$$
\newpage

\chapter{My commentary on the Ashtavakra Gita}

A couple of years ago, my colleague and friend presented me with a verse to verse translation of the celebrated Ashtavakra Gita in English along with a commentary on each verse. I was struck by the grandeur and the simplicity of the Sanskrit used by Ashtavakra for explaining to King Janaka, the true meaning of self-realization, how to attain it immediately without any difficulty and also King Janaka's reply to Ashtavakra about how he has attained immeasurable joy and peace by understanding this profound truth. Ashtavakra was a sage who lived during the time in which the incidents of the Raamayana took place and hence his Gita dates much much before even the Bhagavad Gita spoken by Lord Krisnna to Arjuna was written down by Sage Vyasa through Ganapati the scribe. One can see in fact many of the slokas in the Bhagavad Gita have been adapted from the Ashtavakra Gita and hence I feel that it is imperative for any seeker of the absolute truth and everlasting peace, joy and tranquility to read the Ashtavakra Gita, so relevant in our modern times where there are so many disturbances. Unlike the Bhagavad Gita, Ashtavkra does not bring in the concept of a personal God to explain true knowledge and the process of self-realization. He simply says that you attain self-realization immediately at this moment when you understand that the physical universe that you see around you is illusory and the fact that your soul is the entire universe. Every living being's soul is the same and it fills the entire universe and that all the physical events that take place within our universe as perceived by our senses are false, ie, they do not occur. In the language of quantum mechanics, these events occur only because of the presence of the observer, namely the living individuals with their senses. The absolute truth is that each individual is a soul and not a body with his/her senses and this soul neither acts nor causes to act. Knowledge is obtained immediately at the moment when we understand the truth that the impressions we have from our physical experiences are illusory, ie false and that the truth is that our soul fills the entire universe and that this soul is indestructible and imperishable as even the Bhagavad Gita states. The Bhagavad Gita however distinguishes between two kinds of souls, the jivatma and the paramatma, the former belonging to the living entities in our universe and the latter to the supreme personality of Godhead. The Bhagavad Gita tells us that within each individual entity, both the jivatma and the paramatma reside and when the individual becomes arrogant, he believes that his jivatma is superior to the paramatma and this then results in self conflict owing to which the individual suffers. By serving the paramatma with devotion as outlined in the chapter on Bhakti yoga, the jivatma is able to befriend the paramatma and then the individual lives in peace and tranquility. To explain self-knowledge more clearlh without bringing in the notion of a God, Ashtavakra tells Raja Janaka that there are two persons who are doing exactly the same kind of work every day both in their offices as well as in their family home. However, one of them is very happy and tranquil while the other is depressed and is always suffering. What is the reason for this ? The reason, Ashtavakra says is that the second person believes that his self is doing these works and consequently the reactions generated by his works affect him and make him either very happy or very sad. He is confused and does not know how to escape from this knot of physical experiences. The first person, on the other hand, knows very well that he is compelled to act in the way he is acting owing to effects of his past life's karma's  and that in fact his soul is pure and has nothing at all to do with the results and effects of his actions. He therefore neither rejoices nor does he get depressed. The physical experiences that he goes through do not leave any imprints on his self because he knows that the physical experiences are illusory and that his soul which pure and without any blemish fills the entire universe and is the truth. Ashtavarka further clarifies his point by saying that the person in ignorance will
practice meditation and try to cultivate detachment from the objects of this world thinking that this will give him peace. Momentarily such a person will get joy but when he comes of the meditation, immediately all the desires of this illusory physical world will come back to him causing even more misery. He will try to run to a secluded place like a forest and meditate to forget about his experiences, but again when his meditation is over, misery will come back to him. However, once he realizes that he is actually not his body but his soul and that the soul is free of any physical quality and that it is the immeasurable truth, then he will neither try forcefully to cultivate detachment nor will he try forcefully to acquire possessions. Both the extremes of attachment and detachment will not affect him. In short, he attains peace even living in this world immediately at this moment when true knowledge comes to him. According to Ashtavkara, the seeker of attachement as well as the seeker of detachment are both unhappy because they have not realized that the mind must not be forced to perform an activity. Knowledge is already present within the individual, he just has to remain in his worldly position as it is, performing his duties with the full knowledge that this compulsion is due to his past karmas and then simultaneously be conscious of the fact that his soul is the perfect Brahman and it pervades the entire universe. At one point, Ashtavakra tells Janaka that let even Hari (Vishnu) or Hara (Shiva) be your guru, even then you cannot be happy unless you have realized at this moment the falsity of the physical world around you and the reality of your soul as that which pervades the entire universe. Such a person who has understood this, will according to Ashtavakra, be unmoved if either if death in some form approaches him or if temptations of various kinds appear before him. He will remain neutral to both of these extremes. Such a realized soul has eliminated all the thought processes that occur within his mind. When true knowledge dawns upon a person, his mind will become empty and like a dry leaf in a strong wind, he will appear to be blown from one place to other doing different kinds of karma but always with the full understanding that these karmas have nothing at all to do with his soul which is ever at bliss, imperturbable, imperishable and omnisicient. In the quantum theory of matter and fields, we learn about how the observer perturbs a system in order to measure it and gain knowledge about the system. When the system is thus perturbed, one measures not the true system but only the perturbed system. This is the celebrated Heisenberg uncertainty principle. Likewise, in Ashtavakra's philosophy, if we attempt to gain true knowledge about the universe and its living entities from our sensory experience, we are in effect perturbing the universe and hence we would gain only perturbed knowledge, not true knowledge. True knowledge comes only with the instant realization that we are not these bodies, we are all permanent souls. Even from a scientific standpoint, as we grow older, all our body cells eventually get replaced by new ones, then what it is that in our new body after ten years that we are able to retain our identity ? What scientific theory of the DNA or RNA can explain this fact ? What is the difference between the chemical composition of our body just before it dies and just after it dies ? Absolutely no difference. Then what is it that disinguishes a dead body from a living body ? That something which distinguishes the two must be the soul which is not composed of any chemical. It must be an energy, the life force which our modern science is incapable of explaining. 
\newpage

\chapter{LDP-UKF based controller}

State equations:
$$
x(n+1)=f(x(n),u(n))+w(n+1)
$$
Measurement equations:
$$
z(n)=h(x(n),u(n))+v(n)
$$
$w,v$ are independent white Gaussian processes with mean zero and covariances $Q,R$ respectively. $x_d(n)$ is the desired process to be tracked and it satisfies
$$
x_d(n+1)=f(x_d(n),u(n))
$$
UKF equations: Let $\{\xi(k):1\leq k\leq K\}$ and $\{\eta(k):1\leq k\leq K\}$ be independent iid $N(0,I)$ random vectors. The UKF equations are
$$
\hat x(n+1|n)=K^{-1}\sum_{k=1}^Kf(\hat x(n|n)+\sqrt{P(n|n)}\xi(k),u(n))
$$
$$
\hat x(n+1|n+1)=\hat x(n+1|n)+K(n)(z(n+1)-\hat z(n+1|n))
$$
where
$$
K(n+1)=\Sigma_{XY}\Sigma_{YY}^{-1}
$$
Where
$$
\Sigma_{XY}=K^{-1}\sum_{k=1}^K[\sqrt{P(n+1|n)}\xi(k)(h(\hat x(n+1|n)+\sqrt{P(n+1|n)}\eta(k))-\hat z(n+1|n))^T],
$$
$$
\hat z(n+1|n)=K^{-1}\sum_{k=1}^Kh(\hat x(n+1|n)+\sqrt{P(n+1|n)}\eta(k),u(n+1))
$$
$$
\Sigma_{YY}=K^{-1}.\sum_{k=1}^K(h(\hat x(n+1|n)+\sqrt{P(n+1|n)}\eta(k))(,,)^T
$$
$$
P(n+1|n)=K^{-1}\sum_{k=1}^K[(f(\hat x(n|n)+\sqrt{P(n|n)}\xi(k),u(n))-\hat x(n+1|n))(,,)^T]+Q
$$
$$
P(n+1|n+1)=\Sigma_{XX}-\Sigma_{XY}.\Sigma_{YY}^{-1}\Sigma_{XY}^T
$$
where
$$
\Sigma_{XX}=P(n+1|n)
$$
Tracking error feedback controller: Let
$$
f(n)=x_d(n)-\hat x(n|n), f(n+1|n)=x_d(n+1)-\hat x(n+1|n)
$$
Controlled dynamics:
$$
x(n+1)=f(x(n),u(n))+w(n+1)+K_cf(n)
$$
Aim: To choose the controller coefficient $K_c$ so that the probability
$$
P(max_{1\leq n\leq N}(|e(n)|^2+|f(n)|^2)>\epsilon)
$$
is minimized where
$$
e(n)=x(n)-\hat x(n), e(n+1|n)=x(n+1)-\hat x(n+1|n), \hat x(n)=\hat x(n|n)
$$
Linearized difference equations for $e(n),f(n)$: (Approximate)
$$
e(n+1|n)=x(n+1)-\hat x(n+1|n)=f(x(n),u(n))+w(n+1)+K_cf(n+1|n)-K^{-1}.\sum_kf(\hat x(n)+\sqrt{P(n|n)}\xi(k),u(n))
$$
$$
=f(x(n),u(n))-f(\hat x(n),u(n))-f'(\hat x(n),u(n))\sqrt{P(n|n)}K^{-1}\sum_k\xi(k)+w(n+1)+K_cf(n)
$$
$$
=f'(\hat x(n),u(n))e(n)-f'(\hat x(n),u(n))\sqrt{P(n|n)}K^{-1}\sum_k\xi(k)+w(n+1)+K_cf(n)---(1)
$$
$$
f(n+1|n)=x_d(n+1)-\hat x(n+1|n)=f_(x_d(n),u(n))-K^{-1}.\sum_kf(\hat x(n)+\sqrt{P(n|n)}\xi(k),u(n))
$$
$$
=f'(\hat x(n),u(n))f(n)-f'(\hat x(n),u(n))\sqrt{P(n|n)}K^{-1}\sum_k\xi(k)---(2)
$$
$$
e(n+1)=x(n+1)-\hat x(n+1|n+1)=
$$
$$
e(n+1|n)+K_cf(n+1|n)-K(n)(z(n+1)-\hat z(n+1|n))
$$
$$
=e(n+1|n)+K_cf(n+1|n)-K(n)(h(x(n+1),u(n+1))+v(n+1)-K^{-1}\sum_kh(\hat x(n+1|n)+\sqrt{P(n+1|n)}\eta(k),u(n+1)))
$$
$$
=e(n+1|n)+K_cf(n+1|n)-K(n)(h'(\hat x(n+1|n),u(n+1))e(n+1|n)+v(n+1)
$$
$$
-h'(\hat x(n+1|n),u(n+1)).\sqrt{P(n+1|n)}.K^{-1}.\sum_k\eta(k))
$$
$$
=(I-K(n)h'(\hat x(n+1|n),u(n+1)))e(n+1|n)+K_cf(n)-K(n)v(n+1)
$$
$$
+K(n)h'(\hat x(n+1|n),u(n+1)).\sqrt{P(n+1|n)}K^{-1}\sum_k\eta(k)---(3)
$$
Finally,
$$
f(n+1)=x_d(n+1)-\hat x(n+1|n+1)=
$$
$$
f(n+1|n)-K(n)(z(n+1)-\hat z(n+1|n))
$$
$$
=f(n+1|n)-K(n)(h(x(n+1),u(n+1))+v(n+1)-K^{-1}\sum_kh(\hat x(n+1|n)+\sqrt{P(n+1|n)}\eta(k),u(n+1))
$$
$$
=f(n+1|n)-K(n)(h'(\hat x(n+1|n),u(n+1))e(n+1|n)
$$
$$
-h'(\hat x(n+1|n),u(n+1)).\sqrt{P(n+1|n)}K^{-1}\sum_k\eta(k))-K(n)v(n+1)
$$
$$
=f(n+1|n)-K(n)h'(\hat x(n+1|n),u(n+1)))e(n+1|n)
$$
$$
+K(n)h'(\hat x(n+1|n),u(n+1)).\sqrt{P(n+1|n)}.K^{-1}.\sum_k\eta(k)-K(n)v(n+1)---(4)
$$
Substituting for $e(n+1|n)$ and $f(n+1|n)$ from (1) and (2) respectively into (3) and (4) results in the following recursion for $e(n),f(n)$:
$$
e(n+1)=
$$
$$
f'(\hat x(n),u(n))f(n)-f'(\hat x(n),u(n))\sqrt{P(n|n)}K^{-1}\sum_k\xi(k)-K(n)h'(\hat x(n+1|n),u(n+1)))(f'(\hat x(n),u(n))e(n)
$$
$$
-f'(\hat x(n),u(n))\sqrt{P(n|n)}K^{-1}\sum_k\xi(k)+w(n+1)+K_cf(n))
$$
$$
+K(n)h'(\hat x(n+1|n),u(n+1)).\sqrt{P(n+1|n)}.K^{-1}.\sum_k\eta(k)-K(n)v(n+1)---(5)
$$
and
$$
f(n+1)=
$$
$$
f'(\hat x(n),u(n))f(n)-f'(\hat x(n),u(n))\sqrt{P(n|n)}K^{-1}\sum_k\xi(k)
$$
$$
-K(n)h'(\hat x(n+1|n),u(n+1))).
$$
$$
.(f'(\hat x(n),u(n))e(n)-f'(\hat x(n),u(n))\sqrt{P(n|n)}K^{-1}\sum_k\xi(k)+w(n+1)+K_cf(n))
$$
$$
+K(n)h'(\hat x(n+1|n),u(n+1)).\sqrt{P(n+1|n)}.K^{-1}.\sum_k\eta(k)-K(n)v(n+1)---(6)
$$
If we regard the error processes $e(n),f(n)$ and the noise processes w(n),v(n), $\xi_1(n)=K^{-1}.\sum_{k=1}^K\xi(k)$ and $\eta_1(n)=K^{-1}\sum_k\eta(k)$ to be all of the first order of smallness in perturbation theory, then to the same order, we can replace in
(4) and (5), $\hat x(n)$ appearing on the rhs by $x_d(n)$ and then the error thus incurred will be only of the second order of smallness. Doing so, we can express (4) and (5) in the form after equilibrium has been obtained as
$$
e(n+1)=A_{11}e(n)+(A_{12}+A_{18}K_c)f(n)+A_{13}w(n+1)+A_{14}A_{15}v(n+1)+A_{16}\xi_1(n+1)+A_{17}\eta_1(n+1)
$$
$$
f(n+1)=A_{21}e(n)+(A_{22}+A_{28}K_c)f(n)+A_{23}w(n+1)+A_{24}A_{25}v(n+1)+A_{26}\xi_1(n+1)+A_{27}\eta_1(n+1)
$$
Define the aggregate noise process
$$
W[n]=[w(n)^T,v(n)^T,\xi_1(n)^T,\eta_1(n)^T]^T
$$
It is a zero mean vector valued white Gaussian process and its covariance matrix is block diagonal:
$$
\Bbb E(W(n)W(m)^T)=R_W(n)\delta(n-m)
$$
where
$$
R_W(n)=\Bbb E(W(n)W(n)^T)=diag[Q,R,P[n|n)/K, P(n+1|n)/K]
$$
Let $P$ denote the limiting equilibrium value of $P[n|n]$ as $n\rightarrow\infty$.
The limiting equilibrium value of $P[n+1|n]$ is then given approximately by
$$
P_+=f'(\hat x,u)Pf'(\hat x,u)^T+Q
$$
where $\hat x$ may be replaced by the limiting value of $x_d$, assuming it to be asymptotically d.c and $f'(x,u)$ is the Jacobian matrix of $f(x,u)$ w.r.t $x$. We write
$f'(x_d,u)=F$ and then
$$
P_+=FPF^T+Q
$$
If we desire to use a more accurate value of $P_+$, then we must replace it by
$$
P_+=Cov(f(x_d+P.\xi_1))+Q
$$
where $\xi_1$ is an $N(0,K^{-1}P)$ random vector. Now, let $R_W$ denote the limiting equilibrium value of $R_W(n)$, ie,
$$
R_W=diag[Q,R,P/K, P_+/K]
$$
Then, the above equations for 
$$
\chi[n]=[e[n]^T,f[n]^T]^T
$$ 
can be expressed as
$$
\chi[n+1]=(G_0+G_1K_cG_2)\chi[n]+G_3W[n+1]
$$
and its solution is given by
$$
\chi[n]=\sum_{k=0}^{n-1}(G_0+G_1K_cG_2)^{n-1-k}G_3W[k+1]
$$
This process is zero mean Gaussian with an autocorrelation given by
$$
\Bbb E(\chi[n]\chi[m]^T)=\sum_{k=0}^{min(n,m)-1}(G_0+G_1K_cG_2)^{n-1-k}.R_W.(G_0+G_1K_cG_2)^{T m-1-k}
$$
$$
=R_{\chi}[n,m|K_c]
$$
say. The LDP rate function of the process $\chi[.]$ over the time interval $[0,N]$ is then given by
$$
I[\chi[n]:0\leq n\leq N]=(1/2)\sum_{n,m=0}^N\chi[n]^TQ_{\chi}[n,m|K_c]\chi[m]
$$
where
$$
Q_{\chi}=((R_{\chi}[n,m|K_c]))_{0\leq n,m\leq N}^{-1}
$$
According to the LDP for Gaussian processes, the probability that $\sum_{n=0}^N\parallel\chi[n]\parallel^2$ will exceed a threshold $\delta$ after we scale $\chi$ by a small number $\sqrt{\epsilon}|$ is given approximately by
$$
exp(-\epsilon^{-1}min(I(\chi):\sum_{n=0}^N\parallel\chi[n]\parallel^2>\delta))
$$
and in order to minimize this deviation probability, we must choose $K_c$ so that
the maximum eigenvalue $\lambda_{max}(K_c)$ of $((R_{\chi}[n,m|K_c]))_{0\leq n,m\leq N}^{-1}$ is a minimum.
\bigskip

\chapter{Abstracts for a book on quantum signal processing using quantum field theory}

{\bf Chapter 1}

An introduction to quantum electrodynamics

[1] The Maxwell and Dirac equations.

[2] Quantization of the electromagnetic field using Bosonic creation and annihilation operators.

[3] Second Quantization of the Dirac field using Fermionic creation and annihilation operators.

[4] The interaction term between the Dirac field and the Maxwell field.

[5] Propagators for the photons and electron-positron fields.

[7] The Dyson series and its application to calculation of amplitudes for absorption, emission and scattering processes between electrons and positrons.

[8] Representing the terms in the Dyson series for interactions between photons, electrons and positrons using Feynman diagrams.

[9] The role of propagators in Feynman diagrammatic calculations.

[10] Quantum electrodynamics using the Feynman path integral method.

[11] Calculating the propagator of fields using Feynman path integrals.

[12] Electrons self energy, vacuum polarization, Compton scattering and anomalous magnetic moment of the electron.
\bigskip

{\bf Chapter 2}

{The design of quantum gates using quantum field theory}

[1] Design of quantum gates using first quantized quantum mechanics by controlled modulation of potentials.

[2] Design of quantum gates using harmonic oscillators with charges perturbed by and electric field.

[3] Design of quantum gates using 3-D harmonic oscillators perturbed by control electric and magnetic field.

[4] Representing the Klein-Gordon Hamiltonian with perturbation using truncated spatial Fourier modes within a box.

[5] Design of the perturbation potential in Klein-Gordon quantization for realizing very large sized quantum gates.

[6] Representing the Hamiltonian of the Dirac field plus electromagnetic field in terms of photon and electron-positron creation and annihilation operators.

[7] Design of control classical electromagnetic fields and control classical current sources in quantum electrodynamics for approximating very large sized quantum unitary gates.

[8] The statistical performance of quantum gate design methods in the presence of classical noise.
\bigskip

{\bf Chapter 3}

The quantum Boltzmann equation with applications to the design of quantum noisy gates, ie,
TPCP channels.

[1] The classical Boltzmann equation for a plasma of charged particles interacting with an external electromagnetic field.

[2] Calculating the conductivity and permeability of a plasma interacting with an electromagnetic field using first order perturbation theory.

[3] The quantum Boltzmann equation for the one particle density operator for a system of $N$ identically charged particles interacting with a classical electromagnetic field.
The effect of quantum charge and current densities on the dynamics of the Maxwell field and the effect of the Maxwell field on the dynamics of the one particle density operator.

[4] The quantum nonlinear channel generated by the evolving the one particle density operator under an external field in the quantum Boltzmann equation.
\bigskip

{\bf Chapter 4}

The Hudson-Parthasarathy quantum stochastic calculus with application to quantum gate and quantum TPCP channel design.

[1] Boson and Fermion Fock spaces.

[2] Creation, annihilation and conservation operator fields in Boson and Fermion Fock spaces.

[3] Creation, annihilation and conservation processes in Boson Fock space.

[4] The Hudson-Parthasarathy quantum Ito formulae.

[5] Hudson-Parthasarathy-Schrodinger equation for describing the evolution of a quantum system in the presence of bath noise.

[6] The GKSL (Lindlbad) master equation for open quantum systems.

[7] Design of the Lindlbad operators for an open quantum system for generation of a given 
TPCP map, ie, design of noisy quantum channels.

[8] Schrodinger's equation with Lindlbad noisy operators in the presence of an external electromagnetic field. Design of the electromagnetic field so that the evolution matches a given TPCP map in the least squares sense.

[9] Quantum Boltzmann equation in the presence of an external electromagnetic field with Lindblad noise terms. Design of the Lindblad operators and the electromagnetic field so that the evolution matches a given nonlinear TPCP map.
\bigskip

{\bf Chapter 5}

Non-Abelian matter and gauge fields applied to the design of quantum gates.

[1] Lie algebra of a Lie group.

[2] Gauge covariant derivatives.

[3] The non-Abelian matter and gauge field Lagrangian that is invariant under local gauge transformations.

[4] Application of non-Abelian matter and gauge field theory to the design of large sized quantum gates in the first quantized formalism, gauge fields are assumed to be classical and the control fields and current are also assumed to be classical.
\bigskip

{\bf Chapter 6}

The performance of quantum gates in the presence of quantum noise.
\bigskip

Quantum plasma physics: For an $N$ particle system with identitcal particles
$$
i\rho_{,t}=[\sum_kH_k,\rho]+[\sum_{i<j}V_{ij},\rho]+Lindlbad noise terms.
$$
where
$$
H_k=-(\nabla_k-ieA(t,r_k))^2/2m+e\Phi(t,r_k)
$$
$$
=H_{0k}+V_k
$$
where
$$
H_{0k}=-\nabla_k^2/2m, V_k=(ie/2m)(2.div A_k+(A_k,\nabla_k))+e^2A_k^2/2m+e\Phi_k
$$
Approximate one particle Boltzmann equation obtained by partial tracing over the remaining $N$ particles.
$$
i\rho_{1,t}=[H_1,\rho_1]+(N-1)Tr_2[V_{12},\rho_{12}]+Lindblad noise terms
$$
$$
i\rho_{12,t}=[H_1+H_2+V_{12},\rho_{12}]+(N-2)Tr_3[V_{13}+V_{23},\rho_{123}]+Lindblad noise terms
$$
Writing
$$
\rho_{12}=\rho_1\otimes\rho_1+g_{12}
$$
we get from the above two equations,
$$
i(\rho_{1,t}\otimes\rho_1+\rho_1\otimes\rho_{1,t})+ig_{12,t}=
$$
$$
[H_1,\rho_1]\otimes\rho_1+\rho_1\otimes[H_1,\rho_1]+ig_{12,t}=
$$
$$
[H_1,\rho_1]\otimes\rho_1+\rho_1\otimes[H_1,\rho_1]+[H_1+H_2+V_{12},g_{12}]
$$
$$
+Lindblad noise
$$
Here, we have neglected the cubic terms, ie, the terms involving $\rho_{123}$. After making the appropriate cancellations, we get
$$
ig_{12,t}=[H_1+H_2+V_{12},g_{12}]+Lindblad noise
$$
Thus we get for the one particle density,
$$
i\rho_{1,t}=[H_1,\rho_1]+(N-1)Tr_2[V_{12},\rho_1\otimes\rho_1]+(N-1)Tr_2[V_{12},g_{12}]
$$
$$
+Lindblad noise
$$
We can express this approximate equation as
$$
i\rho_{1,t}=[H_1,\rho_1]+(N-1)Tr_2[V_{12},\rho_1\otimes\rho_1]
$$
$$
+(N-1)Tr_2[V_{12},exp(-itad(H_1+H_2+V_{12}))(g_{12}(0))]
$$
Since the nonlinear term is small, we can approximate the $\rho_1(t)$ occurring in it by
$exp(-itad(H_1))(\rho_1(0))$. Thus,
$$
i\rho_{1,t}(t)=[H_1,\rho_1(t)]+(N-1)Tr_2[V_{12},exp(-it.ad(H_1+H_2))(\rho_1(0)\otimes\rho_1(0))]
$$
$$
+(N-1)Tr_2[V_{12},exp(-itad(H_1+H_2+V_{12}))(g_{12}(0))]+\theta(\rho_1(t))
$$
where $\theta(\rho_1)$ is the Lindblad noise term. The solution is
$$
\rho_1(t)=exp(-itad(H_1))(\rho_1(0))+
$$
$$
(N-1)\int_0^texp(-i(t-s)ad(H_1))(Tr_2[V_{12},exp(-is.ad(H_1+H_2))(\rho_1(0)\otimes\rho_1(0))])ds
$$
$$
+(N-1)\int_0^texp(-i(t-s)ad(H_1))(Tr_2[V_{12},exp(-isad(H_1+H_2+V_{12}))(g_{12}(0))])ds
$$
$$
+Lindblad noise terms.
$$
The first term is the initial condition term, the second term is the collision term and the third term is the scattering term. Note that taking the trace gives us
$$
i\partial_t(Tr(\rho_1(t))=0
$$
which means that the above quantum Boltzmann equation as it evolves the one particle state, determines a nonlinear trace preserving mapping. This suggests that the quantum Boltzmann equation can be used to design nonlinear TPCP maps by controlling the electromagnetic potentials $A,\Phi$ as well as the Lindblad terms.

Now let $\rho_1(t,r,r')$ denote the position space representation of the one particle density operator. Then the charge density corresponding to this operator is given by
$e\rho_1(t,r,r)$ and the current density is
$$
J(t,r)=(e/m)\nabla_{r'}Im(\rho(t,r,r'))|_{r'=r}+(e/m)A(t,r)\rho_1(t,r,r)
$$
This is based on the fact that if $\psi(t,r)$ denotes a pure state in the position representation, ie, a wave function, then the mixed state is an average of
$\psi(t,r)\psi(t,r')^*$ over ensembles of such pure states, ie,
$$
\rho_1(t,r,r')=<\psi(t,r)\psi(t,r')^*>
$$
In particular, the charge density is
$$
e<|\psi(t,r)|^2>=e\rho_1(t,r,r)
$$
Likewise the current density for a pure state is
$$
J(t,r)=(ie/2m)(\psi(t,r)^*(\nabla-ieA(t,r))\psi(t,r)-\psi(t,r)(\nabla+ieA(t,r))\psi(t,r)^*)
$$
$$
=(e/m)A(t,r)|\psi(t,r)|^2-(e/m)Im(\psi(t,r)^*\nabla\psi(t,r))
$$
which on taking average gives
$$
J(t,r)=(e/m)A(t,r)\rho_1(t,r,r)-(e/m)\nabla_{r'}Im(\rho_1(t,r',r)|_{r'=r})
$$
$$
=(e/m)A(t,r)\rho_1(t,r,r)+(e/m)\nabla_{r'}Im(\rho_1(t,'r,r)|_{r'=r})
$$
The classical em field then satisfies the Maxwell equations
$$
\square A(t,r)=\mu_0J(t,r)=\mu_0((e/m)A(t,r)\rho_1(t,r,r)-(e/m)\nabla_{r'}Im(\rho_1(t,r',r)|_{r'=r}))
$$
$$
\square\Phi(t,r)=\rho(t,r)/\epsilon_0=\epsilon_0^{-1}(e\rho_1(t,r,r))
$$
where
$$
\square=(c^{-2}\partial_t^2-\nabla^2)
$$
is the wave operator. The complete system of the quantum Boltzmann or rather quantum Vlasov (quantum for the density operator and classical for the em field) is therefore summarized below:
$$
H_1=-(\nabla_1+ieA(t,r_1))^2/2m+e\Phi(t,r_1)
$$
$$
V_{12}=V(|r_1-r_2|)
$$
$$
\rho_1(t)=exp(-itad(H_1))(\rho_1(0))+
$$
$$
(N-1)\int_0^texp(-i(t-s)ad(H_1))(Tr_2[V_{12},exp(-is.ad(H_1+H_2))(\rho_1(0)\otimes\rho_1(0))])ds
$$
$$
+(N-1)\int_0^texp(-i(t-s)ad(H_1))(Tr_2[V_{12},exp(-isad(H_1+H_2+V_{12}))(g_{12}(0))])ds
$$
$$
+Lindblad noise terms.
$$
$$
\square A(t,r)=\mu_0((e/m)A(t,r)\rho_1(t,r,r)-(e/m)\nabla_{r'}Im(\rho_1(t,r',r)|_{r'=r}))
$$
$$
\square\Phi(t,r)=\epsilon_0^{-1}(e\rho_1(t,r,r))
$$
\bigskip

Second quantization of the quantum plasma equations.

Consider the Dirac field of electrons and positrons. The wave function $\psi(t,r)$ is a Fermionic field operator satisfying the Dirac equation
$$
[\gamma^{\mu}(i\partial_{\mu}+eA_{\mu})-m]\psi(t,r)=0
$$
The unperturbed field $\psi_0(t,r)$, ie, when $A_{\mu}=0$ can be expanded as a superposition of annihilation and creation operator fields in momentum space of electrons and positrons. This is the case when we second quantize the motion of a single electron-positron pair. When however we second quantize the motion of $N$ identical electron-positron pairs and then form the marginal density operator of a single electron-positron pair, we obtain the quantum Boltzmann equation for the one particle density operator field $\rho_1(t,r',r')$ which in the absence of mutual interactions $V(r_1-r_2)$, is simply $\psi_0(t,r)\psi_0(t,r')^*$ which is expressible as a quadratic combination of Fermionic creation and annihilation operator fields in momentum space. The single particle Dirac Hamiltonian is
$$
H_D=(\alpha,-i\nabla+eA)+\beta m-e\Phi
$$
which in the absence of electromagnetic interactions reads
$$
H_{D0}=(\alpha,-i\nabla)+\beta m
$$
The second quantized density operator field in position space representation that corresponds to this unperturbed Dirac Hamiltonian is given by
$$
\rho_0(t,r,r')=\psi_0(t,r)\psi_0(t,r')
$$
where
$$
\psi_0(t,r)=\int[a(P,\sigma)u(P,\sigma)exp(-i(E(P)t-P.r))+b(P,\sigma)^*v(P,\sigma)exp(i(E(P)t-P.r))]d^3P
$$
with $u(P,\sigma)$ and $v(-P,\sigma)$ being eigenvectors for $H_{D0}(P)=(\alpha,P)+\beta m$ with eigenvalues $\pm E(P)$ respectively where $E(P)=\sqrt{P^2+m^2}$. Thus, we can write after discretization in the momentum domain,
$$
\rho_{100}(t,r,r')=\sum_{k,m}[a(k)a(m)f_1(t,r,r'|k,m)+a(k)a(m)^*f_2(t,r,r')]+H.c
$$
where $a(k),a(k)^*$ are respectively Fermionic annihilation and creation operators satisfying the canonical anticommutation relations
$$
[a(k),a(m)^*]_+=\delta(k,m), [a(k),a(m)]_+=0=[a(k)^*,a(m)^*]_+
$$
After interaction with the quantum em field, the density operator $\rho_{10}(t,r,r')$ satisfies the equation
$$
i\partial_t\rho_{10}(t)=[H_{D0}+H_{int}(t),\rho_{10}(t)]
$$
where
$$
H_{int}(t)=e(\alpha,A(t,r))-e\Phi(t,r)
$$
where $A(t,r), \Phi(t,r)$ are now both Bosonic quantum operator fields expressible upto zeroth order approximations as linear combinations of Bosonic annihilation and creation operators $c(k),c(k)^*, k=1,2,...$ that satisfy the canonical commutation relations:
$$
[c(k),c(m)^*]=\delta(k,m), [c(k),c(m)]=[c(k)^*,c(m)^*]=0
$$
We have upto first order perturbation theory,
$$
\rho_{10}=\rho_{100}+\rho_{101}
$$
where
$$
i\partial_t\rho_{101}(t)=[H_{D0},\rho_{101}(t)]+[H_{int}(t),\rho_{100}(t)]
$$
This has the solution
$$
\rho_{101}(t)=\int_0^texp(-i(t-s)ad(H_{D0})([H_{int}(s),\rho_{100}(s)])ds
$$
It is clear that this expression for $\rho_{101}(t)$ is a trilinear combination of the photonic creation and annihilation operators and the electron-positron creation and annihilation operators such that it is linear in the former and quadratic in the latter.
Thus, we can write approximately
$$
\rho_{101}(t,r,r')=\sum_{klm}[c(k)a(l)a(m)g_1(t,r,r'|klm)+c(k)^*a(l)a(m)g_2(t,r,r'|klm)+c(k)a(l)^*a(m)g_3(t,r,r'|klm)+c(k)^*a(l)^*a(m)g_4(t,r,r'|klm)]+H.c
$$
After taking into account self interactions of the plasma of electrons and positrons, the one particle density operator field $\rho_1(t,r,r')$ satisfies the quantum Boltzmann equation
$$
i\rho_{1,t}(t,r_1,r_1')=\int(H_D(t,r_1,r_1'')\rho_1(t,r'',r')-\rho_1(t,r,r'')H_D(t,r'',r'))d^3r''
$$
$$
+(N-1)\rho_{10}(t,r_1,r_1')\int(V(r_1,r_2)-V(r_1`',r_2))\rho_{10}(t,r_2,r_2)d^3r_2
$$
$$
+(N-1)\int(V(r_1,r_2)-V(r_1',r_2))g_{12}(t,r_1,r_2|r_1',r_2)d^3r_2---(a)
$$
where
$$
g_{12}(t,r_1,r_2|r_1',r_2')
$$
Note that in perturbation theory,
$$
\rho_1=\rho_{10}+\rho_{11}
$$
where $\rho_{10}$ as calculated above, satisfies (a) with $V=0$.
\bigskip

\chapter{Quantum neural networks and learning using mixed state dynamics for open quantum systems}

Let $\rho(t)$ be the state of an evolving quantum system according to the dynamics
$$
i\rho'(t)=[H_0+V(t),\rho(t)]+\theta(\rho(t))
$$
where $V(t)$ is the control potential and $\theta(\rho)$ is the Lindblad noise term. Our aim is to design this control potential $V(t)$ so that the probability density
$\rho(t,r,r)$ in the position representation tracks a given probability density function
$p(t,r)$ and the quantum current $(e/m)\nabla_{r'}Im(\rho(t,r,r')|_{r'=r})$ tracks a given current density $J(t,r)$. By analogy with pure state dynamics and fluid dynamical systems, the corresponding problem is formulated as follows: Consider the wave function $\psi(t,r)$
that satisfies Schrodinger's equation for a charged particle in an electromagnetic field:
$$
i\partial_t\psi(t,r)=(-(\nabla+ieA(t,r))^2/2-e\Phi(t,r)+W(t,r)V(t,r))\psi(t,r)
$$
The aim is to control the neural weight $W(t,r)$, so that $p_q(t,r)=|\psi(t,r)|^2$ tracks the fluid density $\rho(t,r)$ that is normalized so that $\int\rho(t,r)d^3r=1$ and further that the probability current density 
$$
J_q(t,r)=Im(\psi(t,r)\nabla\psi(t,r)^*)+A(t,r)|\psi(t,r)|^2
$$
tracks a given current density $J(t,r)$ of the fluid. Note that we are assuming that $p, J$ satisfy the equation of continuity:
$$
\partial_t\rho+div(J)=0
$$
This is an effective QNN algorithm for simulating the motion of a fluid because the corresponding quantum mechanical probability and current density pairs $(p_q,J_q)$ automatically by virtue of Schrodinger's equation satisfy the equation of continuity and simultaneously guarantee that $p_q$ is a probability density, ie, it is non-negative and integrates over space to unity. Thus rather than using a QNN for tracking only the density of a fluid or a probability density, we can in addition use it to track also the fluid current density by adjusting the control weights $W(t,r)$. Note that for open quantum systems, ie, in the presence of Lindblad noise terms, the equation of continuity for the QNN does not hold. This means that when bath noise interacts with our QNN, we would not get an accurate simulation of the fluid mass and current density. However, if there are sources of mass generation in the fluid so that the equation of continuity reads
$$
\partial_t\rho(t,r)+div J(t,r)=g(t,r)
$$
then we can use the Lindblad operator coefficient terms as additional control parameters to simulate $\rho, J$ for such a system with mass generation $g>0$ or mass destruction
$g<0$.
$$
|J(t,r)-J_q(t,r)|^2=|Im(\psi(t,r)\nabla\psi(t,r)^*)+A(t,r)|\psi(t,r)|^2-J(t,r)|^2
$$
is the space-time instantaneous norm error square between the true fluid current density and that predicted by the QNN. Its gradient w.r.t $A(t,r)$ is
$$
\delta|J(t,r)-J_q(t,r)|^2/\delta A(t,r)=
$$
$$
2|\psi(t,r)|^2(J(t,r)-J_q(t,r))
$$
Note that this is a vector. This means that for tracking the current density, the update equation for the control vector potential must be
$$
\partial_tA(t,r)=-\mu\delta|J(t,r)-J_q(t,r)|^2/\delta A(t,r)=
$$
$$
-2\mu|\psi(t,r)|^2(J(t,r)-J_q(t,r))
$$
We can also derive a corresponding update equation for the scalar potential in Schrodinger's equation based on matching the true mass density of the fluid to that predicted by the QNN as follows.
$$
|p(t,r)-p_q(t,r)|^2=|p(t,r)-|\psi(t,r)|^2|^2
$$
so
$$
\partial_t\Phi(t,r)=-\mu\delta|p(t,r)-|\psi(t,r)|^2|^2/\delta\Phi(t,r)=
$$
$$
2\mu[\bar\psi(t,r)\delta\psi(t,r)/\delta\Phi(t,r)+\psi(t,r)\delta\bar\psi(t,r)/\delta\Phi(t,r)][p(t,r)-|\psi(t,r)|^2]
$$
Now using the Dyson series expansion,
$$
\psi(t,r)=U_0(t)\psi(0,r)+\sum_{n\geq 1}(-i)^n\int_{0<t_n<...<t_1<t}U_0(t-t_1)V(t_1)U_0(t_1-t_2)....U_0(t_{n-1}-t_n)\psi(0,r)dt_1...dt_{n-1}
$$
giving
$$
\delta\psi(t,r)/\delta V(t,r)=V(t,r)\sum_{n\geq 1}(-i)^n\int_{0<t_n<...<t_2<t}U_0(t-t_2)V(t_2)U_0(t_2-t_3)...U_0(t_{n-1}-t_n)\psi(0,r)dt_2...dt_n
$$
$$
=-iV(t,r)(\psi(t,r)-\psi(0,r))
$$
Taking the conjugate of this equation gives us
$$
\delta\bar\psi(t,r)/\delta V(t,r)=iV(t,r)(\bar\psi(t,r)-\bar\psi(0,r))
$$
Thus the update equation for $\Phi(t,r)=V(t,r)/e$ becomes
$$
\partial_t\Phi(t,r)=2\mu.\Phi(t,r)Im(\psi(t,r)-\psi(0,r)).(p(t,r)-|\psi(t,r)|^2)
$$
Note that here, we can more generally consider the error energy to be a weighted linear combination of the error energies between the fluid mass density and its QNN estimate and that between the fluid current density and its QNN estimate:
$$
E=w_1|p(t,r)-|\psi(t,r)|^2|^2+w_2|J(t,r)-Im(\psi(t,r)\nabla\psi(t,r)^*)+A(t,r)|\psi(t,r)|^2|^2
$$
while calculating the updates for the vector and scalar potentials using this energy function, we must take into account the dependence of $\psi(t,r)$ and $\bar\psi(t,r)$ on both $A(t,r)$ and $\Phi(t,r)$. For this we write the Hamiltonian as
$$
H(t)=H_0+eV_1(t)+e^2V_2(t)
$$
where
$$
H_0=-\nabla^2/2, V_1(t)=\Phi(t,r)+(1/2)(div A(t,r)+2(A(t,r),\nabla)), V_2(t)=A(t,r)^2/2
$$
Then in terms of variational derivatives, from the Dyson series expansion,
$$
\delta\psi(t,r)/\delta A(t,r)=e((i/2)\nabla\psi(t,r)-i\nabla\psi(t,r))
$$
$$
=(-i/2)\nabla\psi(t,r)
$$
and taking conjugate
$$
\delta\bar\psi(t,r)/\delta A(t,r)=(i/2)\nabla\bar\psi(t,r)
$$
so we get
$$
\delta|\psi(t,r)|^2/\delta A(t,r)=
$$
$$
(-i/2)(\bar\psi(t,r)\nabla\psi(t,r)-\psi(t,r)\nabla\bar\psi(t,r))
$$
$$
=Im(\bar\psi(t,r)\nabla\psi(t,r))
$$
\bigskip

Exercise: Using the above formulas, evaluate
$$
\delta E/\delta A(t,r), \delta E/\delta\Phi(t,r)
$$

QNN in the presence of Hudson-Parthasarathy noise:

Let $|\psi(t)>$ be a pure state in the Hilbert space $\mathfrak h\otimes\Gamma_s(L^2(\Bbb R_+))$ evolving according to the Hudson-Parthasarathy noisy Schrodinger equation where
$\mathfrak h=L^2(\Bbb R_+)$:
$$
d|\psi(t)>=(-(iH+LL^*/2)dt+LdA-L^*dA^*)|\psi(t)>
$$
where $H,L$ are system operators. Specifically, choose
$$
H=-\nabla^2/2+V(t,r), L=(a,\nabla)+(b,r)
$$
where $a,b$ are constant complex 3-vectors. For a specific representation of the creation and annihilation processes $A(t)^*,A(t)$, we choose an onb $|e_n>$ for $L^2(\Bbb R_+)$ and then write
$$
A(t)=\sum_na_n<\chi_{[0,t]}>|e_n>, A(t)^*=\sum_na_n^*<e_n|\chi_{[0,t]}>
$$
where $a_n$ are operators in a Hilbert space satisfying the CCR
$$
[a_n,a_m]=0, [a_n,a_m^*]=\delta[n-m], [a_n^*,a_m^*]=0
$$
Then an easy calculation shows that
$$
[A(t),A(s)^*]=\sum_n<\chi_{[0,t]}|e_n><e_n|\chi_{[0,s]}>=<\chi_{[0,t]}|\chi_{[0,s]}>=min(t,s)
$$
from which we easily deduce the quantum Ito formula of Hudson and Parthasarathy. We can also introduce a conservation process $\Lambda(t)$
$$
\Lambda(t)=\sum_{n,m}a_n^*a_m<e_n|\chi_{[0,t]}|e_m>
$$
and deduce that
$$
[A(t),\Lambda(s)]=\sum_{n,m}a_m<\chi_{[0,t]}|e_n><e_n|\chi_{[0,s]}|e_m>
$$
$$
=\sum_ma_m<\chi_{[0,min(t,s)]}|e_m>=A(min(t,s))
$$
and taking the adjoint of this equation then gives us
$$
[\Lambda(s),A(t)^*]=A(min(t,s))^*
$$
These equations yield the other quantum Ito formulas involving the products of the creation and annihilation operator differentials with the conservation operator differentials. Now suppose that we assume that our wave function on the tensor product of the system and bath space has the form
$$
|\psi(t)>=|\psi_s(t)>\otimes|\phi(u)>
$$
where $|\phi(u)>$ is the bath coherent state and $|\psi_s(t)>$ is the system state. Then
substituting this into the Hudson-Parthasarathy noisy Schrodinger equation gives us
using
$$
dA(t)|\phi(u)>=dt.u(t)|\phi(u)>
$$
$$
dA(t)^*|\phi(u)>=(dB(t)-u(t)dt)|\phi(u)>
$$
where
$$
B(t)=A(t)+A(t)^*
$$
is a classical Brownian motion process and since the bath is in the coherent state, it is should be regarded as a Brownian motion with drift. Thus, the HPS equation assumes the form for such a special class of pure states on system and bath space assumes the form of a stochastic Schrodinger equatiion
$$
i\partial_t|\psi_s(t)>=(-(iH+LL^*/2)dt+u(t)dtL-(dB(t)-u(t)dt)L^*)|\psi_s(t)>
$$
$$
=(((-iH-LL^*/2+u(t)(L+L^*))dt-L^*dB(t))|\psi_s(t)>
$$
This stochastic Schrodinger equation is parametrized by the coherent state parameter function $u\in L^2(\Bbb R_+)$. We can now use this stochastic Schrodinger equation to track a given probability density $p(t,r)$ using control on all $H,L,u(t)$. This would thus define a QNN in the Hudson-Parthasarathy-Stochastic-Schrodinger framework. The idea is that Brownian noise is undesired but one cannot help it enter into the picture and hence one must track the given pdf in spite of it being present.
\bigskip

Using quantum electrodynamics (QED) to track joint probability densities in a QNN.

The Lagrangian of (QED) is given by the sum of a Maxwell term, a Dirac term and an interaction term:
$$
L=L_M+L_D+L_{int}=
$$
$$
(-1/4)F_{\mu\nu}F^{\mu\nu}+\bar\psi(\gamma^{\mu}(i\partial_{\mu}+eA_{\mu})-m)\psi
$$
$$
=(-1/4)F_{\mu\nu}F^{\mu\nu}+\bar\psi(i\gamma^{\mu}(\partial_{\mu}-m)\psi+e\bar\psi\gamma^{\mu}\psi.A_{\mu}
$$
Thus we identify
$$
L_M=(-1/4)F_{\mu\nu}F^{\mu\nu}, L_D=\bar\psi(i\gamma^{\mu}\partial_{\mu}-m)\psi,
$$
$$
L_{int}=e\bar\psi\gamma^{\mu}\psi.A_{\mu}
$$
To proceed further, we write down the Schrodinger equation corresponding to this second quantized system as
$$
iU'(t)=H(t)U(t)
$$
where $H(t)$ is the second quantized Hamiltonian of the system expressed in terms of the creation and annihilation operator fields of the photons and Fermions. To get at this Hamiltonian, we must first calculate to a given order of perturbation, the Dirac wave field and the Maxwell wave field in terms of these Bosonic and Fermionic creation and annihilation operators. Let us as an example, derive these expressions upto second order of perturbation. The Maxwell and  Dirac equations derived from the Lagrangian are
$$
\partial_{\alpha}\partial^{\alpha}A^{\mu}=e\psi^*\alpha^{\mu}\psi,
$$
$$
(i\gamma^{\mu}\partial_{\mu}-m)\psi=-e\gamma^{\mu}\psi.A_{\mu}
$$
where we have applied the Lorentz gauge condition
$$
\partial_{\mu}A^{\mu}=0
$$
in deriving the former. In the absence of interactions, we can express the solutions
as
$$
\psi_0(x)=\sum_k(a(k)f_k(x)+a(k)^*g_k(x)),
$$
$$
A^{\mu}(x)=\sum_kc(k)\phi_k^{\mu}(x)+c(k)^*\bar\phi_k^{\mu}(x)
$$
where $c(k),c(k)^*$ satisfy the Bosonic CCR while $a(k),a(k)^*$ satisfy the Fermionic CAR.
Writing
$$
\psi=\psi_0+\psi_1+\psi_2, A_{\mu}=A_{\mu 0}+A_{\mu 1}+A_{\mu 2},
$$
we get

a.Zeroth order perturbation equations.
$$
(i\gamma^{\mu}\partial_{\mu}-m)\psi_0=0, \partial_{\alpha}\partial^{\alpha}A_{\mu 0}=0,
$$
\bigskip

b.First order perturbation equations.
$$
(i\gamma^{\mu}\partial_{\mu}-m)\psi_1=-e\gamma^{\mu}\psi_0A_{\mu 0},
$$
$$
\partial_{\alpha}\partial^{\alpha}A_{\mu 1}=e\psi_0^*\alpha_{\mu}\psi_0
$$
\bigskip

c.Second order perturbation equations.
$$
(i\gamma^{\mu}\partial_{\mu}-m)\psi_1=-e\gamma^{\mu}(\psi_0A_{\mu 1}+\psi_1A_{\mu 0})
$$
$$
\partial_{\alpha}\partial^{\alpha}A_{\mu 2}=e(\psi_0^*\alpha_{\mu}\psi_1+\psi_1^*\alpha_{\mu}\psi_0)
$$
In general, writing the $n^{th}$ order perturbation series
$$
A_{\mu}=\sum_{k=0}^nA_{\mu k}, \psi=\sum_{k=0}^n\psi_k,
$$
we obtain the $n^{th}$ order perturbation equations as
$$
(i\gamma^{\mu}\partial_{\mu}-m)\psi_n=-e\gamma^{\mu}\sum_{k=0}^{n-1}\psi_kA_{\mu n-1-k},
$$
$$
\partial_{\alpha}\partial^{\alpha}A_{\mu n}=e\sum_{k=0}^{n-1}\psi_k^*\alpha_{\mu}\psi_{n-1-k}
$$
\bigskip

\chapter{Quantum gate design, magnet motion,quantum filtering,Wigner distributions}

\section{Designing a quantum gate using QED with a control c-number four potential} 

Consider the interaction Lagrangian between the em four potential and the Dirac current:
$$
L_{int}=e\bar\psi\gamma^{\mu}\psi.A_{\mu}
$$
The corresponding contribution of this interaction Lagrangian to the interaction Hamiltonian is its negative (in view of the Legendre transformation) integrated over space:
$$
H_{int}=-L_{int}=-e\int\bar\psi\gamma^{\mu}\psi.A_{\mu}d^3x
$$
The contribution of this term upto second order in the em potential to the scattering amplitude between two states $|p_1,\sigma_1,p_2,\sigma_2>$ and $|p_1',\sigma_1',p_2',\sigma_2'>$ in which there are two electrons with the specified four momenta and z-component of the spins or equivalently two positrons or one electron and one positron is given by
$$
A_2(p_1',\sigma_1',p_2',\sigma_2'|p_1,\sigma_1,p_2,\sigma_2)=-e^2<p_1',\sigma_1',p_2',\sigma_2'|T(\int H_{int}(x_1)H_{int}(x_2))d^4x_1d^4x_2|p_1,\sigma_1,p_2,\sigma_2>
$$
Let $u(p,\sigma)$ denote the free particle Dirac wave function in the momentum-spin domain. Thus it satisfies
$$
(\gamma^{\mu}p_{\mu}-m)u(p,\sigma)=0,\sigma=1,2
$$
Taking the adjoint of this equation after premultiplying by $\gamma^0$ and making use of the self-adjointness of the matrices $\gamma^0\gamma^{\mu}$ gives us
$$
\bar u(p,\sigma)(\gamma^{\mu}p_{\mu}-m)=0
$$
where
$$
\bar u(p,\sigma)=u(p,\sigma)^*\gamma^0
$$
Now observe that for one particle momentum-spin states of an electron,
$$
a(p',\sigma')|p,\sigma>=\delta(\sigma',\sigma)\delta^3(p'-p)|0>
$$
and therefore,
$$
\psi(x)|p,\sigma>_e=exp(-ip.x)u(p,\sigma)|0>
$$
and for a one particle positron state
$$
\psi(x)^*|p,\sigma>_p=exp(ip.x)v(p,\sigma)^*|0>
$$
or equivalently,
$$
\bar\psi(x)|p,\sigma>=\psi(x)^*\gamma^0|p,\sigma>_p=exp(ip.x)\bar v(p,\sigma)|0>
$$
Thus one of the terms in $A_2(p_1',\sigma_1',p_2',\sigma_2'|p_1,\sigma_1,p_2,\sigma_2)$ for two electron momentum state transitions is given by
$$
-e^2\int<p_1',\sigma_1',p_2',\sigma_2'|(\bar\psi(x_1)\otimes\bar\psi(x_2))\gamma^{\mu}\otimes\gamma^{\nu})(\psi(x_1)\otimes\psi(x_2))|p_1,\sigma_1,p_2,\sigma_2>D_{\mu\nu}(x_1-x_2)d^4x_1d^4x_2
$$
$$
=\int D_{\mu\nu}(x_1-x_2)(\bar u(p_1',\sigma_1')\otimes\bar u(p_2',\sigma_2'))(\gamma^{\mu}\otimes\gamma^{\nu})(u(p_1,\sigma_1)\otimes u(p_2,\sigma_2))exp(-i(p_1-p_1').x_1-i(p_2-p_2').x_2)d^4x_1d^4x_2
$$
$$
=\hat D_{\mu\nu}(p_1-p_1')\delta^4(p_1+p_2-p_1'-p_2').
$$
$$
.(\bar u(p_1',\sigma_1')\otimes\bar u(p_2',\sigma_2'))(\gamma^{\mu}\otimes\gamma^{\nu})(u(p_1,\sigma_1)\otimes u(p_2,\sigma_2))
$$
This evaluates to
$$
[(\bar u(p_1',\sigma_1')\gamma^{\mu}u(p_1,\sigma_1)).(\bar u(p_2',\sigma_2')\gamma_{\mu}u(p_2,\sigma_2))/(p_1'-p_1)^2]\delta^4(p_1+p_2-p_1-p_2')
$$
Likewise another term having a different structure also comes from the same expression for $A_2$ and that is
$$
[(\bar u(p_2',\sigma_2')\gamma^{\mu}u(p_1,\sigma_1)).(\bar u(p_1',\sigma_1')\gamma_{\mu}u(p_2,\sigma_2))/(p_2'-p_1)^2]\delta^4(p_1+p_2-p_1-p_2')
$$
The total amplitude is a superposition of these two amplitudes. Here, $D_{\mu\nu}(x)$ is the photon propagator and $\hat D_{\mu\nu}(p)$ is its four dimensional space-time Fourier transform and is given by $\eta_{\mu\nu}/p^2$ where $p^2=p^{\mu}p_{\mu}$. This propagator is defined by
$$
D_{\mu\nu}(x_1-x_2)=<0|T(A_{\mu}(x_1)A_{\nu}(x_2))|0>
$$
Now we consider a situation when apart from the quantum em field, there is also an external classical control em field described by the four potential $A_{\mu}^c$.
Then in the $n^{th}$ order term in the Dyson series for the scattering of $M$ electrons, there will be a term of the form
$$
\int<p_k',\sigma_k', k=1,2,..., M|T(H_{int}(x_1)...H_{int}(x_n))|p_k,\sigma_k, k=1,2,..., M>d^4x_1...d^4x_n
$$
where
$$
H_{int}(x)=e\bar\psi(x)\gamma^{\mu}\psi(x)(A_{\mu}(x)+A_{\mu}^c(x))
$$
Here, $A_{\mu}$ is the four potential of the quantum electromagnetic field while $A_{\mu}^c(x)$ is the four potential of the classical c-number control electromagnetic four potential. Expanding these, it is clear that this term will be a sum of terms of the form
$$
\int<p_k',\sigma_k',k=1,2,..., MT(\Pi_{k=1}^n\bar\psi(x_k)\gamma^{\mu_k}\psi(x_k))|p_k,\sigma_k, k=1,2,..., M>
$$
$$
.<0|T(A_{\mu_1}(x_1)...A_{\mu_r}(x_r))|0>\Pi_{m=r+1}^nA_{\mu_m}^c(x_m)d^4x_1...d^4x_n
$$
\bigskip

\section{On the motion of a magnet through a pipe with a coil under gravity and the reaction force exerted by the induced current through the coil on the magnet}
\bigskip

{\bf Problem formulation}

Assume that the coil is wound around a tube with $n_0$ turns per unit length extending from a height of $h$ to $h+b$. The tube is cylindrical and extends from $z=0$ to $z=d$. Taking relativity into account via the retarded potentials, if $m(t)$ denotes the magnetic moment of the magnet, then the vector potential generated by it is approximately obtained from the retarded potential formula as
$$
A(t,r)=\mu m(t-r/c)\times r/4\pi r^3---(1)
$$
where we assume that the magnet is located near the origin. This approximates to
$$
A(t,r)=\mu m(t)\times r/4\pi r^3-\mu m'(t)\times r/4\pi cr^2---(2)
$$
Now Assume that the magnet falls through the central axis with its axis always being oriented along the $-\hat z$ direction. Then, after time $t$, let its $z$ coordinate be
$\xi(t)$. We have the equation of motion
$$
M\xi''(t)=F(t)-Mg---(3)
$$
where $F(t)$ is the $z$ component of the force exerted by the magnetic field generated by the current induced in the coil upon the magnet. Now, the flux of the magnetic field generated by the magnet through the coil is given by
$$
\Phi(t)=n_0\int_{X_2+Y^2\leq R_T^2,h\leq z\leq h+b}B_z(t,X,Y,Z)dXdYdZ---(4)
$$
where
$$
B_z(t,X,Y,Z)=\hat z.B(t,X,Y,Z)---(5a)
$$
$$
B(t,X,Y,Z)=curl A(t,{\bf R}-\xi(t)\hat z)---(5b)
$$
where
$$
{\bf R}=(X,Y,Z)
$$
and where $R_T$ is the radius of the tube. The current through the coil is then
$$
I(t)=\Phi'(t)/R_0---(6)
$$
where $R_0$ is the coil resistance. The force of the magnetic field generated by this current upon the falling magnet can now be computed easily. To do this, we first compute the magnetic field $B_c(t,X,Y,Z)$ generated by the coil using the usual retarded potential formula
$$
B_c(t,X,Y,Z)=curl A_c(t,X,Y,Z)---(7a)
$$
where
$$
A_c(t,X,Y,Z)=(\mu/4\pi)\int_0^{2n_0\pi}(I(t-|{\bf R}-R_T(\hat x.cos(\phi)+\hat y.sin(\phi))-\hat z(h+b/2n_0\pi)\phi|/c).
$$
$$
.(-\hat x.sin(\phi)+\hat y.cos(\phi))/|{\bf R}-R_T(\hat x.cos(\phi)+\hat y.sin(\phi))-\hat z(b/2n_0\pi)\phi|)R_Td\phi---(7a)
$$
Note that $A_c(t,X,Y,Z)$ is a function of $\xi(t)$ and also of $m(t)$ and $m'(t)$. It follows that $I(t)$ is a function of $\xi(t),\xi'(t),m(t),m'(t),m''(t)$.  Then we calculate the interaction energy between the magnetic fields generated by the magnet and by the coil as
$$
U_{int}(\xi(t),\xi'(t),t)=(2\mu)^{-1}\int(B_c(t,X,Y,Z),B(t,X,Y,Z))dXdYdZ---(8)
$$
and then we get for the force exerted by the coil's magnetic field upon the falling magnet as
$$
F(t)=-\partial U_{int}(\xi(t),\xi'(t),t)/\partial\xi(t)---(9)
$$

Remark: Ideally since the interaction energy between the two magnetic fields appears in the Lagrangian $\int(E^2/c^2-B^2)d^3R/2\mu$, we should write down the Euler-Lagrange equations using the Lagrangian
$$
L(t,\xi'(t),\xi'(t))=M\xi'(t)^2/2-U_{int}(\xi(t),\xi'(t),t)-Mg\xi(t)--(10)
$$
in the form
$$
\frac{d}{dt}\partial L/\partial\xi'-\partial L/\partial\xi=0---(11)
$$
which yields
$$
M\xi''(t)-\frac{d}{dt}\partial U/\partial\xi'+\partial U_{int}/\partial\xi+Mg=0--(12)
$$
\bigskip

{\bf 3. Numerical simulations}

Define the force function as in (9):
$$
F(\xi(t),\xi'(t)=-\partial U_{int}(\xi(t),\xi'(t))/\partial\xi(t)
$$
Neglecting the term involving the gradient of $U_{int}$ w.r.t $\xi'$ (this is justified because the dominant reaction force of the coil field on the magnet is due to the spatial gradient of the dipole interaction energy), we can discretize (2) with a time step of $\$\delta$ to get
$$
M(\xi[n+1]-2\xi[n]+\xi[n-1])/\delta^2+F_{int}(\xi[n],(\xi[n]-\xi[n-1])/\delta)+Mg=0
$$
or equivalently in recursive form,
$$
\xi[n+1]==2\xi[n]-\xi[n-1]-(\delta^2/M)F_{int}(\xi[n],(\xi[n]-\xi[n-1])/\delta)-\delta^2g
$$
Fig.1 shows plot of $\xi[n]$ obtained using a  "for loop". The current through the coil
as a function of time has also been plotted in fig.2 using the formula
$$
I(t)=R_0^{-1}\frac{d}{dt}(n_0\int_{X_2+Y^2\leq R_T^2,h\leq z\leq h+b}B_z(t,X,Y,Z)dXdYdZ)
$$
$$
=I(t|\xi(t),\xi'(t))
$$
where
$$
B_z(t,X,Y,z)=\hat z.curl A(t,X,Y,Z-\xi(t))=\partial A_y(X,Y,Z)\partial X-\partial A_x(X,Y,Z)/\partial Y
$$
with
$$
A(X,Y,Z)=\mu_0.{\bf m}\times{\bf R}/4\pi R^3, {\bf R}=(X,Y,Z)
$$
Here relativistic corrections have been neglected. The formula for the interaction energy
between the magnet and the field generated by the coil discussed in the previous section can be simplified after making appropriate approximations to $-{\bf m}.{\bf B}_c=mB_{cz}$ where
$B_{cz}$, the $z$-component of the reaction magnetic field produced by the coil is evaluated at the site of the magnet, ie, at $(0,0,\xi(t))$. Thus, the formula for the force of the field on the coil is approximated as
$$
F_{int}(\xi,\xi')=F_{int}(\xi)=-m\partial B_{cz}(t,0,0,\xi|\xi,\xi')/\partial\xi
$$
The coil magnetic field is evaluated using the approximate non-relativistic Biot-Savart formula
$$
B_c(t,X,Y,Z|\xi,\xi')=(\mu_0I(t|\xi,\xi')/4\pi))\int_0^{2n_0\pi}R_Td\phi(-\hat x.sin(\phi)+\hat y.cos(\phi))\times((X-R_T.cos(\phi))\hat x+(Y-R_T.sin(\phi))
$$
$$
.\hat y+(Z-h-b\phi/2n_0\pi)\hat z)/[(X-R_T.cos(\phi))^2+(Y-R_T.sin(\phi))^2+(Z-h-b\phi/2n_0\pi)^2]^{3/2}
$$
These approximations are shown to produce excellent agreement with hardware experiments as a comparison of figs.() shows.
\bigskip

Taking into account relativistic corrections: The magnetic field generated by the coil has the form $B_c(t,X,Y,Z|\xi(t),\xi'(t))$. More precisely, if $I_0(t)$ is the current of the electrons within the magnet, the magnetic vector potential generated by this current is given by 
$$
A(t,R)=(\mu/4\pi)\int_{magnet}I_0(t-|R-\xi(t)-R'|/c)dR'/|R-\xi(t)-R'|
$$
In this expression, $\xi(t)$ is the position vector of the centre of the magnet at time $t$ and $R'$ is the position on the magnet electron loop relative to its centre. Since the magnet current is a constant in the case of a permanent magnet, the above simplifies to
$$
A(t,R)=(\mu I_0/4\pi)\int_{magnet}dR'/|R-\xi(t)-R'|
$$
and hence by making the infinitesimal loop approximation, this becomes
$$
A(t,R)=(\mu/4\pi)m\times(R-\xi(t))/|R-\xi(t)|^3
$$
In other words, relativistic corrections to the magnetic field produced by the magnet will arise only if the magnet is an electromagnet so that the current through it varies with time. However, the current through the coil wound on the tube surface varies with time and the magnetic field produced by it will have to be calculated using the retarded potential.
Hence relativistic corrections will arise in this case.
\bigskip

Estimating the magnetic moment $m$ of the magnet using the following scheme: The magnetic field through the coil and hence the magnetic flux and hence the induced current in the coil are all functions of $m$. We've already noted that the induced current can be expressed in the form given by equations (2b),(4),(5) and (6). We write this as
$$
I(t)=I(m,\xi(t),\xi'(t))
$$
in the special case $m$ is a constant. It is also clear from the above formulas that
for given $\xi(t),\xi'(t)$, the coil current is a linear function of $m$. This is because the magnetic field produced by the magnet is a linear function of $m$. Hence, we can write
$$
I(t)=P(\xi(t),\xi'(t))m
$$
where $P(\xi(t),\xi'(t))$ is an easily calculable function. Thus, if we take measurements
of the coil current and the displacement of the magnet at a sequence of times $t_1<t_2<..<t_N$, then $m$ may be estimated by the least squares method:
$$
\hat m=argmin_m\sum_{k=1}^N(I(t_k)-P(\xi(t_k),(\xi(t_{k+1})-\xi(t_k))/(t_{k+1}-t_k))m)^2
$$
Carrying out this minimization results in the estimate
$$
\hat m=\sum_kI_kP_k/\sum_kP_k^2
$$
where
$$
I_k=I(t_k), P_k=P(\xi(t_k),(\xi(t_{k+1})-\xi(t_k))/(t_{k+1}-t_k))
$$
The estimates of $m$ obtained in this way for different measurements are shown in Table(.).

Remark: $m$ is the z component of the magnetic moment which equals in magnitude the total magnetic moment since the magnet is assumed to be oriented along the $z$ direction. The 
flux through the coil can be expressed as
$$
\Phi(t)=m.(\mu n_0/4\pi)\int_{X^2+Y^2\leq R_T^2, h\leq Z\leq h+b}curl(\hat z\times{\bf R}-\xi(t)\hat z)/|{\bf R}-\xi(t)\hat z|^3).\hat zdXdYdZ
$$
so this provides the immediate identification
$$
P(\xi(t),\xi'(t))=
$$
$$
\frac{d}{dt}[(\mu n_0/4\pi R_0)\int_{X^2+Y^2\leq R_T^2, h\leq Z\leq h+b}curl(\hat z\times{\bf R}/|{\bf R}-\xi(t)\hat z|^3).\hat zdXdYdZ
$$
\bigskip

Quantization of the motion of a magnet in the vicinity of the coil. The magnetic dipole
is replaced by a constrained sea of Dirac electrons and positrons within the boundary of the magnet. This Dirac field can be expressed in the form
$$
\psi(x)=\sum_k(a(k)f_k(x)+a(k)^*g_k(x))
$$
where the functions $f_k(x),g_k(x)$ are eigen functions of the Dirac equation with boundary conditions determined by the vanishing of the Dirac wave function on the boundary of the magnet. These are four component wave functions. In other words, they satisfy
$$
(i\gamma^{\mu}\partial_{\mu}-m)(f_k(x),g_k(x))=0
$$
for all $x$ within the bounding walls of the magnet and they vanish on the boundary, or rather the charge density corresponding to these functions as well as the normal component of the spatial current density vanishes on the boundary. The magnetic dipole moment of the magnet after second quantization is given by
$$
{\bf m}(t)=\int_M\psi(x)^*(eJ/2m)\psi(x)d^3x
$$
where 
$$
{\bf J}={\bf L}+g\sigma/2
$$
is the total orbital plus spin angular momentum of an electron taking the $g$ factor into account, normally we put $g=1$. It is a conserved vector operator for the Dirac Hamiltonian. The $a(k),a(k)^*$ are the annihilation and creation operators of the electrons and positrons. They satisfy the CAR
$$
[a(k),a(m)^*]_+=\delta(k,m)
$$
It is easily noted that ${\bf m}(t)$ is a quadratic function of the electron-positron creation and annihilation operators.

The magnetic field produced by this magnet at the site of the coil is given by the usual formula
$$
B(t,{\bf R})=(\mu/4\pi)curl({\bf m}(t)\times({\bf R}-\xi(t))/|{\bf R}-\xi(t)|^3)
$$
When integrated over the cross-section of the coil and differentiated w.r.t time and then divided by the coil resistance, this gives the induced current as a function of the position $\xi(t)$ and velocity $\xi'(t)$ of the magnet and this expression for the current is a quadratic form in the Fermion creation and annihilation operators. The magnetic field produced by the coil current at the site of the magnet can be computed and hence the force exerted by the coil's magnetic field on the magnet obtained by taking the dot product of
${\bf m}(t)$ with this magnetic field followed by the spatial gradient will be a fourth degree polynomial in the Fermionic creation and annihilation operators. The Heisenberg equation of motion of the coil will therefore have the form
$$
\xi''(t)=F(t,\xi(t),\xi'(t),\{a(k),a(k)^*\})-g
$$
where $g$ is the acceleration due to gravity and the rhs is a homogeneous fourth degree polynomial in $a,a^*$. We can thus express this operator equation as
$$
\xi''(t)=F_1(t,\xi(t),\i'(t)|kmpq)a(k)a(m)a(p)a(q)+F_2(t,\xi(t),\i'(t)|km)a(k)^*a(m)a(p)a(q)+F_3(t,\xi(t),\i'(t)|kmpq)a(k)^*a(m)^*a(p)a(q)+F_4(t,\xi(t),\i'(t)|km)a(k)^*a(m)...+sixteen terms in all-g
$$
These equations are solved approximately using perturbation theory as follows. We express the solution as
$$
\xi(t)=d-gt^2/2+\delta\xi(t)=\xi_0(t)+\delta\xi(t), \xi_0(t)=d-gt^2/2
$$
where $\delta\xi(t)$ satisfies the first order perturbed equation
$$
\delta\xi''(t)=F(t,\xi_0(t),\xi_0'(t),a,a^*)
$$
giving
$$
\delta\xi(t)=\int_0^t(t-s)F(s,\xi_0(s),\xi_0'(s),a,a^*)ds
$$
Statistical properties of the quantum fluctuations $\delta\xi(t)$ of the magnet's position in a coherent state of the Fermions within the magnet can now be evaluated. For example if
$|\psi(\gamma)>$ is such a coherent state, then
$$
a(k)|\psi(\gamma)>=\gamma(k)|\psi(\gamma)>
$$
$$
a(k)^*|\psi(\gamma)>=\partial/\partial\gamma|\psi(\gamma)>
$$
where $\gamma(k),\gamma(k)^*$ are Grassmannian numbers, ie, Fermionic parameters. They are assumed to all mutually anticommute, ie, 
$$
[\gamma(k),\gamma(l)]_+=0=[\gamma(k)^*,\gamma(l)^*]_+,
$$
$$
[\gamma(k),\gamma(l)^*]_+=0
$$
and further they anticommute with the $a(m),a(m)^*$. To check the consistency of these operations, we have
$$
\gamma(m)a(k)|\psi(\gamma)>=\gamma(m)\gamma(k)|\psi(\gamma)>
$$
on the one hand and on the other,
$$
\gamma(k)\gamma(m)|\psi(\gamma)>=-\gamma(m)\gamma(k)|\psi(\gamma)>=-\gamma(m)a(k)|\psi(\gamma)>
$$
in agreement with
$$
\gamma(m)\gamma(k)=-\gamma(k)\gamma(m)
$$
Further,
$$
a(k)a(m)^*|\psi(\gamma)>=(\delta(k,m)-a(m)^*a(k))|\psi(\gamma)>=
$$
$$
(\delta(k,m)-a(m)^*\gamma(k))|\psi(\gamma)>=(\delta(k,m)+\gamma(k)a(m)^*)|\psi(\gamma)>
$$
while on the other hand,
$$
a(m)^*a(k)|\psi(\gamma)>=a(m)^*\gamma(k)|\psi(\gamma)>
$$
$$
=-\gamma(k)a(m)^*|\psi(\gamma)>
$$
which is a consistent picture.

Let us for example take a single CAR pair $a,a^*$ so that $a^2=a^{*2}=0, aa^*+a^*a=1$.
Let $|0>$ be the vacuum state and $|1>$ the single particle state, so that
$$
a|0>=0, a^*|1>=0, a^*|0>=|1>, a|1>=|0>
$$
Then, the number operator $N=a^*a$ obviously satisfies
$$
N|0>=0, N|1>=|1>
$$
Suppose we assume that
$$
|\psi(\gamma)>=(c_1\gamma+c_2\gamma^*+c_3)|0>+(c_4\gamma+c_5\gamma^*+c_6)|1>
$$
Then since $a\gamma=-\gamma a, a\gamma^*=-\gamma^*a$, we get
$$
a|\psi(\gamma)>=(-c_4\gamma|0>-c_5\gamma^*+c_6)|0>
$$
and for this to equal $\gamma|\psi(\gamma)>$, we require that $c_2=c_5=c_6=0$ and
$c_3=-c_4$. In other words,
$$
|\psi(\gamma)>=(c_1\gamma+c_3)|0>-c_3\gamma|1>
$$
We then get
$$
a^*|\psi(\gamma)>=(-c_1\gamma+c_3)|1>
$$
Note that we require that
$$
\gamma\gamma^*+\gamma^*\gamma=1
$$
We have
$$
\partial/\partial\gamma|\psi(\gamma)>=c_1|0>-c_3|1>
$$
which cannot equal $a^*|\psi(\gamma)>$. So to get agreement with this equation also, we must suppose that
$$
|\psi(\gamma)>=(c_1\gamma+c_2\gamma^*+c_3\gamma^*\gamma+c_4)|0>+(d_1\gamma+d_2\gamma^*+d_3\gamma^*\gamma+d_4)|1>
$$
Note that by hypothesis,
$$
\gamma^*\gamma+\gamma\gamma^*=0
$$
Then,
$$
a|\psi(\gamma)>=(-d_1\gamma-d_2\gamma^*+d_3\gamma^*\gamma+d_4)|0>
$$
$$
\gamma|\psi(\gamma)>=(c_2\gamma\gamma^*+c_4\gamma)|0>+(d_2\gamma\gamma^*+d_4\gamma)|1>
$$
For equality of the two expressions, we must thus assume that
$$
d_2=d_4=0, d_3=-c_2, d_1=-c_4
$$
This gives
$$
|\psi(\gamma)>=(c_1\gamma+c_2\gamma^*+c_3\gamma^*\gamma+c_4)|0>+(-c_4\gamma-c_2\gamma^*\gamma)|1>
$$
Now we also require that
$$
a^*|\psi(\gamma)>=-\partial|\psi(\gamma)>/\partial\gamma
$$
and this gives
$$
-(c_1-c_3\gamma^*)|0>+(c_4-c_2\gamma^*)|1>=(-c_1\gamma-c_2\gamma^*+c_3\gamma^*\gamma+c_4)|1>
$$
Thus,
$$
c_1=c_3=0
$$
and we get
$$
|\psi(\gamma)>=(c_2\gamma^*+c_4)|0>+(-c_4\gamma-c_2\gamma^*\gamma)|1>
$$
As a check, we compute
$$
a|\psi(\gamma)>=(c_4\gamma-c_2\gamma^*\gamma)|0>,
$$
$$
\gamma|\psi(\gamma)>=(c_2\gamma\gamma^*+c_4\gamma)|0>,
$$
$$
a^*|\psi(\gamma)>=(-c_2\gamma^*+c_4)|1>,
$$
$$
\partial|\psi(\gamma)>/\partial\gamma=(-c_4+c_2\gamma^*)|1>
$$
\bigskip

Now suppose that we have a term for example as $a(k)^*a(m)a(n)^*a(r)$ and we wish to evaluate its expectation value in the coherent state $|\psi(\gamma)>$. It will be given by
$$
<\psi(\gamma)|a(k)^*a(m)a(n)^*a(r)|\psi(\gamma)>=
$$
$$
-\gamma(r)<\psi(\gamma)|a(k)^*a(m)a(n)^*|\psi(\gamma)>
$$
$$
=-\gamma(r)\gamma(k)^*<\psi(\gamma)|a(m)a(n)^*|\psi(\gamma)>=
$$
$$
=-\gamma(r)\gamma(k)^*<\psi(\gamma)|\delta(m,n)-a(n)^*a(m)|\psi(\gamma)>
$$
$$
=-\delta(m,n)\gamma(r)\gamma(k)+\gamma(r)\gamma(k)^*\gamma(n)^*\gamma(m)
$$
The problem is how to interpret this in terms of real numbers ?
\bigskip

\section{Quantum filtering theory applied to quantum field theory for tracking a given probability density function}

Consider for example the Klein-Gordon field $\phi(x)$ in the presence of an external potential $V(\phi)$ whose dynamics after it gets coupled to the electron-positron Dirac field is given by the Lagrangian
$$
L=\bar\psi(i\gamma^{\mu}\partial_{\mu}-e\phi-m_e)\psi+(1/2)\partial_{\mu}\phi\partial^{\mu}\phi-m_k^2\phi^2/2-V(\phi)
$$
$$
=L_e+L_{kg}+L_{int}
$$
where $L_e$, the electronic Lagrangian, is given by
$$
L_e=\bar\psi(i\gamma^{\mu}\partial_{\mu}-m_e)\psi,
$$
$L_k$ the KG Lagrangian is given by
$$
L_{kg}=(1/2)\partial_{\mu}\phi.\partial^{\mu}\phi-m_k^2\phi^2/2-V(\phi)
$$
and $L_{int}$, the interaction Lagrangian between the KG field and the electron field is given by
$$
L_{int}=-e\bar\psi\psi.\phi
$$
Note that we could also include interaction terms of the form given by the real and imaginary parts of
$$
\bar\psi\gamma^{\mu_1}...\gamma^{\mu_k}\psi.\partial_{\mu_1}...\partial_{\mu_k}\phi
$$
The Schrodinger evolution equation for such a system of quantum fields is obtained by introducing Fermionic and Bosonic creation and annihilation operators to describe the non-interacting fields in the presence of zero potential $V$, then applying perturbation theory to approximately solve for the interacting fields in the presence of the potential $V$, then use this approximate solution to write down an expression for the total second quantized field Hamiltonian in terms of the creation and annihilation operators.
\bigskip

The electron-positron field interacting with a KG field and the photon field:
$$
L=(-1/4)F_{\mu\nu}F^{\mu\nu}+\bar\psi(\gamma^{\mu}(i\partial_{\mu}+e_1A_{\mu})+e_2\phi-m_1)\psi
$$
$$
+(1/2)\partial_{\mu}\phi.\partial^{\mu}\phi-m_2\phi^2/2
$$
The interaction Hamiltonian derived from this expression is
$$
H_{int}(t)=e_1\int\bar\psi(x)\gamma^{\mu}\psi(x)A_{\mu}(x)d^3x
$$
$$
+e_2\int\bar\psi(x)\psi(x)\phi(x)d^3x
$$
One of the terms for the scattering amplitude of two electrons in the Dyson series formulation is
$$
e_1^2<p_1',p_2'|\int T(\bar\psi(x_1)\gamma^{\mu}\psi(x_1)\bar\psi(x_2)\gamma^{\nu}\psi(x_2))D^A_{\mu\nu}(x_1-x_2)d^4x_1d^4x_2|p_1,p_2>
$$
Another term is
$$
e_2^2<p_1',p_2'|\int T(\bar\psi(x_1)\psi(x_1)\bar\psi(x_2)\psi(x_2))D^{\phi}(x_1-x_2)d^4x_1d^4x_2|p_1,p_2>
$$
Yet another term is obtained from
$$
e_1^2e_2^2\int\bar\psi(x_1)\gamma^{\mu}\psi(x_1)\bar\psi(x_2)\gamma^{\nu}\psi(x_2)\bar\psi(x_3)\psi(x_3)\bar\psi(x_4)\psi(x_4)D^A_{\mu\nu}(x_1-x_2)D^{\phi}(x_3-x_4)dx_1dx_2dx_3dx_4
$$
This last term leads to a term of the form
$$
e_1^2e_2^2\int<p_1',p_2'|\bar\psi(x_4)S_e(x_4-x_1)\gamma^{\mu}S_e(x_1-x_2)\gamma^{\nu}\psi(x_2)\bar\psi(x_3)\psi(x_3)|p_1,p_2>D^A_{\mu\nu}(x_1-x_2)D^{\phi}(x_3-x_4)dx_1dx_2dx_3dx_4
$$
This term can be evaluated easily by expanding in terms of matrix elements or equivalently by using the Kronecker tensor product:
$$
e_1^2e_2^2\int <p_1',p_2'|(\bar\psi(x_4)\otimes\bar\psi(x_3))(S_e(x_4-x_1)\gamma^{\mu}S_e(x_1-x_2)\otimes I)(\psi(x_2)\otimes\psi(x_3))|p_1,p_2>
$$
$$
.D^A_{\mu\nu}(x_1-x_2)D^{\phi}(x_3-x_4)dx_1dx_2dx_3dx_4
$$
\bigskip

One of the terms in the scattering amplitude of two electrons in the presence of an external electromagnetic potential is
$$
A(p_1',p_2'|p_1,p_2)=<p_1',p_2'|\int A_{\mu}^c(x_1)T(\bar\psi(x_1)\gamma^{\mu}T(\psi(x_1)\bar\psi(x_2))\gamma^{\nu}\psi(x_2)\bar\psi(x_3))\gamma^{\rho}\psi(x_3))D_{\nu\rho}(x_2-x_3)dx_1dx_2dx_3|p_1,p_2>
$$
$$
=\int A_{\mu}^c(x_1)<p_1',p_2'|T(\bar\psi(x_1)\gamma^{\mu}S_e(x_1-x_2)\gamma^{\nu}\psi(x_2)\bar\psi(x_3)\gamma^{\rho}\psi(x_3)|p_1,p_2>D_{\nu\rho}(x_2-x_3)dx_1dx_2dx_3
$$
This can be expressed as
$$
\int A_{\mu}^c(x_1)<p_1',p_2',|T((\bar\psi(x_1)\otimes\bar\psi(x_3))(\gamma^{\mu}S_e(x_1-x_2)\gamma^{\nu}\otimes\gamma^{\rho})(\psi(x_2)\otimes\psi(x_3))|p_1,p_2>D_{\nu\rho}(x_2-x_3)dx_1dx_2dx_3
$$
which evaluates to
$$
\int A_{\mu}^c(x_1)exp(i(p_1'.x_1+p_2'.x_3-p_1.x_2-p_2.x_3))\bar u(p_1')\gamma^{\mu}S_e(x_1-x_2)\gamma^{\nu}u(p_1).\bar u(p_2')\gamma^{\rho}u(p_2)D_{\nu\rho}(x_2-x_3)dx_1dx_2dx_3
$$
By the change of variables
$$
x_1-x_2=x, x_2-x_3=y
$$
and by expressing $A_{\mu}^c(x_1)$ in terms of its momentum domain Fourier integral, this integral can be expressed as
\bigskip

\section{Lecture on the role of statistical signal processing in general relativity and quantum mechanics}

[1] Using the extended Kalman filter to calculate the estimate of the fluid velocity field when the fluid dynamical equations are described by the general relativisitic versions of the Navier-Stokes equation and the equation of continuity and external noise is present.
The fluid equations are
$$
((\rho+p)v^{\mu}v^{\nu}-pg^{\mu\nu})_{:\nu}=f^{\mu}
$$
where $f^{\mu}$ is a random field. From these equations, we can derive
$$
((\rho+p)v^{\nu})_{:\nu}v^{\mu}+(\rho+p)v^{\nu}v^{\mu}_{:\nu}-p^{,\mu}=f^{\mu}
$$
which results in on using $v_{\mu}v^{\mu}=1$,
$$
((\rho+p)v^{\nu})_{:\nu}-p_{,\mu}v^{\mu}-f_{\mu}v^{\mu}=0
$$
This represents the general relativistic equation of continuity and on substituting this into the previous equation, we get
$$
(\rho+p)v^{\nu}v^{\mu}_{:\nu}-p^{,\mu}-f^{\mu}+v^{\mu}(p_{,\nu}v^{\nu}+f_{\nu}v^{\nu})=0
$$
This equation is the general relativistic version of the Navier-Stokes equation.
This system of equations can be expressed as a system of four differential equations for
$v^r,r=1,2,3,\rho$ assuming that the equation of state $p=p(\rho)$ is known. Then, the measurement process consists of measuring the velocity field $v^r$ at a set of $N$ discrete spatial points, taking into account measurement noise and then applying the EKF to estimate the fields $v^r,\rho$ at all spatial points on a real time basis.
\bigskip

Applications of quantum statistical signal processing to quantum fluid dynamics. 

Consider the Navier-Stokes equation along with the equation of continuity in the form
$$
v_{,t}(t,r)={\bf F}_1(t,r,v(t,r),\nabla v(t,r),\rho(t,r),\nabla\rho(t,r))+w(t,r)
$$
$$
\rho_{,t}(t,r)=F-2(t,r,v(t,r),\nabla\rho(t,r),\nabla v(t,r))
$$
Denoting the four component vector time varying vector field $[\rho(t,r),v(t,r)^T]^T$ by $\xi(t,r)$, these equations can be expressed in the form
$$
\xi_{,t}(t,r)={\bf F}(t,r,\xi(t,r),\nabla\xi(t,r))+w(t,r)
$$
where $w$ is the noise field. Now to quantize this system of equations, there are two available routes. The first is to introduce an auxiliary four vector field $\eta(t,r)$ which acts as a Lagrange multiplier and write down the Lagrangian density for the pair of fields $\xi,\eta$ as
$$
L(\xi,\eta,\xi_{,t},\nabla\xi,\nabla\eta)=\eta(t,r)^T(\xi_{,t}(t,r)-F(t,r,\xi(t,r),\nabla\xi(t,r)))
$$
The equations of motion of $\eta$ are derived from
$$
\delta_{\xi}L=0
$$
which result in
$$
\eta_{,t}(t,r)+\eta(t,r)^T(\partial F/\partial\xi)-div(\eta(t,r)^T\partial F/\partial\nabla\xi)=0
$$
Now we can express the Hamiltonian density of this system of two fields as follows.
The Hamiltonian density of this system of fields is first computed: The canonical momenta are
$$
P_{\eta}=\partial L/\partial\eta_{,t}=0
$$
$$
P_{\xi}=\partial L/\partial\xi_{,t}=\eta
$$
$$
\mathcal H=\eta\xi_{,t}-L
$$
It is impossible to bring this to Hamiltonian form owing to the strange nature of of the constraints involved. Indeed, to bring this to Hamiltonian form, $\xi_{,t}$ must be expressed in terms of the canonical momenta, $\xi,\eta$ and their spatial gradients which is not possible. Therefore, we add a regularizing term to give the Lagrangian density the following form:
$$
L=\eta^T(\xi_{,t}-F)+\epsilon.\xi_{,t}^2/2+\mu.\eta_{,t}^2/2
$$
The canonical momenta now are
$$
P_{\xi}=\partial L/\partial\xi_{,t}=\eta+\epsilon.\xi_{,t},
$$
$$
P_{\eta}=\partial L/\partial\eta_{,t}=\mu\eta_{,t}
$$
Thus,
$$
\mathcal H=P_{\xi}.\xi_{,t}+P_{\eta}.\eta_{,t}-L=
$$
\bigskip

\section{Problems in classical and quantum stochastic calculus}

[1] Let $P:0=t_0<t_1<...<t_n=T$ be a partition of $[0,T]$. Let $B$ be standard one dimensional Brownian motion. Let $f$ be adapted to the Brownian filtration and bounded, continuous over $[0,T]$. Let $Q:0=s_0<s_1<...<s_m=T$ be another partition of $[0,T]$.
Define
$$
I(P,f)=\sum_{k=0}^{n-1}f(t_k)(B(t_{k+1})-B(t_k))
$$
Show that if $R=P\cup Q$, then
$$
\Bbb E(I(P,f)-I(P\cup Q,f))^2\leq T.max_ksup_{t_k<s<t_{k+1}}\Bbb E(f(s)-f(t_k))^2
$$
and that this quantity converges to zero as $|P|\rightarrow 0$. Deduce using the inequality
$$
\Bbb E(I(P,f)-I(Q,f))^2\leq 2\Bbb E(I(P,f)-I(P\cup Q,f))^2+2\Bbb E(I(Q,f)-I(P\cup Q,f))^2
$$
that as $|P|,|Q|\rightarrow 0$, we get
$$
\Bbb E(I(P,f)-I(Q,f))^2\rightarrow 0
$$
and hence that 
$$
lim_{|P|\rightarrow 0}I(P,f)=\int_0^Tf(t)dB(t)
$$
exists.
\bigskip

[2] Derive the quantum Fokker-Planck or Master equation or GKSL equation from Hamiltonian quantum mechanics along the following lines: Let $H(t)=H_0+V(t)$ be the Hamiltonian
where $H_0$ is non-random and $\delta H(t)$ is an operator valued Hermitian stochastic process with $dt^2.\Bbb E(V(t)\otimes V(t))$ being of $O(dt)$ just as in Ito's formula for Brownian motion. Here $V(t)$ is to be regarded as operator valued white noise so that
$V(t)dt$ becomes an operator valued Brownian differential. Then consider the Schrodinger equation
$$
i\rho'(t)=[H_0+V(t),\rho(t)]=[H(t),\rho(t)]
$$
Show that this implies approximately for small $\tau$ that
$$
i(\rho(t+\tau)-\rho(t))=\int_t^{t+\tau}[H(s),\rho(s)]ds=\int_t^{t+\tau}[H(s),\rho(t)+\rho'(t)(s-t)]ds+O(\tau^3)
$$
$$
=[H(t),\rho(t)]\tau-i[H(t),[H(t),\rho(t)]]\tau^2/2
$$
which gives on taking average over the probability distribution of $V(t)$,
$$
\rho(t+\tau)-\rho(t)=-i[H_0,\rho(t)]\tau-(\tau^2/2)<[V(t),[V(t),\rho(t)]]>
$$
$$
=-i[H_0,\rho(t)]\tau-(\tau^2/2)(<V(t)^2\rho(t)+\rho(t)V(t)^2-2V(t)\rho(t)V(t)>)
$$
Now, write
$$
V(t)=\sum_k\xi(k)L_k
$$
where $\xi(k)$ are iid r.v's with zero mean and variance $1/\tau$ in accordance with the assumption that $V(t)$ is operator valued white noise. $L_k's$ are assumed to be self-adjoint operators in the Hilbert space. Then show that the above equation gives
$$
\rho'(t)=-i[H_0,\rho(t)]-(1/2)\sum_k(L_k^2\rho(t)+\rho(t)L_k^2-2L_k\rho(t)L_k)
$$
which is a special case of the master equation for open quantum systems.
\bigskip

[3] Quantum fluctuation dissipation theorem: The general Lindblad term in the master equation has the form
$$
\sum_k(L_k^*L_k\rho+\rho L_k^*L_k-2L_k\rho L_k^*]
$$
For $\rho=C(\beta)exp(-\beta H_0)$ to satisfy the GKSL as an equilibrium state, we require the quantum fluctuation-dissipation theorem to hold, ie,
$$
\sum_k(L_k^*L_kexp(-\beta H)+exp(-\beta H)L_k^*L_k-2L_k.exp(-\beta H).L_k^*)=0
$$

Problem: let
$$
H=P^2/2m+U(Q)
$$
in $L^2(\Bbb R)$ where $P=-id/dQ$ and put
$$
L_k=g_k(Q,P)=\alpha_kQ+\beta_kP
$$
Evaluate the conditions of the quantum fluctuation-dissipation theorem. 

hint:
$$
exp(-\beta H).F(Q,P).exp(\beta H)=exp(-\beta ad H)(F)
$$
$$
=F+\sum_{n\geq 1}((-\beta)^n/n!)(ad H)^n(F)
$$
Now,
$$
[H,Q]=-iP/m, [H,P]=iU'(Q)
$$
Make the induction hypothesis
$$
ad(H)^n(Q)=\sum_{k=0}^nF_{nk}(Q)P^k, ad(H)^n(P)=\sum_{k=0}^nG_{nk}(Q)P^k
$$
Deduce recursion formulae for $\{F_{nk}(Q)\}_k$ and $\{G_{nk}(Q)\}_k$.
\bigskip

[4] Evaluate $exp(-\beta P^2/2m).exp(-\beta U(Q))$ in terms of $exp(-\beta(P^2/2m+U(Q))$ and vice versa and multiple commutators between $P$ and $Q$.

hint: Let $A,B$ be two operators. Put
$$
exp(tA).exp(tB)=exp(Z(t))
$$
Then,
$$
exp(Z)g(ad Z)(Z')=Aexp(Z)+exp(Z)B
$$
where
$$
g(z)=(1-exp(-z))/z
$$
Thus,
$$
g(ad Z)(Z')=exp(-ad Z)(A)+B
$$
or equivalently
$$
Z'=f_1(ad Z)(A)+f_2(ad Z)(B)
$$
where
$$
f_1(z)=g(z)^{-1}, f_2(z)=exp(-z)g(z)^{-1}
$$
\bigskip

\section{Some remarks on the equilibrium Fokker-Planck equation for the Gibbs distribution} 

The quantum fluctuation-dissipation theorem for a single Lindlbad operator is
$$
L^*Lexp(-\beta H)+exp(-\beta H)L^*L-2Lexp(-\beta H)L^*=0
$$
This is the condition on the Lindblad operator $L$ so that $\rho=exp(-\beta H)/Tr(exp(-\beta H))$ satisfies the steady state GKSL equation:
$$
0=i\partial_t\rho=[H,\rho]-(1/2)(L^*L\rho+\rho L^*L-2L\rho L^*)
$$
Here
$$
H=P^2/2m+U(Q), P=-i\partial/\partial Q
$$
The classical version of this is
$$
0=\partial_tf(Q,P)=-(P/m)\partial_Qf+U'(Q)\partial_Pf+\gamma\partial_P(Pf)+(\sigma^2/2)\partial_P^2f
$$
with 
$$
f=Cexp(-\beta H), H=P^2/2m+U(Q)
$$
Note that this classical Fokker-Planck equation is derived from the Ito sde
$$
dQ=Pdt/m, dP=-\gamma Pdt-U'(Q)dt+\sigma dB(t)
$$
Now,
$$
\partial_Pf=-\beta Pf/m, \partial_P^2f=-\beta f/m+\beta^2P^2f/m^2
$$
$$
\partial_Qf=-\beta U'(Q)f
$$
So the above classical equilibrium condition becomes
$$
0=(\beta P/m)U'(Q)+U'(Q)(-\beta P/m)+\gamma(1-\beta P^2/m)+(\sigma^2/2)(-\beta/m+\beta^2P^2/m^2)
$$
This is satisfied iff
$$
\gamma-\sigma^2\beta/2m=0
$$
or equivalently,
$$
2kTm\gamma=\sigma^2
$$
which is the classical fluctuation-dissipation theorem as first derived by Einstein.
In the quantum context, the dissipation coefficient $\gamma$ and the diffusion/fluctuation coefficient $\sigma^2$ are both represented by the Lindblad operator $L$ as follows from the derivation the master equation from the Hudson-Parthasarathy-Schrodinger equation 
$$
dU(t)=(-(iH+L^*L/2)dt+LdA-L^*dA^*)U(t)
$$
by partial tracing of the density evolution over the bath state. Now consider the Heisenberg quantum equations of motion for $Q,P$ for the master equation: For any Heisenberg observable $X$, the dual of the master equation is
$$
X'=i[H,X]-(1/2)(XL^*L+L^*LX-2L^*XL)=i[H,X]-(1/2)([X,L^*]L+L^*[L,X])
$$
In particular for $X=Q$,
$$
Q'=P/m-(1/2)([Q,L^*]L+L^*[L,Q])
$$
and for $X=P$,
$$
P'=-U'(Q)-(1/2)([P,L^*]L+L^*[L,P])
$$
Now to get an equation that describes the motion of a particle in a potential $U(Q)$ with linear velocity damping, we assume that
$$
L=aQ+bP, L^*=\bar aQ+\bar bP
$$
Then,
$$
[L,Q]=[aQ+bP,Q]=-ib, [Q,L^*]=[Q,\bar aQ+\bar bP]=i\bar b
$$
$$
[Q,L^*]L+L^*[L,Q]=i\bar b(aQ+bP)+(\bar aQ+\bar bP)(-ib)=2.Im(\bar ab)Q
$$
$$
[P,L^*]L+L^*[L,P]=(-i\bar a)(aQ+bP)+(\bar aQ+\bar bP)(ia)=2.Im(\bar ab)P
$$
Thus the Heisenberg-Lindblad equations read
$$
Q'=P/m-Im(\bar ab)Q,
$$
$$
P'=-U'(Q)-Im(\bar ab)P
$$
Eliminating $P$ between these two equations gives us
$$
Q''=(-U'(Q)-m.Im(\bar ab)(Q'+Im(\bar ab)Q))/m-Im(\bar ab)Q'
$$
or
$$
mQ''=-(U'(Q)+(Im(\bar ab))^2)Q-2Im(\bar ab)Q'
$$
which is the velocity damped equation in a potential $U(Q)+\gamma^2Q^2/8$ and with velocity damping coefficient $\gamma=2Im(\bar ab)$. The harmonic correction to the potential $U(Q)$ is actually a quantum correction as can be verified by not assuming
Planck's constant $h=1$ as we have done here. (Here $h$ stands for Planck's constant divided by $2\pi$). Now, the quantum fluctuation-dissipation theorem can also be expressed as
$$
L^*L+exp(-\beta ad H)(L^*L)-2Lexp(-\beta ad H)(L^*)=0
$$
For high temperatures or equivalently small $\beta$, this reads
$$
[L^*,L]-\beta([H,L^*L]-L[H,L^*])=0
$$
where $O(\beta^2)$ terms have been neglected. Now for $L=aQ+bP$, we get
$$
[L^*,L]=[\bar aQ+\bar bP,aQ+bP]=i\bar ab-ia\bar b=2Im(\bar ab)
$$
$$
[H,L^*L]=[H,L^*]L+L^*[H,L]
$$
so
$$
[H,L^*L]-L[H,L^*]=[[H,L^*],L]+L^*[H,L]
$$
\bigskip

\section{Wigner-distributions in quantum mechanics}

Since $Q,P$ do not commute in quantum mechanics, we cannot talk of a joint probability density of these two at any time unlike the classical case. So how to interpret the Lindblad equation
$$
i\partial_t\rho=[H,\rho]-(1/2)(L^*L\rho+\rho L^*L-2L\rho L^*)
$$
as a quantum Fokker-Planck equation for the joint density of $(Q,P)$ ? Wigner suggested the following method: Let $\rho(t,Q,Q')=<Q|\rho(t)|Q>$ denote $\rho(t)$ in the position representation. Then $\rho(t)$ in the momentum representation is given by
$$
\hat\rho(t,P,P')=<P|\rho(t)|P'>=\int <P|Q><Q|\rho(t)|Q'><Q'|P'>dQdQ'
$$
$$
=\int\rho(t,Q,Q')exp(-iQP+iQ'P')dQdQ'
$$
$\rho(t,Q,Q)$ is the probability density of $Q(t)$ and $\hat\rho(t,P,P)$ is the probability density of $P(t)$. Both are non-negative and integrate to unity. Now Wigner defined the function
$$
W(t,Q,P)=\int\rho(t,Q+q/2,Q-q/2)exp(-iPq)dq/(2\pi)
$$
This is in general a complex function of $Q,P$ and therefore cannot be a probability density. However,
$$
\int W(t,Q,P)dP=\rho(t,Q,Q),
$$
is the marginal density of $Q(t)$
Further, we can write
$$
W(t,Q,P)=\int\hat\rho(t,P_1,P_2)exp(iP_1(Q+q/2)-iP_2(Q-q/2))exp(-iPq)dP_1dP_2dq/(2\pi)^2
$$
$$-
=\int\hat\rho(t,P_1,P_2)exp(i(P_1-P_2)Q)\delta((P_1+P_2)/2-P)dP_1dP_2/(2\pi)
$$
so that
$$
\int W(t,Q,P)dQ=\int\hat\rho(t,P_1,P_2)\delta(P_1-P_2)\delta((P_1+P_2)/2-P)dP_1dP_2
$$
$$
=\hat\rho(t,P,P)
$$
is the marginal density of $P(t)$. This suggest strongly that the complex valued function$W(t,Q,P)$ should be regarded as the quantum analogue of the joint probability density of $(Q(t),P(t))$. Our aim is to derive the pde satisfied by $W$ when $\rho(t)$ satisfies the quantum Fokker-Planck equation. To this end, observe that
$$
[P^2,\rho]\psi(Q)=(-\int\rho_{,11}(Q,Q')\psi(Q')dQ'+\int\rho(Q,Q')\psi''(Q')dQ'
$$
$$
=\int(-\rho_{,11}+\rho_{,22})(Q,Q')\psi(Q')dQ'
$$
which means that the position space kernel of $[P^2,\rho]$ is
$$
[P^2,\rho](Q,Q')=(-\rho_{,11}+\rho_{,22})(Q,Q')
$$
Also
$$
[U,\rho](Q,Q')=(U(Q)-U(Q'))\rho(Q,Q')
$$
$$
L^*L\rho(Q,Q')=(\bar aQ+\bar bP)(aQ+bP)\rho(Q,Q')
$$
$$
=|a|^2Q^2\rho(Q,Q')-|b|^2\rho_{,11}(Q,Q')-i\bar abQ\rho_{,1}(Q,Q')
$$
$$
-ia\bar b(\rho(Q,Q')+Q\rho_{,1}(Q,Q'))
$$
How to transform from $W(Q,P)$ to $\rho(Q,Q')?$ 
$$
W(Q,P)=\int\rho(Q+q/2,Q-q/2)exp(-iPq)dq
$$
gives
$$
\int W(Q,P)exp(iPq)dP/(2\pi)=\rho(Q+q/2,Q-q/2)
$$
or equivalently,
$$
\int W((Q+Q')/2,P)exp(iP(Q-Q'))dP/(2\pi)=\rho(Q,Q')
$$
So
$$
\rho_{,1}(Q,Q')=\int((1/2)W_{,1}((Q+Q')/2,P)+iPW((Q+Q')/2,P))exp(iP(Q-Q'))dP
$$
Also note that
$$
Q\rho(Q,Q')=\int Q W((Q+Q')/2,P)exp(iP(Q-Q'))dP=
$$
$$
=\int ((Q+Q')/2)+(Q-Q')/2)W((Q+Q')/2,P)exp(iP(Q-Q'))dP
$$
$$
\int(((Q+Q')/2)W((Q+Q')/2,P)+(i/2)W_{,2}((Q+Q')/2,P))exp(iP(Q-Q'))dP
$$
on using integration by parts. Consider now the term
$$
(U(Q)-U(Q'))\rho(Q,Q')
$$
In the classical case, $\rho(Q,Q')$ is diagonal and so such a term contributes only when
$Q'$ is in the vicinity of $Q$. For example,  if we have
$$
\rho(Q,Q')=\rho_1(Q)\delta(Q'-Q)+\rho_2(Q)\delta'(Q'-Q)
$$
then,
$$
(U(Q')-U(Q))\rho(Q,Q')=\rho_2(Q)(U(Q')-U(Q))\delta'(Q'-Q)
$$
or equivalently,
$$
\int(U(Q')-U(Q))\rho(Q,Q')f(Q')dQ'=
$$
$$
-\partial_{Q'}(\rho_2(Q)(U(Q')-U(Q))f(Q'))|_{Q'=Q}
$$
$$
=-\rho_2(Q)U'(Q)f(Q)
$$
which is equivalent to saying that
$$
(U(Q')-U(Q))\rho(Q,Q')=-\rho_2(Q)U'(Q)\delta(Q'-Q)
$$
just as in the classical case. In the general quantum case, we can write
$$
\int\rho(Q,Q')f(Q')dQ'=\sum_{n\geq 0}\rho_n(Q)\partial_Q^nf(Q)
$$
This can be seen by writing
$$
f(Q')=\sum_nf^{(n)}(Q)(Q'-Q)^n/n!
$$
and then defining
$$
\int\rho(Q,Q')(Q'-Q)^ndQ'/n!=\rho_n(Q)
$$
Now, this is equivalent to saying that the kernel of $\rho$ in the position representation is given by
$$
\rho(Q,Q')=\sum_n\rho_n(Q)\delta^{(n)}(Q-Q')
$$
and hence
$$
\int(U(Q')-U(Q))\rho(Q,Q')f(Q')dQ'=\sum_n\rho_n(Q)\int\delta^{(n)}(Q-Q')(U(Q')-U(Q))f(Q')dQ'
$$
$$
=\sum_n\rho_n(Q)(-1)^n\partial_{Q'}^n((U(Q')-U(Q))f(Q'))|_{Q'=Q}
$$
$$
=\sum_{n\geq 0,1\leq k\leq n}\rho_n(Q)(-1)^n{n\choose k}U^{(k)}(Q)f^{(n-k)}(Q)
$$
Thus,
$$
(U(Q')-U(Q))\rho(Q,Q')=
$$
$$
\sum_{n\geq 0, 1\leq k\leq n}\rho_n(Q)(-1)^k{n\choose k}U^{(k)}(Q)\delta^{(n-k)}(Q'-Q)
$$
This is the quantum generalization of the classical term
$$
-U(Q)\partial_Pf(Q,P)
$$
which appears in the classical Fokker-Planck equation. Another way to express this kernel is to directly write
$$
(U(Q')-U(Q))\rho(Q,Q')=\sum_{n\geq 1}U^{(n)}(Q)(Q'-Q)^n\rho(Q,Q')/n!
$$
From the Wigner-distribution viewpoint, it is more convenient to write
$$
U(Q')-U(Q)=U((Q+Q')/2-(Q-Q')/2)-U((Q+Q')/2+(Q-Q')/2)
$$
$$
=-\sum_{k\geq 0}U^{(2k+1)}((Q+Q')/2)(Q-Q')^{2k+1)}/(2k+1)!2^{2k}
$$
Consider a particular term $U^{(n)}((Q+Q')/2)(Q-Q')^n\rho(Q,Q')$ in this expansion:
$$
U^{(n)}((Q+Q')/2)(Q-Q')^n\rho(Q,Q')=U^{(n)}((Q+Q')/2)(Q-Q')^n\int W((Q+Q')/2,P)exp(iP(Q-Q'))dP
$$
$$
=i^n\int U^{(n)}((Q+Q')/2)\partial_P^n(W((Q+Q')/2,P))exp(iP(Q-Q'))dP
$$
and thus
$$
(U(Q')-U(Q))\rho(Q,Q')=
$$
$$
-2\int exp(iP(Q-Q'))\sum_{k\geq 0}(i/2)^{2k+1}((2k+1)!)^{-1}(U^{(2k+1)}((Q+Q')/2)\partial_P^{2k+1}W((Q+Q')/2,P))dP
$$
Also note that
$$
\rho_{,11}(Q,Q')=
$$
$$
\partial_Q^2\int W((Q+Q')/2,P)exp(iP(Q-Q'))dP
$$
$$
=\int((1/4)W_{,11}((Q+Q')/2,P)-P^2W((Q+Q')/2,P)+iPW_{,1}((Q+Q')/2,P))exp(iP(Q-Q'))dP
$$
and likewise,
$$
\rho_{,22}(Q,Q')=\int((1/4)W_{,11}((Q+Q')/2,P)-P^2W((Q+Q')/2,P)-iPW_{,1}((Q+Q')/2,P))exp(iP(Q-Q'))dP
$$
Thus, the noiseless quantum Fokker-Planck equation (or equivalently, the Liouville-Schrodinger-Von-Neumann-equation)
$$
i\rho'(t)=[(P^2/2m+U(Q)),\rho(t)]
$$
is equivalent to
$$
i\partial_t(W(t,(Q+Q')/2,P))=
$$
$$
(-iP/m)W_{,1}(t,(Q+Q')/2,P)
$$
$$
+2\sum_{k\geq 0}(i/2)^{2k+1}((2k+1)!)^{-1}(U^{(2k+1)}((Q+Q')/2)\partial_P^{2k+1}W(t,(Q+Q')/2,P))
$$
or equivalently,
$$
i\partial_tW(t,Q,P)=
$$
$$
(-iP/m)W_{,1}(t,Q,P)+U'(Q)\partial_PW(t,Q,P)+
$$
$$
+2\sum_{k\geq 0}(i/2)^{2k+1}((2k+1)!)^{-1}(U^{(2k+1)}(Q)\partial_P^{2k+1}W(t,Q,P))
$$
This can be expressed in the form of a classical Liouville equation with quantum correction terms:
$$
\partial_tW(t,Q,P)=(-P/m)\partial_QW(t,Q,P)+U'(Q)\partial_PW(t,Q,P)
$$
$$
+\sum_{k\geq 1}((-1)^k/(2^{2k}(2k+1)!))U^{(2k+1)}(Q)\partial_P^{2k+1}W(t,Q,P)
$$
The last term is the quantum correction term.

Remark: To see that the last term is indeed a quantum correction term, we have to use general units in which Planck's constant is not set to unity. Thus, we define
$$
W(Q,P)=C\int\rho(Q+q/2,Q-q/2).exp(-iPq/h)dq
$$
For $\int W(Q,P)dP=\rho(Q,Q)$, we must set 
$$
2\pi hC=1
$$
Then,
$$
\rho(Q+q/2,Q-q/2)=\int W(Q,P)exp(iPq/h)dP
$$
Thus, with $P=-ih\partial_Q$, we get
$$
[P^2/2m,\rho](Q,Q')=(h^2/2m)(-\rho_{,11}(Q,Q')+\rho_{,22}(Q,Q'))
$$
$$
=(h^2/2m)(-\partial_Q^2+\partial_{Q'}^2)\int W((Q+Q')/2,P)exp(iP(Q-Q')/h)dP
$$
$$
=(-ih/m)\int PW_{,1}((Q+Q')/2,P)exp(iP(Q-Q')/h)dP
$$
Further,
$$
(U(Q)-U(Q'))\rho(Q,Q')=\sum_{k\geq 0}2^{-2k}((2k+1)!)^{-1}U^{(2k+1)}((Q+Q')/2)(Q-Q')^{2k+1}\rho(Q,Q')
$$
$$
=\int\sum_k2^{-2k}((2k+1)!)^{-1}(ih)^{2k+1}U^{(2k+1)}((Q+Q')/2)\partial_P^{2k+1}W((Q+Q')/2,P)exp(iP(Q-Q')/h)dP
$$
and hence its we get from the Schrodinger equation
$$
ih\partial_t\rho=[H,\rho]=[P^2/2m,\rho]+[U(Q),\rho],
$$
the equation
$$
ih\partial_tW(t,Q,P)=(-ih/m)PW_{,1}((Q+Q')/2,P)
$$
$$
+ihU'(Q)\partial_PW(t,Q,P)+ih\sum_{k\geq 1}(-1)^k2^{-2k}h^{2k}((2k+1)!)^{-1}U^{(2k+1)}(Q)\partial_P^{2k+1}W(t,Q,P)
$$
or equivalently,
$$
\partial_tW(t,Q,P)=(-P/m)\partial_QW(t,Q,P)+U'(Q)\partial_PW(t,Q,P)
$$
$$
+\sum_{k\geq 1}(-h^2/4)^k((2k+1)!)^{-1}U^{(2k+1)}(Q)\partial_P^{2k+1}W(t,Q,P)
$$
which clearly shows that the quantum correction to the Liouville equation is a power series in $h^2$ beginning with the first power of $h^2$. Note that the quantum corrections involve $\partial_P^{2k+1}W(t,Q,P), k=1,2,...$ but a diffusion term of the form
$\partial_P^2W$ is absent. It will be present only when we include the Lindblad terms for it is these terms that describe the effect of a noisy bath on our quantum system.
\bigskip

{\bf Construction of the quantum Fokker-Planck equation for a specific choice of the Lindblad operator}

Here,
$$
H=P^2/2m+U(Q), L=g(Q)
$$
Assume that $g$ is a rea; function. Then, the Schrodinger equation for the mixed state
$\rho$ is given by
$$
\partial_t\rho=-i[H,\rho]-(1/2)\theta(\rho)
$$
where
$$
\theta(\rho)=L^*L\rho+\rho L^*L-2L\rho L^*
$$
Thus, the kernel of this in the position representation is given by
$$
\theta(\rho)(Q,Q')=(g(Q)^2+g(Q')^2-2g(Q)g(Q'))\rho(Q,Q')
$$
$$
=(g(Q)-g(Q'))^2\rho(Q,Q')
$$
More generally, if
$$
\theta(\rho)=\sum_k(L_k^*L_k\rho+\rho L_k^*L_k-2L_k\rho L_k)
$$
where
$$
L_k=g_k(Q)
$$
are real functions, then
$$
\theta(\rho)(Q,Q')=F(Q,Q')\rho(Q,Q')
$$
where
$$
F(Q,Q')=\sum_k(g_k(Q)-g_k(Q'))^2
$$
We define
$$
G(Q,q)=F(Q+q/2,Q-q/2)
$$
or equivalently,
$$
F(Q,Q')=G((Q+Q')/2,Q-Q')
$$
Then,
$$
\theta(\rho)(Q,Q')=G((Q+Q')/2,(Q-Q')\rho(Q,Q')=
$$
$$
\sum_{n\geq 0}G_n((Q+Q')/2)(Q-Q')^n\int W((Q+Q')/2,P)exp(iP(Q-Q'))dP
$$
$$
=\sum_{n\geq 0}G_n((Q+Q')/2)i^n\int\partial_P^nW((Q+Q')/2,P)exp(iP(Q-Q'))dP
$$
and this contributes a factor of
$$
-(1/2)\sum_{n\geq 0}i^nG_n(Q)\partial_P^nW(Q,P)
$$
$$
=(-1/2)G_0(Q)W(Q,P)-(i/2)G_1(Q)\partial_PW(Q,P)+(1/2)G_2(Q)\partial_P^2W(Q,P)
$$
$$
-(1/2)\sum_{n>2}i^nG_n(Q)\partial_P^nW(Q,P)
$$
The term $(1/2)G_2(Q)\partial_P^2W(Q,P)$ is the quantum analogue of the classical diffusion term in the Fokker-Planck equation. Here, we are assuming an expansion
$$
G(Q,q)=\sum_nG_n(Q)q^n
$$
ie,
$$
G_n(Q)=n!^{-1}\partial_q^nG(Q,q)|_{q=0}
$$
\bigskip

{\bf Problems in quantum corrections to classical theories in probability theory and in mechanics}

Consider the Lindblad term in the density evolution problem:
$$
\theta(\rho)=L^*L\rho+\rho L^*L-2L\rho L^*)
$$
where
$$
L=g_0(Q)+g_1(Q)P
$$
We have
$$
L^*=\bar g_0(Q)+P\bar g_1(Q)
$$
$$
L^*L\psi(Q)=\bar g_0+P\bar g_1)(g_0+g_1P)\psi(Q)=|g_0(Q)|^2\psi(Q)+P|g_1(Q)|^2P\psi(Q)+\bar g_0g_1P\psi(Q)+P\bar g_1g_0\psi(Q)
$$
Now, defining
$$
|g_1(Q)|^2=f_1(Q), |g_0(Q)|^2=f_0(Q),
$$
we get
$$
Pf_1(Q)P\psi(Q)=-\partial_Qf_1(Q)\psi'(Q)=-f_1'(Q)\psi'(Q)-f_1(Q)\psi''(Q)
$$
$$
P\bar g_1(Q)g_0(Q)\psi(Q)=-i(\bar g_1g_0)'(Q)\psi(Q)-\bar g_1(Q)g_0Q)\psi'(Q)
$$
Likewise for the other terms. Thus, $L^*L$ has the following kernel in the position representation:
$$
L^*L(Q,Q')=-f_1(Q)\delta''(Q-Q')+f_2(Q)\delta'(Q-Q')+f_3(Q)\delta(Q-Q')
$$
where $f_1(Q)$ is real positive and $f_2,f_3$ are complex functions. Thus,
$$
L^*L\rho(Q,Q')=-f_1(Q)\partial_Q^2\rho(Q,Q')+f_2(Q)\partial_Q\rho(Q,Q')+f_3(Q)\rho(Q,Q')
$$
$$
\rho L^*L(Q,Q')=\int\rho(Q,Q'')L^*L(Q'',Q')dQ''=-\partial_{Q'}^2(\rho(Q,Q')f_1(Q'))-
$$
$$
\partial_{Q'}(\rho(Q,Q')f_2(Q'))+\rho(Q,Q')f_3(Q')
$$
Likewise,
$$
L\rho L^*(Q,Q')=g_1(Q)\bar g_1(Q')\partial_Q\partial_{Q'}\rho(Q,Q')+T
$$
where $T$ contains terms of the form a function of $Q,Q'$ times first order partial derivatives in $Q,Q'$. Putting all the terms together, we find that tbe Lindblad contribution to the master equation has the form
$$
(-1/2)\theta(\rho)(Q,Q')=(f_1(Q)\partial_Q^2+f_1(Q')\partial_{Q'}^2)\rho(Q,Q')-g_1(Q)\bar g_1(Q')\partial_Q\partial_{Q'}\rho(Q,Q')+h_1(Q,Q')\partial_Q\rho(Q,Q')+
$$
$$
h_2(Q,Q')\partial_{Q'}\rho(Q,Q')+h_3(Q,Q')\rho(Q,Q')
$$
\bigskip

\section{Motion of an element having both an electric and a magnetic dipole moment in an external electromagnetic field plus a coil}

Let $m, p$ denote respectively the magnetic and electric dipole moments of the element at time $t=0$. Let $E(t,r),B(t,r)$ denote the external electric and magnetic fields. After time $t$, the element undergoes a rotation
$S(t)\in SO(3)$ so that its magnetic moment becomes
$$
m(t)=S(t)m,
$$
and its electric dipole moment becomes
$$
p(t)=S(t)p
$$
Assume that after time $t$, the centre of the element is located at the position $R(t)$. It should be noted that if $n(t)$ denotes the unit vector along the length of the element at time $t$, then we can write
$$
m(t)=K_1n(t), p(t)=K_2n(t)
$$
where $K_1,K_2$ are constants. Thus, we can write
$$
m(t)=K_1S(t)n(0), p(t)=K_2S(t)n(0)
$$
The angular velocity tensor $\Omega(t)$ in the lab frame can be expressed using
$$
S'(t)=\Omega(t)S(t)
$$
or equivalently,
$$
\Omega(t)=S'(t)S(t)^{-1}=S'(t)S(t)^T
$$
The angular velocity tensor in the rest frame of the element is
$$
\Omega_r(t)=S(t)^{-1}\Omega(t)S(t)=S(t)^{-1}S'(t)=S(t)^TS'(t)
$$
The total kinetic energy of the element due to translational moment of its cm and rotational motion around its cm is given by
$$
K(t)=(1/2)Tr(S'(t)JS'(t)^T)+MR'(t)^2/2
$$
the interaction potential energy between the magnetic moment of the element and the magnetic field plus that between the electric dipole moment and the electric field is given by
$$
V(t)=-(m(t),B(t,R(t)))-(p(t),E(t,R(t))=-m^TS(t)^TB(t,R(t))-p^TS(t)^TE(t,R(t))
$$
and hence taking into account the gravitational potential energy, the total Lagrangian of the element is given by
$$
L(R(t),R'(t),S(t),S'(t))=(1/2)Tr(S'(t)^TJS'(t))+MR'(t)^2/2+m^TS(t)^TB(t,R(t))+p^TS(t)^TE(t,R(t))-Mge_3^TR(t)
$$
where $e_3$ is the unit vector along the $z$ direction. While writing the Euler-Lagrange equations of motion from this Lagrangian, we must take into account the constraint $S(t)^TS(t)=I$ by adding a Lagrange multiplier term
$-Tr(\Lambda(t)(S(t)^TS(t)-I))$ to the above Lagrangian.
\bigskip

Remark: The magnetic vector potential produced by a current loop $\Gamma$ carrying a current $I(t)$ is given by the retarded potential formula in the form of a line integral
$$
A(t,r)=(\mu/4\pi)\int_{\Gamma}I(t-|r-r'|/c)dr'/|r-r'|
$$
This approximates in the far field zone $|r|>>|r'|$ to
$$
\int_{\Gamma}I(t-r/c+\hat r.r'/c)(1+\hat r.r'/r)dr'/r
$$
which further approximates to
$$
(I(t-r/c)/r^2)\int_{\Gamma}\hat r.r'dr'+(I'(t-r/c)/cr)\int_{\Gamma}(\hat r.r')dr'
$$
$$
=(I(t-r/c)/r^3)\int_{\Gamma}r.r'dr'+(I'(t-r/c)/cr^2)\int_{\Gamma}r.r'dr'
$$
which can be expressed in terms of the magnetic moment
$$
m(t)=I(t)\int_{\Gamma}r'\times dr'
$$
as
$$
A(t,r)=(\mu/4\pi)[m(t-r/c)\times r/r^3+m'(t-r/c)\times r/cr^2]
$$

Remark: The interaction energy between a system of charged particles in motion and a magnetic field is derived from the Hamiltonian term
$$
\sum_i(p_i-e_iA(t,r_i))^2/2m_i
$$
The interaction term in this expression is
$$
-\sum_ie_i(p_i,A(t,r_i))/m_i
$$
However, noting the relationship between momentum and velocity in a magnetic field: $p_i=m_iv_i+e_iA(t,r_i)$, it follows that the interaction energy should be taken as
$$
-\sum_ie_i(v_i,A(t,r_i))
$$
Now if the system of charged particles spans a small region of space around the point $r$ and in this region, the magnetic field does not vary too rapidly, then we can approximate the above interaction energy term by
$$
U=-\sum_ie_i(v_i,A(t,r+r_i'))=-\sum_ie_i(v_i,(r_i',\nabla)A(t,r))
$$
where $r_i'=r_i-r$. Now assuming $A$ to be time independent,
$$
d/dt(r_i',(r_i',\nabla)A)=(v_i,(r_i',\nabla)A)+(r_i',(v_i,\nabla)A)
$$
For bounded motion, the time average of the rhs is zero and we get denoting time averages by $<.>$,
$$
<(v_i,(r_i',\nabla)A)>=-<(r_i',(v_i,\nabla)A)>
$$
Hence,
$$
<(v_i,(r_i',\nabla)A)>=(1/2(<(v_i,(r_i',\nabla)A)-(r_i',(v_i,\nabla)A)>)
$$
On the other hand,
$$
(r_i'\times v_i).(nabla\times A)=(v_i,(r_i',\nabla)A)-(r_i',(v_i,\nabla)A)
$$
and hence we deduce that
$$
<U>=-\sum_i(e_ir_i'\times v_i/2).curl A(t,r)
$$
provided that $A$ does not vary too rapidly with time. This formula can be expressed as
$$
U=-m.B
$$
where
$$
m=(1/2)\sum_ie_ir_i'\times v_i
$$
is the magnetic moment of the system of particles.
\bigskip

\chapter{Large deviations for Markov chains, quantum master equation, Wigner distribution for the Belavkin filter}

\section{Large deviation rate function for the empirical distribution of a Markov chain}

Assume that $X(n),n\geq 0$ is a discrete time Markov chain with finite state space
$E=\{1,2,..., K\}$ and with transition probabilities
$$
P(X(n+1)=y|X(n)=x)=\pi(x,y), x,y\in E
$$
let $\mu$ be the distribution of $X(1)$. The empirical distribution of the chain based on data collected upto time $N$ is given by
$$
\nu_N=N^{-1}\sum_{n=1}^N\delta_{X(n)}
$$
Then the logarithmic generating function of $\nu_N$ is given by
$$
\Lambda_N(f)=log\Bbb Eexp(\int f(x)d\nu_N(x))=log\Bbb E[exp(N^{-1}\sum_{n=1}^Nf(X(n)))]
$$
and hence,
$$
\Lambda_N(Nf)=log\Bbb E[exp(\sum_{n=1}^Nf(X(n)))]=
$$
$$
log(\mu^T\pi_f^N(1))
$$
where
$$
\pi_f=((exp(f(x))\pi(x,y)))_{1\leq x,y\leq K}
$$
and 
$$
\pi_fu(x)=\sum_y\pi_f(x,y)u(y)
$$
It follows from the spectral theorem for matrices that
$$
lim_{N\rightarrow\infty}N^{-1}\Lambda_N(Nf)=log(\lambda(f))
$$
where $\lambda(f)$ is the maximum eigenvalue of $\pi_f$. By the Perron-Frobenius theory, this is real. Equivalently, this fact follows from the above relation since the limit is real because each term is real and the limit equals the eigenvalue of $\pi_f$ having maximum magnitude. Thus if $q$ is a probability distribution on $E$, then the LDP rate function of the process evaluated at $q$ is given by
$$
I(q)=sup_f(\sum_xf(x)q(x)-log\lambda(f))
$$
We claim that
$$
I(q)=J(q)=sup_{u>0}\sum_xq(x)log(u(x)/\pi u(x))
$$
where
$$
\pi u(x)=\sum_y\pi(x,y)u(y)
$$
To see this, we first observe that for any $f,u$, we have
$$
\sum_xf(x)q(x)-\sum_xq(x)log(u(x)/\pi u(x))=-\sum_xq(x)log(u(x)/\pi_fu(x))
$$
Now fix $f$ and choose $u=u_0$ so that
$$
\pi_fu_0(x)=\lambda(f)u_0(x), x\in E
$$
ie $u_0$ is an eigenvector of $\pi_f$ corresponding to its maximum eigenvalue. Then, 
$$
\sum_xf(x)q(x)-J(q)=inf_u(\sum_xf(x)q(x)-\sum_xq(x)log(u(x)/\pi u(x))
$$
$$
=inf_u-\sum_xq(x)log(u(x)/\pi_fu(x))\leq -\sum_xlog(u_0(x)/\pi_fu_0(x))=log\lambda(f)
$$
Thus, for any $f$,
$$
\sum_xf(x)q(x)-log\lambda(f)\leq J(q)
$$
and taking supremum over all $f$, gives us
$$
I(q)\leq J(q)
$$
In order to show that
$$
J(q)\leq I(q)
$$
it suffices to show that for every $u$, there exists an $f$ such that
$$
\sum_xq(x)log(u(x)/\pi u(x))\leq\sum_xf(x)q(x)-log\lambda(f)
$$
or equivalently, such that
$$
-\sum_xq(x)log(u(x)/\pi_fu(x))=log\lambda(f)---(a)
$$
To show this fix $u$ and define
$$
f(x)=log(u(x)/\pi u(x))
$$
Then, clearly
$$
\pi_fu(x)=u(x)
$$
and hence for every $n\geq 1$,
$$
\pi_f^nu=u
$$
and hence
$$
log(\lambda(f))=lim_nn^{-1}log(\mu^T\pi_f^nu)=0
$$
ie
$$
\lambda(f)=1
$$
But then, both sides of (a) equal zero and the proof is complete.
\bigskip

\section{Quantum Master Equation in general relativity}

Assume that the metric of space-time is restricted to have the form
$$
d\tau^2=(1+\delta.V_0(x))dt^2-\sum_{k=1}^3(1+\delta.V_k(x))(dx^k)^2
$$
where $V_0,V_k,k=1,2,3$ are some functions of $x=(t,x^k,k=1,2,3)$. The corresponding metric is given by
$$
g_{00}=1+\delta.V_0, g_{kk}=-(1+\delta.V_k), k=1,2,3
$$
and
$$
g_{\mu\nu}=0,\mu\neq\nu
$$
We substitute this into the Einstein-Hilbert Lagrangian density
$$
L_g=g^{\mu\nu}\sqrt{-g}(\Gamma^{\alpha}_{\mu\nu}\Gamma^{\beta}_{\alpha\beta}-\Gamma^{\alpha}_{\mu\beta}\Gamma^{\beta}_{\nu\alpha})
$$
and retain terms only upto $O(\delta^2)$. Likewise, we compute the Lagrangian density of the electromagnetic field 
$$
L_{em}=(-1/4)F_{\mu\nu}F^{\mu\nu}
$$
and retain terms only upto $O(\delta)$, ie, terms linear in the $V_{\mu}$. The resulting Lagrangian of the gravitational field interacting with the external em field will have the form
$$
L=L_g+L_{em, int}=\delta^2C_1(\mu\nu\alpha\beta)V_{\mu,\alpha}V_{\nu,\beta}+
$$
$$
+\delta.C_2(\mu\alpha\beta\rho\sigma)V_{\mu}F_{\alpha\beta}F_{\rho\sigma}
$$
where we regard the covariant components $A_{\mu}$ of the electromagnetic four potential as being fundamental and the contravariant components being derived from the covariant components after interaction with the gravitational field. Specifically, with neglect of $O(\delta^2)$ terms,
$$
A^0=g^{0\mu}A_{\mu}=g^{00}A_0=(1-\delta.V_0)A_0
$$
$$
A^r=g^{rr}A_r=-(1-\delta.V_r)A_r
$$
We regard the field
$$
B_{\mu}=C_2(\mu\alpha\beta\rho\sigma)F_{\alpha\beta}F_{\rho\sigma}
$$
as a new c-number external field with which the gravitational field interacts and then write down our gravitational field Lagrangian as
$$
L=\delta^2C_1(\mu\nu\alpha\beta)V_{\mu,\alpha}V_{\nu,\beta}+\delta.V_{\mu}B_{\mu}
$$
Keeping this model in mind, it can be abstracted into the Lagrangian of a set of Klein-Gordon field interacting with an external c-number field. The canonical momenta are
$$
P_{\mu}=\partial L/\partial V_{\mu,0}=2\delta^2C_1(\mu\nu 0\beta)V_{\nu,\beta}
$$
$$
=2\delta^2C_1(\mu\nu 00)V_{\nu,0}+2\delta^2C_1(\mu\nu 0r)V_{\nu,r}
$$
which can be easily solved to express $V_{\mu,0}$ as a linear combination of
$P_{\nu}$ and $V_{\nu,r}$. In this way our Hamiltonian density has the form
$$
H=C_1(\mu\nu)P_{\mu}P_{\nu}+C_2(\mu\nu r)P_{\mu}V_{\nu,r}
$$
$$
+C_3(\mu\nu rs)V_{\mu,r}V_{\nu,s}+C_4(\mu\nu)V_{\mu}B_{\nu}
$$
Abstracting this to one dimension, we get a Hamiltonian of the form
$$
H=P^2/2+\omega^2Q^2/2+\beta(PQ+QP)+\gamma E(t)Q
$$
where $E(t)$ is an external c-number function of time. Application of a canonical transformation then gives us a Hamiltonian of the form
$$
H=P^2/2+\omega^2Q^2/2+E(t)(aP+bQ)
$$
If $E(t)$ is taken as quantum noise coming from a bath, then we can derive the following Hudson-Parthasarathy noisy Schrodinger equation for the evolution operator $U(t)$:
$$
dU(t)=(-(iH+LL^*/2)dt+LdA(t)-L^*dA(t)^*)
$$
where
$$
L=aQ+bP
$$
\bigskip

\section{Belavkin filter for the Wigner distirbution function}

The HPS dynamics is
$$
dU(t)=(-(iH+LL^*/2)dt+LdA(t)-L^*dA(t)^*)U(t)
$$
where
$$
L=aP+bQ, H=P^2/2+U(Q)
$$
The coherent state in which the Belavkin filter is designed to operate is 
$$
|\phi(u)>=exp(-|u|^2/2)|e(u)>, u\in L^2(\Bbb R_+)
$$
The measurement process is
$$
Y_o(t)=U(t)^*Y_i(t)U(t), Y_i(t)=cA(t)+\bar cA(t)^*
$$
Then,
$$
dY_o(t)=dY_i(t)-j_t(\bar cL+cL^*)dt
$$
where
$$
j_t(X)=U(t)^*XU(t)
$$
Note that
$$
\bar cL+cL^*=\bar c(aP+bQ)+c(\bar aQ+\bar bP)=2Re(\bar ac)Q+2Re(\bar bc)P
$$
Thus, our measurement model corresponds to measuring the observable
$2Re(\bar ac)Q(t)+2Re(\bar bc)P(t)$ plus white Gaussian noise. Let
$$
\pi_t(X)=\Bbb E(j_t(X)|\eta_o(t)), \eta_o(t)=\sigma(Y_o(s):s\leq t)
$$
Let
$$
d\pi_t(X)=F_t(X)dt+G_t(X)dY_o(t)
$$
with
$$
F_t(X),G_t(X)\in\eta_o(t)
$$
Let
$$
dC(t)=f(t)C(t)dY_o(t), C(0)=1
$$
The orthogonality principle gives
$$
\Bbb E[(j_t(X)-\pi_t(X))C(t)]=0
$$
and hence by quantum Ito's formula and the arbitrariness of the complex valued function
$f(t)$, we have
$$
\Bbb E[(dj_t(X)-d\pi_t(X))|\eta_o(t)]=0,
$$
$$
\Bbb E[(j_t(X)-\pi_t(X))dY_o(t)|\eta_o(t)]+\Bbb E[(dj_t(X)-d\pi_t(X))dY_o(t)|\eta_o(t)]=0
$$
Note that
$$
dj_t(X)=j_t(\theta_0(X))dt+j_t(\theta_1(X))dA(t)+j_t(\theta_2(X))dA(t)^*
$$
where
$$
\theta_0(X)=i[H,X]-(1/2)(LL^*X+XLL^*-2LXL^*),
$$
$$
\theta_1(X)=-LX+XL=[X,L], \theta_2(X)=L^*X-XL^*=[L^*,X]
$$
Then the above conditions give
$$
\pi_t(\theta_0(X))+\pi_t(\theta_1(X))u(t)+\pi_t(\theta_2(X))\bar u(t)=
$$
$$
F_t(X)+G_t(X)(cu(t)+\bar c\bar u(t)-\pi_t(\bar cL+cL^*))
$$
and
$$
-\pi_t(X(\bar cL+cL^*))+\pi_t(X)\pi_t(\bar cL+cL^*)+\pi_t(\theta_1(X))\bar c
$$
$$
-G_t(X)|c|^2=0
$$
This second equation simplifies to
$$
G_t(X)=|c|^{-2}(-\pi_t(cXL^*+\bar cLX)+\pi_t(\bar cL+cL^*)\pi_t(X))
$$
Then,
$$
d\pi_t(X)=(\pi_t(\theta_0(X))+\pi_t(\theta_1(X))u(t)+\pi_t(\theta_2(X))\bar u(t))dt
$$
$$
+G_t(X)(dY_o(t)+\pi_t(\bar cL+cL^*-cu(t)-\bar c\bar u(t))dt)
$$
In the special case when $c=-1$, we get
$$
d\pi_t(X)=(\pi_t(\theta_0(X))+\pi_t(\theta_1(X))u(t)+\pi_t(\theta_2(X))\bar u(t))dt
$$
$$
+(\pi_t(XL^*+LX)-\pi_t(L+L^*)\pi_t(X))(dY_o(t)-(\pi_t(L+L^*)-2Re(u(t)))dt)
$$
\bigskip

Consider the simplified case when $u=0$, ie, filtering is carried out in the vacuum coherent state. Then, the above Belavkin filter simplifies to
$$
d\pi_t(X)=\pi_t(\theta_0(X))dt+(\pi_t(XL^*+LX)-\pi_t(L+L^*)\pi_t(X))(dY_o(t)-\pi_t(L+L^*)dt)
$$
Equivalently in the conditional density domain, the dual of the above equation gives
the Belavkin filter for the conditional density
$$
\rho'(t)=\theta_0^*(\rho(t))dt+(L^*\rho(t)+\rho(t)L-Tr(\rho(t)(L+L^*))\rho(t))(dY_o(t)-Tr(\rho(t)(L+L^*))dt)
$$
where
$$
\theta_0^*(\rho)=-i[H,\rho]-(1/2)(LL^*\rho+\rho LL^*-2L^*\rho L)
$$
Now let us take
$$
L=aQ+bP, a,b\in\Bbb C
$$
and
$$
H=P^2/2+U(Q)
$$
and translate this Belavkin equation to the Wigner-distribution domain and then compare the resulting differential equation for $W$ with the  Kushner-Kallianpur filter for the classical conditional density. We write
$$
\rho(t,Q,Q')=\int W(t,(Q+Q')/2,P)exp(iP(Q-Q'))dP
$$
Note that the Kushner-Kallianpur filter for the corresponding classical probabilistic problem
$$
dQ=Pdt, dP=-U'(Q)dt-\gamma Pdt+\sigma dB(t),
$$
$$
dY(t)=(2Re(a)Q(t)+2Re(b)P(t))dt+\sigma dV(t)
$$
where $B,V$ are independent identical Brownian motion processes is given by
$$
dp(t,Q,P)=L^*p(t,Q,P)+((\alpha Q+\beta P)p(t,Q,P)-(\int(\alpha Q+\beta P)p(t,Q,P)dQdP)p(t,Q,P))(dY(t)-dt\int (\alpha Q+\beta P)p(t,Q,P)dQdP)
$$
where
$$
\alpha=2Re(a), \beta=2Re(b)
$$
where
$$
L^*p=-P\partial_Qp+U'(Q)\partial_Pp+\gamma.\partial_P(Pf)+(\sigma^2/2)\partial_P^2p
$$
Let us see how this Kushner-Kallianpur equation translates in the quantum case using the Wigner distribution $W$ in place of $p$. As noted earlier, the equation
$$
\rho'(t)=-i[H,\rho(t)]
$$
is equivalent to
$$
\partial_tW(t,Q,P)=(-P/m)\partial_QW(t,Q,P)+U'(Q)\partial_PW(t,Q,P)
$$
$$
+\sum_{k\geq 1}(-h^2/4)^k((2k+1)!)^{-1}U^{(2k+1)}(Q)\partial_P^{2k+1}W(t,Q,P)---(a)
$$
while the Lindblad correction term to (a) is $(-1/2)$ times
$$
LL^*\rho+\rho LL^*-2L^*\rho L
$$
$$
=(aQ+bP)(\bar aQ+\bar bP)\rho+\rho(aQ+bP)(\bar aQ+\bar bP)-2(\bar aQ+\bar bP)\rho(aQ+bP)
$$
which in the position representation has a kernel of the form
$$
-|b|^2(\partial_Q^2+\partial_{Q'}^2+2\partial_Q\partial_{Q'})-|a|^2(Q-Q')^2
$$
$$
+(c_1Q+c_2Q'+c_3)\partial_Q+(c_4Q+c_5Q'+c_6)\partial_{Q'}]\rho(t,Q,Q')
$$
This gives a Lindblad correction to (a) of the form
$$
(\sigma_1^2/2)\partial_Q^2W(t,Q,P)+(\sigma_2^2/2)\partial_P^2W(t,Q,P)
$$
$$
+d_1\partial_Q\partial_PW(t,Q,P)+(d_3Q+d_4P+d_5)\partial_QW(t,Q,P)
$$
Thus, the master/Lindblad equation (in the absence of measurements) has the form
$$
\partial_tW(t,Q,P)=(-P/m)\partial_QW(t,Q,P)+U'(Q)\partial_PW(t,Q,P)
$$
$$
+\sum_{k\geq 1}(-h^2/4)^k((2k+1)!)^{-1}U^{(2k+1)}(Q)\partial_P^{2k+1}W(t,Q,P)
$$
$$
+(\sigma_1^2/2)\partial_Q^2W(t,Q,P)+(\sigma_2^2/2)\partial_P^2W(t,Q,P)
$$
$$
+d_1\partial_Q\partial_PW(t,Q,P)+(d_3Q+d_4P+d_5)\partial_QW(t,Q,P)
$$

Remark: Suppose $b=0$ and $a$ is replaced by $a/h$. Then, taking into account the fact that in general units, the factor $exp(iP(Q-Q'))$ must be replaced by $exp(iP(Q-Q'))/h)$ in the expression for $\rho(t,Q,Q')$ in terms of $W(t,Q,P)$, we get the Lindblad correction as
$$
(|a|^2/2)(Q-Q')^2\rho(t,Q,Q')
$$
or equivalently in the Wigner-distribution domain as
$$
(|a|^2/2)\partial_P^2W(t,Q,P)
$$
and the master equation simplifies to
$$
\partial_tW(t,Q,P)=(-P/m)\partial_QW(t,Q,P)+U'(Q)\partial_PW(t,Q,P)+(|a|^2/2)\partial_P^2W(t,Q,P)
$$
$$
+\sum_{k\geq 1}(-h^2/4)^k((2k+1)!)^{-1}U^{(2k+1)}(Q)\partial_P^{2k+1}W(t,Q,P)
$$
This equation clearly displays the classical terms and the quantum correction terms.
\bigskip

Exercise:Now express the terms $L^*\rho(t), \rho(t)L^*$ and $Tr(\rho(t)(L+L^*)$ in terms of the Wigner distribution for $\rho(t)$ assuming $L=aQ+bP$ and complete the formulation of the Belavkin filter in terms of the Wigner distribution. Compare with the corresponding classical Kushner-Kallianpur filter for the conditional density of $(Q(t),P(t))$ given noisy measurements of $aQ(s)+bP(s),s\leq t$ by identifying the precise form of the quantum correction terms.

\chapter{String and field theory with large deviations,stochastic algorithm convergence}

\section{String theoretic corrections to field theories}

Let 
$$
X^{\mu}(\tau,\sigma)=x^{\mu}+p^{\mu}\tau-i\sum_{n\neq 0}a^{\mu}(n)exp(in(\tau-\sigma))/n
$$
be a string field where
$$
a^{\mu}(-n)=a^{\mu*}(n)
$$
and
$$
[a^{\mu}(n),a^{\nu}(m)]=\eta^{\mu\nu}\delta(n+m)
$$
When this string interacts with an external c-number gauge field described by the antisymmetric potentials 
$$
B_{\mu\nu}(X(\tau,\sigma))
$$
The total string Lagrangian is given by
$$
L=(1/2)\eta^{\alpha\beta}X^{\mu}_{,\alpha}X_{\mu,\beta}-(1/2)B_{\mu\nu}\epsilon^{\alpha\beta}X^{\mu}_{,\alpha}X^{\nu}_{,\beta}
$$
The string equations are given by
$$
\eta^{\alpha\beta}X_{\mu,\alpha\beta}=-\epsilon^{\alpha\beta}(B_{\mu\nu}(X)X^{\nu}_{,\beta})_{,\alpha})=-\epsilon^{\alpha\beta}B_{\mu\nu,\rho}(X)X^{\rho}_{,\alpha}X^{\nu}_{,\beta}
$$
This is the string theoretic version of the following point particle theory for a charged particle in an electromagnetic field:
$$
MX_{\mu,\tau\tau}=q.F_{\mu\nu}(X)X^{\nu}_{,\tau}
$$
In the case when the field $B_{\mu\nu,\rho}$ is a constant, this gives corrections to the string field that are quadratic functions of the $\{a^{\mu}(n)\}$:
$$
X_{\mu}=X_{\mu}^{(0)}+X_{\mu}^{(1)}
$$
where
$$
X_{\mu}^{(0)}=x^{\mu}+p^{\mu}\tau-i\sum_{n\neq 0}a^{\mu}(n)exp(in(\tau-\sigma))/n
$$
$$
-i\sum_{n\neq 0}b^{\mu}(n).exp(in(\tau+\sigma))/n
$$
and
$$
\eta_{\alpha\beta}X^{(1)}_{\mu,\alpha\beta}=-\epsilon^{\alpha\beta}B_{\mu\nu,\rho}(X)X^{(0)\rho}_{,\alpha}X^{(0)\nu}_{,\beta}
$$
$$
=\epsilon^{\alpha\beta}B_{\mu\nu,\rho}(a^{\rho}(n)b^{\nu}(m)u_{\alpha}v_{\beta}exp(i(n(\tau-\sigma)+m(\tau+\sigma)))
$$
$$
+b^{\rho}(n)a^{\nu}(m)v_{\alpha}u_{\beta}exp(i(n(\tau+\sigma)+m(\tau-\sigma))))
$$
where
$$
(u_{\alpha})_{\alpha=0,1}=(1,-1), (v_{\alpha})_{\alpha=0,1}=(1,1)
$$
\bigskip

\section{Large deviations problems in string theory}

Consider first the motion of a point particle with Poisson noise. Its equation of motion is
$$
dX(t)=V(t)dt, dV(t)=-\gamma V(t)dt-U'(X(t))dt+\sigma\epsilon dN(t)
$$
The rate function of the Poisson process is computed as follows. Let $\lambda(\epsilon)$ be the mean arrival rate of the Poisson process. Then
$$
M_{\epsilon}(f)=\Bbb E[exp(\epsilon\int_0^Tf(t)dN(t))]=
$$
$$
exp((\lambda/\epsilon)\int_0^T(exp(\epsilon f(t))-1)dt)
$$
Thus the Gartner-Ellis logarithmic moment generating function of the family of processes
$\epsilon.N(.)$ over the time interval $[0,T]$ is given by
$$
\Lambda(f)=\epsilon.log(M_{\epsilon}(f/\epsilon))=\lambda\int_0^T(exp(f(t))-1)dt
$$
So the rate function of this family of processes is given by
$$
I_T(\xi)=sup_f(\int_0^Tf(t)\xi'(t)dt-\Lambda(f))
$$
$$
=\int_0^Tsup_x(\xi'(t)x-\lambda(exp(x)-1))dt=\int_0^T\eta(\xi'(t))dt
$$
where
$$
\eta(y)=sup_x(xy-\lambda(exp(x)-1))
$$
By the contraction principle, the rate function of the process $\{X(t):0\leq t\leq T\}$ is given by
$$
J_T(X)=\int_0^T\eta(X''(t)+\gamma X'(t)+U'(X(t)))dt
$$
assuming $\sigma=1$. 

Application to control: let the potential $U(x|\theta)$ have some control parameters $\theta$. We choose $\theta$ so that
$$
inf_{\parallel X-X_d\parallel>\delta}J_T(X)
$$
is as large as possible. This would guarantee that the probability of deviation of $X$
from desired noiseless trajectory $X_d$ obtained with $\theta=\theta_0$ by an amount more than $\delta$ is as small as possible. To proceed further, we write
$$
e(t)=X(t)-X_d(t)
$$
and then with $\theta=\theta_0+\delta\theta$,
$$
\eta(X''+\gamma X'+U'(X|\theta))=\eta(e''+\gamma e'+U'(X|\theta)-U'(X_d|\theta_0))
$$
$$
\approx\eta(e''+\gamma e'+U''(X_d|\theta_0)e+U_1(X_d|\theta_0)^T\delta\theta+\delta\theta^TU_2(X_d|\theta_0)\delta\theta)
$$
where
$$
U_1(X_d|\theta_0)=\partial_{\theta}U''(X_d|\theta_0),
$$
$$
U_2(X_d|\theta_0)=(1/2)\partial_{\theta}\partial_{\theta}^TU''(X_d|\theta_0)
$$
In particular, consider the string field equations in the presence of a Poisson source:
$$
\partial_0^2X^{\mu}-\partial_1^2X^{\mu}=f^{\mu}_k(\tau,\sigma)dN^k(\tau)/d\tau
$$
Here
$$
\sigma^0=\tau, \sigma^1=\sigma
$$
By writing down the Fourier series expansion
$$
X^{\mu}(\tau,\sigma)=\sum_na^{\mu}(n)exp(in(\tau-\sigma))+\delta X^{\mu}(\tau,\sigma)
$$
where $\delta X^{\mu}$ is the Poisson contribution, evaluate the rate function of the field $\delta X^{\mu}$ and hence calculate the approximate probability that the quantum average of the string field in a coherent state will deviate from its unforced value by an amount more than a threshold over a given time duration with the spatial region extending over the entire string.
\bigskip

\section{Convergence analysis of the LMS algorithm}

$$
X(n)\in\Bbb R^N, d(n), n\in\Bbb Z
$$
are jointly iid Gaussian processes, ie, $(X(n),d(n)), n\in\Bbb Z$ is a Gaussian process with zero mean. Let
$$
\Bbb E(X(n)X(n)^T)={\bf R}, \Bbb E(X(n)d(n))={\bf r}
$$
The LMS algorithm is
$$
h(n+1)=h(n)-\mu.\partial_{h(n)}(d(n)-h(n)^TX(n))^2
$$
$$
=(I-2\mu X(n)X(n)^T)h(n)+2\mu d(n)X(n)
$$
$h(n)$ is a function of $h(0),(X(k),d(k)), k\leq n-1$ and is thus independent of
$(X(n),d(n))$. Then taking means with $\lambda(n)=\Bbb E(h(n))$, we get
$$
\lambda(n+1)=(I-2\mu R)\lambda(n)+2\mu r
$$
and hence
$$
\lambda(n)=(I-2\mu R)^n\lambda(0)+2\mu\sum_{k=0}^{n-1}(I-2\mu R)^{n-k-1}2\mu r
$$
$$
=(I-2\mu R)^n\lambda(0)+(I-(I-2\mu R)^n)h_0
$$
where
$$
h_0=R^{-1}r
$$
is the optimal Wiener solution. It follows that if all the eigenvalues of $I-2\mu R$ are smaller than unity in magnitude, then
$$
lim_{n\rightarrow\infty}\lambda(n)=h_0
$$
However, we shall show that $Cov(h(n))$ does not converge to zero. It converges to a non-zero matrix. We write
$$
\delta h(n)=h(n)-\lambda(n)
$$
and then get
$$
\lambda(n+1)+\delta h(n+1)=(I-2\mu X(n)X(n)^T)(\lambda(n)+\delta h(n))+2\mu d(n)X(n)
$$
or equivalently,
$$
\delta h(n+1)=((I-2\mu X(n)X(n)^T)-(I-2\mu R))\lambda(n)+(I-2\mu X(n)X(n)^T)\delta h(n)+2\mu(d(n)X(n)-r)
$$
$$
=-2\mu(X(n)X(n)^T-R)\lambda(n)+(I-2\mu X(n)X(n)^T)\delta h(n)+2\mu(d(n)X(n)-r)
$$
and hence
$$
\Bbb E(\delta h(n+1)\otimes\delta h(n+1))=
$$
$$
4\mu^2\Bbb E[(X(n)X(n)^T-R)\otimes(X(n)X(n)^T-R)](\lambda(n)\otimes\lambda(n))
$$
$$
+\Bbb E[(I-2\mu X(n)X(n)^T)\otimes(I-2\mu X(n)X(n)^T)]\Bbb E(\delta h(n)\otimes\delta h(n))
$$
$$
+4\mu^2\Bbb E[(d(n)X(n)-r)\otimes(d(n)X(n)-r)]
$$
$$
-4\mu^2(\Bbb E[(X(n)X(n)^T-R)\otimes(d(n)X(n)-r)]+\Bbb E[d(n)X(n)-r)\otimes(X(n)X(n)^T-R)])\lambda(n)
$$
Note that so far no assumption regarding Gaussianity of the process $(d(n),X(n))$ has been made. We've only assumed that this process is iid. Now making the Gaussianity assumption gives us
$$
\Bbb E(X_i(n)X_j(n)X_k(n)X_m(n))=R_{ij}R_{km}+R_{ik}R_{jm}+R_{im}R_{jk}
$$

Exercise: Using this assumption, evaluate
$$
lim_{n\rightarrow\infty}\Bbb E(\delta h(n)\delta h(n)^T)
$$
Note that in the general non-Gaussian iid case, in case that $\Bbb E(\delta h(n)\otimes\delta h(n))$ converges, say to $V$, then the above equation implies that
$$
V=(I-\Bbb E[(I-2\mu X(0)X(0)^T)\otimes(I-2\mu X(0)X(0)^T)])^{-1}
$$
$$
.[4\mu^2\Bbb E[(X(0)X(0)^T-R)\otimes(X(0)X(0)^T-R)](h_0\otimes h_0)
$$
$$
+4\mu^2\Bbb E[(d(0)X(0)-r)\otimes(d(0)X(0)-r)]
$$
$$
-4\mu^2(\Bbb E[(X(0)X(0)^T-R)\otimes(d(0)X(0)-r)]+\Bbb E[d(0)X(0)-r)\otimes(X(0)X(0)^T-R)])h_0]
$$
It should be noted that a sufficient condition for this convergence to occur is that
all the eigenvalues of the real symmetric matrix
$$
I-\Bbb E[(I-2\mu X(0)X(0)^T)\otimes(I-2\mu X(0)X(0)^T)]
$$
be smaller than unity in magnitude.

\section{String interacting with scalar field, gauge field and a gravitational field}

The total action is
$$
S[X,\Phi,g]=\int(1/2)\eta^{\alpha\beta}g_{\mu\nu}(X)X^{\mu}_{,\alpha}X^{\nu}+_{,\beta}d^2\sigma
$$
$$
+\int\Phi(X)d^2\sigma+\int\epsilon_{\alpha\beta}X^{\mu}_{,\alpha}X^{\nu}_{,\beta}B_{\mu\nu}(X)d^2\sigma
$$
The string field equations are
$$
-\eta_{\alpha\beta}(g_{\mu\nu}(X)X^{\nu}_{,\beta})_{,\alpha}+\Phi_{,\mu}(X)-\epsilon_{\alpha\beta}B_{\mu\nu,\rho}(X)X^{\rho}_{,\alpha}X^{\nu}_{,\beta}=0
$$
In flat space-time, $g_{\mu\nu}=\eta_{\mu\nu}$ and the approximate string equations simplify to
$$
\eta_{\alpha\beta}X_{\mu,\alpha\beta}=-\Phi_{,\mu}(X)+\epsilon_{\alpha\beta}B_{\mu\nu,\rho}(X)X^{\rho}_{,\alpha}X^{\nu}_{,\beta}
$$
The propagator of the string
$$
D_{\mu\nu}(\sigma|\sigma')=<T(X_{\mu}(\sigma)X_{\nu}(\sigma'))>
$$
thus depends on $\Phi,B_{\mu\nu,\rho}$. Let us evaluate this propagator approximately:
$$
\partial_0D_{\mu\nu}(\sigma|\sigma')=\delta(\sigma^0-\sigma_{0'})<[X_{\mu}(\sigma),X_{\nu}(\sigma')]>
$$
$$
+<T(X_{\mu,0}(\sigma)X_{\nu}(\sigma'))>
$$
$$
=<T(X_{\mu,0}(\sigma)X_{\nu}(\sigma'))>
$$
Further,
$$
\partial_0^2D_{\mu\nu}(\sigma|\sigma')=\delta(\sigma^0-\sigma^{0'})<[X_{\mu,0}(\sigma),X_{\nu}(\sigma')]>
$$
$$
+<T(X_{\mu,00}(\sigma)X_{\nu}(\sigma'))>
$$
$$
=-i\delta^2(\sigma-\sigma')+\partial_1^2<T(X_{\mu}(\sigma)X_{\nu}(\sigma'))>
$$
$$
+<T(-\Phi_{,\mu}(X(\sigma))+\epsilon_{\alpha\beta}B_{\mu\gamma,\rho}(X(\sigma))X^{\rho}_{,\alpha}(\sigma)X^{\gamma}_{,\beta}(\sigma)X_{\nu}(\sigma'))>
$$
Thus our exact string propagator equation in the presence of the scalar and gauge fields becomes
$$
\square D_{\mu\nu}(\sigma|\sigma')=-i\delta^2(\sigma-\sigma')
$$
$$
-<T(\Phi_{,\mu}(X(\sigma))X_{\nu}(\sigma'))>+
$$
$$
\epsilon_{\alpha\beta}<T(B_{\mu\gamma,\rho}(X(\sigma))X^{\rho}_{,\alpha}(\sigma)X^{\gamma}_{,\beta}(\sigma)X_{\nu}(\sigma'))>
$$
We now make approximations by expanding the scalar and gauge fields around the centre of of the string position $x^{\mu}$. This approximation gives
$$
<T(\Phi_{,\mu}(X(\sigma))X_{\nu}(\sigma'))>=
$$
$$
\Phi_{,\mu\rho}(x)<T(X^{\rho}(\sigma)X_{\nu}(\sigma'))>
$$
$$
=\Phi_{,\mu\rho}(x)D^{(0)\rho}_{\nu}(\sigma|\sigma')
$$
where $D^{(0)\mu\nu}$ is the unperturbed string propagator and is given by
$\eta^{\mu\nu}ln(|\sigma-\sigma'|)$ where
$$
|\sigma-\sigma'|=\sqrt{(\sigma^0-\sigma^{0'})^2-(\sigma^1-\sigma^{1'})^2}
$$
Likewise,
$$
<T(B_{\mu\gamma,\rho}(X(\sigma))X^{\rho}_{,\alpha}(\sigma)X^{\gamma}_{,\beta}(\sigma)X_{\nu}(\sigma'))>
$$
$$
=B_{\mu\gamma,\rho\delta}(x)(<T(X^{\delta}(\sigma)X^{\rho}_{,\alpha}(\sigma)X^{\gamma}_{,\beta}(\sigma)X_{\nu}(\sigma'))>
$$
Since three out of four of the arguments of $X$ or its partial derivatives in this last expression have the same arguments, it is clear that it cannot contribute anything to the propagator except singularities. So we neglect this term. Thus our corrected propagator equation becomes
$$
\square D_{\mu\nu}(\sigma|\sigma')=-i\delta^2(\sigma-\sigma')
$$
$$
-\Phi_{,\mu\rho}(x)D^{(0)\rho}_{\nu}(\sigma|\sigma')
$$
Note that this is a local propagator equation since it depends on the coordinates $x$ of the centre of the string. We solve this to get
$$
D_{\mu\nu}(\sigma-\sigma')=D_{\mu\nu}^{(0)}(\sigma-\sigma')-\Phi_{,\mu\nu}(x)\int D^{(0)}(\sigma-\sigma'')D^{(0)}(\sigma''-\sigma')d^2\sigma''
$$
Where
$$
D^{(0)}(\sigma)=ln|\sigma|
$$
Now we raise the question about what field equations should be satisfied by $B_{\mu\nu}$ and $\Phi$ so that the string action has conformal invariance. It
For solving this problem, we must replace the above action by
$$
S[X,\Phi,g]=\int(1/2)\eta^{\alpha\beta}exp(\phi(\sigma))g_{\mu\nu}(X)X^{\mu}_{,\alpha}X^{\nu}_{,\beta}d^2\sigma
$$
$$
+\int exp(\phi(\sigma))\Phi(X)d^2\sigma+\int\epsilon_{\alpha\beta}exp(\phi(\sigma))X^{\mu}_{,\alpha}X^{\nu}_{,\beta}B_{\mu\nu}(X)d^2\sigma
$$
since this corresponds to using the string sheet metric $exp(\phi(\sigma))\eta_{\alpha\beta}$ and correspondingly replacing the string sheet area element $d^2\sigma$ by the invariant area element $exp(\phi(\sigma))d^2\sigma$. For small $\phi$, conformal invariance then requires the quantum average of the above action to be independent of $\phi$. Taking into account the larger singularity involved in the product of four string field terms at the same point on the string world sheet as compared with the product of two string field terms, this means that the quantity
$$
\eta^{\alpha\beta}g_{\mu\nu,\rho\gamma}(x)<X^{\rho}(\sigma)X^{\gamma}(\sigma)X^{\mu}_{,\alpha}(\sigma)X^{\nu}_{,\beta}(\sigma)>
$$
$$
+\Phi_{,\mu\nu}(x)<X^{\mu}(\sigma)X^{\nu}(\sigma)>
$$
$$
+\epsilon_{\alpha\beta}B_{\mu\nu,\rho}(x)<X^{\rho}(\sigma)X^{\mu}_{,\alpha}(\sigma)X^{\nu}_{,\beta}(\sigma)>
$$
$$
+\epsilon_{\alpha\beta}B_{\mu\nu,\rho\gamma}(x)<X^{\rho}(\sigma)X^{\gamma}(\sigma)X^{\mu}_{,\alpha}(\sigma)X^{\nu}_{,\beta}(\sigma)>
$$
must vanish.
\bigskip

Another way to approach this problem is to start with the string equations of motion:
$$
\square X_{\mu}=-\Phi_{,\mu}(X)+\epsilon_{\alpha\beta}B_{\mu\nu,\rho}(X)X^{\rho}_{,\alpha}X^{\nu}_{,\beta}
$$
and express it in terms of perturbation theory by writing
$$
X^{\mu}=x^{\mu}+\delta X^{\mu}=x^{\mu}+Y^{\mu}
$$
as
$$
\square Y^{\mu}=-\Phi_{,\mu\nu}(x)Y^{\nu}+\epsilon_{\alpha\beta}(B_{\mu\nu,\rho}(x)+B_{\mu\nu,\rho\gamma}(x)Y^{\gamma})Y^{\rho}_{,\alpha}Y^{\nu}{,\beta}
$$
giving
$$
Y_{\mu}=Z_{\mu}-\Phi_{,\mu\nu}(x)D^{(0)}.Z^{\nu}+\epsilon_{\alpha\beta}B_{\mu\nu,\rho}(x)D^{(0)}.(Z^{\rho}_{,\alpha}Z^{\nu}_{,\beta})
+\epsilon_{\alpha\beta}B_{\mu\nu,\rho\gamma}(x)D^{(0)}.(Z^{\gamma}Z^{\rho}_{,\alpha}Z^{\nu}_{,\beta})
$$
where $x+Z$ is the string field in the absence of any interactions. With this expansion in mind, we then require that
$$
\eta^{\alpha\beta}g_{\mu\nu,\rho\gamma}(x)<Y^{\rho}(\sigma)Y^{\gamma}(\sigma)Y^{\mu}_{,\alpha}(\sigma)Y^{\nu}_{,\beta}(\sigma)>
$$
$$
+\Phi_{,\mu\nu}(x)<Y^{\mu}(\sigma)Y{\nu}(\sigma)>
$$
$$
+\epsilon_{\alpha\beta}B_{\mu\nu,\rho}(x)<Y^{\rho}(\sigma)Y^{\mu}_{,\alpha}(\sigma)Y^{\nu}_{,\beta}(\sigma)>
$$
$$
+\epsilon_{\alpha\beta}B_{\mu\nu,\rho\gamma}(x)<Y^{\rho}(\sigma)Y^{\gamma}(\sigma)Y^{\mu}_{,\alpha}(\sigma)Y^{\nu}_{,\beta}(\sigma)>
$$
should vanish.
\bigskip

Another problem in information theory: Let $p_i,i=1,2,..., N$ be a probability distribution. Let $p_i^{\downarrow}, i=1,2,..., N$ denote the permutation of these probabilities in decreasing order. Consider the set
$$
E=\{i:p_i>e^b\}
$$
Clearly the size $|E|$ of this set satisfies for $0\leq s\leq 1$
$$
|E|\leq \sum_i(p_i/e^b)^{1-s}=exp(-(1-s)b+\psi(s))=exp(sR)\leq exp(R)
$$
where
$$
\psi(s)=log\sum_ip_i^{1-s}
$$
and where
$$
b=b(s,R)=(\psi(s)-sR)/(1-s)
$$
and it is clear that
$$
lim_{s\rightarrow 0}\psi(s)/s=H(p)=-\sum_ip_ilog(p_i)
$$
Hence, if $R>H(s)$, by choosing $s$ positive and sufficiently close to zero, it follows that
if $p^{(n)}$ denotes the product probability measure of the $p_i$ on $\{1,2,...,N\}^n$, then with
$$
b_n(s,R)=n(\psi(s)-sR))/(1-s)
$$
it follows that $b_n(s,R)\rightarrow-\infty$ and hence $exp(b_n(s,R))\rightarrow 0$ and further,
$$
|\{p^{(n)}>exp(b_n(s,R))\}\leq exp(nsR)\leq exp(nR)
$$
Now define
$$
P(p,L)=\sum_{i=1}^Lp_i^{\downarrow}
$$
Then, it is clear that since
$$
|\{p>exp(b(s,R))\}|\leq exp(R)
$$
for all $s\in[0,1]$, we have
$$
1-P(p,exp(R))\leq exp(b(s,R))
$$
and this implies for the product probability distribution,
$$
1-P(p^{(n)},exp(nR))\leq exp(nb(s,R))=exp(n(\psi(s)-sR)/(1-s))\rightarrow 0, n\rightarrow\infty
$$
provided that $s$ is chosen sufficiently close to zero and that $R>H(p)$. In other words, we have proved that if $R>H(p)$, then
$$
P(p^{(n)},exp(nR))\rightarrow 1
$$
\bigskip

Motion of a string field in the presence of an external gauge field with potentials that are linear in the space-time coordinates. Evaluation of the low energy field action in terms of the statistics of the string field.
$$
L=(1/2)\partial^{\alpha}X^{\mu}\partial_{\alpha}X_{\mu}+\epsilon_{\alpha\beta}B_{\mu\nu}(X)\partial_{\alpha}X^{\mu}\partial_{\beta}X^{\nu}
$$
The equations of motion are
$$
\square X_{\mu}=\partial^{\alpha}\partial_{\alpha}X_{\mu}=\epsilon^{\alpha\beta}B_{\mu\nu,\rho}(X)\partial_{\beta}X^{\nu})\partial_{\alpha}X^{\rho}
$$
Writing
$$
B_{\mu\nu,\rho}(X)=\sum A_n(\mu\nu\rho|\mu_1,...,\mu_n)X^{\mu_1}X^{\mu_2}...X^{\mu_n}
$$
we find that for the expansion
$$
X^{\mu}=x^{\mu}-i\sum_na(n)^{\mu}exp(inu^a\sigma_a)/n-ib(n)^{\mu}exp(inv^a\sigma_a)/n
$$
where
$$
(u^a)=(1,-1), (v^a)=(1,1)
$$
we find that
$$
\epsilon^{\alpha\beta}B_{\mu\nu,\rho}(X)\partial_{\beta}X^{\nu})\partial_{\alpha}X^{\rho}=
$$
$$
\sum_{r,s,n_1,...,n_r,m_1,...,m_s,l_1,l_2}\epsilon_{\alpha\beta}(-i)^{n_1+...+n_r+m_1+...+m_s}C_{rs}(\mu\nu\rho|\mu_1,...,\mu_r,\nu_1,...,\nu_s)[\Pi_{k=1}^r(a^{\mu_k}(n_k)/n_k)\Pi_{k=1}^s(b^{\nu_k}(m_k)/m_k)]
$$
$$
(a^{\nu}(l_1)b^{\rho}(l_2)exp(i(l_1u+l_2v).\sigma)-b^{\nu}(l_1)a^{\rho}(l_2))exp(i(l_1v+l_2u).\sigma))u^T\epsilon v.exp(i((n_1+...+n_r)u+(m_1+...+m_s)v).\sigma)
$$
where the coefficients $C_{rs}(.)$ are determined in terms of $A_n(\mu\nu\rho|\mu_1,...\mu_n)$.
In particular if $B_{\mu\nu\rho}$ are constants, then this gauge term contributes only quadratic terms in the creation and annihilation operators and hence can be expressed as
$$
B_{\mu\nu\rho}=\sum_{l_1,l_2}B_{\mu\nu,\rho}(u^T\epsilon v)(a^{\nu}(l_1)b^{\rho}(l_2)exp(i(l_1u+l_2v).\sigma)-b^{\nu}(l_1)a^{\rho}(l_2))exp(i(l_1v+l_2u).\sigma)
$$
In short, in this special case, we can express the gauge field interaction corrected string field upto first order perturbation terms as
$$
X^{\mu}(\sigma)=x^{\mu}+\sum_na^{\mu}(n)f_n(\sigma)+F^{\mu}_{\nu\rho}\sum_{n,m}a^{\nu}(n)a^{\rho}(m)g_{nm}(\sigma)
$$
$$
=x^{\mu}+\delta X^{\mu}(\sigma)
$$
where
$$
[a^{\mu}(n),a^{\nu}(m)]=\eta^{\mu\nu}\delta(n+m)
$$
Now consider a point field $\phi(x)$. Assume that its Lagrangian density is given by $L(\phi(x),\phi_{,\mu}(x))$. The average action of this string field when string theoretic corrections are taken into account in a coherent state $|\phi(u)>$ of the string is given by
$$
<\phi(u)|\int L(\phi(x+\delta X(\sigma)),\phi_{,\mu}(x+\delta X(\sigma)))d\sigma d^4x|\phi(u)>
$$
We can evaluate this action upto quadratic terms in the string perturbation by using the Taylor approximations
$$
\phi(x+\delta X(\sigma))=\phi(x)+\phi_{,\mu}(x)\delta X^{\mu}(\sigma)+(1/2)\phi_{,\mu\nu}(x)\delta X^{\mu}(\sigma)\delta X^{\nu}(\sigma)
$$
$$
\phi_{,\mu}(x)(x+\delta X(\sigma))=\phi_{,\mu}(x)+\phi_{,\mu\nu}(x)\delta X^{\mu}(\sigma)+(1/2)\phi_{,\mu\nu}(x)\delta X^{\mu}(\sigma)\delta X^{\nu}(\sigma)
$$
and then us the following identities:
$$
<\phi(u)|\delta X^{\mu}(\sigma)|\phi(u)>=
$$
$$
\sum_n<\phi(u)|a^{\mu}(n)|\phi(u)>f_n(\sigma)+F^{\mu}_{\nu\rho}\sum_{n,m}<\phi(u)|a^{\nu}(n)a^{\rho}(m)|\phi(u)>g_{nm}(\sigma)
$$
$$
<\phi(u)|\delta X^{\mu}(\sigma)\delta X^{\nu}(\sigma)|\phi(u)>
$$
is expressible as a linear combination of
$$
<\phi(u)|a^{\mu_1}(n_1)a^{\mu_2}(n_2)|\phi(u)>,
$$
$$
<\phi(u)|a^{\mu_1}(n_1)a^{\mu_2}(n_2)a^{\mu_3}(n_3)|\phi(u)>,
$$
and
$$
<\phi(u)|a^{\mu_1}(n_1)a^{\mu_2}(n_2)a^{\mu_3}(n_3)a^{\mu_4}(n_4)|\phi(u)>
$$
Finally,
$$
<\phi(u)|a^{\mu}(n)|\phi(u)>=u^{\mu}(n), u^{\mu}(-n)=\bar u^{\mu}(n)
$$
For $n+m\neq 0$,
$$
<\phi(u)|a^{\mu}(n)a^{\nu}(m)|\phi(u)>=u^{\mu}(n)u^{\nu}(m)
$$
For $n>0$
$$
<\phi(u)|a^{\mu}(n)a^{\nu}(-n)|\phi(u)>=\eta^{\mu\nu}+u^{\mu}(n)u^{\nu}(-n)
$$
and so on. 
\bigskip

{\bf Problems on the Fokker-Planck equation}

[1] Derive the quantum Fokker-Planck equation for the Wigner distribution of the position-momentum pair of a particle
moving in a potential with quantum noise. The HPS evolution equation is
$$
dU(t)=(-(iH+LL^*/2)dt+LdA-L^*dA^*)U(t)
$$
so that the master equation obtained as the equation for $\rho_s(t)$ where
$$
\rho_s(t)=Tr_2(U(t)(\rho_s(0)\otimes|\phi(0><\phi(0)|)U(t)^*)
$$
is given by
$$
\rho_s'(t)=-i[H,\rho_s(t)]-(1/2)(LL^*\rho_s(t)+\rho_s(t)LL^*-2L^*\rho_s(t)L)
$$
\bigskip

[2] Given a master equation of the above type, how would you adapt the parameters $\theta_k(t), k=1,2,..., d$ where
$$
H=H_0+\sum_{k=1}^d\theta_k(t)V_k, L=L_0+\sum_{k=1}^d\theta_k(t)L_k
$$
so that the probability density $\rho(t,q,q)$ in the position representation tracks a given pdf $p_0(t,q)$ ?
\bigskip

hint: Express the master equation as
$$
\rho'(t)=T_0(\rho(t))+\sum_k\theta_k(t)T_k(\rho(t))+\sum_{k,m}\theta_k(t)\theta_m(t)T_{km}(\rho(t))
$$
where $T_0, T_k, T_{km}$ are linear operators on the space of mixed states in the underlying Hilbert space. It follows then that
\bigskip

A remark about supersymmetry, large deviations and stochastic control: Suppose we are given a set of Bosonic and Fermionic fields and a Lagrangian functional of these fields such that this Lagrangian functional is invariant under global supersymmetry transformations. Suppose we now break the supersymmetry of this Lagrangian by adding terms to it that can be represented as a coupling between stochastic c-number fields and the Bosonic and Fermionic fields appearing in the original Lagrangian. Theh, we write down the equations of motion corresponding to this stochastically perturbed Lagrangian. The equations of motion will be stochastic partial differential equations which are non-supersymmetric because of the stochastic perturbations. Now we calculate the change in the modified action functional under an infinitesimal supersymmetry transformation and require that this change be bounded by a certain small number in the sense of expected value of modulus square calculated by assuming that the Bosonic and Fermionic fields are solutions to the unperturbed supersymmetric equations of motion. This condition will impose a restriction on the amplitude of the c-number stochastic fields. Keeping this restriction, we then try to calculate using the theory of large deviations, the probability that the solution of the stochasically perturbed field equations which are non-supersymmetric will deviate from the supersymmetric unperturbed solutions by an amount more than a certain threshold. We then repeat this calculation by adding error feedback control terms to the perturbed dynamical equations and choose the the control coefficients so that the resulting deviation probability is minimized. This in effect, amounts to attempting to restore supersymmetry by dynamic feedback control when the supersymmetry has been broken.

\end{document}